\documentclass[11pt]{article}
\pdfoutput=1
\usepackage{jheppub,amsmath,amssymb,amsfonts,amsxtra, mathrsfs, makeidx,graphics,graphicx,amsthm,epsfig, youngtab,bm,longtable,float}
\usepackage[margin=1cm]{caption}
\newcommand{\ba}{\begin{array}}
\newcommand{\ea}{\end{array}}
\newcommand{\bi}{\begin{itemize}}
\newcommand{\ei}{\end{itemize}}
\def\vec#1{\bm{#1}}
\def\bea#1\eea{\allowdisplaybreaks \begin{align}#1\end{align}}
 \newcommand{\ben}{\begin{enumerate}}
\newcommand{\een}{\end{enumerate}}
\newcommand{\bean}{\begin{eqnarray*}}
\newcommand{\eean}{\end{eqnarray*}}
\newcommand{\eref}[1]{(\ref{#1})}

\newcommand{\tref}[1]{Table~\ref{#1}}
\newcommand{\fref}[1]{Figure~\ref{#1}}
\newcommand{\nn}{\nonumber}

\newcommand{\tr}{\mathrm{Tr}}
\newcommand{\PE}{\mathop{\rm PE}}
\newcommand{\PL}{\mathop{\rm PL}}

\newcommand{\tQ}{\widetilde{Q}}
\newcommand{\tA}{\widetilde{A}}
\newcommand{\tS}{\widetilde{S}}
\newcommand{\tX}{\widetilde{X}}
\newcommand{\tB}{\widetilde{B}}

\newcommand{\BC}{\mathbb{C}}
\newcommand{\BR}{\mathbb{R}}

\newcommand{\BZ}{\mathbb{Z}}

\newcommand{\comment}[1]{}
\newcommand{\CF}{{\cal F}}

\newcommand{\CT}{{\cal T}}

\newcommand{\CM}{{\cal M}}

\newcommand{\CN}{{\cal N}}

\newcommand{\CZ}{{\cal Z}}
\newcommand{\CR}{{\cal R}}

\newcommand{\ie}{{\it i.e.}}
\newcommand{\eg}{{\it e.g.}}

\newcommand{\ud}{\mathrm{d}}

\newcommand{\fflat}{\mathcal{F}^\flat}

\title{Hilbert Series for Moduli Spaces of Instantons on $\BC^2/\BZ_n$}
\author[a]{Anindya Dey,} 
\author[b]{Amihay Hanany,} 
\author[c,d]{Noppadol Mekareeya,}
\author[e]{Diego Rodr\'iguez-G\'omez,}
\author[b,f]{and Rak-Kyeong Seong}

\affiliation[a]{
Theory Group and Texas Cosmology Center, Department of Physics,\\
University of Texas at Austin,\\
Austin, TX 78712, USA
}
\affiliation[b]{
Theoretical Physics Group, The Blackett Laboratory, \\
Imperial College London, Prince Consort Road, \\
London SW7 2AZ, United Kingdom
}
\affiliation[c]{
Max-Planck-Institut f\"ur Physik (Werner-Heisenberg-Institut),\\
F\"ohringer Ring 6, 80805 M\"unchen, Deutschland
}
\affiliation[d]{
Theory Group, Physics Department, CERN, CH-1211, Geneva 23, Switzerland
}
\affiliation[e]{
Department of Physics, Universidad de Oviedo, \\
Avda. Calvo Sotelo 18, 33007, Oviedo, Spain
}
\affiliation[f]{
School of Physics, Korea Institute for Advanced Study, \\
85 Hoegi-ro, Seoul 130-722, South Korea
}

\emailAdd{anindya@physics.utexas.edu, a.hanany@imperial.ac.uk, noppadol@mpp.mpg.de,d.rodriguez.gomez@uniovi.es, rak-kyeong.seong@imperial.ac.uk}

\preprint{
{\scriptsize
\begin{flushright}
UTTG-12-13\\
TCC-009-13\\
Imperial/TP/13/AH/02\\
MPP-2013-150\\
KIAS-P13051
\end{flushright}
}
}

\abstract{We study chiral gauge-invariant operators on moduli spaces of $G$ instantons for any classical group $G$ on A-type ALE spaces using Hilbert Series (HS). Moduli spaces of instantons on an ALE space can be realized as Higgs branches of certain quiver gauge theories which appear as world-volume theories on D$p$ branes in a D$p$-D$(p+4)$ system with the D$(p+4)$ branes (with or without O$(p+4)$ planes) wrapping the ALE space. We study in detail a list of quiver gauge theories which are related to $G$-instantons of arbitrary ranks and instanton numbers on a generic $A_{n-1}$ ALE space  and discuss the corresponding brane configurations. For a large class of theories, we explicitly compute  the Higgs branch HS which reveals various algebraic/geometric aspects of the moduli space such as the dimension of the space, generators of the moduli space and relations connecting them.  In a large number of examples involving lower rank instantons, we demonstrate that HS for equivalent instantons of isomorphic gauge groups but very different quiver descriptions do indeed agree, as expected.
}

\begin{document}
\setcounter{tocdepth}{2}
\maketitle
\section{Introduction and Main Results}
Asymptotically locally Euclidean (ALE) spaces have generated a lot of interest since their discovery in 1978 by Gibbons and Hawking \cite{Gibbons1978430, Gibbons:1979xm} as a family of self-dual and positive-definite solutions to Einstein's equations with zero cosmological constant.  As pointed out later by Hitchin {\it et al.} \cite{PSP:2086068, Hitchin:1986ea} and Kronheimer \cite{Kronheimer:1989zs}, such spaces can also be realised as resolutions of the complex quotient singularities $\BC^2/\Gamma$, where $\Gamma$ is a finite subgroup of $SU(2)$.  We focus on the special case $\Gamma = \BZ_n$ in this paper.

One interesting feature of ALE spaces is the construction of self-dual/anti-self-dual solutions of the Yang-Mills equations, \ie~instanton solutions, on such spaces. It was shown by Kronheimer and Nakajima (KN) \cite{kronheimer1990yang} in 1990 that given an ALE space, the problem of constructing an instanton solution can be translated into a problem of linear algebra in a way similar to the flat space analogue proposed by Atiyah, Drinfeld, Hitchin and Manin (ADHM) \cite{Atiyah:1978ri}.  In particular, the aforementioned linear algebra problem involves solving a set of quadratic equations for certain finite-dimensional matrices; see \cite{Christ:1978jy, king1989instantons} for  ADHM construction and \cite{kronheimer1990yang, Bianchi:1996zj} for  KN construction. In case of ADHM construction for instantons on $\BC^2$, such equations can be identified with the vacuum equations of a supersymmetric gauge theory with eight real supercharges living on the world-volume of  D$p$ branes probing a set of D$(p+4)$ branes wrapping the $\BC^2$ in a Type II theory \cite{Witten:1995gx, Douglas:1995bn}. In fact, the moduli space of instantons on $\BC^2$ can  be identified with the Higgs branch of theories described above.
For KN construction of instantons on ALE spaces, the corresponding Type II picture generically involves D$p$ branes probing a set of D$(p+4)$ branes which wrap the ALE space in presence of orientifold planes (also wrapping the ALE). The moduli space of instantons on an ALE space can be identified, as before, with the Higgs branch of the quiver gauge theory with eight real supercharges living on the world-volume of  D$p$ branes in the Type II picture described above.\\
The quiver gauge theories related to KN construction and their D-brane descriptions were first discussed in \cite{Douglas:1996sw} and subsequent papers such as \cite{Cherkis:2008ip, Witten:2009xu}. Some of these quiver gauge theories have been studied in various physical contexts  \cite{Hanany:1999sj, Erlich:1999rb, Dey:2011pt, Bergman:2012kr, Bergman:2012qh, Dey:2013nf}. However, in this paper, we present a complete catalogue of quivers associated to instantons on an $A-$type ALE space for any classical group $G$, with a precise dictionary relating a given instanton to the corresponding quiver.\\


One mathematical tool that proves to be useful for studying moduli spaces of supersymmetric gauge theories and, in particular, instanton moduli spaces is Hilbert series \cite{Nakajima:2003pg, Benvenuti:2010pq, Hanany:2012dm,Benvenuti:2006qr}.  It is a generating function that counts chiral gauge-invariant operators on the moduli space of vacua with respect to a certain global $U(1)$ charge. Moreover, it is closely related to the instanton contribution to the partition function of five dimensional pure supersymmetric Yang-Mills theory on $\BR^4 \times S^1$ \cite{Nekrasov:2002qd, Nakajima:2003pg, Keller:2011ek, RGZ2013}.  When the radius of $S^1$ is taken to zero, the four dimensional Nekrasov instanton partition function \cite{Nekrasov:2002qd, Nekrasov:2003af} can be obtained from the Hilbert series \cite{Nakajima:2003pg, Keller:2011ek, RGZ2013} and, as a consequence, the non-perturbative parts of the prepotential (at each instanton number) can be derived (see \eg~ \cite{Fucito:2004ry}).

In this paper, we use the correspondence between instanton moduli spaces and Higgs branches of quiver gauge theories to study instantons with {\it special unitary}, {\it special orthogonal} and {\it symplectic} gauge groups on an A-type ALE space . The quiver diagrams in each case are presented explicitly.  We compute the Hilbert series on the Higgs branch for several illustrative examples in each family of quivers and associate these to the appropriate moduli space of instantons.  We show that Hilbert series corresponding to equivalent instantons of various isomorphic gauge groups do match, thereby providing a non-trivial consistency check for our approach.

Our study of instantons on $A_{n-1}$ ALE spaces is a natural follow-up of the previous works \cite{Benvenuti:2010pq,Hanany:2012dm} on the moduli spaces of one and two instantons on the flat space $\mathbb{R}^4$. In this paper, we will also focus on cases where the instanton number is small. In addition, we will mainly consider cases with low $n$.\\

\noindent The organisation of this paper is as follows:
\bi
\item The moduli spaces of $SU(N)$ instantons on $\BC^2/\BZ_n$ are studied in sections \ref{sec:quivdata}- \ref{moreex:loc}. We further study  the particular case of $k$ pure instantons of rank 0 through holography focusing on the $n=2$ case in appendix \ref{holo}.
\bi
\item In section \ref{sec:quivdata}, we discuss the KN quiver and summarize the geometrical data corresponding to $SU(N)$ instantons on $\BC^2/\BZ_n$. 
\item In section \ref{sec:SUNC2Z2}, we discuss the computation of Hilbert series using Molien integral formula for 
$SU(N)$ instantons on $\BC^2/\BZ_n$ focussing on the special case of $n=2$.  Using several examples, such as \eref{HS:C2Z2SU1N11}, we demonstrate that the Hilbert series can be explicitly written in terms of a  $G$-invariant character expansion in a closed form similar to those obtained for instantons on $\BR^4$ \cite{Benvenuti:2010pq, Hanany:2012dm}. 
The generators of the instanton moduli space are spelt out explicitly in section \ref{sec:genSUNC2Zn} using the Plethystic Logarithm (PL) analysis of the Hilbert Series.

\item In section \ref{sec:localisationSU}, we present a localisation method similar to \cite{Fucito:2004ry, Ito:2011mw, Alfimov:2011ju, Bonelli:2011kv, Bonelli:2012ny, Ito:2013kpa}, that can be used to compute the Hilbert series of $SU(N)$ instantons on $\BC^2/\BZ_n$.  In section \ref{sec:pureinstfeatures}, we address the special case of pure instantons. We discuss certain interesting features of the moduli space of such instantons and present several illustrative examples of the Hilbert Series using localisation method. Several examples of Hilbert series for instantons computed via localisation can be found in section \ref{moreex:loc}.
 \ei

\item The moduli spaces of $SO(N)$ and $Sp(N)$ instantons on $\BC^2/\BZ_n$ are studied in sections \ref{sec:SOSpC2Zn} and \ref{sec:SONC2Z2}.
\bi
\item In section \ref{sec:SOSpC2Zn}, we present six infinite families of quiver gauge theories associated with $SO(N)$ and $Sp(N)$ instantons on $\BC^2/\BZ_n$.   The brane configurations of these six quivers are discussed in Appendix \ref{app:SOSpbrane}. Among these quivers, two infinite families correspond to odd $n$ and can be distinguished by boundary conditions at the two ends of the quiver as follows:
\ben
\item {\bf Unitary gauge group with an antisymmetric hyper at one end and symplectic gauge group at the other end} as shown in \fref{fig:SO_N_oddn}. This quiver corresponds to $SO(N)$ instantons.
\item {\bf Unitary gauge group with a symmetric hyper at one end and orthogonal gauge group at the other end} as shown in  \fref{fig:Sp_N_oddn}. This quiver corresponds to $Sp(N)$ instantons.
\een
 
 The remaining four families correspond to $n=2m$ and may also be  distinguished by boundary conditions in the following manner:
 \ben
 \item {\bf Each end of the quiver has (special) orthogonal global symmetry group.} We refer to such a quiver as the {\it O/O quiver}.   This is depicted in \fref{fig:SO_N_withVSevenn}.  It correspond to $SO(N)$ instantons on $\BC^2/\BZ_{2m}$.
\item {\bf Each end of the quiver has symplectic global symmetry group.} We refer to such a quiver as the {\it S/S quiver}. This is depicted in \fref{fig:Sp_N_withVSevenn}.  It correspond to $Sp(N)$ instantons on $\BC^2/\BZ_{2m}$.
\item {\bf Each end has unitary gauge group with one antisymmetric hypermultiplet.} We refer to such a quiver as the {\it AA quiver}.   This is depicted in \fref{fig:SO_N_withNOVSevenn}.  It correspond to $SO(2N)$ instantons on $\BC^2/\BZ_{2m}$.
 \item {\bf Each end has unitary gauge group with one symmetric hypermultiplet.} We refer to such a quiver as the {\it SS quiver}.   This is depicted in \fref{fig:Sp_N_withNOVSevenn}.  It correspond to $Sp(N)$ instantons on $\BC^2/\BZ_{2m}$.
\een
 
 For each category of quivers with $n \geq 3$ , we compute Hilbert series for several examples involving lower rank instantons. We discuss in detail the generators of the instanton moduli spaces and the relations connecting them in each case.

\item  In section \ref{sec:SONC2Z2}, we discuss the degenerate cases associated with $n=2$. For each of the four category of quivers one can have for $n=2$, we compute Hilbert Series for several examples involving lower rank instantons. We discuss some of the general features of generators of the instanton moduli spaces and their relations. 

\item  We find agreement between Hilbert series of different quivers corresponding to instantons of gauge groups that have isomorphic Lie algebras; these are presented in Tables \ref{tab:match1} and \ref{tab:match2}.  As an immediate consequence, many correspondences between different quiver gauge theories can be established.
\ei
\item In section \ref{sec:hybrids}, we discuss a family of quiver diagrams that can be realised from the brane configurations in a way similar to those presented in section \ref{sec:SOSpC2Zn}.  These quivers arise in the Type IIB picture by taking boundary conditions associated with quivers for $SO$ instantons  on one end and that for $Sp$ instantons on the other. We therefore refer to such theories as `hybrid quivers'.  It is found that the symmetry present on the Higgs branch of each of these quivers is similar to that present in the string backgrounds studied in \cite{Chaudhuri:1995fk, Chaudhuri:1995bf,Dabholkar:1996pc}.
\ei

\paragraph{Quiver Notation.} In this paper, we use the following convention for drawing quiver diagrams:
\begin{center}
\includegraphics[scale=0.7]{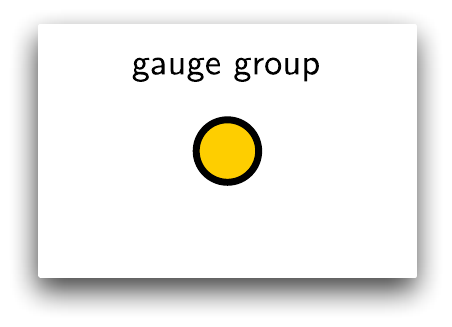}
\includegraphics[scale=0.7]{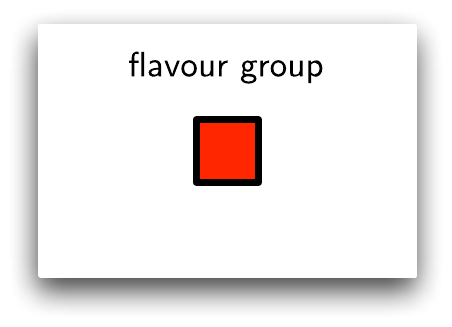}
\\
\includegraphics[scale=0.7]{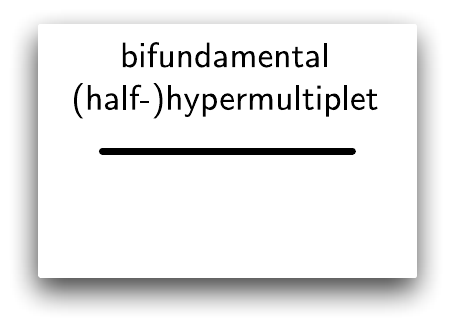}
\includegraphics[scale=0.7]{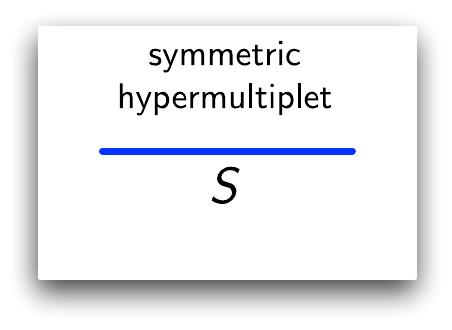}
\includegraphics[scale=0.7]{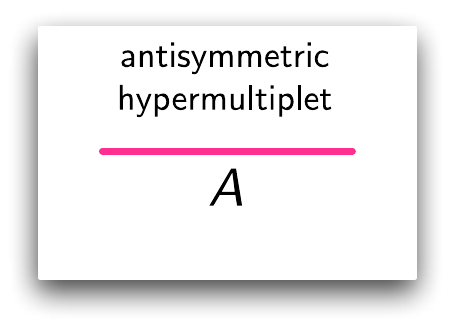}
\end{center}

\section{$SU(N)$ instantons on $\mathbb{C}^2/\mathbb{Z}_n$ \label{U(N)inst}}

This section introduces the quiver and geometrical data for $SU(N)$ instantons on $\mathbb{C}^2/\mathbb{Z}_n$, following which we discuss the Molien integral as well as a localisation method for the Hilbert series computation. We mostly focus on one-instanton examples with $n \leq 4$. Generalisations for Hilbert series of $k$ $SU(N)$ instantons on $\mathbb{C}^2/\mathbb{Z}_n$ (for generic $k$ and $n$) are also presented.
\\

\subsection{Quiver and Geometrical Data} \label{sec:quivdata}

The quiver data to specify the moduli space for $SU(N)$ instantons on $\mathbb{C}^2/\mathbb{Z}_n$ are presented in this subsection. As reviewed in appendix \ref{classification}, this information is related to the geometrical data of the instanton bundle.

\paragraph{Quiver data.} For $SU(N)$ instantons on $\mathbb{C}^2$, the quiver data are given by the instanton number $k$ and the rank $N$. In the case of $SU(N)$ instantons on $\mathbb{C}^2/\mathbb{Z}_n$, the quiver data are given as follows:
\ben
\item {\bf A $k$-tuple analogue of the instanton number.}  We denote this by 
$\vec k =(k_1, \ldots, k_n)$ 
with $k_i \in \BZ_{\geq 0}$ for all $i =1, \ldots, n$.

\begin{figure}[ht!]
\begin{center}
\includegraphics[scale=0.7]{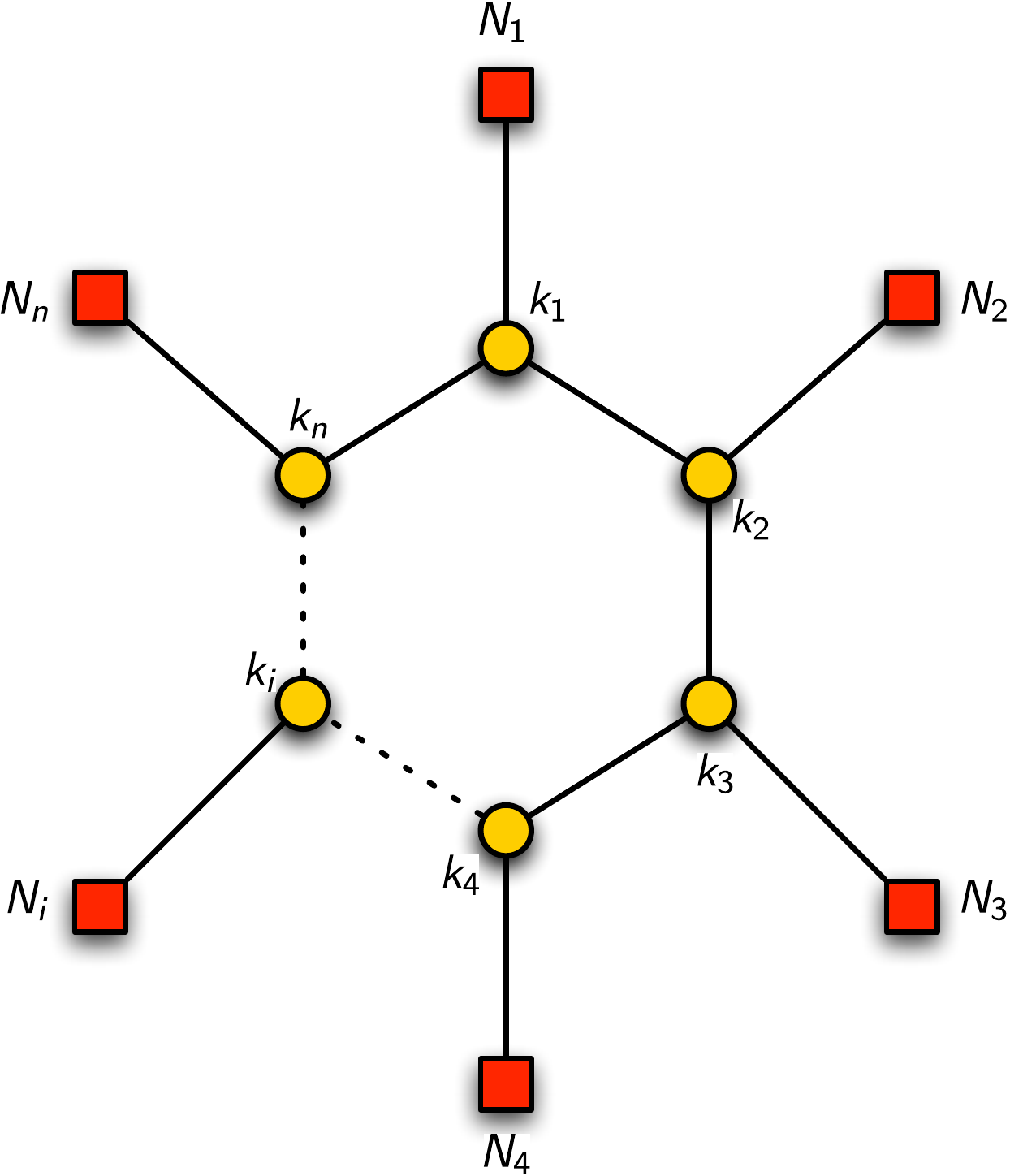}
\caption{The Kronheimer-Nakajima quivers for $SU(N)$ instantons on $\BC^2/\BZ_n$. Each node denotes the unitary group with the labelled rank. The circular nodes denote gauge groups and the square nodes denote the flavour symmetries.}
\label{fig:necklace}
\end{center}
\end{figure}

\item {\bf A $n$-partition of $N$.}  We denote this by $\vec N =(N_1, \ldots, N_n)$ such that 
\bea N =  N_1+\ldots + N_n~,
\eea with $N_i \in \BZ_{\geq 0}$ for all $i =1, \ldots, n$.
\een
These pieces of information are collected in a four dimensional $\CN=2$ quiver diagram, also known as the {\it Kronheimer-Nakajima (KN) quiver} \cite{kronheimer1990yang}. The quiver diagram is depicted in \fref{fig:necklace}.  The instanton moduli space corresponds to the Higgs branch of the quiver theory. Therefore, the instanton moduli space is uniquely specified by the vectors $\vec k$ and $\vec N$ (see \eg, \cite{kronheimer1990yang, Bianchi:1996zj, Fucito:2004ry, Cherkis:2008ip, Witten:2009xu, Cherkis:2009jm, Cherkis:2010bn}).  It is known \cite{Hanany:1996ie} that the KN quiver can also be realised from a system of D3 branes on a circle, D5 and NS5 branes. Such a configuration and relevant geometical data are reviewed in Appendix \ref{classification}.

For each gauge group $U(k_i)$, the one loop beta function coefficient $\beta_i$ is given by
\bea
\beta_i = N_i + k_{i-1}+ k_{i+1} -2k_i~, \label{betafn}
\eea
 and the index $i$ runs from $1, \ldots, n$ and is taken modulo $n$.   Note that $\beta_i$ is also equal to the relative linking numbers of the $(i+1)$-th and the $i$-th NS5-branes, and that
\bea 
 \sum_{i=1}^n \beta_i = N~. \label{sumbeta}
\eea

\paragraph{Geometrical Data.} The relation between the quiver data, namely $\vec k$ and $\vec N$, and the geometrical data of the instanton bundle can be summarised as follows \cite{kronheimer1990yang, Bianchi:1996zj, Fucito:2004ry, Cherkis:2008ip, Witten:2009xu, Cherkis:2009jm, Cherkis:2010bn}:
\bi
\item {\bf Rank of the instanton bundle} is given by $\sum_{i=1}^n N_i =N$.
\item The {\bf monodromy at infinity} is specified by the $n$-partition of $N$ : $\vec N =(N_1,N_2, \ldots, N_n)$.
\item The {\bf first Chern class} of the instanton bundle is 
\bea
c_1=\sum_{i =1}^n \beta_i c_1(\mathcal{T}_i)~, \label{firstChernclass}
\eea 
where $\CT_i$ are the line bundles defined in \eref{deflinebundT}.  According to \cite{Bianchi:1996zj}, $c_1$ has a physical interpretation as a {\it monopole charge} coming from an integral of the field strength corresponding to the instanton configuration.
\item The {\bf second Chern class} of the instanton bundle is 
\bea
c_2=\sum_{i=1}^n \beta_i c_2(\mathcal{T}_i)+\frac{1}{n}\,\sum_{i=1}^n k_i~.
\eea
It is possible to make a choice so that one of $\CT_1, \ldots, \CT_n$ is a trivial line bundle.  Henceforth, we choose $\mathcal{T}_n$ to be trivial, so that 
\bea
c_1(\mathcal{T}_n)=c_2(\mathcal{T}_n)=0~. \label{choiceoftautbundle}
\eea 
\item Given the vector $\vec k =(k_1, \ldots, k_n)$, we define an {\bf instanton number} as
\begin{equation}
k=\frac{1}{n} \sum_{i=1}^n k_i~.
\end{equation}
We stress that this does not necessarily equal to the second Chern class $c_2$. 
\item In a special case that $\beta_1 = \cdots = \beta_{n-1}=0$, the monopole charge $c_1 =0$ due to \eref{firstChernclass} and \eref{choiceoftautbundle}.  We thus refer to this special case as the {\bf pure instanton} configuration.  As an immediate consequence, the instanton number is equal to the second Chern class of the instanton bundle.
\ei

\subsection{The Hilbert series via the Molien integral\label{sec:SUNC2Z2}}

\begin{figure}[H]
\centering
\includegraphics[scale=0.7]{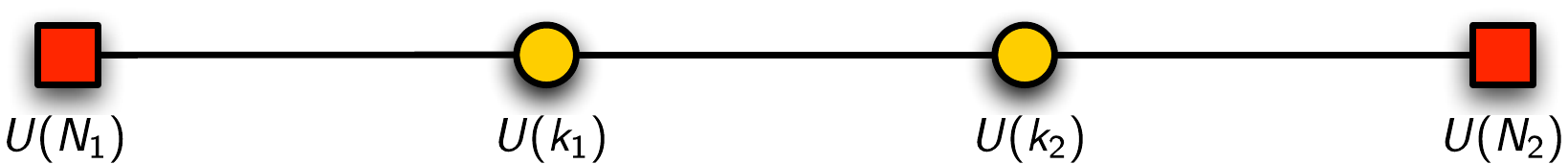}
\caption{The quiver for $SU(N)$ instantons on $\mathbb{C}^2/\mathbb{Z}_2$: $\vec k=(k_1,k_2)$, $\vec N =(N_1,N_2)$ with $N=N_1+N_2$. The square nodes represent the flavor symmetries, whereas the circular nodes represent gauge symmetries.  Each line between the groups $U(r_1)$ and $U(r_2)$ represent $r_1 r_2$ hypermultiplets in the bi-fundamental representation.}
\label{UNC2modZ2}
\end{figure}

In this subsection, we focus on the computation of the Hilbert series with a particular emphasis on the $SU(N)$ instanton on $\BC^2/\BZ_2$. The quiver data is given by $\vec k=(k_1,k_2)$ and $\vec N =(N_1,N_2)$ with $N=N_1+N_2$.  The Higgs branch of this theory is identified with the moduli space of the corresponding instanton. 

\begin{figure}[H]
\centering
\includegraphics[scale=0.7]{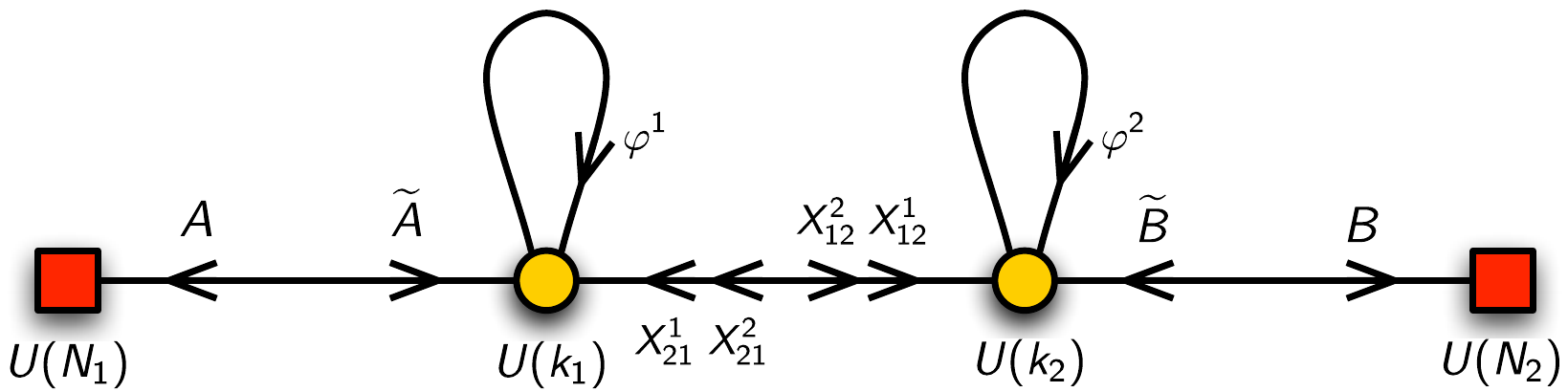}
\caption{Quiver diagram in 4d $\CN=1$ notation for $SU(N)$ instantons on $\mathbb{C}^2/\mathbb{Z}_2$: $\vec k=(k_1,k_2)$, $\vec N =(N_1,N_2)$ with $N=N_1+N_2$. The superpotential is given by \eref{WSUNC2Z2}.}
\label{N1C2modZ2}
\end{figure}

\paragraph{Quiver and Superpotential.} Using a similar method presented in  \cite{Benvenuti:2010pq, Hanany:2012dm}, we can compute the Hilbert series associated to the instanton moduli space in question.  We translate the 4d $\CN=2$ quiver into $\CN=1$ language with the corresponding quiver diagram depicted in \fref{N1C2modZ2}. The $\CN=1$ superpotential is
\bea \label{WSUNC2Z2}
W&=\epsilon_{\alpha_1 \alpha_2} \left[ (X^{\alpha_1}_{12})^{a_1}_{~b_1}(\varphi_2)^{b_1}_{~b_2} (X^{\alpha_2}_{21})^{b_2}_{~a_1} -(X^{\alpha_1}_{21})^{b_1}_{~a_1}(\varphi_1)^{a_1}_{~a_2} (X^{\alpha_2}_{12})^{a_2}_{~b_1}  \right] +  \nn \\
& \quad \tA_{a_1}^{i_1} (\varphi_1)_{~a_2}^{a_1} A_{i_1}^{a_2}+\tB^{j_1}_{b_1}(\varphi_2)_{~b_2}^{b_1} B^{b_2}_{j_1}~,
\eea
where $a, a_1, a_2, \ldots =1, \ldots, k_1$ are the indices for the gauge group $U(k_1)$, $b, b_1,b_2, \ldots=1, \ldots, k_2$ are the indices for the gauge group $U(k_2)$, $i, i_1, i_2, \ldots =1, \ldots, N_1$ are the fundamental indices for the flavor symmetry group $U(N_1)$ and $j, j_1, j_2, \ldots=1, \ldots, N_2$ are the fundamental indices for the flavor symmetry group $U(N_2)$.  Note that there is an $SU(2)$ global symmetry under which each of $X^{\alpha}_{12}$ and $X^{\alpha}_{21}$ transforms as a doublet, where $\alpha, \alpha_1, \alpha_2, \ldots =1,2$ are the fundamental indices for this global symmetry which we will refer to as $SU(2)_x$.

Given that the theory has $\mathcal{N}=2$ supersymmetry, the $SU(2)$ $R$-symmetry transforms each of the following pairs of chiral fields as doublets,
\begin{equation}
\left\{ \left(\begin{array}{c} X_{12}^1 \\ (X_{21}^2)^{\dagger}\end{array}\right), \qquad \left(\begin{array}{c} X_{12}^2 \\ (X_{21}^1)^{\dagger}\end{array}\right) \right\}, \qquad  \left(\begin{array}{c} A_{i_1} \\ (\tA^{i_1})^{\dagger}\end{array}\right)\qquad \left(\begin{array}{c} B_{j_1} \\ (\tB^{j_1})^{\dagger}\end{array}\right)~,
\end{equation}
where the pair in the curly bracket transforms as a doublet under $SU(2)_x$ and the others transform as singlets under $SU(2)_x$.
The scalar in each vector multiplet transforms as a singlet under $SU(2)$ $R$-symmetry.

In $\CN=1$ formalism, only one generator of $SU(2)_R$, which we will call $U(1)_t$, is manifest.\footnote{This $U(1)$ global symmetry is referred to as $U(1)_J$ in, for example, \cite{Seiberg:1994rs}. Note that, when thinking of the theory as a 4d $\mathcal{N}=2$ one, the R-symmetry is $U(1)_r\times SU(2)_R$. On the Higgs branch, up to a normalization, the action of the $U(1)_r$ and the $U(1)_J\in SU(2)_R$ is identical, \textit{i.e.} assigns the same charges to all fields in hypermultiplets.}  Therefore, each of the chiral fields $X^\alpha_{12}$, $X^\alpha_{21}$, $A_i$, $\tA^i$, $B_j$, $\tB^j$, coming from the hypermultiplets, carries charge $+1$ under $U(1)_t$, whereas the chiral fields $\varphi_1$ and $\varphi_2$, coming from the vector multiplets, carries charge $0$ under $U(1)_t$. 

\paragraph{F-terms.} Since we are interested in the Higgs branch of this theory, the relevant $F$-terms arising from differentiating the superpotential \eref{WSUNC2Z2} with respect to the scalars in the vector multiplets, $\varphi_1$ and $\varphi_2$, are given by
\bea
0 &= (\mathcal{F}_1)^{a_1}_{~a_2}:= \partial_{(\varphi_1)^{a_2}_{a_1}} W= -(X_{21}^{\alpha_1})^{b_1}_{~a_2} (X_{12}^{\alpha_2})^{a_1}_{~b_1}\epsilon_{\alpha_1 \alpha_2}+A_{i_1}^{a_1}  \tA_{a_2}^{i_1}~, \nn \\
0 &= (\mathcal{F}_2)^{b_1}_{~b_2} :=  \partial_{(\varphi_2)^{b_2}_{b_1}} W=(X_{12}^{\alpha_1})^{a_1}_{~b_2} (X_{21}^{\alpha_2})^{b_1}_{~a_1}\epsilon_{\alpha_1 \alpha_2}+B_{j_1}^{b_1} \tB_{b_2}^{j_1}~.
\eea

\paragraph{Gauge and Global Symmetries.} The transformation rule of the chiral fields and the $F$-terms under the gauge and global symmetries are summarised in \tref{tcharges}. 

\begin{table}
\centering
$
\begin{array}{|c | c c | c c c c c |c }
\hline
 & U(k_1) & U(k_2) & U(1)_t & SU(2)_x &  U(N_1) & U(N_2) &  \\ \hline
 \hline
 X_{12}^\alpha &\Box_{+1}              &  \overline{\Box}_{-1}  & 1   &   \Box   &                       \textbf{1}_0  &  \textbf{1}_0 &  \\
 X_{21}^\alpha &   \overline{\Box}_{-1}                     &   \Box_{+1}   &   1  &   \Box            &    \textbf{1}_0 &  \textbf{1}_0 &  \\
\varphi_1   &    \mathbf{Adj}      &   \textbf{1}   &   0   &   \textbf{1}         &  \textbf{1}_0   & \textbf{1}_0 &   \\
\varphi_2  & \textbf{1}               & \mathbf{Adj}   &   0  &   \textbf{1}     &    \textbf{1}_0 &  \textbf{1}_0 &  \\
  A      &   \Box_{+1}                  &   \textbf{1}_0    &   1    &   \textbf{1}           &  \overline{\Box}_{-1}  &  \textbf{1}_0 & \\
  {\tA}&  \overline{\Box}_{-1}            &   \textbf{1}_0   &  1   &   \textbf{1}         &  \Box_{+1}  &   \textbf{1}_0& \\
B          &    \textbf{1}_0            &  \Box_{+1}   &   1   &    \textbf{1}               &  \textbf{1}_0    &  \overline{\Box}_{-1}& \\ 
 {\tB} &   \textbf{1}_0             & \overline{\Box}_{-1}   &   1   &     \textbf{1}      &  \textbf{1}_0   & \Box_{+1}  &  \\
 \hline
\mathcal{F}_1     &   \mathbf{Adj}   &   \textbf{1}_0          &   2   &   \textbf{1}            &  \textbf{1}_0   &   \textbf{1}_0&\\
  \mathcal{F}_2     &  \textbf{1}_0             & \mathbf{Adj}        &   2   &    \textbf{1}   &   \textbf{1}_0  &   \textbf{1}_0 &  \\
  \hline
 \end{array}
 $
 \caption{The transformation rule of the chiral fields and the $F$-terms under the gauge and global symmetries. \label{tcharges}}
\end{table}

\paragraph{Moduli Space Data.} When $N_1$ and $N_2$ are sufficiently large compared to $k_1$ and $k_2$, the gauge symmetry is completely broken on the hypermultiplet moduli space.  Upon Higgsing, $k_1 N_1+k_2 N_2 +2k_1k_2 - k_1^2-k_2^2$ quaternionic gauge invariant degrees of freedom remain. Hence, the quaternionic dimension of the Higgs branch is given by
\begin{equation}
{\rm dim}_{\mathbb{H}}\,\mathcal{M}^{(N_1,N_2)}_{(k_1,k_2)}=k_1\,(N_1+k_2-k_1)+k_2\,(N_2+k_1-k_2)
\end{equation}


\subsubsection{Hilbert Series for one $SU(N)$ instanton on $\BC^2/\BZ_2$: $\vec{k}=(1,1)$ and $\vec{N}=(0,N)$.\label{(0,N)}}
The Molien integral formula for Hilbert Series is most conveniently expressed in terms of plethystic exponential (PE) of functions. For a multi-variable function $f(x_1,x_2,....,x_n)$ which vanishes at the origin, this is defined as,
\bea
\PE\left[f(x_1,x_2,....,x_n)\right]= \exp\left(\sum^{\infty}_{r=1} \frac{f(x^r_1,x^r_2,....,x^r_n)}{r}\right)
\eea
The inverse function to the plethystic exponential is known as plethystic logarithm (PL). Character expansion of PL of the Hilbert Series in terms of the global symmetry group is a direct way to read off generators of the chiral ring at every level including their relations. For more details, the reader is referred to the earlier body of work in \cite{Benvenuti:2010pq}-\cite{Benvenuti:2006qr}.\\
\noindent In the present example, let $z_1$ and $z_2$ be the fugacities for the gauge symmetry $U(1) \times U(1)$.  Furthermore, let $x$ be fugacity of the global symmetry $SU(2)_x$ and $(u,\vec y)$ be the fugacities of $U(N)=U(1) \times SU(N)$.\\

The Hilbert series is then given by the Molien integral 
\bea
g_{(1,\,1)}^{(0,\,N)} (t,x,u, \vec y)=\oint_{|z_1|=1}\frac{\ud z_1}{(2 \pi i )z_1}\,\oint_{|z_2|=1}\frac{\ud z_2}{(2 \pi i )z_2} \frac{\chi_B (t, u, \vec y,z_2) \chi_X (t,x,z_1,z_2)}{\chi_F(t)}
\eea
where the contributions from $B$, $X$'s and $F$-terms are given respectively by
\bea
\chi_B (t, u, \vec y) &=  \PE\Big[[0,\ldots,0,1]_{\vec y}u^{-1}  z_2 t+[1,0,\ldots,\,0]_{\vec y} u z_2^{-1} t\Big]~, \nn \\
\chi_X(t, x, z_1, z_2) &= \PE \Big[ [1]_x\left(z_1 z_2^{-1}+z_1^{-1} z_2 \right)t \Big]~, \nn \\
\chi_F(t) &= (1-t^2)^{-2}~.
\eea

The $z_1$ integral can be easily done. Letting $z'_2 = z_2 u^{-1}$, we have 
\bea
\label{C2Z2SUk110N}
g_{(1,\,1)}^{(0,\,N)}(t, x, \vec y) &=g_{\BC^2/\BZ_2} (t,x) \Big\{(1-t^2)\,\oint_{|z_2'|=1} \frac{\ud z'_2}{z'_2}\,\PE\Big[ [0,\ldots,\,1]_{\vec y}  z'_2 t+[1,\ldots,0,0]_{\vec y} {z'_2}^{-1} t \Big]\Big\}\nn\\
&=\sum_{n_1=0}^{\infty} \sum_{n_2=0}^{\infty} [2n_1;n_2,0,\dots,0,n_2]_{x,\vec y}\,\, t^{2n_1+2n_2} \nn \\
&= g_{\BC^2/\BZ_2} (t,x) \times \widetilde{g}_{1, SU(N), \BC^2} (t,\vec y)~,
\eea
where 
\bea
g_{\BC^2/\BZ_2} (t,x) &= (1-t^4) \PE[ [2]_x t^2] = \frac{(1+t^2)}{(1-t^2\,x^2)\,(1-t^2\,x^{-2})} = \sum_{n=0}^\infty [2n]_x t^{2n}~, \\ 
\widetilde{g}_{1, SU(N), \BC^2} (t,\vec y) &= \sum_{n=0}^\infty [n,0,\ldots,0,n]_{\vec y} t^{2n}~. \label{C2Z2SUk110N-2}
\eea
From \cite{Benvenuti:2010pq}, $\widetilde{g}_{1, SU(N), \BC^2} (t,\vec b)$ can be identified as the Hilbert series for the reduced instanton moduli space of one $SU(N)$ instanton on $\mathbb{C}^2$. On the other hand, $g_{\BC^2/\BZ_2} (t,x)$ is the Hilbert series for $\mathbb{C}^2/\mathbb{Z}_2$. This infinite family of quivers therefore gives a class of instantons where the moduli space is a product of $\BC^2/\BZ_2$ and a reduced moduli space of instantons on $\BC^2$.
In the next subsection, we will encounter a class of instantons where the moduli space does not factorise into such simple components as above.

The plethystic logarithm of the Hilbert series \eref{C2Z2SUk110N} is
\bea
\PL \left[ g_{(1,\,1)}^{(0,\,N)}(t, x, \vec y) \right] &= ([2]_{\vec x}t^2 - t^4) + [1,0,\ldots, 0,1]_{\vec y}t^2 -([0, 1, 0, \ldots, 0, 1, 0]_{\vec y} +
[1, 0, \ldots, 0, 1]_{\vec y} \nn \\
& \qquad + [0, \ldots, 0]_{\vec y}) t^4 +\ldots~.
\eea

In order to check that the order of the pole of the function $g_{(1,\,1)}^{(0,\,N)}(t, x, \vec y)$ at $t=1$ gives the correct complex dimension of the instanton moduli space, one can re-sum the series in equation (\ref{C2Z2SUk110N-2}) at $\vec y=\left(1,1,.....,1\right)$. Using the dimension formula for the representations of $SU(N)$ \cite{Benvenuti:2010pq}, we have
\bea
\mbox{dim} \left[n,0,0,....,0,n\right]=\left(\frac{(N-2+n)!}{n!(N-2)!}\right)^2 \left(\frac{2n+N-1}{N-1}\right)
\eea
The unrefined index for the HS will therefore be given as 
\bea
g_{(1,\,1)}^{(0,\,N)}(t)&= \frac{1+t^2}{(1-t^2)^2} \times \left[\sum_{n=0}^\infty \mbox{dim}[n,0,\ldots,0,n] t^{2n}\right]\nn \\
&
=\frac{1+t^2}{(1-t^2)^2}\times \left[ {}_2F_1\left(N-1,N-1,1;t^2\right)+ 2t^2 (N-1){}_2F_1\left(N,N,2;t^2\right) \right]
\eea
where ${}_2F_1\left(a,b,c;x\right)$ is a hypergeometric function of the indicated type.\\
We enumerate the unrefined HS for the first few cases of $(0,N)$ instantons of instanton number one:
\bea
N=1\;\;:\; \; g_{(1,\,1)}^{(0,\,N=1)}(t)=&\frac{1+t^2}{\left(1-t^2\right)^2} \nn \\
N=2\;\;:\; \; g_{(1,\,1)}^{(0,\,N=2)}(t)=&\frac{\left(1+t^2\right)^2}{\left(1-t^2\right)^4}\nn \\
N=3\;\;:\; \; g_{(1,\,1)}^{(0,\,N=3)}(t)=&\frac{\left(1+t^2\right) \left(1+4 t^2+t^4\right)}{\left(1-t^2\right)^6}\nn \\
N=4\;\;:\; \; g_{(1,\,1)}^{(0,\,N=4)}(t)=& \frac{\left(1+t^2\right)^2 \left(1+8 t^2+t^4\right)}{\left(1-t^2\right)^8}~.
\eea
Note that the order of the pole at $t=1$ in each case matches with the complex dimension of the instanton moduli space.\\
For a generic $N$, the order of the pole can be obtained by considering the asymptotic form of the hypergeometric functions in the formula for $g_{(1,\,1)}^{(0,\,N)}(t)$ as $t \to 1$.
\bea
g_{(1,\,1)}^{(0,\,N)}(t \to 1) \sim \frac{1}{(1-t^2)^2}\times \frac{1}{(1-t^2)^{2N-2}}=\frac{1}{(1-t^2)^{2N}}
\eea
The order of the pole gives the correct complex dimension of the instanton moduli space as expected.\\
One can obviously unrefine the Molien integral formula directly and evaluate the residue using Leibniz's derivative rule to obtain the following equivalent formula for the unrefined HS,
\bea
g_{(1,\,1)}^{(0,\,N)}(t)=\frac{1+t^2}{(1-t^2)^2} \times \frac{\sum^{N-1}_{n=0} {N-1 \choose n}^2 t^{2n}}{(1-t^2)^{2(N-1)}}
\eea
The order of the $t=1$ pole for a generic theory of the class $(0,N)$ can be directly read off from the above formula and matches with the moduli space dimension of the corresponding instanton, as expected. \\

\subsubsection{Hilbert Series for one $SU(N+1)$ instanton on $\BC^2/\BZ_2$: $\vec{k}=(1,1)$ and $\vec{N}=(1,N)$.}
The Hilbert series is given by the Molien integral
\bea
&g_{(1,\,1)}^{(1,\,N)} (t,x,u_1, u_2, \vec y) \nn \\
& =
\oint_{|z_1|=1}\frac{\ud z_1}{(2 \pi i )z_1} \oint_{|z_2|=1}\frac{\ud z_2}{(2 \pi i )z_2} 
\frac{\chi_A (t, u_1,z_1) \chi_B (t, u_2, \vec y,z_2) \chi_X (t,x,z_1,z_2)}{\chi_F(t)}~,
\eea
where the contributions from $A$, $B$, $X$'s and $F$-terms are given respectively by
\bea
\chi_A (t, u_1,z_1) &=  \PE\Big[u_1^{-1}  z_1 t+ u_1 z_1^{-1} t\Big]~, \nn \\
\chi_B (t, u_2, \vec y,z_2) &=  \PE\Big[[0,\ldots,0,1]_{\vec y}u_2^{-1}  z_2 t+[1,0,\ldots,\,0]_{\vec y} u_2 z_2^{-1} t\Big]~, \nn \\
\chi_X(t, x, z_1, z_2) &= \PE \Big[ [1]_x\left(z_1 z_2^{-1}+z_1^{-1} z_2 \right)t \Big]~, \nn \\
\chi_F(t) &= (1-t^2)^{-2}~.
\eea

\subsubsection*{Example: $N=1$}

Let us consider the case for $N=1$. We find the Hilbert series
\begin{equation}
\label{Hilbert series(1,N)}
g_{(1,\,1)}^{(1,\,1)}(t,x,q)=\frac{P(t,x,q)}{(1-\frac{t^2}{x^2})\,(1-t^2\,x^2)\,(1-\frac{t^3\,q}{x})\,(1-\frac{t^3}{x\,q})\,(1-{t^3\,x\,q})\,(1-\frac{t^3\,x}{q})}~,
\end{equation}
where $P(t,x,q)$ is a polynomial
\bea
P(t,x,q)&=1+2\,t^2+2\,t^4-(x+x^{-1})\,(q+q^{-1})\,t^5-(x^2+x^{-2})\,t^6 \nn \\
&\qquad -(x+x^{-1})\,(q+q^{-1})\,t^7+2\,t^8+2\,t^{10}+t^{12}~,
\eea
with $q=u_1 u_2^{-1}$. The unrefined Hilbert series is
\begin{equation}
g_{(1,\,1)}^{(1,\,1)}(t,1,1)=\frac{1+2\,t^2+2\,t^3+2\,t^4+t^6}{(1-t)^4\,(1+t)^2\,(1+t+t^2)^2}
\label{C2Z2SUk11N11}
\end{equation}
The order of the pole at $t=1$ matches with the complex dimension of the moduli space. We will observe to be true in subsequent examples below as well.

Note that \eref{Hilbert series(1,N)} is invariant under $q \rightarrow q^{-1}$ due to the reflection symmetry of the quiver.  Even though the $U(1)_q$ fugacities in the refined Hilbert series can be arranged into $SU(2)$ characters, there does not seem to be such an enhancement in the Lagrangian. There may be such an enhancement at infinite coupling, but the methods which are used in this paper cannot provide the answer. This is left for future investigations.



\subsubsection*{Example: $N=2$}



The HS can be expanded in terms of the characters of $SU(2)_x \times SU(2)_y \times U(1)_q$ as follows: 
\bea
 g_{(1,\,1)}^{(1,\,2)} (t,x,q,y) &=1+t^2\left(\left[2\right]_x + [2]_y +1 \right) + t^3 \left[1\right]_x [1]_y (q+q^{-1}) + t^4 (\left[4\right]_x+[4]_y+[2]_x+[2]_b  \nn \\ 
&\qquad +[2]_y [2]_x +1 ) +\ldots~.
\eea
where $q= u_1 u_2^{-1}$ with $u_1$ and $u_2$ fugacities of the $U(1)$'s coming from each flavour node.  

The plethystic logarithm of this Hilbert series is given by
\bea
\PL[ g_{(1,\,1)}^{(1,\,2)} (t,x,q,y)] &=t^2\left(\left[2\right]_x + [2]_y +1 \right) + t^3 \left[1\right]_x [1]_y \left[1\right]_q- 2t^4 -\ldots~.
\eea

The Hilbert series can also be partially unrefined by setting $q=1$ or $u_1=u_2$ and the resultant formula can be written as an $SU(2)_x\times SU(2)_y$ character expansion in a nice closed form as follows,
\bea
g_{(1,\,1)}^{(1,\,2)} (t,x,1,y) &=\frac{1}{(1-t^2)} \sum_{n_1,n_2,n_3=0}^\infty \Big\{  [2 n_2+ n_3; 2 n_1 + n_3]_{x,y} t^{2n_1+2n_2+3n_3}\nn\\
&
+ [2 n_2+ n_3 + 1; 2 n_1+ n_3 + 1]_{x,y} t^{2n_1+2n_2+3n_3 + 3}
\Big\}
\eea
Upon unrefining completely, we find that
\bea
g_{(1,\,1)}^{(1,\,2)}(t,1,1,1) &=\frac{1+t+4\,t^2+9\,t^3+13\,t^4+12\,t^5+13\,t^6+9\,t^7+4\,t^8+t^9+t^{10}}{(1-t)^6\,(1+t)^4\,(1+t+t^2)^3} \nn \\
&= 1 + 7 t^2 + 8 t^3 + 26 t^4 + 40 t^5 + 88 t^6 + 120 t^7 + 233 t^8 + 
 312 t^9 +\ldots~.
\eea

\subsubsection*{Example: $N=3$}
The Hilbert series can be partially  unrefined by setting $q=1$ or $u_1=u_2$ and the resultant formula can be written as an $SU(2)_x\times SU(3)_{\vec y}$ character expansion:
\bea
g_{(1,\,1)}^{(1,\,3)}(t,x,1,\vec y) &=
\frac{1}{(1-t^2)}
\sum_{n_1, n_2, n_3=0}^\infty \Big\{
[2 n_2+n_3; n_1+n_3,n_1]_{x,\vec y} t^{2n_1+2n_2+3n_3} \nn\\
&
+  [2 n_2 + n_3 +1;n_1,n_1+n_3+1]_{x,\vec y} t^{2n_1+2n_2+3n_3 +3} \Big\}
\eea
The corresponding unrefined Hilbert series is
\bea
g_{(1,\,1)}^{(1,\,3)}(t,1, 1,\vec 1)&=\frac{1 + 2 t + 9 t^2 + 24 t^3 + 50 t^4 + 76 t^5 + 108 t^6 + 120 t^7 + 
 108 t^8 + \text{palindrome} +t^{14}}{(1-t)^8\,(1+t)^6\,(1+t+t^2)^4} ~. \label{HS:C2Z2SU4N13}
\eea

\subsubsection*{The Case of General $N$}
One can now summarise the case of general $N$.  Unrefining partially by setting $q=1$ or $u_1=u_2$, the Hilbert series can be written as an $SU(2)_x\times SU(N)_y$ character expansion for any $N > 1$,
\bea \label{HS:C2Z2SU1N11}
g_{(1,\,1)}^{(1,\,N)} (t,x,1,\vec y)&=
\frac{1}{(1-t^2)}
\sum_{n_1, n_2, n_3=0}^\infty \Big\{
[2 n_2+n_3; n_1+n_3,0,\dots,0,n_1]_{x,\vec y} t^{2n_1+2n_2+3n_3}
\nn\\
&
+ [2 n_2 + n_3 +1; n_1,0,\dots,0,n_1+n_3+1]_{x,\vec y} t^{2n_1+2n_2+3n_3 +3}
\Big \}~.
\eea
The plethystic logarithm for the corresponding fully refined Hilbert series is
\bea \label{PL1Ngen}
&\PL[ g^{(1, N)}_{(1, 1)} (t, x;q ; \vec y)] \nn\\
&= ([2;0,\ldots,0]+[0;1,0, \ldots,0,1]+[0;0, \ldots,0])t^2 + \nn \\
&\qquad + (q^{-1}[1;1,0, \ldots,0] + q[1;0,\ldots,0,1]) t^3 - \ldots~,
\eea
where the notation $[a;b_1, \ldots, b_n]$ denotes the representation of the global symmetry $SU(2)_x \times SU(N)$, where the fugacity $q=u_1 u_2^{-1}$ where $u_1$ and $u_2$ correspond to the $U(1)$ coming from each flavour node.  

The above character expansion can be used to obtain a formula for the unrefined index for a generic $N$. We define a function $K\left(x,y,N\right)$ such that,
\bea \label{dimform2}
K\left(a_1,a_{N-1},N\right) =&\mbox{dim} \left[a_1,0,...,0,a_{N-1}\right]_{SU(N)}\nn \\
=&(a_1+a_{N-2}+N-1) \frac{G(N-1)}{G(N+1)} \left(a_1+1\right)_{N-2} \left(a_{N-2}+1\right)_{N-2}
\eea
\\
where $G(z)$ is a Barnes $G$-function and $\left(a\right)_n=a(a-1)(a-2)....(a-n+1)$ is the Pochhammer symbol.\\
The above formula may be directly obtained from the following dimension formula of $SU(N)$ irreps in terms of Dynkin labels \cite{Fulton1991}:
\bea
\mbox{dim}\left[a_1,a_2,....,a_{N-1}\right]=\prod_{1\leq i<j \leq N} \frac{(a_i+.....+a_{j-1})+j-i}{j-i}
\eea

One can therefore write the following formula for the unrefined HS:
\bea 
g_{(1,\,1)}^{(1,\,N)} (t,1,1,\vec 1)&=
\frac{1}{(1-t^2)}
\sum_{n_1, n_2, n_3=0}^\infty \Big\{
\left(2 n_2+n_3+1\right) K\left(n_1+n_3,n_1,N\right) t^{2n_1+2n_2+3n_3}
\nn\\
&
+ \left(2 n_2 + n_3 +2\right) K\left(n_1,n_1+n_3+1,N\right) t^{2n_1+2n_2+3n_3 +3}
\Big \}~.
\eea
One cannot re-sum the above series to a closed form for a generic $N$, but one can quickly check that the result agrees with the unrefined HS for $N=2,3$. One can use it to obtain the unrefined series for the $N=4$ case as well.
\bea 
&g_{(1,\,1)}^{(1,\,N=4)} (t,1,1,\vec 1)\nn \\
&=\frac{\left(1+3 t+17 t^2+54 t^3+143 t^4+293 t^5+533 t^6+798 t^7+1018 t^8+1088 t^9+\mbox{palindrome}+t^{18}\right)}{(1-t)^{10} (1+t)^8 \left(1+t+t^2\right)^5}
\eea
with the order of the pole at $t=1$ again matching the complex dimension of the instanton moduli space.\\

Note that Hilbert series for the class of $SU(N+1)$ instantons  discussed above does not factorize into Hilbert series for the center of mass motion on $\mathbb{C}^2/\mathbb{Z}_2$ times Hilbert series of the reduced moduli space. This shows that, on the contrary to the $\mathbb{C}^2$ case in \cite{Benvenuti:2010pq,Hanany:2012dm}, generically the moduli space for instantons on ALE space does not have a factorisation property.

\begin{figure}[H]
\begin{center}
\includegraphics[scale=0.5]{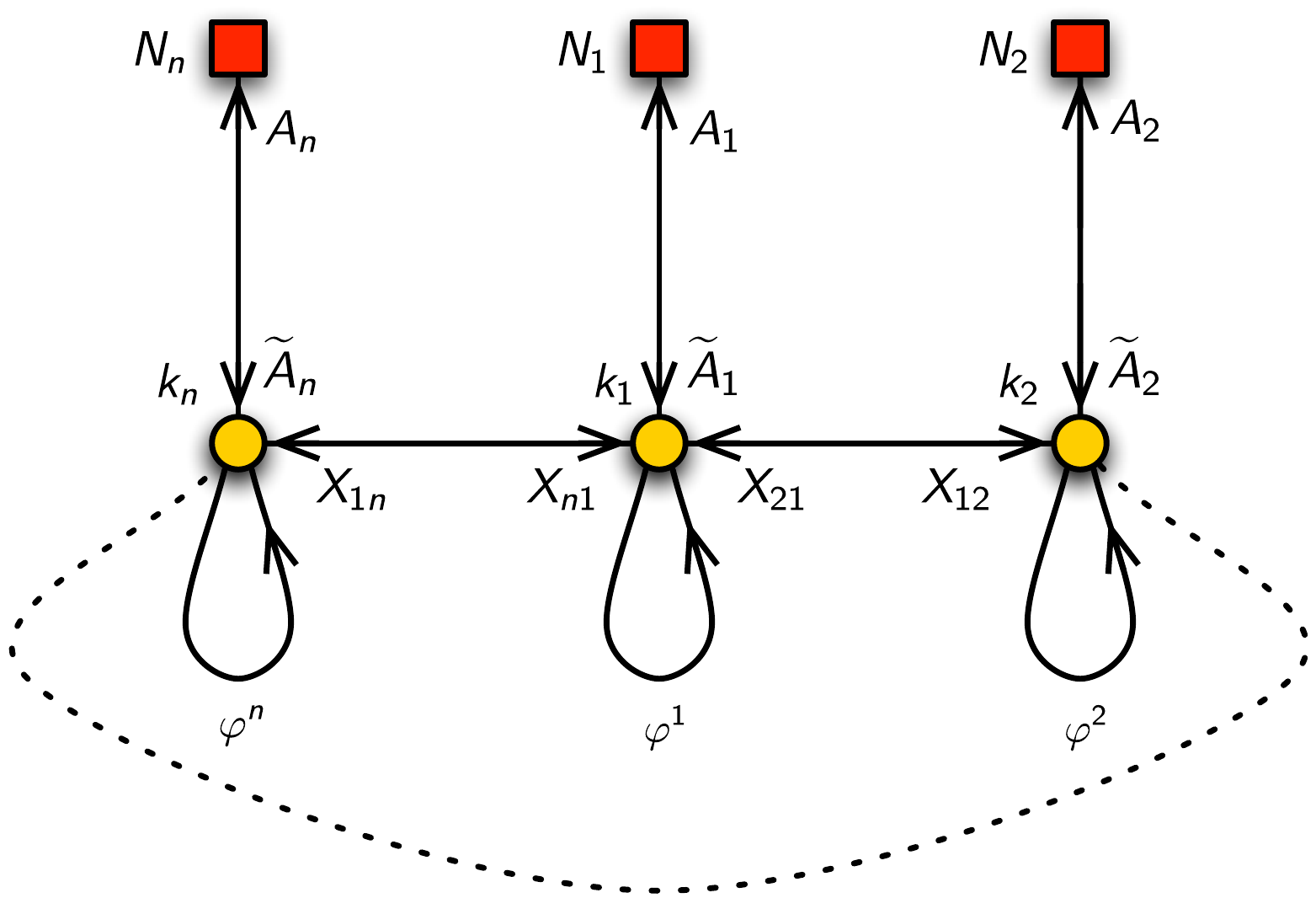}
\caption{The $\mathcal{N}=1$ quiver for $SU(N)$ instantons on $\BC^2/\BZ_n$. }
\label{fig:necklacen1}
\end{center}
\end{figure}

\subsection*{One instanton in a Generic Necklace Quiver: $\vec{k}=(1,\dots,1)$ and $\vec{N}=(N_1,\dots,N_n)$.} 
The Molien integral formula for the Hilbert series can be generalised for a cyclic quiver of the form as shown in \fref{fig:necklacen1}. For $n>2$, there is no analogue of the $SU(2)_x$ global symmetry.
The generalisation is for $\vec{k}=(1,\dots,1)$ and $\vec{N}=(N_1,\dots,N_n)$ and takes the following form,
\bea\label{ehsgeneral}
&g^{(N_1,\dots,N_n)}_{(1,\dots,1)}(t,\vec{y}_j) 
\nn\\
&= \oint_{|z_1|=1} \frac{\ud z_1}{(2\pi i)z_1} \dots \oint_{|z_n|=1} \frac{\ud z_n}{(2\pi i)z_n} 
\frac{
\prod_{j=1}^{n}\chi_{A_{j}}(t,u_j,\vec{y}_j,z_j)~
\chi_X(t,z_j)
}{
\chi_F(t)
}~~,
\eea
where the contributions from $A_j$ (with $j=1, \ldots, n$) , $X$'s and $F$-terms are given by
\bea
\chi_{A_j}(t,u_j,\vec{y}_j,z_j) &= \PE\Big[
[0,\dots,0,1]_{\vec{y}_j} u_j^{-1} z_j t +
[1,0,\dots,0]_{\vec{y}_j} u_j z_j^{-1} t
\Big]
~~,\nn\\
\chi_{X}(t,z_j) &= \PE\Big[
\sum_{j=1}^{n} (z_j^{-1} z_{j+1} t + z_{j+1}^{-1} z_{j} t)
\Big]
~~,\nn\\
\chi_{F}(t)&=(1-t^2)^{-n}~~,
\eea
where $[1,0,\dots,0]_{\vec{y}_j}$ and $[0,\dots,0,1]_{\vec{y}_j}$ are characters of the global symmetry $SU(N_j)$.\\

We discuss the generators of the chiral ring of operators associated to the moduli spaces of instantons presented above and the relations connecting them in \ref{sec:genSUNC2Zn}, using the Plethystic Logarithm (PL) analysis.\\

\subsection{The Hilbert series via localisation} \label{sec:localisationSU}

The Hilbert series can be computed using the Molien integral formula as in the preceding section.  However, larger the rank of the gauge group, greater is computational power required to compute the multi-contour integrations corresponding to Haar measures of the gauge groups.  Nevertheless,  the poles and their residues can be arranged combinatorially so that the problem becomes that of an enumeration of restricted coloured partitions.  In this way, the Hilbert series can be computed more efficiently.  The computation of the Hilbert series from the contributions from such restricted coloured partitions is referred to as the ``localisation method". This method was applied in  \cite{Fucito:2004ry, Ito:2011mw, Alfimov:2011ju} (see also \cite{Bonelli:2012ny, Ito:2013kpa}) to compute the Nekrasov partition functions for $SU(N)$ instantons on $\mathbb{C}^2/\mathbb{Z}_n$.  In this section, we adapt the above technique to compute the Hilbert series.

\subsubsection{The Hilbert series of $k$ $SU(N)$ instantons on $\BC^2$ via localisation}
Let us start out by summarising the localisation method for the Hilbert series of $SU(N)$ instantons on $\BC^2$.  This will turn out to be very useful for the subsequent computation for $\BC^2/\BZ_n$. Here we follow the prescription given in Theorem 2.11 and Proposition 4.1 of \cite{Nakajima:2003pg}. The idea is also similar to Section 3.5 of \cite{Nekrasov:2002qd}.  

Let $t_1 , t_2$ be the fugacities correspond to the $U(2)$ isometry of the space $\BC^2$.  It is sometimes convenient to rewrite these as
\bea
t_1 = t x~, \qquad t_2 = t x^{-1}~,
\eea
where $t$ is the fugacity of the $U(1)$ subgroup of $U(2)$, and $x$ is the fugacity of the $SU(2)$ subgroup.
Let $e_1, \ldots, e_N$ be the fugacities corresponding to the internal symmetry $SU(N)$ of instantons such that $e_1+\ldots+e_N$ is the character of the fundamental representation of $SU(N)$ and 
\bea
\prod_{\alpha=1}^N e_\alpha =1~.
\eea  

Localization fixes the gauge field configuration such that the instantons are located at the origin. Such fixed instantons are labelled by an $N$-tuple of Young diagrams, denoted by $\vec Y = (Y_1, Y_2, \ldots, Y_N)$.  We refer to each element by $Y_\alpha$, with $\alpha=1, \ldots, N$, and we allow the cases in which there exist empty diagrams. The instanton number $k$ is given by the total number of boxes
\bea
k = |\vec Y| := \sum_{\alpha=1}^N | Y_\alpha|~.
\eea

Let us define
\bea \label{eq:NC2}
N_{\alpha ,\beta}^{\vec{Y}}(t_1,t_2, \vec e) 
&=
\sum_{s \in Y_\alpha } \left(
\frac{e_\alpha}{e_\beta} t_1^{-l_{Y_\beta}(s)}t_2^{1+a_{Y_\alpha }(s)}+ \frac{e_\beta}{e_\alpha} t_1^{1+l_{Y_\beta }(s)}t_2^{-a_{Y_\alpha}(s)} \right) ~,
\eea
where for a given box $s$ at the $a$-th column and $b$-th row, \ie~ at the coordinates $(a,b)$, of a given Young diagram $Y = (\lambda_1 \geq \lambda_2 \geq \ldots)$, one can define $a_Y(s)$ and $l_Y(s)$, known as the arm length and the leg length, as follows:
\bea
a_Y(s) = \lambda_a- b~, \qquad l_Y(s) = \lambda'_b - a~,
\eea
where $\lambda'_b$ corresponds to the transpose diagram of $Y$, namely $Y^T = (\lambda'_1 \geq \lambda'_2 \geq \ldots)$.  Note that in \eref{eq:NC2}, if $s \in Y_\alpha$ but $s \notin Y_\beta$, then $l_{Y_\beta}(s)$ is negative.

The Hilbert series of $k$ $SU(N)$ instantons on $\BC^2$ is then given by
\bea
H^{N}_{k, \BC^2} (t_1, t_2, \vec e) = \sum_{\vec Y: |\vec Y| =k}  \PE \left[ \sum_{\alpha, \beta=1}^N N_{\alpha ,\beta}^{\vec{Y}}(t_1,t_2, \vec e) \right ] ~, \label{eq:HSNkC2}
\eea
where the summation outside the $\PE$ runs over all possible $N$-tuples of the Young diagrams whose total number of boxes equal to the instanton number $k$.  Explicit computations can be found, for example, in (4.3) of \cite{Nakajima:2003pg} and Section 3.1 of \cite{Hanany:2012dm}.

\subsubsection{The Hilbert series of $SU(N)$ instantons on $\BC^2/\BZ_n$ via localisation}
In this section, we turn to $SU(N)$ instantons on $\BC^2/\BZ_n$ and use the quiver data as stated in Section \ref{sec:quivdata}. The $\BZ_n$ orbifold acts on the fugacities $t_1$, $t_2$ and $e_\alpha$ (with $\alpha=1, \ldots, N$) as follows.  Let $\omega = \exp(2 \pi i/n)$, the actions are
\bea \label{orbactionZn}
t_1 \rightarrow \omega t_1~, \qquad t_2 \rightarrow \omega^{-1} t_2~, \qquad e_\alpha \rightarrow \omega^{r_\alpha} e_\alpha~,
\eea
where $r_\alpha$ (with $0\leq  r_\alpha \leq  n-1$) are related to the monodromy $\vec N = (N_1, \ldots, N_n)$ as follows:
\bea
N_i = \sum_{\alpha=1}^N \delta_{r_\alpha, i~(\mathrm{mod}~n)}~, \label{Nandr}
\eea
where $i~(\mathrm{mod}~n)$ runs over $0, \ldots, n-1$.  Therefore, $N_i$ is the number of times that $i~(\mathrm{mod}~n)$ appears in the vector $\vec r = (r_1, \ldots, r_N)$.  In order that the orbifold action is compatible with the $SU(N)$ symmetry, we need to impose that
\bea
\sum_{\alpha=1}^N r_\alpha = 0 ~(\mathrm{mod}~n)~. \label{constraintr}
\eea
Let us now describe the method to compute the Hilbert series of instantons on $\BC^2/\BZ_n$ for given $\vec k$ and $\vec r$ (or $\vec N$).

\paragraph{The algorithm.}  Let us define a quantity as analogous to \eref{eq:NC2}:
\bea
& \widetilde{N}_{\alpha ,\beta}^{\vec{Y}, \vec r}(t_1,t_2,\vec e) \nn \\
&=
\sum_{s \in Y_\alpha } \left(
\frac{e_\alpha}{e_\beta} t_1^{-l_{Y_\beta}(s)}t_2^{1+a_{Y_\alpha }(s)}+ \frac{e_\beta}{e_\alpha} t_1^{1+l_{Y_\beta }(s)}t_2^{-a_{Y_\alpha}(s)} \right) \delta_{a_{Y_\alpha}(s)+l_{Y_\beta}(s)+1-r_\alpha+r_\beta ~(\mathrm{mod}~n)}~,
\eea
Let us comment on this formula:
\bi
\item  The quantity in the round brackets is the same as that of instantons on $\BC^2$, see \eref{eq:NC2}.
\item  The delta symbol selects only the terms in the summation that carry charge zero modulo $n$, with respect to the charge assignment \eref{orbactionZn}.  Other terms carrying non-zero charge (modulo $n$) do not contribute to the Hilbert series and are thrown away.
\ei
In addition to the above restriction, one needs to select only certain $N$-tuples of Young diagrams in the summation of \eref{eq:HSNkC2}.  In particular, the required Hilbert series is
\bea \label{eq:Hilbert serieskSUN}
H^{\vec N}_{\vec k} (t_1, t_2, \vec e) = \sum_{\substack{\vec Y\in \CR}}  \PE \left[ \sum_{\alpha, \beta=1}^N \widetilde{N}_{\alpha ,\beta}^{\vec{Y},  \hat{\vec r}}(t_1,t_2, \vec e) \right ] ~,
\eea
where $\hat{\vec r}$ denotes a solution $(r_1, \ldots, r_N)$ of \eref{Nandr} for a given $(N_1, \ldots, N_n)$, and $\CR$ is a set of $N$-tuples of Young diagrams such that all of the following conditions are satisfied:
\ben
\item The total number of boxes in $\vec Y$ is given by $|\vec Y| := \sum_{\alpha=1}^N Y_\alpha = \sum_{i=1}^n k_i$.
\item Upon assigning the numbers $\hat{r}_\alpha+a-b~(\mathrm{mod}~n)$ to all $(a,b)$ boxes of every non-trivial Young diagram $Y_\alpha \neq \emptyset$ for all $\alpha=1, \ldots, N$, there must be precisely $k_j$ boxes in total that are labelled by the number $j~(\mathrm{mod}~n)$ for all $j=1, \ldots, n$.
\een

\paragraph{Notation.} It is conventional to make the following change of variables 
\bea
& t_1 = t x~, \quad t_2 = t x^{-1}~, \nn \\
& e_1 = y_1~, \quad e_2 = y_2 y_1^{-1}~, \ldots,~ e_{N-1} = y_{N-1} y_{N-2}^{-1}~, \quad e_N = y_{N-1}^{-1}~,
\eea
 Let us denote 
\bea
g^{\vec N}_{\vec k} (t, x, \vec y) := H^{\vec N}_{\vec k}  (t_1, t_2, \vec e)
\eea
written in these new variables.  

\paragraph{Nekrasov partition function.}
We mention in passing that the $4d$ Nekrasov partititon function $\CZ^{\vec N}_{\vec k}(\epsilon_1, \epsilon_2, a_1, \ldots, a_N)$ for a given $\vec k$ and $\vec N$ can be computed from the Hilbert series $g^{\vec N}_{\vec k} (t, x, y_1, \ldots, y_N)$ by setting
\bea
t = e^{-\frac{1}{2}\beta(\epsilon_1+\epsilon_2)}, \quad  x = e^{-\frac{1}{2}\beta(\epsilon_1-\epsilon_2)}, \quad y_i = e^{-\beta a_i}~,
\eea
and taking the following limit \cite{Nekrasov:2002qd, Nakajima:2003pg, Keller:2011ek, RGZ2013}:
\bea
\CZ^{\vec N}_{\vec k}(\epsilon_1, \epsilon_2, a_1, \ldots, a_N) = \lim_{\beta \rightarrow 0} \beta^{2kN} g^{\vec N}_{\vec k} (t, x, y_1, \ldots, y_N)~.
\eea
This is indeed a computation of the volume of the base of the hyperK\"ahler cone corresponding to the instanton moduli space \cite{Martelli:2006yb}. In Section \ref{C2Z4pureinst}, we compute explicitly the Nekrasov partition function for a given instanton number $k$ and compare the results with those in \cite{Wyllard:2011mn}.
\\

\subsection{Generators of the moduli space \label{sec:genSUNC2Zn}}

\paragraph{Plethystic logarithm.} The plethystic logarithm of the Hilbert series for instantons on $\BC^2/\BZ_n$ given $\vec {k}$ and $\vec N$ (see \fref{fig:necklace} and \fref{fig:necklacen1}) has the general form
\bea \label{PLkSUnC2Zn}
& \PL[ g^{\vec N}_{\vec k} (t, x; {\vec \xi}_1, \ldots, {\vec \xi}_n)] = \left( \sum_{\substack{1 \leq i  \leq n \\ N_i \neq 0}} \underbrace{{\bf fund} (\vec{\xi}_i){\bf fund} (\vec{\xi}_i^{-1})}_{{\bf Adj}(\vec{\xi}_i)} \right)t^2 +  (x^n + x^{-n}) t^n \nn \\
& \qquad +  \sum_{\ell = 1}^{n-1} \sum_{\substack{1 \leq i  \leq n \\ N_i\neq 0}}  \Big( x^{\ell} {\bf fund}({\vec \xi}_i) {\bf fund}({\vec \xi}_{ \ell+i }^{-1}) + x^{-\ell} {\bf fund}({\vec \xi}_i^{-1}) {\bf fund}({\vec \xi}_{ \ell+i}) \Big) t^{\ell+2} \nn \\
& \qquad - \text{terms corresponding to relations from $t^4$ onwards}~. 
\eea
where $\vec{\xi}_1, \ldots, \vec{\xi}_n$ are the global fugacities for $U(N_1)$, $U(N_2)$, $\ldots$, $U(N_n)$ and $x$ is a fugacity which measures clockwise-anticlockwise fields in the cyclic quiver (see more explanation below). The index $\ell+i$ is taken modulo $n$ and it takes values $1, 2, \ldots, n$.  Since the symmetry of the Higgs branch of the theory is $S(\prod_{\alpha=1}^n U(N_\alpha))$, we impose the condition
\bea
\prod_{\alpha=1}^n \prod_{i=1}^{N_\alpha} \xi_{\alpha,i} =1~.
\eea
Note from \eref{PLkSUnC2Zn} that the total number of generators is
\bea
\sum_{i=1}^n N_i^2 + 2 + 2\sum_{1\leq i < j \leq n} N_i N_j = 2N^2 -\sum_{i=1}^n N_i^2 +2~.
\eea

\paragraph{Generators.} The generators of the moduli space for $\vec k =(k_1, \ldots, k_n)$ and $\vec N =(N_1, \ldots N_n)$ can be read off from \eref{PLkSUnC2Zn} as follows:
\ben
\item At order 2, there are generators in the adjoint representation of $U(N_i)$ for each $i=1, \ldots, n$ with $N_i \neq 0$.  For each of these $i$, these generators arise as gauge invariant combinations (``mesons'') of the chiral fields between the node $U(k_i)$ and $U(N_i)$.
\item At order $n$, there are two generators coming from $\BC^2/\BZ_n$.\footnote{In fact, the space $\BC^2/\BZ_n$ has 3 generators: one, denoted by $G_1$, at order 2 and two, denoted by $G_2, G_3$, at order $n$; they are subject to the relation $G_1^{n} =G_2 G_3$ at order $2n$. The generators $G_2$ and $G_3$ corresponds to the ones appearing at order $t^n$ in \eref{PLkSUnC2Zn}.  On the other hand, the generator $G_1$ can be traded (using the $F$ term condition) with one generator at order $t^2$ that is a singlet under $SU(N)$.}  These arise as gauge invariant combinations of the chiral fields in the cyclic quiver from any node $U(k_i)$ back to itself in clockwise and anti-clockwise directions.
\item At order $\ell+2$, with $\ell =1, \ldots, n-1$, there are two sets of generators: one in the representation $( \Box_{+1}; \overline{\Box}_{-1})$ of $U(N_i) \times U(N_{\ell+i })$ and the other in $( \overline{\Box}_{-1}; \Box_{+1})$ of $U(N_i) \times U(N_{\ell+i })$, where $i=1, \ldots, n$ and $\ell+i$ is taken modulo $n$.  These arise as gauge invariant combinations of the chiral fields going from the node $U(N_i)$ to a different node $U(N_{\ell+i })$ and back, via $\ell$ edges in the cyclic quiver.
\een

Note that, in case of $\BC^2/\BZ_2$, one can rewrite \eref{PLkSUnC2Zn} in such a way that the $SU(2)_x$ isometry is manifest, see \eg, \eref{PL1Ngen}.


\subsection{Special case of pure instantons: $\vec \beta=(0, \ldots, 0, N)$}  \label{sec:pureinstfeatures}

Let us consider the case of pure instantons, namly those with $\beta_1 = \beta_2 = \ldots = \beta_{n-1}=0$.\footnote{The simplest such case is that of rank zero instantons, which we consider in appendix \ref{holo}. Interestingly, this case can also be studied from a holographic perspective.} Given $\vec \beta=(0, \ldots, 0, N)$, the components of $\vec k$ cannot take arbitrary values, but are constrained by \eref{betafn}.   In the special case we are considering, \eref{betafn} can be rewritten as
\bea
k_i = m + \sum_{j=1}^n C_{ij} N_j~, \quad \text{with}~\sum_{j=1}^n C_{ij} N_j \in \BZ~\text{for all}~ i=1, \ldots, n~,
\eea
where $m = 0, 1, 2, \ldots$ and $C$ is the (augmented) inverse Cartan matrix of $A_{n-1}$,
\bea
C_{i_1 i_2} = \min(i_1, i_2)- \frac{i_1 i_2}{n}~, \qquad 1\leq i_1, i_2 \leq n~.
\eea
Below we present examples in which all possible $\vec k$ and $\vec N$ are listed, given $(N,n)$.  One of the easiest ways to enumerate all possibilities is to start from all possible $\vec r$ subject to the contraint \eref{constraintr}. 

\subsubsection*{Examples for $SU(2)$ pure instantons}
For $(N,n)=(2,2)$, \ie~ $SU(2)$ pure instantons on $\BC^2/\BZ_2$, the possibilities are
\bea
\begin{array}{llll}
\vec r= (0,0) \quad &\Rightarrow \quad \vec N = (0,2) \quad &\Rightarrow \quad \vec k = (m,m) \qquad &\Rightarrow \qquad k =m~,  \\
\vec r= (1,1)  \quad &\Rightarrow \quad \vec N = (2,0) \quad &\Rightarrow \quad \vec k = (m+1, m) \qquad &\Rightarrow \qquad k =m+\frac{1}{2}~. \label{C2Z2}
\end{array}
\eea
For $(N,n)=(2,3)$, \ie~ $SU(2)$ pure instantons on $\BC^2/\BZ_3$, the possibilities are
\bea
\begin{array}{llll}
\vec r= (0,0) \quad &\Rightarrow \quad \vec N = (0,0,2) \quad &\Rightarrow \quad \vec k = (m,m,m) \qquad &\Rightarrow \qquad k =m~,  \\
\vec r= (1,2) ~\text{or} ~(2,1) \quad &\Rightarrow \quad \vec N = (1,1,0) \quad &\Rightarrow \quad \vec k = (m+1, m+1, m) \qquad &\Rightarrow \qquad k =m+\frac{2}{3}~. \label{C2Z3}
\end{array}
\eea
For $(N,n)=(2,4)$, \ie~ $SU(2)$ pure instantons on $\BC^2/\BZ_4$, the possibilities are
\bea
\begin{array}{llll}
\vec r= (0,0) \quad &\Rightarrow \quad \vec N = (0,0,0,2) \quad &\Rightarrow \quad \vec k = (m,m,m,m) \qquad &\Rightarrow \qquad k =m~, \\
\vec r= (1,3)~\text{or} ~(3,1) \quad &\Rightarrow \quad \vec N = (1,0,1,0) \quad &\Rightarrow \quad \vec k = (m+1, m+1, m+1, m) \qquad &\Rightarrow \qquad k =m+\frac{3}{4}~,  \\
\vec r= (2,2)  \quad &\Rightarrow \quad \vec N = (0,2,0,0) \quad &\Rightarrow \quad \vec k = (m+1, m+2, m+1, m) \qquad &\Rightarrow \qquad k =m+1~. \label{C2Z4}
\end{array}
\eea

\subsubsection*{Examples for $SU(3)$ pure instantons}
For $(N,n)=(3,2)$, \ie~ $SU(3)$ pure instantons on $\BC^2/\BZ_2$, the possibilities are
\bea
\begin{array}{llll}
\vec r= (0,0,0)  \quad &\Rightarrow \quad \vec N = (0,3) \quad &\Rightarrow \quad \vec k = (m,m) \qquad &\Rightarrow \qquad k =m~,  \\
\vec r= (0,1,1) ~\text{or perms}  \quad &\Rightarrow \quad \vec N = (2,1) \quad &\Rightarrow \quad \vec k = (m+1, m) \qquad &\Rightarrow \qquad k =m+\frac{1}{2}~. 
\end{array}
\eea
For $(N,n)=(3,3)$, \ie~ $SU(3)$ pure instantons on $\BC^2/\BZ_3$, the possibilities are
\bea
\begin{array}{llll}
\vec r= (0,0,0)  \quad &\Rightarrow \quad \vec N = (0,0,3) \quad &\Rightarrow \quad \vec k = (m,m,m) \qquad &\Rightarrow \qquad k =m~,  \\
\vec r= (0,1,2)~\text{or perms}  \quad &\Rightarrow \quad \vec N = (1,1,1) \quad &\Rightarrow \quad \vec k = (m+1, m+1, m) \qquad &\Rightarrow \qquad k =m+\frac{2}{3}~,  \\
\vec r= (1,1,1)  \quad &\Rightarrow \quad \vec N = (3,0,0) \quad &\Rightarrow \quad \vec k = (m+2, m+1, m) \qquad &\Rightarrow \qquad k =m+1~. 
\end{array}
\eea

\subsubsection*{Examples for $SU(4)$ pure instantons}
For $(N,n)=(4,2)$, \ie~ $SU(4)$ pure instantons on $\BC^2/\BZ_2$, the possibilities are
\bea \label{SU4pureC2Z2}
\begin{array}{llll}
\vec r= (0,0,0,0)  \quad &\Rightarrow \quad \vec N = (0,4) \quad &\Rightarrow \quad \vec k = (m,m) \qquad &\Rightarrow \qquad k =m~,  \\
\vec r= (0,0,1,1)~\text{or perms}  \quad &\Rightarrow \quad \vec N = (2,2) \quad &\Rightarrow \quad \vec k = (m+1, m) \qquad &\Rightarrow \qquad k =m+\frac{1}{2}~,  \\
\vec r= (1,1,1,1)  \quad &\Rightarrow \quad \vec N = (4,0) \quad &\Rightarrow \quad \vec k = (m+2, m) \qquad &\Rightarrow \qquad k =m+1~. 
\end{array}
\eea

\subsubsection{Certain features of the moduli space of pure instantons} 
Subsequently, we provide several examples, including explicit Hilbert series for the pure instantons.  Recall that the Kronheimer--Nakajima constructions for pure $SU(N)$ instantons correspond to $\vec \beta =(0, \ldots,0, N)$.  

There are important features that can be drawn from these examples:
\ben
\item The moduli space of $(n-1)/n$ pure $SU(N)$ instantons on $\BC^2/\BZ_n$, corresponding to $\vec k =(1, \ldots, 1,0)$, is $\BC^2/\BZ_n$.
\item The moduli space of {\it one} $SU(N)$ pure instanton on $\BC^2/\BZ_n$, corresponding to $\vec k =(1, \ldots, 1)$, is 
\bea
\BC^2/\BZ_n \times \widetilde{\CM}_{1,SU(N),\BC^2}~, 
\eea
where $\widetilde{\CM}_{1,SU(N),\BC^2}$ is the reduced moduli space of $1$ $SU(N)$ instanton on $\BC^2$.
\een
We demonstrate these claims in several examples below.

\subsubsection{$SU(4)$ instantons on $\BC^2/\BZ_2$ with $\vec \beta = (0, 4)$}
\section*{$k = 1$}
Given $k=1$, from \eref{SU4pureC2Z2} there are two options in this case.
\paragraph{The first option.} Consider $\vec k =(1,1)$ and $\vec N =(0,4)$; we choose $\vec r = (0,0,1,1)$. The Hilbert series is
\bea \label{g0114}
g^{(0,4)}_{(1,1)} (t,x, \vec y) = \sum_{m_2, n_2 \geq 0} [2m_2; n_2,0,n_2] t^{2m_2+2n_2}~.
\eea
The moduli space is therefore
\bea
\CM^{(0,4)}_{(1,1)} = (\BC^2/\BZ_2) \times \widetilde{\CM}_{1,SU(4), \BC^2}~.
\eea
The unrefined Hilbert series is
\bea \label{g0114ur}
g^{(0,4)}_{(1,1)} (t,x, \{ y_i=1\}) = \frac{\left(1+t^2\right)^2 \left(1+8 t^2+t^4\right)}{\left(1-t^2\right)^8}~.
\eea

\paragraph{The second option.} Consider $\vec k =(2,0)$ and $\vec N =(4,0)$; we choose $\vec r = (1,1,1,1)$.  This is actually $U(2)$ gauge theory with 4 flavours.  The Higgs branch Hilbert series is given by
\bea \label{g4200}
 g^{(4,0)}_{(2,0)} (t, x, \vec y) = \sum_{m_2 \geq 0} [0; m_2,0,m_2] t^{2m_2} + \sum_{n_2,n_4 \geq 0} [0; n_2,2n_4+2,n_2] t^{2n_2+4n_4+4} ~.
\eea
The PL of this Hilbert series is
\bea
\PL \left[ g^{(4,0)}_{(2,0)} (t, x, \vec y) \right] = [0;1,0,1] t^2 - ([0;1,0,1]+[0;0,0,0]) t^4 + \ldots~.
\eea
The unrefined Hilbert series is
\bea \label{g4200ur}
g^{(4,0)}_{(2,0)} (t,x, \{ y_i=1\}) = \frac{\left(1+t^2\right)^2 \left(1+5 t^2+t^4\right)}{\left(1-t^2\right)^8}~.
\eea

\subsubsection{$SU(2)$ instantons on $\BC^2/\BZ_3$ with $\vec \beta = (0, 0, 2)$}
\section*{$k = 2/3$}
From \eref{C2Z3}, this corresponds to $\vec k = (1,1,0)$ with $\vec N=(1,1,0)$.  Let us first  choose $\vec r =(1,2)$.  The set $\CR$ contains three elements, namely
\bea
\CR = \{ (\emptyset, \tiny \yng(1,1)), \quad (\yng(1),\yng(1)), \quad (\yng(1,1), \emptyset) \}~.
\eea
The Hilbert series is given by
\bea \label{g110110}
g^{(1,1,0)}_{(1,1,0)}(t, x, y) &= \frac{1}{\left(1-\frac{t^3}{x y^2}\right) \left(1-\frac{x y^2}{t}\right)}+\frac{1}{\left(1-\frac{t}{x y^2}\right) \left(1-t x y^2\right)}+\frac{1}{\left(1-\frac{1}{t x y^2}\right) \left(1-t^3 x y^2\right)} \nn \\
&= \frac{1}{3} \left[ \sum_{p=0}^2 \frac{1}{(1- \omega^p \chi^{1/3} t ) (1- \omega^{-p} \chi^{-1/3} t)} \right]~, \quad \chi := x y^{2}~,  \nn \\
&= g_{\BC^2/\BZ_3} (t, \chi^{1/3})~, 
\eea
where the last two lines indicate that the Hilbert series $g^{(1,1,0)}_{(1,1,0)}(t, x, y)$ is in fact the Hilbert series of $\BC^2/\BZ_3$, with
\bea \label{C2Z3HS}
g_{\BC^2/\BZ_3}(t,x) &= (1-t^6) \PE \left[ t^2 + t^3(x^3 +x^{-3})\right]~, \nn \\
g_{\BC^2/\BZ_3}(t,x=1) &= \frac{1-t^6}{(1-t^2)(1-t^3)^2} = \frac{1-t+t^2}{(1-t)^2 \left(1+t+t^2\right)}~.
\eea
Note that if we choose $\vec r =(2,1)$, the result is still \eref{g110110} with a redefinition of $x$ to $x^{-1}$.

\section*{$k=1$}
From \eref{C2Z3}, this corresponds to $\vec k = (1,1,1)$ with $\vec N=(2,0,0)$; one may choose $\vec r =(0,0)$.
In this case, the set $\CR$ contains six elements, namely
\bea
\CR = \{ (\emptyset, \tiny \yng(1,1,1)), \quad(\emptyset, \tiny \yng(1,2)), \quad(\emptyset, \tiny \yng(3)), (\tiny \yng(1,1,1), \emptyset), \quad( \tiny \yng(1,2), \emptyset), \quad( \tiny \yng(3), \emptyset) \}~.
\eea
The Hilbert series is
\bea
&g^{(2,0,0)}_{(1,1,1)}(t, x, y) \nn \\
&= \frac{1}{\left(1-\frac{t^3}{x^3}\right) \left(1-\frac{x^3}{t}\right) \left(1-\frac{t^2}{y^2}\right) \left(1-y^2\right)}+\frac{1}{\left(1-\frac{t}{x^3}\right) \left(1-t x^3\right) \left(1-\frac{t^2}{y^2}\right) \left(1-y^2\right)} \nn \\
& \quad +\frac{1}{\left(1-\frac{1}{t x^3}\right) \left(1-t^3 x^3\right) \left(1-\frac{t^2}{y^2}\right) \left(1-y^2\right)} +\frac{1}{\left(1-\frac{t^3}{x^3}\right) \left(1-\frac{x^3}{t}\right) \left(1-\frac{1}{y^2}\right) \left(1-t^2 y^2\right)} \nn \\
& \quad +\frac{1}{\left(1-\frac{t}{x^3}\right) \left(1-t x^3\right) \left(1-\frac{1}{y^2}\right) \left(1-t^2 y^2\right)}+\frac{1}{\left(1-\frac{1}{t x^3}\right) \left(1-t^3 x^3\right) \left(1-\frac{1}{y^2}\right) \left(1-t^2 y^2\right)} \nn \\
&=(1-t^4)(1-t^6) \PE \left[ ([2]_y +1)t^2 + [1]_x t^3\right] \nn \\
&= g_{\BC^2/\BZ_3} (t, x) \sum_{m=0}^\infty [2m]_{y} t^{2m}~, \label{HS200111ref}
\eea
where $g_{\BC^2/\BZ_3} (t, x)$ is given by \eref{C2Z3HS}.
Thus, the instanton moduli space is $\BC^2/\BZ_3 \times \widetilde{\CM}_{1,SU(2),\BC^2}$, where $ \widetilde{\CM}_{1,SU(2),\BC^2}$ is the reduced moduli space of one $SU(2)$ instanton on $\BC^2$.
The unrefined Hilbert series is
\bea
g^{(2,0,0)}_{(1,1,1)}(t, 1, 1) = \frac{(1-t^4)(1-t^6)}{(1-t^2)^4 (1-t^3)^2} = \frac{\left(1+t^2\right) \left(1-t+t^2\right)}{(1-t)^4 (1+t)^2 \left(1+t+t^2\right)}~.\label{HS200111unref}
\eea

\section*{$k=5/3$}
From \eref{C2Z3}, this corresponds to $\vec k = (2,2,1)$ with $\vec N=(1,1,0)$.  We choose $\vec r =(1,2)$.
The unrefined Hilbert series is
\bea
g^{(1,1,0)}_{(2,2,1)}(t, x=1, y=1) = \frac{1-2 t+3 t^2-2 t^3+3 t^4+3 t^8-2 t^9+3 t^{10}-2 t^{11}+t^{12}}{(1-t)^6 (1+t)^2 \left(1+t^2\right) \left(1-t+t^2\right) \left(1+t+t^2\right)^3}~.
\eea



\subsubsection{$SU(2)$ instantons on $\BC^2/\BZ_4$ with $\vec \beta = (0, 0, 0, 2)$} \label{C2Z4pureinst}
Let us now turn to $SU(2)$ pure instantons on $\BC^2/\BZ_4$.  Note that the Nekrasov partition function of such instantons for various second Chern class $k$ is explicitly presented in Eq. (2.10) of \cite{Wyllard:2011mn}.  Subsequently, we demonstrate that the latter can be derived from the Hilbert series in a certain limit.

\section*{$k = 3/4$}
From \eref{C2Z4}, we consider $\vec k = (1, 1, 1, 0),~ \vec N = (1,0,1,0)$ and choose $\vec r= (1,3)$.  The elements of $\CR$ are
\bea
\CR = \{ (\emptyset,\tiny \yng(1,1,1) ), \quad (\yng(1), \yng(1,1)), \quad (\yng(2), \yng(1)), \quad (\yng(3), \emptyset) \}~. 
\eea
The Hilbert series is given by
\bea
g^{(1,0,1,0)}_{(1,1,1,0)} (t,x,y) &= \frac{1}{\left(1-\frac{t^2}{x^2 y^2}\right) \left(1-x^2 y^2\right)}+\frac{1}{\left(1-\frac{t^4}{x^2 y^2}\right) \left(1-\frac{x^2 y^2}{t^2}\right)} \nn \\
&\quad +\frac{1}{\left(1-\frac{1}{x^2 y^2}\right) \left(1-t^2 x^2 y^2\right)}+\frac{1}{\left(1-\frac{1}{t^2 x^2 y^2}\right) \left(1-t^4 x^2 y^2\right)} \nn \\
&=\frac{1}{4} \left[ \sum_{p=0}^3 \frac{1}{(1- \omega^p \chi^{1/4} t ) (1- \omega^{-p} \chi^{-1/4} t)} \right]~, \quad \chi := x^2 y^2 \nn \\
&= g_{\BC^2/\BZ_4}(t,\chi^{1/4})~,
\eea
where the last two lines indicate that the Hilbert series $g^{(1,0,1,0)}_{(1,1,1,0)} (t,x,y)$ is in fact the Hilbert series of $\BC^2/\BZ_4$, with
\bea \label{C2Z4HS}
g_{\BC^2/\BZ_4}(t,x) &= (1-t^8) \PE \left[ t^2 + t^4(x^4 +x^{-4})\right]~, \nn \\
g_{\BC^2/\BZ_4}(t,x=1) &= \frac{1-t^8}{(1-t^2)(1-t^4)^2}~.
\eea
On the other hand, if we choose $\vec r= (3,1)$, the result is still the same up to the redefinition of $x$ to be $x^{-1}$.

\paragraph{The Nekrasov partition function.} The coefficient of $y^3$ in Eq. (2.10) of \cite{Wyllard:2011mn} can be derived from the above Hilbert series as follows.  The case of $k=3/4$ corresponds to two possible choices of $\vec r$, namely $\vec r= (1,3)$ and $\vec r= (3,1)$.  As can be seen above, the contributions from these two choices are related to each other by exchanging $x$ with $x^{-1}$.  
Setting
\bea
t = e^{-\frac{1}{2}\beta( \epsilon_1+\epsilon_2)}~, \quad x = e^{-\frac{1}{2}\beta( \epsilon_1-\epsilon_2)}~, \quad y= e^{-\beta a}~,
\eea
we obtain the Nekrasov partition function from the following limit
\bea
\CZ_{k=3/4}(\epsilon_1, \epsilon_2, a) &= \lim_{\beta \rightarrow 0} \beta^2 \left[ g^{(1,0,1,0)}_{(1,1,1,0)} (t,x,y) + g^{(1,0,1,0)}_{(1,1,1,0)} (t,x^{-1},y) \right]_{\substack{t = e^{-\frac{1}{2}\beta( \epsilon_1+\epsilon_2)} \\ x = e^{-\frac{1}{2}\beta( \epsilon_1- \epsilon_2)}, y = e^{-\beta a}}} \nn \\
&= -\frac{4}{\left(2 a-\epsilon _1-3 \epsilon _2\right) \left(2 a+3 \epsilon _1+\epsilon _2\right)}-\frac{4}{\left(2 a-3 \epsilon _1-\epsilon _2\right) \left(2 a+\epsilon _1+3 \epsilon _2\right)} \nn \\
&= \frac{8 \left(-4 a^2+3 \epsilon _1^2+10 \epsilon _1 \epsilon _2+3 \epsilon _2^2\right)}{\left(2 a-\epsilon _1-3 \epsilon _2\right) \left(2 a-3 \epsilon _1-\epsilon _2\right) \left(2 a+3 \epsilon _1+\epsilon _2\right) \left(2 a+\epsilon _1+3 \epsilon _2\right)}~.
\eea

\section*{$k = 1$}
From \eref{C2Z4}, there are two choices for this case.  

\paragraph{The first option.} We consider $\vec k = (1, 1, 1, 1)$ and $\vec N = (0,0,0,2)$ or $\vec r= (0,0)$. The Hilbert series is
\bea
g^{(0,0,0,2)}_{(1,1,1,1)} (t,x,y)  = g_{\BC^2/\BZ_4} (t, x) \sum_{m=0}^\infty [2m]_{y} t^{2m}~, \label{C2Z4SU2N0002}
\eea
where $g_{\BC^2/\BZ_4} (t, x)$ is given by \eref{C2Z4HS}.
The unrefined Hilbert series is 
\bea
g^{(0,0,0,2)}_{(1,1,1,1)} (t,x=1,y=1)  =  \frac{1 + t^4}{(1 - t^2)^4}~. \label{C2Z4SU2N0002ur}
\eea

\paragraph{The second option.}  Now let us consider $\vec k = (1, 2, 1, 0),~ \vec N = (0,2,0,0)$ or $\vec r= (2,2)$.  The Hilbert series is
\bea
g^{(0,2,0,0)}_{(1, 2, 1, 0)} (t,x,y) = \frac{1}{1-t^4} \left( [0; 2m_2 +2m_4]_{x;y} t^{2m_2+4m_4} + [0; 2m_2 +2m_4+2]_{x;y} t^{2m_2+4m_4+6}  \right)~.
\eea
Observe that this does not depend on the fugacity $x$.
The PL of this Hilbert series is 
\bea
\PL \left[ g^{(0,2,0,0)}_{(1, 2, 1, 0)} (t,x,y)  \right] = [0;2]_{x;y} t^2 + [0;2]_{x;y} t^4 - t^6 - t^8~.
\eea
The unrefined Hilbert series is 
\bea
g^{(0,2,0,0)}_{(1, 2, 1, 0)} (t,x=1,y=1)  = \frac{1+t^2+2 t^4+t^6+t^8}{\left(1-t^2\right)^4 \left(1+t^2\right)^2}~.
\eea

\paragraph{The Nekrasov partition function.} The coefficient of $y^4$ in Eq. (2.10) of \cite{Wyllard:2011mn} comes from both options described above:
\bea
\CZ_{k=1}(\epsilon_1, \epsilon_2, a) &= \lim_{\beta \rightarrow 0} \beta^4 \left[g^{(0,0,0,2)}_{(1,1,1,1)} (t,x,y)+g^{(0,2,0,0)}_{(1, 2, 1, 0)} (t,x,y)  \right]_{\substack{t = e^{-\frac{1}{2}\beta( \epsilon_1+\epsilon_2)} \\ x = e^{-\frac{1}{2}\beta( \epsilon_1- \epsilon_2)}, y = e^{-\beta a}}}  \nn \\
&= \frac{1}{2 \epsilon _1 \epsilon _2 \left(-2 a+\epsilon _1+\epsilon _2\right) \left(2 a+\epsilon _1+\epsilon _2\right)} \nn \\
& \quad -\frac{3}{\left(-2 a-\epsilon _1-\epsilon _2\right) \left(-2 a+\epsilon _1+\epsilon _2\right) \left(a+\epsilon _1+\epsilon _2\right) \left(-2 a+2 \epsilon _1+2 \epsilon _2\right)} \nn \\
&=\frac{-a^2+\epsilon _1^2+5 \epsilon _1 \epsilon _2+\epsilon _2^2}{2 \epsilon _1 \epsilon _2 \left(-2 a+\epsilon _1+\epsilon _2\right) \left(-a+\epsilon _1+\epsilon _2\right) \left(a+\epsilon _1+\epsilon _2\right) \left(2 a+\epsilon _1+\epsilon _2\right)}~.
\eea

\subsubsection*{$k=7/4$}
From \eref{C2Z4}, let us consider $\vec k = (2, 2, 2, 1)$ and $\vec N = (1,0,1,0)$.  We can choose either $\vec r= (1,3)$ or $\vec r =(3,1)$.  The unrefined Hilbert series is
\bea
g^{(1,0,1,0)}_{(2,2,2,1)} (t,x=1,y=1) =\frac{1-t^2+5 t^4-t^6+6 t^8-t^{10}+5 t^{12}-t^{14}+t^{16}}{(1-t^2)^6 \left(1+t^2\right)^3 \left(1+t^4\right)}~.
\eea
\paragraph{The Nekrasov partition function.} The coefficient of $y^7$ in Eq. (2.10) of \cite{Wyllard:2011mn} comes from $\vec r = (1,3)$ and $\vec r =(3,1)$.  The contributions from these two choices of $\vec r$ are related to each other by exchanging $x$ with $x^{-1}$. 
\bea
\CZ_{k=7/4}(\epsilon_1, \epsilon_2, a) &= \lim_{\beta \rightarrow 0} \beta^6 \left[ g^{(1,0,1,0)}_{(2,2,2,1)} (t,x,y) + g^{(1,0,1,0)}_{(2,2,2,1)} (t,x^{-1},y) \right]_{\substack{t = e^{-\frac{1}{2}\beta( \epsilon_1+\epsilon_2)} \\ x = e^{-\frac{1}{2}\beta( \epsilon_1- \epsilon_2)}, y = e^{-\beta a}}} \nn \\
&= \frac{2 \left(-4 a^2-8 a \epsilon _1+21 \epsilon _1^2+8 a \epsilon _2+70 \epsilon _1 \epsilon _2+21 \epsilon _2^2\right)}{\epsilon _1 \epsilon _2 A_{1,3} A_{3,1} A_{1,7} A_{7,1}}  + \left( \epsilon_1 \leftrightarrow \epsilon_2 \right) \nn \\
&= \frac{1}{\epsilon _1 \epsilon _2 A_{1,3} A_{3,1} A_{1,7} A_{7,1}} \times 4 \Big[16 a^4-16 a^2 \left(13 \epsilon _1^2+18 \epsilon _1 \epsilon _2+13 \epsilon _2^2\right) \nn \\
& \quad +7 \left(21 \epsilon _1^4+220 \epsilon _1^3 \epsilon _2+542 \epsilon _1^2 \epsilon _2^2+220 \epsilon _1 \epsilon _2^3+21 \epsilon _2^4\right)\Big]~,
\eea
where 
\bea
A_{r,s} = (2a -r \epsilon_1 -s \epsilon_2)(2a + r \epsilon_1 +s \epsilon_2)~. \label{Ars}
\eea

\subsubsection*{$k = 2$}
There are two possibilities for $k=2$. The first one is $\vec k = (2, 2, 2, 2)$ and $\vec N = (0,2,0,0)$ or $\vec r= (0,0)$.  The unrefined Hilbert series is
\bea
g^{(0,2,0,0)}_{(2,2,2,2)} (t,x=1,y=1) &= \frac{1}{\left(1-t^2\right)^8 \left(1+t^2\right)^4 \left(1+t^2+t^4\right)^3} \Big(1+3 t^2+8 t^4+20 t^6+41 t^8+61 t^{10} \nn \\
& \quad +78 t^{12}+84 t^{14}+78 t^{16}+\text{palindrome}+t^{28} \Big)~.
\eea

The second possibility is $\vec k = (2, 1, 2, 3)$ and $\vec N = (0,0,0,2)$ or $\vec r= (2,2)$.  The unrefined Hilbert series is
\bea
g^{(0,0,0,2)}_{(2,1,2,3)} (t,x=1,y=1) &= \frac{1}{\left(1-t^2\right)^8 \left(1+t^2\right)^2 \left(1+t^2+t^4\right)^3} \left(1+t^4\right) \times \nn\\
&\quad   \Big(1+t^2+3 t^4+9 t^6+11 t^8+8 t^{10}+11 t^{12}+9 t^{14} +3 t^{16}+t^{18}+t^{20}\Big)~.
\eea

\paragraph{The Nekrasov partition function.} The coefficient of $y^8$ in Eq. (2.10) of \cite{Wyllard:2011mn} comes from both possibilities discussed above.
\bea
\CZ_{k=2}(\epsilon_1, \epsilon_2, a) &= \lim_{\beta \rightarrow 0} \beta^8 \left[ g^{(0,0,0,2)}_{(2,2,2,2)} (t,x,y) + g^{(0,0,0,2)}_{(2,1,2,3)} (t,x,y) \right]_{\substack{t = e^{-\frac{1}{2}\beta( \epsilon_1+\epsilon_2)} \\ x = e^{-\frac{1}{2}\beta( \epsilon_1- \epsilon_2)}, y = e^{-\beta a}}} \nn \\
&= \frac{1}{\epsilon _1^2 \epsilon _2^2 A_{1,1} A_{2,2} A_{1,5} A_{5,1}} \Big[8 a^4+50 \epsilon _1^4+645 \epsilon _1^3 \epsilon _2+1526 \epsilon _1^2 \epsilon _2^2+645 \epsilon _1 \epsilon _2^3 \nn \\
& \quad +50 \epsilon _2^4-2 a^2 \left(29 \epsilon _1^2+60 \epsilon _1 \epsilon _2+29 \epsilon _2^2\right) \Big]~,
\eea
where $A_{r,s}$ is defined as in \eref{Ars}.

\subsection{Further examples} \label{moreex:loc}

\subsubsection{$SU(4)$ instanton on $\BC^2/\BZ_2$: $\vec k =(1,1)$ and $\vec N = (2,2)$}
The Hilbert series can be obtained using the above procedure by choosing, \eg~ $\vec r = (0,0,1,1)$. Let $(\xi_{1,1}, \xi_{1,2}), (\xi_{2,1},\xi_{2,2})$ be the fugacities corresponding to the global symmetry $U(2)$, $U(2)$ in the KN quiver.
The unrefined Hilbert series is
\bea
&g^{(2,2)}_{(1,1)} (t,x=1; \{\xi_{\alpha,i} =1 \}) \nn\\
&= \frac{1}{(1- t)^8 (1 + t)^4 (1 + t + t^2)^4}   \Big(1 + 6 t^2 + 12 t^3 + 18 t^4 + 24 t^5 + 34 t^6 +24 t^7 \nn \\
&\quad +\text{palindrome} +t^{12}\Big)~. \label{HS:C2Z2SU4N22}
\eea
Notice that the numerator is a palindromic polynomial and the moduli space is 8 complex dimensional, as expected.  The refined plethystic logarithm is
\bea
&\PL \left[ g^{(2,2)}_{(1,1)} (t,x;  \xi_{1,1}, \xi_{1,2}; \xi_{2,1}; \xi_{3,1}) \right] \nn \\
&=  \left[ x^2+x^{-2} + \left( \sum_{\alpha=1}^2 \sum_{i_1, i_2=1}^2   \frac{\xi_{\alpha,i}}{\xi_{\alpha,j}}  \right)  \right] t^2 + \Bigg[  x  \sum_{i_1, i_2=1}^2 \left( \frac{\xi_{1,i}}{\xi_{2,j}} +\frac{\xi_{2,i}}{\xi_{1,j}}  \right) + \nn \\
& \quad + x^{-1}  \sum_{i_1, i_2=1}^2 \left( \frac{\xi_{2,i}}{\xi_{1,j}} +\frac{\xi_{1,i}}{\xi_{2,j}} \right)  \Bigg] t^3 -3 t^4 - \ldots \nn \\
&= ([2;0;0;0]+[0;2;0;0]+[0;0;2;0]+[0;0;0;0])t^2 + ([1;1;1;1])t^3 - 3 t^4 - \ldots~.
\eea

\subsubsection{$SU(4)$ instanton on $\BC^2/\BZ_3$: $\vec k =(1,1,1)$ and $\vec N = (2,1,1)$}
The Hilbert series can be obtained using the above procedure by choosing, \eg~ $\vec r = (0,0,2,1)$. Let $(\xi_{1,1}, \xi_{1,2}), (\xi_{2,1}), (\xi_{3,1})$ be the fugacities corresponding to the global symmetry $U(2)$, $U(1)$, $U(1)$ in the KN quiver.
The unrefined Hilbert series is
\bea
&g^{(2,1,1)}_{(1,1,1)} (t,x=1; \{\xi_{\alpha,i} =1 \}) \nn\\
&= \frac{1}{(1- t)^8 (1 + t)^4 (1 + t^2)^2 (1 + t + t^2)^3}   \Big(1 - t + 4 t^2 + 5 t^3 + 8 t^4 + 11 t^5 + 19 t^6 \nn \\
&\quad +14 t^7+19 t^8+\text{palindrome} +t^{14}\Big)~.
\eea
Notice that the numerator is a palindromic polynomial and the moduli space is 10 complex dimensional, as expected.
The refined plethystic logarithm is
\bea
&\PL \left[ g^{(2,1,1)}_{(1,1,1)} (t,x;  \xi_{1,1}, \xi_{1,2}; \xi_{2,1}; \xi_{3,1}) \right] \nn \\
&=  \left[ 2+( \xi_{1,1}+ \xi_{1,2})( \xi_{1,1}^{-1}+ \xi_{1,2}^{-1})  \right] t^2 + \Bigg[ x^3+x^{-3}+ x  \Big\{ (\xi_{1,1}+ \xi_{1,2} )\xi_{2,1}^{-1} + \xi_{2,1} \xi_{3,1}^{-1}  \nn \\
& \quad +\xi_{3,1} (\xi_{1,1}^{-1}+ \xi_{1,2}^{-1} ) \Big\} +x^{-1}  \Big\{ (\xi^{-1}_{1,1}+ \xi^{-1}_{1,2} )\xi_{2,1}+ \xi_{2,1}^{-1} \xi_{3,1}+\xi^{-1}_{3,1} (\xi_{1,1}+ \xi_{1,2} ) \Big \} \Bigg] t^3  \nn \\
& \quad + \Bigg[ x^{2}  \Big\{  (\xi_{1,1}+ \xi_{1,2} ) \xi^{-1}_{3,1}+\xi_{3,1} \xi_{2,1}^{-1} +\xi_{2,1} (\xi^{-1}_{1,1}+ \xi^{-1}_{1,2} ) \Big \}+x^{-2}  \Big\{ (\xi^{-1}_{1,1}+ \xi^{-1}_{1,2} ) \xi_{3,1}+\xi^{-1}_{3,1} \xi_{2,1}   \nn \\
& \quad +\xi^{-1}_{2,1} (\xi_{1,1}+ \xi_{1,2} ) \Big\} -1 \Bigg] t^4 - O(t^5)~. \label{PL111211}
\eea
This is in agreement with the expression \eref{PLkSUnC2Zn}.

\subsubsection{$SU(4)$ instantons on $\BC^2/\BZ_3$: $\vec k =(1,2,2)$ and $\vec N = (2,1,1)$}
The Hilbert series can be obtained using the above procedure by choosing, \eg~ $\vec r = (0,0,2,1)$. Let $(\xi_{1,1}, \xi_{1,2}), (\xi_{2,1}), (\xi_{3,1})$ be the fugacities corresponding to the global symmetry $U(2)$, $U(1)$, $U(1)$ in the KN quiver.
The unrefined Hilbert series is
\bea
&g^{(2,1,1)}_{(1,2,2)} (t,x=1; \{\xi_{\alpha,i} =1 \}) \nn\\
&= \frac{1}{(1-t)^{10} (1+t)^6 \left(1+t^2\right)^3 \left(1-t+t^2\right) \left(1+t+t^2\right)^4}   \Big( 1-t+4 t^2+5 t^3+8 t^4+15 t^5+27 t^6 \nn \\
&\quad +26 t^7+54 t^8+44 t^9+62 t^{10}+58 t^{11}+62 t^{12}+ \text{palindrome} +t^{22}\Big)~.
\eea
Notice that the numerator is a palindromic polynomial and the moduli space is 10 complex dimensional, as expected.
The refined plethystic logarithm is
\bea
&\PL \left[ g^{(2,1,1)}_{(1,2,2)} (t,x;  \xi_{1,1}, \xi_{1,2}; \xi_{2,1}; \xi_{3,1}) \right] \nn \\
&=  \left[ 2+( \xi_{1,1}+ \xi_{1,2})( \xi_{1,1}^{-1}+ \xi_{1,2}^{-1})  \right] t^2 + \Bigg[ x^3+x^{-3}+ x  \Big\{ (\xi_{1,1}+ \xi_{1,2} )\xi_{2,1}^{-1} + \xi_{2,1} \xi_{3,1}^{-1}  \nn \\
& \quad +\xi_{3,1} (\xi_{1,1}^{-1}+ \xi_{1,2}^{-1} ) \Big\} +x^{-1}  \Big\{ (\xi^{-1}_{1,1}+ \xi^{-1}_{1,2} )\xi_{2,1}+ \xi_{2,1}^{-1} \xi_{3,1}+\xi^{-1}_{3,1} (\xi_{1,1}+ \xi_{1,2} ) \Big \} \Bigg] t^3  \nn \\
& \quad + \Bigg[ x^{2}  \Big\{  (\xi_{1,1}+ \xi_{1,2} ) \xi^{-1}_{3,1}+\xi_{3,1} \xi_{2,1}^{-1} +\xi_{2,1} (\xi^{-1}_{1,1}+ \xi^{-1}_{1,2} ) \Big \}+x^{-2}  \Big\{ (\xi^{-1}_{1,1}+ \xi^{-1}_{1,2} ) \xi_{3,1}+\xi^{-1}_{3,1} \xi_{2,1}   \nn \\
& \quad +\xi^{-1}_{2,1} (\xi_{1,1}+ \xi_{1,2} ) \Big\}  \Bigg] t^4 - O(t^5)~.
\eea
This is in agreement with the expression \eref{PLkSUnC2Zn}.  Observe that, up to order $t^4$, this expression differs from \eref{PL111211} by only the term $-1$; this comes from the relation at order $t^4$ that exists in the latter but not in the former case.

\subsubsection{$SU(6)$ instanton on $\BC^2/\BZ_3$: $\vec k =(1,1,1)$ and $\vec N = (2,2,2)$}
The Hilbert series can be obtained using the above procedure by choosing, \eg~ $\vec r = (0,1,0,2,2,1)$.
Let $\xi_{\alpha,1}$ and $\xi_{\alpha,2}$, with $\alpha=1,2,3$, be the fugacities of the global symmetry $U(2)_\alpha$.
The unrefined Hilbert series is
\bea
&g^{(2,2,2)}_{(1,1,1)} (t,x=1; \{\xi_{\alpha,i} =1 \}) \nn\\
&= \frac{1}{(1-t)^{12} (1+t)^6 \left(1+t^2\right)^3 \left(1+t+t^2\right)^5}  \times \Big( 1-t+9 t^2+12 t^3+42 t^4+78 t^5+170 t^6 \nn \\
&+250 t^7+405 t^8+485 t^9+613 t^{10}+612 t^{11}+613 t^{12}+ \text{palindrome} +t^{22}\Big)~.
\eea
Notice that the numerator is a palindromic polynomial and the moduli space is 12 complex dimensional, as expected.
The refined plethystic logarithm is
\bea
&\PL \left[ g^{(2,2,2)}_{(1,1,1)} (t,x;  \xi_{1,1}, \xi_{1,2}; \xi_{2,1}, \xi_{2,2}; \xi_{3,1}, \xi_{3,2}) \right] \nn \\
&=  \left( \sum_{\alpha=1}^3 \sum_{i_1, i_2=1}^2   \frac{\xi_{\alpha,i}}{\xi_{\alpha,j}}  \right) t^2 + \Bigg[ x^3+x^{-3}+ x  \sum_{i_1, i_2=1}^2 \left( \frac{\xi_{1,i}}{\xi_{2,j}} +\frac{\xi_{2,i}}{\xi_{3,j}} +\frac{\xi_{3,i}}{\xi_{1,j}} \right) + \nn \\
& \quad + x^{-1}  \sum_{i_1, i_2=1}^2 \left( \frac{\xi_{2,i}}{\xi_{1,j}} +\frac{\xi_{3,i}}{\xi_{2,j}} +\frac{\xi_{1,i}}{\xi_{3,j}} \right)  \Bigg] t^3 
+ \Bigg[ x^{2}  \sum_{i_1, i_2=1}^2 \left( \frac{\xi_{1,i}}{\xi_{3,j}} +\frac{\xi_{3,i}}{\xi_{2,j}} +\frac{\xi_{2,i}}{\xi_{1,j}} \right) \nn \\
& \quad +x^{-2}  \sum_{i_1, i_2=1}^2 \left( \frac{\xi_{3,i}}{\xi_{1,j}} +\frac{\xi_{2,i}}{\xi_{3,j}} +\frac{\xi_{1,i}}{\xi_{2,j}} \right) -3 \Bigg] t^4 - O(t^5)~.
\eea
This is in agreement with the expression \eref{PLkSUnC2Zn}.

\subsubsection{$SU(4)$ instanton on $\BC^2/\BZ_4$: $\vec k =(1,1,1,1)$ and $\vec N = (1,1,1,1)$}
The Hilbert series can be obtained using the above procedure by choosing, \eg~ $\vec r = (0,1,2,3)$.
Let $\xi_\alpha$ be the fugacities for the global symmetry $U(1)_\alpha$, with $\alpha =1,\ldots, 4$.
The unrefined Hilbert series is
\bea \label{HS:C2Z4SU(4)N1111}
&g^{(1,1,1,1)}_{(1,1,1,1)} (t,x=1; \{\xi_{\alpha} =1 \}) 
= \frac{1}{(1-t)^8 (1+t)^4 \left(1+t^2\right)^2 \left(1+t+t^2\right)^4 \left(1+t+t^2+t^3+t^4\right)}  \times \nn\\
&\quad \Big( 1+t+3 t^2+7 t^3+18 t^4+33 t^5+51 t^6+69 t^7+93 t^8+110 t^9+120 t^{10} \nn \\
& \quad +110 t^{11}+ \text{palindrome} +t^{20}\Big)~.
\eea
Notice that the numerator is a palindromic polynomial and the moduli space is 8 complex dimensional, as expected.
The refined plethystic logarithm is
\bea
&\PL \left[ g^{(1,1,1,1)}_{(1,1,1,1)} (t,x;  \xi_{1}, \ldots, \xi_4) \right] \nn \\
&=  4 t^2 + \left[ x \left( \xi_1\xi_2^{-1} + \xi_2\xi_3^{-1}+\xi_3\xi_4^{-1}+\xi_4\xi_1^{-1} \right) +x^{-1} \left(\xi_\alpha \rightarrow \xi^{-1}_\alpha \right) \right] t^3 \nn \\
& \quad +\left[x^4+x^{-4}+ x^2 \left( \xi_1\xi_3^{-1} + \xi_2\xi_4^{-1}+\xi_3\xi_1^{-1}+\xi_4\xi_2^{-1} \right) +x^{-2} \left(\xi_\alpha \rightarrow \xi^{-1}_\alpha \right) \right] t^4 \nn \\
& \quad +\left[x^3 \left( \xi_1\xi_4^{-1} + \xi_2\xi_1^{-1}+\xi_3\xi_2^{-1}+\xi_4\xi_3^{-1} \right) +x^{-3} \left(\xi_\alpha \rightarrow \xi^{-1}_\alpha \right) \right] t^5- O(t^6)~.
\eea
This is in agreement with the expression \eref{PLkSUnC2Zn}.

\subsubsection{$SU(8)$ instanton on $\BC^2/\BZ_4$: $\vec k =(1,1,1,1)$ and $\vec N = (2,2,2,2)$}
Let $\xi_{\alpha,1},\xi_{\alpha,2}$ be the fugacities of the global symmetry $U(2)_\alpha$, with $\alpha=1,\ldots,4$.
The unrefined Hilbert series is
\bea
&g^{(2,2,2,2)}_{(1,1,1,1)} (t,x=1; \{\xi_{\alpha,i} =1 \}) 
= \frac{1}{(1-t)^{16} (1+t)^8 \left(1+t^2\right)^4 \left(1+t+t^2\right)^8 \left(1+t+t^2+t^3+t^4\right)^3}  \times \nn\\
&\quad \Big( 1+3 t+18 t^2+70 t^3+263 t^4+850 t^5+2511 t^6+6682 t^7+16350 t^8+36715 t^9+76420 t^{10} \nn \\
& \quad +147737 t^{11}+266766 t^{12}+451026 t^{13}+716737 t^{14}+1072562 t^{15} \nn \\
& \quad + 1515293 t^{16}+2023950 t^{17}+2560234 t^{18}+3069965 t^{19}+3493295 t^{20} \nn \\
& \quad +3773632 t^{21}+3872056 t^{22}+3773632 t^{23}+ \text{palindrome} +t^{44}\Big)~.
\eea
Notice that the numerator is a palindromic polynomial and the moduli space is 16 complex dimensional, as expected.
The refined plethystic logarithm is
\bea
&\PL \left[ g^{(2,2,2,2)}_{(1,1,1,1)} (t,x; \{ \xi_{\alpha,1}, \xi_{\alpha,2}\}) \right] \nn \\
&=  \left( \sum_{\alpha=1}^4 \sum_{i_1, i_2=1}^2   \frac{\xi_{\alpha,i}}{\xi_{\alpha,j}}  \right) t^2 + \underbrace{\Bigg[  x  \sum_{i_1, i_2=1}^2 \left( \frac{\xi_{1,i}}{\xi_{2,j}} +\frac{\xi_{2,i}}{\xi_{3,j}} +\frac{\xi_{3,i}}{\xi_{4,j}} +\frac{\xi_{4,i}}{\xi_{1,j}} \right) + x^{-1}  \left( \xi_{\alpha,i} \rightarrow \xi_{\alpha,i}^{-1}  \right)  \Bigg]}_{(*)} t^3 \nn \\
& \quad  
+ \Bigg[ x^4+x^{-4}+x^{2} \sum_{i_1, i_2=1}^2 \left( \frac{\xi_{1,i}}{\xi_{3,j}} +\frac{\xi_{2,i}}{\xi_{4,j}} +\frac{\xi_{3,i}}{\xi_{1,j}} +\frac{\xi_{4,i}}{\xi_{2,j}} \right) + x^{-2}  \left( \xi_{\alpha,i} \rightarrow \xi_{\alpha,i}^{-1}  \right)   -4 \Bigg] t^4  \nn \\
& \quad + \Bigg[ x^{3} \sum_{i_1, i_2=1}^2 \left( \frac{\xi_{1,i}}{\xi_{4,j}} +\frac{\xi_{2,i}}{\xi_{1,j}} +\frac{\xi_{3,i}}{\xi_{2,j}} +\frac{\xi_{4,i}}{\xi_{3,j}} \right) + x^{-3}  \left( \xi_{\alpha,i} \rightarrow \xi_{\alpha,i}^{-1}  \right)   -2 (*) \Bigg] t^5 - O(t^6)~,
\eea
where (*) represents the terms in the square brackets in the second line.  Notice that the generators are in agreement with the expression \eref{PLkSUnC2Zn}.

\section{Quivers for $SO(N)$ and $Sp(N)$ instantons on $\BC^2/\BZ_n$: \\
Constructions with orientifold planes} \label{sec:SOSpC2Zn}
In this section, we present the KN quivers for the constructions of $SO(N)$ and $Sp(N)$ instantons on $\BC^2/\BZ_n$.  These quivers can be obtained by considering various orientifold actions on the brane configurations of the KN quiver.  We refer the reader to Appendix \ref{app:SOSpbrane} for details on the derivations of such quivers; here, we only present the final results.\\
The class of quivers corresponding to instantons on $\BC^2/\BZ_n$ with odd $n$ consists of two infinite families:
\ben
\item {\bf Unitary gauge group with an antisymmetric hyper at one end and symplectic gauge group at the other end} as shown in \fref{fig:SO_N_oddn}. This quiver corresponds to $SO(N)$ instantons.
\item {\bf Unitary gauge group with a symmetric hyper at one end and orthogonal gauge group at the other end} as shown in  \fref{fig:Sp_N_oddn}. This quiver corresponds to $Sp(N)$ quivers.
\een

On the other hand, the quivers for $SO(N)$ and $Sp(N)$ instanton on $\BC^2/\BZ_n$, with $n=2m$ even, come in four types which can be classified according to their ends or their boundary conditions as follows:
\ben
\item The quivers that correspond to the brane configurations in which the NS5-branes are not stuck on the orientifold 5-plane; see \fref{fig:SOSpNevenorbnoVS}.  There are two options for the boundary conditions of such quivers: 
\ben
\item {\bf Each end has (special) orthogonal global symmetry group.} We refer to such a quiver as the {\it O/O quiver}.   This is depicted in \fref{fig:SO_N_withVSevenn}.  It correspond to $SO(N)$ instantons on $\BC^2/\BZ_{2m}$.
\item {\bf Each end has symplectic global symmetry group.} We refer to such a quiver as the {\it S/S quiver}. This is depicted in \fref{fig:Sp_N_withVSevenn}.  It correspond to $Sp(N)$ instantons on $\BC^2/\BZ_{2m}$.
\een
\item The quivers that correspond to the brane configurations in which the NS5-branes are stuck on the orientifold 5-plane; see \fref{fig:SOSpNevenorbVS}. There are two options for the boundary conditions of such quivers: 
\ben
\item {\bf Each end has unitary gauge group with one antisymmetric hypermultiplet.} We refer to such a quiver as the {\it AA quiver}.   This is depicted in \fref{fig:SO_N_withNOVSevenn}.  It correspond to $SO(2N)$ instantons on $\BC^2/\BZ_{2m}$.
\item {\bf Each end has unitary gauge group with one symmetric hypermultiplet.} We refer to such a quiver as the {\it SS quiver}.   This is depicted in \fref{fig:Sp_N_withNOVSevenn}.  It correspond to $Sp(N)$ instantons on $\BC^2/\BZ_{2m}$.
\een
\een

These quivers were studied in various contexts; see \eg~ \cite{Intriligator:1997kq, Hanany:1997gh, Feng:2001rh, Dey:2011pt, Bergman:2012qh, Bergman:2012kr, Dey:2013nf}.  In \cite{Intriligator:1997kq}, the $O/O$ and $S/S$ quivers are referred to as the cases {\it with vector structure}, whereas the $AA$ and $SS$ quiver are referred to as the cases {\it without vector structure}.

It should be pointed out that, in the quivers for $SO(N)$ instantons with odd $n$ ({\it resp.} the $AA$ and $SS$ quivers for even $n$), only the subgroup $SO(N_1) \times \prod_i U(N_i)$ ({\it {\it resp.}} $\prod_i U(N_i)$)  of $SO(N)$ is realised on the instanton moduli space.  This scenario is related to the string backgrounds studied in \cite{Gimon:1996rq,Bianchi:1990tb, Dabholkar:1996zi, Dabholkar:1996pc}.

Note that, in addition to the six quivers mentioned above, one can form quivers with mixed boundary conditions; for example, using the above notation, it is clear that there are also the $SA$ and the $O/S$ quivers (see also \eg, \cite{Chaudhuri:1995fk, Chaudhuri:1995bf, Hanany:2001iy}). These cases are generically related to the Dabholkar-Park orientifold \cite{Dabholkar:1996pc}. We refer to such quivers as the {\bf hybrids}; they will be discussed in detail in Section \ref{sec:hybrids}.

Subsequently, we provide following checks for the above quiver constructions of $SO(N)$ and $Sp(N)$ instantons:
\bi
\item  The dimension of the hypermultiplet moduli space of a given quiver is equal to that of corresponding instanton moduli space.
\item We compute the Hilbert series of the hypermultiplet moduli space for different quivers and find expected matches for equivalent instantons of isomorphic groups as summarised in \tref{tab:match1} and \tref{tab:match2}.\\

In the following subsections, the moduli spaces of $SO(N)$ and $Sp(N)$ instantons on $\BC^2/\BZ_n$, with $n \geq 3$, are considered.  We turn to the degenerate case with $n=2$ in Section \ref{sec:SONC2Z2}. 

\subsection{Quiver for $SO(N)$ instantons on $\BC^2/\BZ_{2m+1}$}
\begin{figure}[H]
\begin{center}
\includegraphics[scale=0.7]{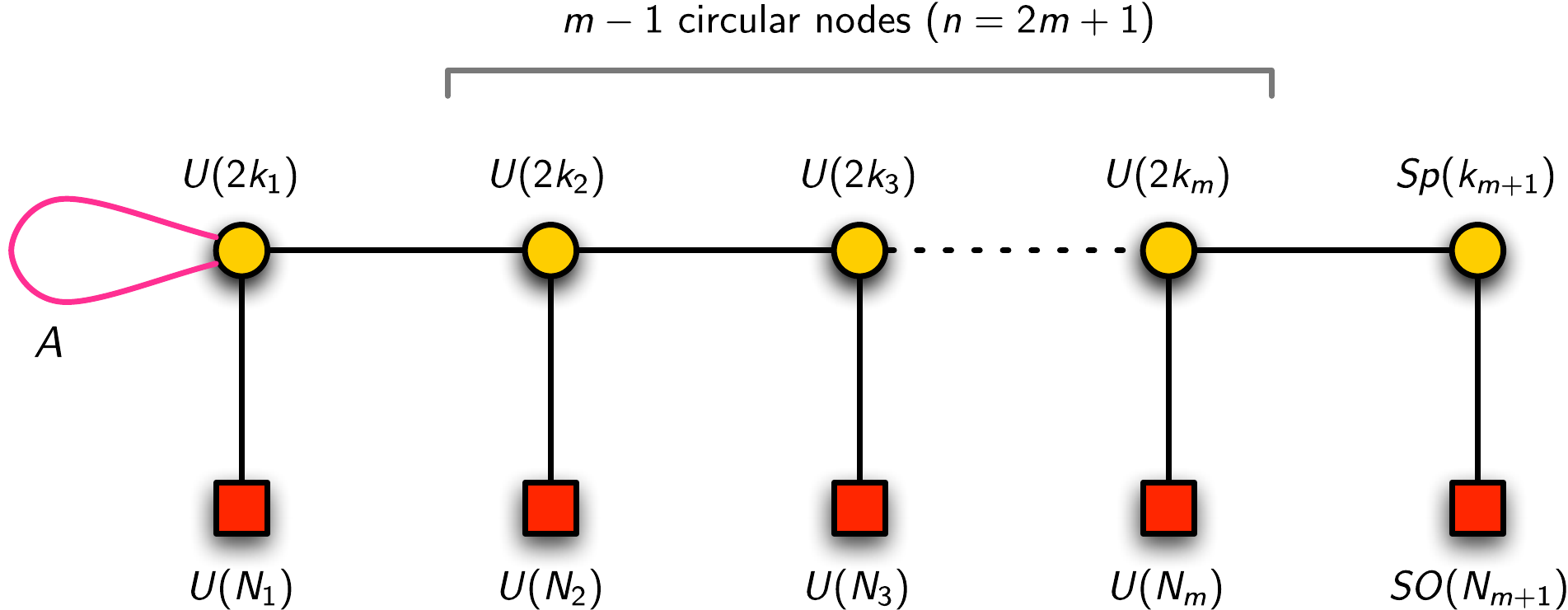}
\caption{The quiver for $SO(N)$ instantons on $\mathbb{C}^2/\mathbb{Z}_{2m+1}$.  Here, each of $k_1, \ldots, k_{m}$ can take either integral or half-integral value and $N=2N_1+ \ldots+2N_m + N_{m+1}$.  The loop labelled by $A$ denotes the rank 2-antisymmetric hypermultiplet in $Sp(k_{m+1})$ gauge group. Each line between $Sp(r_1)$ and $U(r_2)$ denotes $2r_1 r_2$ hypermultiplets, each line between $U(r_1)$ and $U(r_2)$ denotes $r_1 r_2$ hypermultiplets, and each line between $Sp(r_1)$ and $SO(r_2)$ denotes $2r_1 r_2$ half-hypermultiplets.}
\label{fig:SO_N_oddn}
\end{center}
\end{figure}
If $k_1 = \cdots =k_{m+1} =k$, the quaternionic dimension of the Higgs branch of \fref{fig:SO_N_oddn} is given by
\bea
kN+ m(4 k^2)+ \frac{1}{2} (2k)(2k-1) - m(4k^2)- \frac{1}{2}(2k)(2k+1) = k(N-2)~, \label{dimkSONoddn}
\eea
where $N=2N_1+2N_2+ \ldots+2N_m+N_{m+1}$.  This is to be expected for $k$ $SO(N)$ instantons on $\mathbb{C}^2/\mathbb{Z}_{2m+1}$, for $N\geq 5$.

\subsubsection{$SO(2)$ instanton on $\BC^2/\BZ_3$: $\vec k =(1,1)$ with $\vec N=(0,2)$ or $\vec N= (1,0)$}
\begin{figure}[H]
\begin{center}
\includegraphics[scale=0.7]{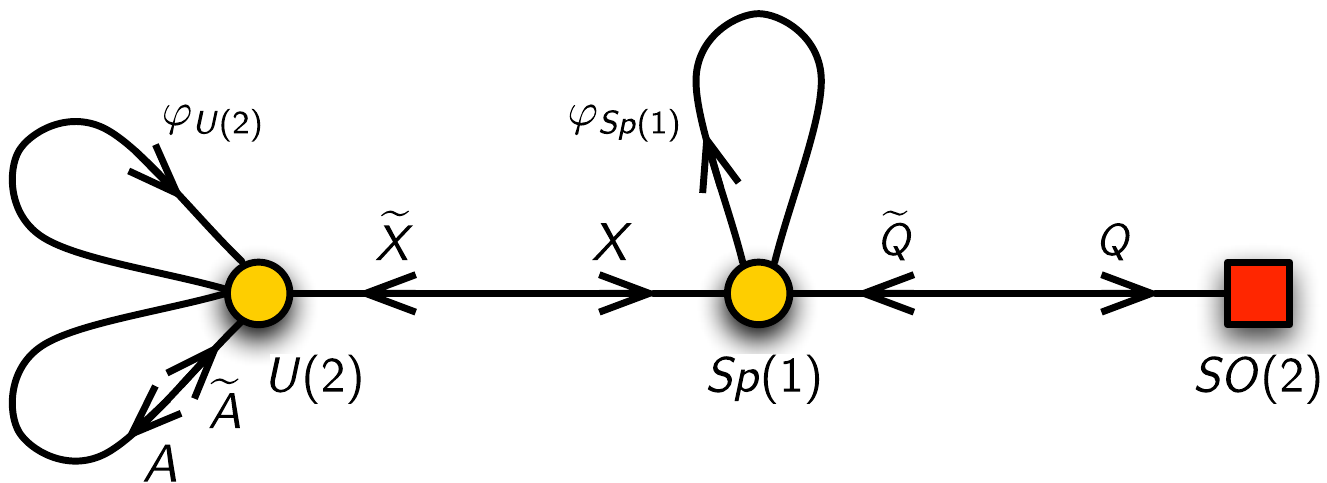}
\caption{Quiver in $\CN=1$ notation for $\vec k =(1,1)$ with $\vec N=(0,2)$.  The superpotential is \eref{Wk11N02}.}
\label{fig:C2Z3SO_2_wVSN1}
\end{center}
\end{figure}

Let us first consider $\vec k =(1,1)$ and $\vec N=(0,2)$. The quiver is given by
\begin{figure}[H]
\begin{center}
\includegraphics[scale=0.7]{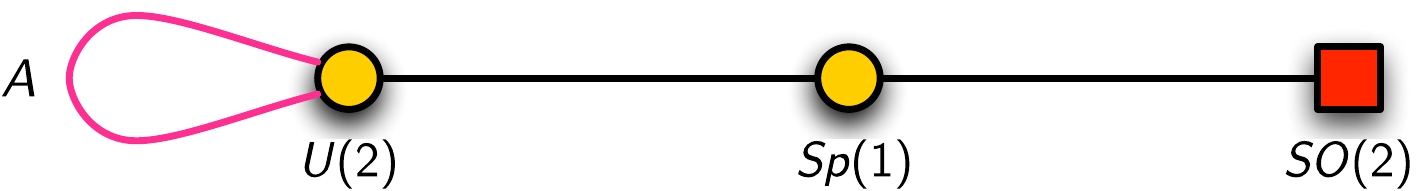}
\caption{The quiver for $SO(2)$ instantons on $\mathbb{C}^2/\mathbb{Z}_{3}$: $\vec k =(1,1)$ with $\vec N=(0,2)$}
\label{fig:C2Z3SO_2_wVS}
\end{center}
\end{figure}
In $\CN=1$ language, let us denote the chiral fields in the bi-fundamental representation of $U(2) \times Sp(1)$ by $X^{a_1}_{~b_1}$ and $\widetilde{X}^{b_1}_{~a_1}$, those of $Sp(1) \times SO(2)$ by $Q_{b_1}$ and $\tQ^{b_1}$, where $a_1, a_2,\ldots=1,2$ are the indices for $U(2)$, and $b_1, b_2, \ldots =1,2$ are those for $Sp(1)$.  The chiral multiplets coming from the antisymmetric adjoint hypermultiplet $A$ are denoted by $A^{a_1a_2}$ and $\widetilde{A}_{a_1a_2}$, where $\alpha=1,2$ is an $SU(2)$ global index and the indices $a,b$ are antisymmetrised.  We emphasise that $A$ and $\tA$ transform in complex conjugate representation of each other. The superpotential is given by
\bea \label{Wk11N02}
W &= \left[\widetilde{X}^{b_1}_{~a_1} (\varphi_{U(2)})^{a_1}_{~a_2} X^{a_2}_{~b_1} - X^{a_1}_{~b_1} (\varphi_{Sp(1)})^{b_1}_{~b_2} \tX^{b_2}_{~a_1}   \right]  \nn \\
& \qquad + \widetilde{A}_{a_1 a_2} (\varphi_{U(2)})^{a_2}_{~a_3} A^{a_3 a_1} + \tQ^{b_1} (\varphi_{Sp(1)})^{a_1}_{~ b_1} Q_{a_1}~,
\eea
where $\varphi_{U(2)}$ and $\varphi_{Sp(1)}$ are the adjoint scalars in the vector multiplets in the gauge groups $U(2)$ and $Sp(1)$ respectively.  The $\CN=1$ quiver is depicted in \fref{fig:C2Z3SO_2_wVSN1}.

In this case, we expect the moduli space of the instanton to be $\BC^2/\BZ_3$.  Notice that formula \eref{dimkSONoddn} cannot be applied to this case, since the gauge symmetry is not completely broken.
This hypermultiplet moduli space of the quiver depicted in \fref{fig:SO_N_oddn} contains more than one branches. In order to obtain the correct Hilbert series for the instanton moduli space, one needs to perform the following steps in order: 
\ben
\item Find the primary decomposition of the $F$-term conditions: $0 =\partial_{\varphi_{U(2)}} W =\partial_{\varphi_{Sp(1)}} W$.
\item Select the relations that gives the $F$-flat space of complex dimension $8$.  In particular, these relations can be chosen as
\bea \label{FflatrelC2Z3SO2}
0 = \tQ^{b_1}~, \qquad 0 =\partial_{\varphi_{U(2)}} W~, \qquad 0 =\partial_{\varphi_{Sp(1)}} W
\eea
The corresponding unrefined Hilbert series of such an $F$-flat space is
\bea
\frac{(1+t) (1+3 t)}{(1-t)^8}~.
\eea
\item Integrating the $F$-flat Hilbert series over gauge groups, we obtain the Hilbert series of $\BC^2/\BZ_3$ as expected:
\bea
g^{(0,2)}_{(1,1)} (t,x) = (1-t^6) \PE [ t^2 +t^3(x+x^{-1}) ] = g_{\BC^2/\BZ_3} (t, x^{1/3})~, \label{C2Z3SO_2_N02HS}
\eea
where $g_{\BC^2/\BZ_3} (t, x)$ is given by \eref{C2Z3HS}, and the fugacitiy $x$ are given such that the global fugacities for $A$ and $\widetilde{A}$ are $tx$ and $tx^{-1}$ respectively.
\een
The same result can be obtained from the case of $\vec k =(1,1)$ with $\vec N= (1,0)$.

\subsubsection*{Generators of the moduli space}
The $F$-flat relations \eref{FflatrelC2Z3SO2} imply that
\bea
X^{a_1}_{~b_1} \tX^{b_1}_{~a_1} + A^{a_1 a_2} \tA_{a_2 a_1}=0~, \qquad \tX^{b_1}_{~a_1} X^{a_1}_{~b_2} =0 \quad({b_1} \neq {b_2})~, \qquad \tX^{1}_{~a_1} X^{a_1}_{1} =  \tX^{2}_{~a_1} X^{a_1}_{2} ~.
\eea
The generator at order $t^2$ can therefore be taken as
\bea
G_2 = A^{a_1b_1} \tA_{b_1a_1}~.
\eea
The two generators at order $t^3$ are
\bea
(G_3)_1=\epsilon^{b_1 b_2} \tA_{a_1 a_2} X^{a_1}_{~ b_1} X^{a_2}_{~ b_2}  ~, \qquad (G_3)_2 = \epsilon_{b_1 b_2} A^{a_1 a_2} \tX^{b_1}_{~ a_1}\tX^{b_2}_{~ a_2}~.
\eea
The relation of order $t^6$ can then be written as
\bea
G_2^3 + 2 (G_3)_1 (G_3)_2 = 0~,
\eea
after the $F$-flat relations have been taken into account.

\subsubsection{$SO(5)$ instanton on $\BC^2/\BZ_3$: $\vec k =(1,1)$ with $\vec N=(1,3)$}
\begin{figure}[H]
\begin{center}
\includegraphics[scale=0.7]{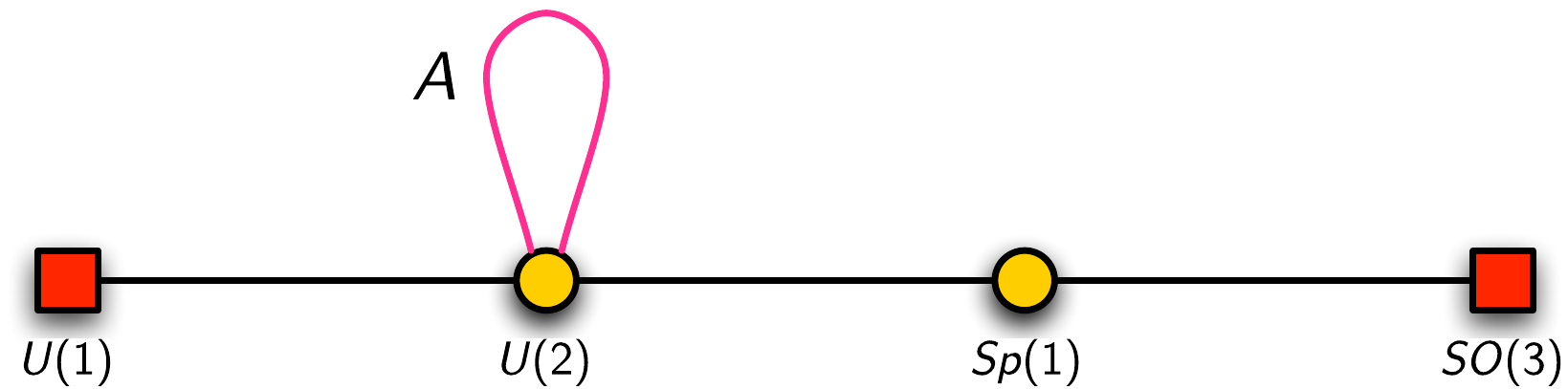}
\caption{The quiver for $SO(5)$ instantons on $\mathbb{C}^2/\mathbb{Z}_{3}$: $\vec k =(1,1)$ with $\vec N=(1,3)$}
\label{fig:C2Z3SO_5_wVS}
\end{center}
\end{figure}
Let $(q,z_1)$ be the fugacities for the gauge group $U(2) = U(1) \times SU(2)$, $z_2$ be that of the gauge group $Sp(1)$, $a$ be that of the global symmetry $U(1)$ and $b$ be that of the global symmetry $SO(3)$.  The Hilbert series is given by
\bea
g^{(1,3)}_{(1,1)} (t, x, a, b)  &= \frac{1}{(2 \pi i)^3} \oint_{|q|=1} \frac{\ud q}{q} \oint_{|z_1|=1} \frac{1-z_1^2}{z_1} \ud z_1 \oint_{|z_2|=1} \frac{1-z_2^2}{z_2} \ud z_2 \times \nn \\
& \quad \frac{\chi_{Q_1}(a,q,z_1) \chi_X(q,z_1,z_2)  \chi_{A} (x,q) \chi_{Q_2} (z_2,b)}{\chi_{F}(z_1, z_2)}~.
\eea
where the subscripts $Q_1$, $X$, $Q_2$ and $A$ denote the contribution from the hypermultiplets in the bi-fundamental representation of $U(1) \times U(2)$, those of $U(2) \times Sp(1)$, those of $Sp(1)\times SO(3)$ and those in the antisymmetric representation of $U(2)$.  The function $\chi_{F}$ takes into account of the $F$-terms.  Explicitly,
\bea
\begin{array}{lll}
\chi_{Q_1}(a,q,z_1) &= \PE \left[ t (a q^{-1}+a^{-1} q) [1]_{z_1} \right] &= \PE \left[ t (a q^{-1}+a^{-1} q) (z_1+z_1^{-1}) \right]    \\
\chi_X(q,z_1,z_2) &=  \PE \left[ t (q+q^{-1}) [1]_{z_1} [1]_{z_2} \right] &=  \PE \left[ t (q+q^{-1}) (z_1+z_1^{-1}) (z_2+z_2^{-1}) \right]  \\
\chi_{Q_2} (b,z_2) &= \PE \left[ t [2]_b [1]_{z_2} \right] &= \PE \left[ t(b^2+1+b^{-2}) (z_2 +z_2^{-1})\right] \\
\chi_{A} (q,x) &= \PE \left[ t (x q^2 + x^{-1} q^{-2} ) \right]  & \\
\chi_{F} (z_1,z_2) &= \PE \left[ t^2 ([2]_{z_1}+1) + t^2 [2]_{z_2} \right] & ~.
\end{array}
\eea
After integration, the Hilbert series is given by
\bea
g^{(1,3)}_{(1,1)} (t, x, a, b)  &= 1+([2]_b +2)t^2+ \left \{ (a+a^{-1})[2]_b + (x+x^{-1}) \right \} t^3 + \nn \\
& \quad + \left \{ [4]_b + (a x + a^{-1} x^{-1} +2) [2]_b +3 \right \} t^4 + \ldots~.
\eea
The plethystic logarithm of this Hilbert series is
\bea
\PL \left[ g^{(1,3)}_{(1,1)} (t, x, a, b)  \right] &= ([2]_b +2)t^2+ \left \{ (a+a^{-1})[2]_b + (x+x^{-1}) \right \} t^3 \nn \\
&\quad + \left\{(a x +a^{-1} x^{-1}) [2]_b -1 \right \} t^4 - \ldots~.
\eea
The corresponding unrefined Hilbert series is
\bea \label{SO51311}
g^{(1,3)}_{(1,1)} (t, 1,1,1)  &= \frac{1}{(1-t)^6 (1+t)^4 \left(1+t^2\right)^2 \left(1+t+t^2\right)^3} \Big( 1+t+4 t^2+9 t^3+18 t^4+ \nn \\
& \quad 25 t^5+33 t^6+30 t^7+33 t^8+ \text{palindrome}+t^{14} \Big) \nn \\
&= 1+5 t^2+8 t^3+20 t^4+32 t^5+70 t^6+96 t^7+183 t^8+ \ldots~.
\eea
It can be seen that this Hilbert series agrees with that of $Sp(2)$ instanton on $\BC^2/\BZ_3$ with $\vec k=(1,1)$ and $\vec N=(2,0)$; see \eref{Sp22011}.  In both cases, the symmetry of the instanton moduli space is $U(2)$.

\subsubsection*{Generators of the moduli space}
\begin{figure}[H]
\centering
\includegraphics[scale=0.7]{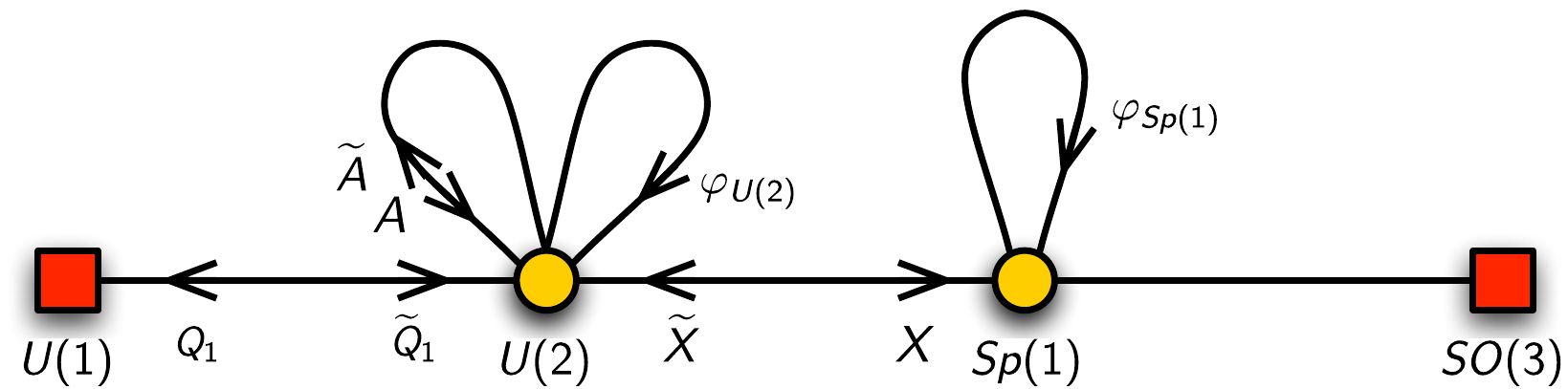}
\caption{Quiver in $\CN=1$ notation for $\vec k =(1,1)$ with $\vec N=(2,1)$. The superpotential is given by \\
$W = (\tQ_1)^{b_1} (\varphi_{U(2)})^{a_1}_{~ b_1} (Q_1)_{a_1} + \left[\widetilde{X}^{b_1}_{~a_1} (\varphi_{U(2)})^{a_1}_{~a_2} X^{a_2}_{~b_1} - X^{a_1}_{~b_2} (\varphi_{Sp(1)})^{b_2}_{~ b_3} \tX^{b_3}_{~a_1}   \right] + \widetilde{A}_{a_1a_2} (\varphi_{U(2)})^{a_2}_{~a_3} A^{a_3 a_1} +  M^{SO(3)}_{j_1 j_2} (\varphi_{Sp(1)})^{b_1 b_2} (Q_2)^{j_1}_{b_1} (Q_2)^{j_2}_{b_2}.$
}
\label{fig:C2Z3SO_2_N21N1}
\end{figure}

Below we use the notation as depicted in \fref{fig:C2Z3SO_2_N21N1}.  The gauge indices $a_1,a_2, \ldots=1,2$ and $b_1, b_2, \ldots =1,2$ correspond to the gauge group $U(2)$ and $SU(2)$ respectively.  The flavour index $j_1,j_2, \ldots=1,2,3$ corresponds to the gauge group $SO(3)$.
\noindent At order 2, there are $5$ generators: 
\bea 
A^{a_1 a_2} \tA_{a_2 a_2}~, \qquad  (Q_1)_{a_1} (\tQ_1)^{a_1}, \qquad \epsilon^{b_1 b_2} (Q_2)^{j_1}_{b_1} (Q_2)^{j_2}_{b_2} 
\eea
At order 3, there are $8$ generators: 
\bea
\begin{array}{ll}
\epsilon^{b_1 b_2} \tA_{a_1 a_2} X^{a_1}_{~ b_1} X^{a_2}_{~ b_2}~,  & \qquad \epsilon_{b_1 b_2} A^{a_1 a_2} \tX^{b_1}_{~ a_1}\tX^{b_2}_{~ a_2}~,  \\
(\tQ_1)^{a_1} \tX^{b_2}_{~a_1} (Q_2)^{j_1}_{b_2}~, & \qquad (Q_1)_{a_1} X^{a_1}_{~ b_2} (Q_2)^{j_1}_{b_3} \epsilon^{b_2 b_3}~. 
\end{array}
\eea
At order 4, there are $6$ generators: 
\bea
(\tQ_1)^{a_1} \tA_{a_1 a_2} X^{a_2}_{~b_1} (Q_2)^{j_1}_{b_2} \epsilon^{b_1b_2}~, \qquad (Q_1)_{a_1} A^{a_1 a_2} \tX^{b_2}_{~ a_2} (Q_2)^{j_1}_{b_2}~.
\eea

\subsubsection{$SO(5)$ instanton on $\BC^2/\BZ_3$: $\vec k =(1,1)$ with $\vec N=(2,1)$}
The quiver is given by
\begin{figure}[H]
\begin{center}
\includegraphics[scale=0.7]{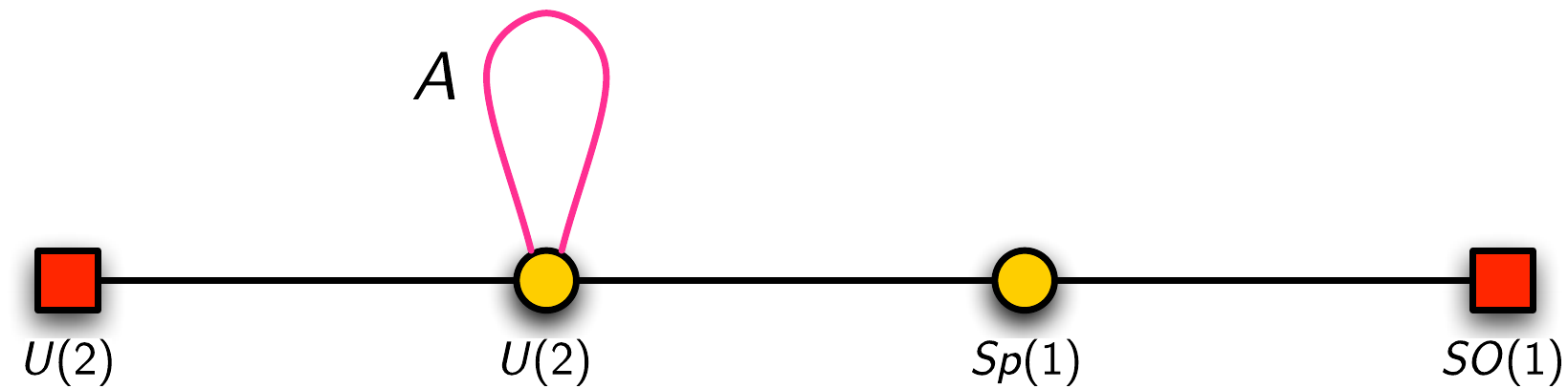}
\caption{The quiver for $SO(5)$ instantons on $\mathbb{C}^2/\mathbb{Z}_{3}$: $\vec k =(1,1)$ with $\vec N=(2,1)$}
\label{fig:C2Z3SO_5_N21}
\end{center}
\end{figure}
Let $(q,z_1)$ be the fugacities for the gauge group $U(2) = U(1) \times SU(2)$, $z_2$ be that of the gauge group $Sp(1)$, and $(u,a)$ be that of the global symmetry $U(2)=U(1) \times SU(2)$.  The Hilbert series is given by
\bea
g^{(2,1)}_{(1,1)} (t, x, a, b)  &= \frac{1}{(2 \pi i)^3} \oint_{|q|=1} \frac{\ud q}{q} \oint_{|z_1|=1} \frac{1-z_1^2}{z_1} \ud z_1 \oint_{|z_2|=1} \frac{1-z_2^2}{z_2} \ud z_2 \times \nn \\
& \quad \frac{\chi_{Q_1}(u,a,q,z_1) \chi_X(q,z_1,z_2)  \chi_{A} (x,q) \chi_{Q_2} (z_2)}{\chi_{F}(z_1, z_2)}~.
\eea
where the subscripts $Q_1$, $X$, $Q_2$ and $A$ denote the contribution from the hypermultiplets in the bi-fundamental representation of $U(2) \times U(2)$, those of $U(2) \times Sp(1)$, those of $Sp(1)\times SO(1)$ and those in the antisymmetric representation of $U(2)$.  The function $\chi_{F}$ takes into account of the $F$-terms.  Explicitly,
\bea
\begin{array}{lll}
\chi_{Q_1}(a,q,z_1) &= \PE \left[ t [1]_a (u q^{-1}+u^{-1} q) [1]_{z_1} \right]    \\
\chi_X(q,z_1,z_2) &=  \PE \left[ t (q+q^{-1}) [1]_{z_1} [1]_{z_2} \right]  \\
\chi_{Q_2} (b,z_2) &= \PE \left[ t  [1]_{z_2} \right] \\
\chi_{A} (q,x) &= \PE \left[ t (x q^2 + x^{-1} q^{-2} ) \right]  & \\
\chi_{F} (z_1,z_2) &= \PE \left[ t^2 ([2]_{z_1}+1) + t^2 [2]_{z_2} \right] & ~.
\end{array}
\eea
After integration, the Hilbert series is given by
\bea
g^{(2,1)}_{(1,1)} (t, x, u, a, b)  &= 1+([2]_a +2)t^2+ \left \{ u^2 x +u^{-2} x^{-1} +(x+x^{-1}) + (u+u^{-1})[1]_a\right \} t^3 + \nn \\
& \quad + \left \{ [4]_a+ 2[2]_a + (u x + u^{-1} x^{-1} ) [1]_a +[2]_u +2 \right \} t^4 + \ldots~.
\eea
The plethystic logarithm of this Hilbert series is
\bea
\PL \left[ g^{(2,1)}_{(1,1)} (t, x, u, a, b)  \right] &= ([2]_a +2)t^2+ \left \{ u^2 x +u^{-2} x^{-1} +(x+x^{-1}) + (u+u^{-1})[1]_a\right \} t^3 \nn \\
&\quad + \left\{(u x +u^{-1} x^{-1}) [1]_a +u^2 + u^{-2} -1 \right \} t^4 - \ldots~.
\eea
The corresponding unrefined Hilbert series is
\bea \label{C2Z3SO5N21HS}
g^{(2,1)}_{(1,1)} (t, 1,1,1,1)  &= \frac{1-2 t+5 t^2-2 t^3+6 t^4-2 t^5+5 t^6-2 t^7+t^8}{(1- t)^6 (1 + t^2) (1 + 2 t + 2 t^2 + t^3)^2}  \nn \\
&= 1+5 t^2+8 t^3+20 t^4+36 t^5+70 t^6+112 t^7+187 t^8 \ldots~.
\eea
It can be seen that this Hilbert series agrees with that of $Sp(2)$ instanton on $\BC^2/\BZ_3$ with $\vec k=(1,1)$ and $\vec N=(1,1)$; see \eref{HS:C2Z3Sp2N11}.

\subsection{Quiver for $Sp(N)$ instantons on $\BC^2/\BZ_{2m+1}$}
\begin{figure}[H]
\begin{center}
\includegraphics[scale=0.7]{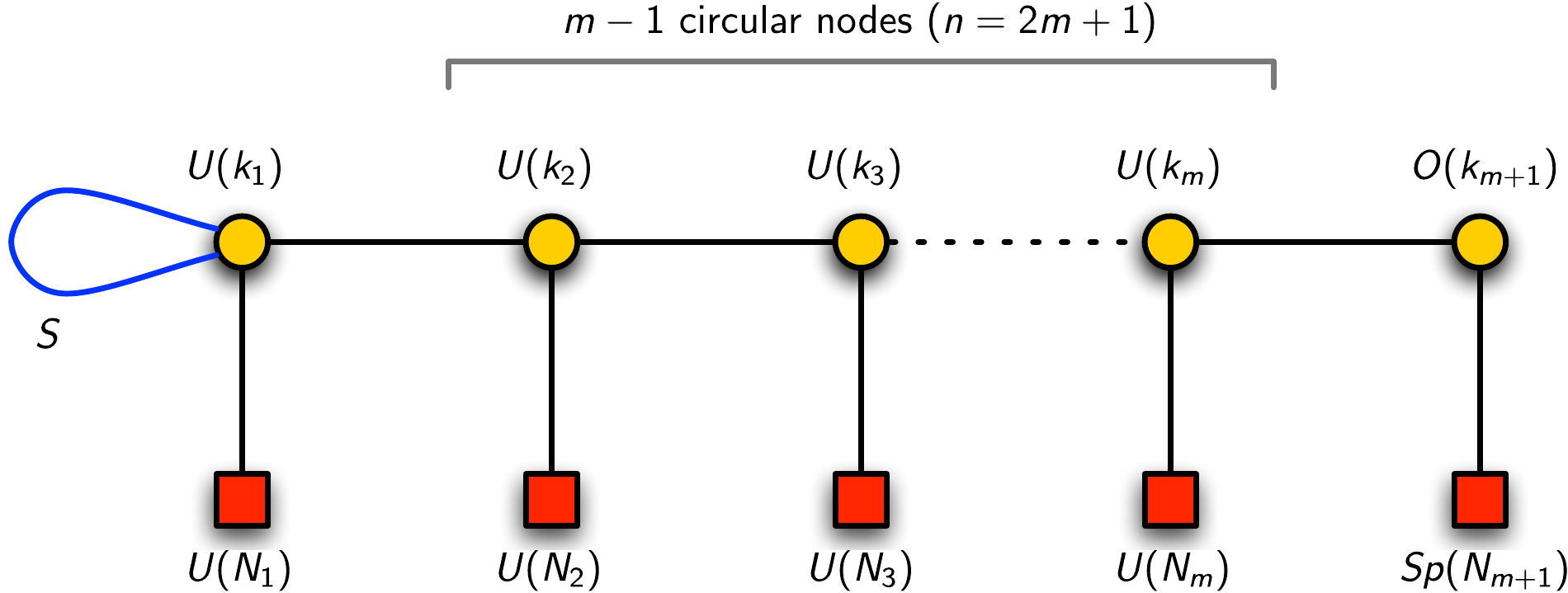}
\caption{The quiver for $Sp(N)$ instantons on $\mathbb{C}^2/\mathbb{Z}_{2m+1}$. Here, $N=N_1+ \ldots+N_{m+1}$.  The loop labelled by $S$ denotes the rank 2-symmetric hypermultiplet in $O(k_{m+1})$ gauge group. Each line between $O(r_1)$ and $U(r_2)$ denotes $r_1 r_2$ hypermultiplets, each line between $U(r_1)$ and $U(r_2)$ denotes $r_1 r_2$ hypermultiplets, and each line between $Sp(r_1)$ and $O(r_2)$ denotes $2r_1 r_2$ half-hypermultiplets.}
\label{fig:Sp_N_oddn}
\end{center}
\end{figure}
If $k_1 =k_2 = \cdots =k_{m+1} =k$, the quaternionic dimension of the Higgs branch of \fref{fig:Sp_N_oddn} is given by
\bea
k \left(\sum_{i=1}^{m+1} N_i \right)+ m k^2 + \frac{1}{2} k(k+1) - m k^2 - \frac{1}{2}k(k-1) = k(N+1)~,
\eea
where $N=\sum_{i=1}^{m+1} N_i $.  This is to be expected for $k$ $Sp(N)$ instantons on $\mathbb{C}^2/\mathbb{Z}_{2m+1}$.

\subsubsection{$Sp(1)$ instanton on $\BC^2/\BZ_3$: $\vec k =(1,1)$ and $\vec N=(0,1)$}
\begin{figure}[H]
\begin{center}
\includegraphics[scale=0.7]{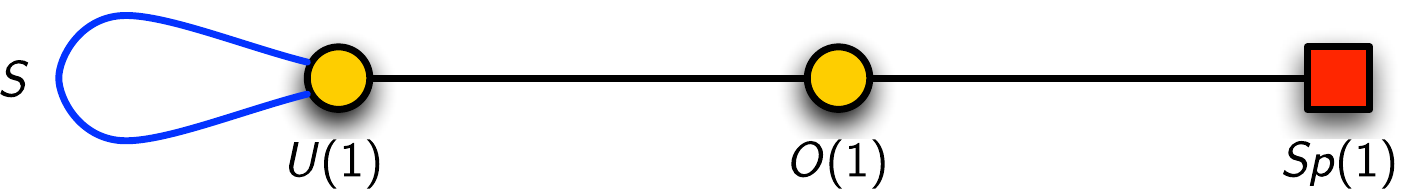}
\caption{The quiver for $Sp(1)$ instantons on $\mathbb{C}^2/\mathbb{Z}_{3}$: $\vec k =(1,1)$ and $\vec N=(0,1)$.}
\label{fig:C2Z3Sp_1_N01.pdf}
\end{center}
\end{figure}

The quiver of our question is depicted in \fref{fig:C2Z3Sp_1_N01.pdf}.
Let $z_1$ be the fugacity for the gauge group $U(1)$.  We take the gauge group $O(1)$ to be $\BZ_2$, and denote the discrete action as $\omega$, with $\omega^2=1$.  Let $b$ be the fugacity of the global $Sp(1)$.
The Hilbert series is given by
\bea
g^{(0,1)}_{(1,1)} (t, x,  b)  &=\frac{1}{2} \sum_{\omega = \pm1}\oint_{|z_1|=1} \frac{\ud z_1}{(2 \pi i)z_1}  \frac{ \chi_X(z_1, \omega)  \chi_{Q} (b, \omega) \chi_{S} (x,z_1)}{\chi_{F}}~.
\eea
where the subscripts $X$, $Q$ and $S$ denote the contribution from the hypermultiplets in the bi-fundamental representation of $U(1) \times O(1)$, those of $O(1)\times Sp(1)$ and those in the symmetric representation of $U(1)$.  The function $\chi_{F}$ takes into account of the $F$-terms.  Explicitly,
\bea
\begin{array}{lll}
\chi_X(z_1, \omega) &=  \PE \left[ t (\omega z_1^{-1} + \omega^{-1} z_1)\right]   \\
\chi_{Q} (b, \omega) &= \PE \left[ t \omega [1]_b \right] \\
\chi_{S} (x,z_1) &= \PE \left[ t (x z_1^2 + x^{-1} z_1^{-2} ) \right]  & \\
\chi_{F} &= \PE \left[ t^2 \right] &= (1-t^2)^{-1} ~.
\end{array}
\eea
After integration and summation, the Hilbert series is given by
\bea
g^{(0,1)}_{(1,1)} (t, x,  b)  &= (1-t^4)(1-t^6) \PE \left[ ([2]_b +1)t^2 + [1]_x t^3\right]
\eea
The plethystic logarithm of this Hilbert series is
\bea
\PL \left[ g^{(0,1)}_{(1,1)} (t, x, b)  \right] &=  ([2]_b +1)t^2 + [1]_x t^3 -t^4 -t^6~.
\eea
The corresponding unrefined Hilbert series is
\bea
g^{(0,1)}_{(1,1)} (t, 1,1)  = \frac{\left(1+t^2\right) \left(1-t+t^2\right)}{(1-t)^4 (1+t)^2 \left(1+t+t^2\right)}~. \label{HS:C2Z3Sp_1_N01.pdf}
\eea
It can be seen that this Hilbert series agrees with that of $SU(2)$ instanton on $\BC^2/\BZ_3$ with $\vec k=(1,1,1)$ and $\vec N=(2,0,0)$; see \eref{HS200111ref} and \eref{HS200111unref}.

\subsubsection{$Sp(1)$ instanton on $\BC^2/\BZ_3$: $\vec k =(1,1)$ and $\vec N=(1,0)$}
\begin{figure}[H]
\begin{center}
\includegraphics[scale=0.7]{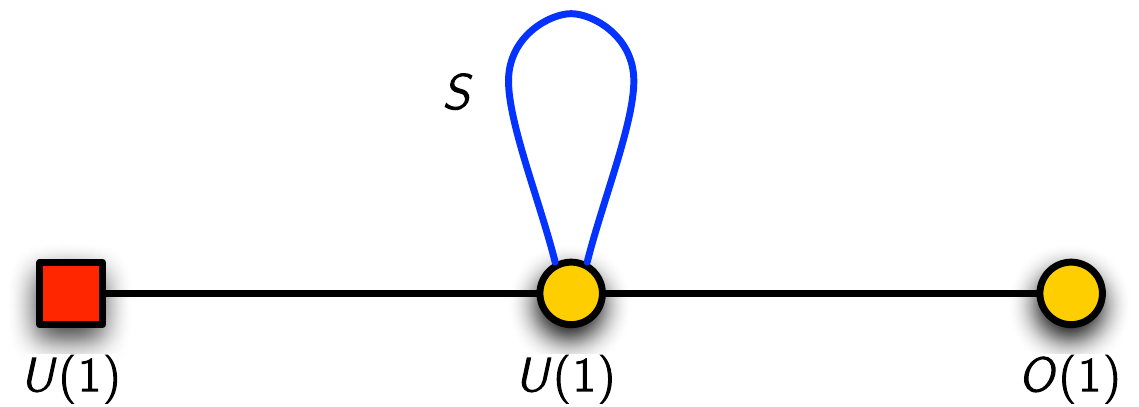}
\caption{The quiver for $Sp(1)$ instantons on $\mathbb{C}^2/\mathbb{Z}_{3}$: $\vec k =(1,1)$ and $\vec N=(1,0)$.}
\label{fig:C2Z3Sp_1_N10.pdf}
\end{center}
\end{figure}

The quiver of our question is depicted in \fref{fig:C2Z3Sp_1_N10.pdf}.
Let $z_1$ be the fugacity for the gauge group $U(1)$.  We take the gauge group $O(1)$ to be $\BZ_2$, and denote the discrete action as $\omega$, with $\omega^2=1$.  Let $a$ be the fugacity of the global $U(1)$.
The Hilbert series is given by
\bea
g^{(1,0)}_{(1,1)} (t, x, a)  &=\frac{1}{2} \sum_{\omega = \pm1}\oint_{|z_1|=1} \frac{\ud z_1}{(2 \pi i)z_1}  \frac{ \chi_X(z_1, \omega)  \chi_{Q} (a, \omega) \chi_{S} (x,z_1)}{\chi_{F}}~.
\eea
where the subscripts $X$, $Q$ and $S$ denote the contribution from the hypermultiplets in the bi-fundamental representation of $U(1) \times O(1)$, those of $U(1)\times U(1)$ and those in the symmetric representation of $U(1)$.  The function $\chi_{F}$ takes into account of the $F$-terms.  Explicitly,
\bea
\begin{array}{lll}
\chi_X(z_1, \omega) &=  \PE \left[ t (\omega z_1^{-1} + \omega^{-1} z_1)\right]   \\
\chi_{Q} (a, z_1) &= \PE \left[ t (a z_1^{-1}+a^{-1} z_1) \right] \\
\chi_{S} (x,z_1) &= \PE \left[ t (x z_1^2 + x^{-1} z_1^{-2} ) \right]  & \\
\chi_{F} &= \PE \left[ t^2 \right] &= (1-t^2)^{-1} ~.
\end{array}
\eea
After integration and summation, the Hilbert series is given by
\bea
g^{(1,0)}_{(1,1)} (t, x, a)  &= 1+2t^2+\left( x a^{2} +x^{-1} a^{-2}  + x+x^{-1} \right) t^3 +(3+a^{-2} +a^2) t^4+\ldots~.
\eea
The plethystic logarithm of this Hilbert series is
\bea
\PL \left[ g^{(1,0)}_{(1,1)} (t, x, a)  \right] &=  2t^2+\left( x a^{2} +x^{-1} a^{-2}   + x+x^{-1} \right) t^3 + (a^2 +a^{-2})t^4 \nn \\
& \quad - (a^2+2+a^{-2}) t^6 + \ldots~.
\eea
The corresponding unrefined Hilbert series is
\bea
g^{(1,0)}_{(1,1)} (t, 1,1)  = \frac{1+t^2+2 t^3+2 t^4+2 t^5+t^6+t^8}{(1-t)^4 (1+t)^2 \left(1+t^2\right) \left(1+t+t^2\right)^2}~. \label{HS:C2Z3Sp_1_N10.pdf}
\eea
It can be seen that this Hilbert series agrees with that of $SU(2)$ instanton on $\BC^2/\BZ_3$ with $\vec k=(1,1,1)$ and $\vec N=(1,1,0)$.

\subsubsection{$Sp(2)$ instanton on $\BC^2/\BZ_3$: $\vec k =(1,1)$ and $\vec N=(2,0)$}
\begin{figure}[H]
\begin{center}
\includegraphics[scale=0.7]{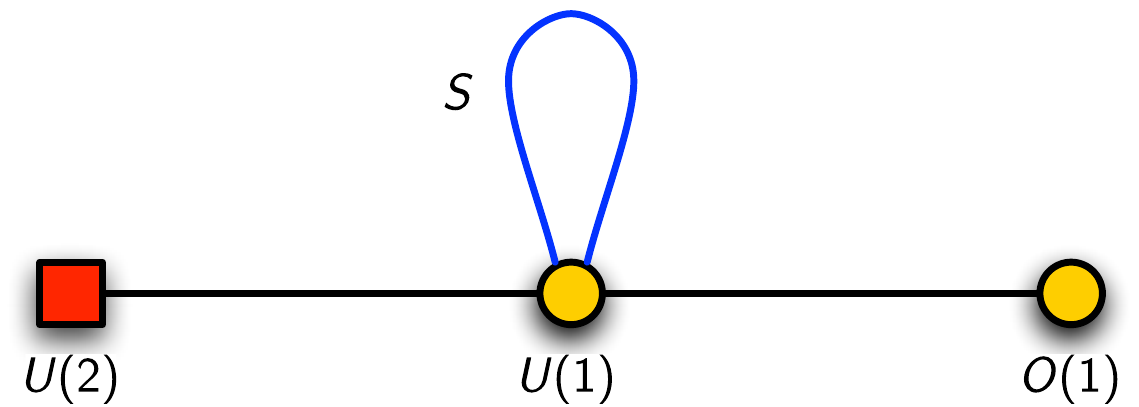}
\caption{The quiver for $Sp(2)$ instantons on $\mathbb{C}^2/\mathbb{Z}_{3}$: $\vec k =(1,1)$ and $\vec N=(2,0)$.}
\label{fig:C2Z3Sp_2_N20.pdf}
\end{center}
\end{figure}

The quiver of our question is depicted in \fref{fig:C2Z3Sp_2_N20.pdf}.
Let $z_1$ be the fugacity for the gauge group $U(1)$.  We take the gauge group $O(1)$ to be $\BZ_2$ and denote the discrete action as $\omega$, with $\omega^2=1$.  Let $(u,a)$ be the fugacity of the global $U(2)=U(1) \times SU(2)$.
The Hilbert series is given by
\bea
g^{(2,0)}_{(1,1)} (t, x, u, a)  &=\frac{1}{2} \sum_{\omega = \pm1}\oint_{|z_1|=1} \frac{\ud z_1}{(2 \pi i)z_1}  \frac{\chi_{Q_1} (u, a, z_1) \chi_X(z_1, \omega)  \chi_{S} (x,z_1)}{\chi_{F}}~.
\eea
where the subscripts $X$, $Q$ and $S$ denote the contribution from the hypermultiplets in the bi-fundamental representation of $U(1) \times O(1)$, those of $U(2)\times U(1)$ and those in the symmetric representation of $U(1)$.  The function $\chi_{F}$ takes into account of the $F$-terms.  Explicitly,
\bea
\begin{array}{lll}
\chi_X(z_1, \omega) &=  \PE \left[ t (\omega z_1^{-1} + \omega^{-1} z_1)\right]   \\
\chi_{Q} (u,a, z_1) &= \PE \left[ t (u z_1^{-1}+u^{-1} z_1) [1]_a\right] \\
\chi_{S} (x,z_1) &= \PE \left[ t (x z_1^2 + x^{-1} z_1^{-2} ) \right]  & \\
\chi_{F} &= \PE \left[ t^2 \right] &= (1-t^2)^{-1} ~.
\end{array}
\eea
After integration and summation, the Hilbert series is given by
\bea
g^{(2,0)}_{(1,1)} (t, x, u,a)  &= 1+([2]_a +2)t^2+ \left \{ (x u^2+x^{-1} u^{-2})[2]_b + (x+x^{-1}) \right \} t^3 + \nn \\
& \quad + \left \{ [4]_a + (u^2 + u^{-2}) [2]_a +3 \right \} t^4 + \ldots~.
\eea
The plethystic logarithm of this Hilbert series is
\bea
\PL \left[ g^{(2,0)}_{(1,1)} (t, x, u,a)  \right] &=  ([2]_a+2)t^2+\left[ (x u^{2} +x^{-1} u^{-2})[2]_a  + x+x^{-1} \right] t^3 \nn \\
& \quad + \left[ (u^2 +u^{-2})[2]_a -1 \right]t^4 - \ldots~.
\eea
The corresponding unrefined Hilbert series is
\bea \label{Sp22011}
g^{(2,0)}_{(1,1)} (t, 1,1,1)  &= \frac{1}{(1- t)^6 (1 + t)^4 (1 + t^2)^2 (1 + t + t^2)^3} \Big(1+t + 4 t^2 + 9 t^3 + 18 t^4 + 25 t^5 \nn \\
& \quad + 33 t^6 + 30 t^7 + 33 t^8+ \text{palindrome} +t^{14} \Big) \nn \\
&= 1 + 5 t^2 + 8 t^3 + 20 t^4 + 32 t^5 + 70 t^6 + 96 t^7 + 183 t^8+ \ldots~.
\eea
It can be seen that this Hilbert series agrees with that of $SO(5)$ instanton on $\BC^2/\BZ_3$ with $\vec k=(1,1)$ and $\vec N=(1,2)$; see \eref{SO51311}.  In both cases, the symmetry of the instanton moduli space is $U(2)$.

\subsubsection{$Sp(2)$ instanton on $\BC^2/\BZ_3$: $\vec k =(1,1)$ and $\vec N=(1,1)$}
\begin{figure}[H]
\begin{center}
\includegraphics[scale=0.7]{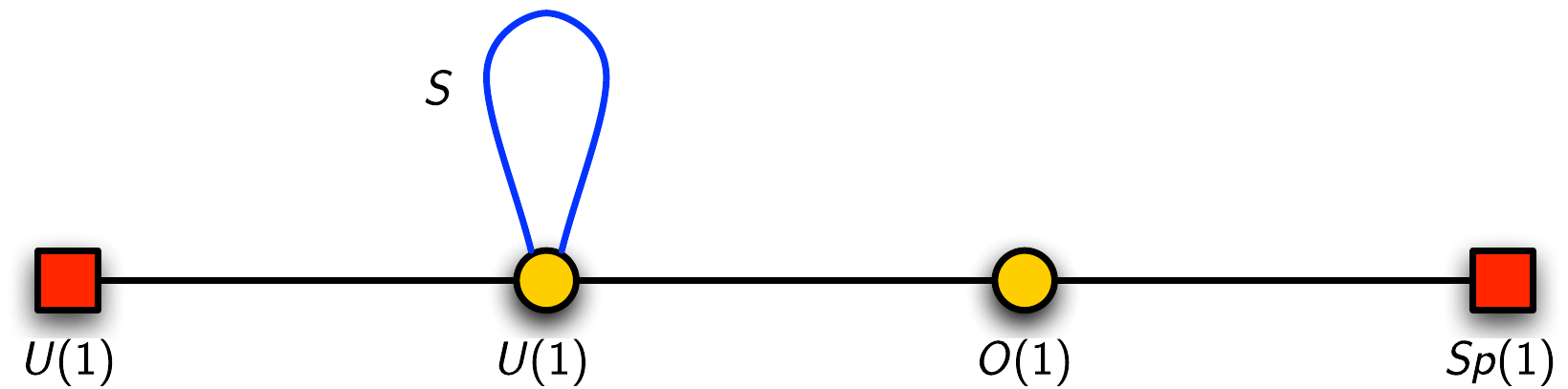}
\caption{The quiver for $Sp(2)$ instantons on $\mathbb{C}^2/\mathbb{Z}_{3}$: $\vec k =(1,1)$ and $\vec N=(1,1)$.}
\label{fig:C2Z3Sp_2_N11.pdf}
\end{center}
\end{figure}

The quiver of our question is depicted in \fref{fig:C2Z3Sp_2_N11.pdf}.
Let $z_1$ be the fugacity for the gauge group $U(1)$.  We take the gauge group $O(1)$ to be $\BZ_2$ and denote the discrete action as $\omega$, with $\omega^2=1$.  Let $(u,a)$ be the fugacity of the global $U(2)=U(1) \times SU(2)$ and $b$ be that of the global $Sp(1)$.
The Hilbert series is given by
\bea
g^{(1,1)}_{(1,1)} (t, x, u, a,b)  &=\frac{1}{2} \sum_{\omega = \pm1}\oint_{|z_1|=1} \frac{\ud z_1}{(2 \pi i)z_1}  \frac{\chi_{Q_1} (u, a, z_1) \chi_X(z_1, \omega)   \chi_{S} (x,z_1) \chi_{Q_2} (b,\omega) }{\chi_{F}}~.
\eea
where the subscripts $Q_1$ $X$, $Q_2$ and $S$ denote the contribution from the hypermultiplets in the bi-fundamental representation of $U(2)\times U(1)$, those of $U(1) \times O(1)$,  those of $O(1) \times Sp(1)$ and those in the symmetric representation of $U(1)$.  The function $\chi_{F}$ takes into account of the $F$-terms.  Explicitly,
\bea
\begin{array}{lll}
\chi_{Q_1} (u,a, z_1) &= \PE \left[ t (u z_1^{-1}+u^{-1} z_1) [1]_a\right] \\
\chi_X(z_1, \omega) &=  \PE \left[ t (\omega z_1^{-1} + \omega^{-1} z_1)\right]   \\
\chi_{S} (x,z_1) &= \PE \left[ t (x z_1^2 + x^{-1} z_1^{-2} ) \right]  & \\
\chi_{Q_2} (b, \omega) &= \PE \left[ t \omega [1]_b\right] \\
\chi_{F} &= \PE \left[ t^2 \right] = (1-t^2)^{-1} ~.
\end{array}
\eea
The corresponding unrefined Hilbert series is
\bea \label{HS:C2Z3Sp2N11}
g^{(1,1)}_{(1,1)} (t, 1, 1, 1,1) &=  \frac{1-2 t+5 t^2-2 t^3+6 t^4-2 t^5+5 t^6-2 t^7+t^8}{(1-t)^6 \left(1+t^2\right) \left(1+2 t+2 t^2+t^3\right)^2} \nn \\
&= 1+5 t^2+8 t^3+20 t^4+36 t^5+70 t^6+112 t^7+187 t^8+ \ldots~.
\eea
It can be seen that this Hilbert series agrees with that of $SO(5)$ instanton on $\BC^2/\BZ_3$ with $\vec k=(1,1)$ and $\vec N=(2,1)$; see \eref{C2Z3SO5N21HS}. 

\subsection{The $O/O$ quiver for $SO(N)$ instantons on $\BC^2/\BZ_{2m}$}
\begin{figure}[H]
\begin{center}
\includegraphics[scale=0.7]{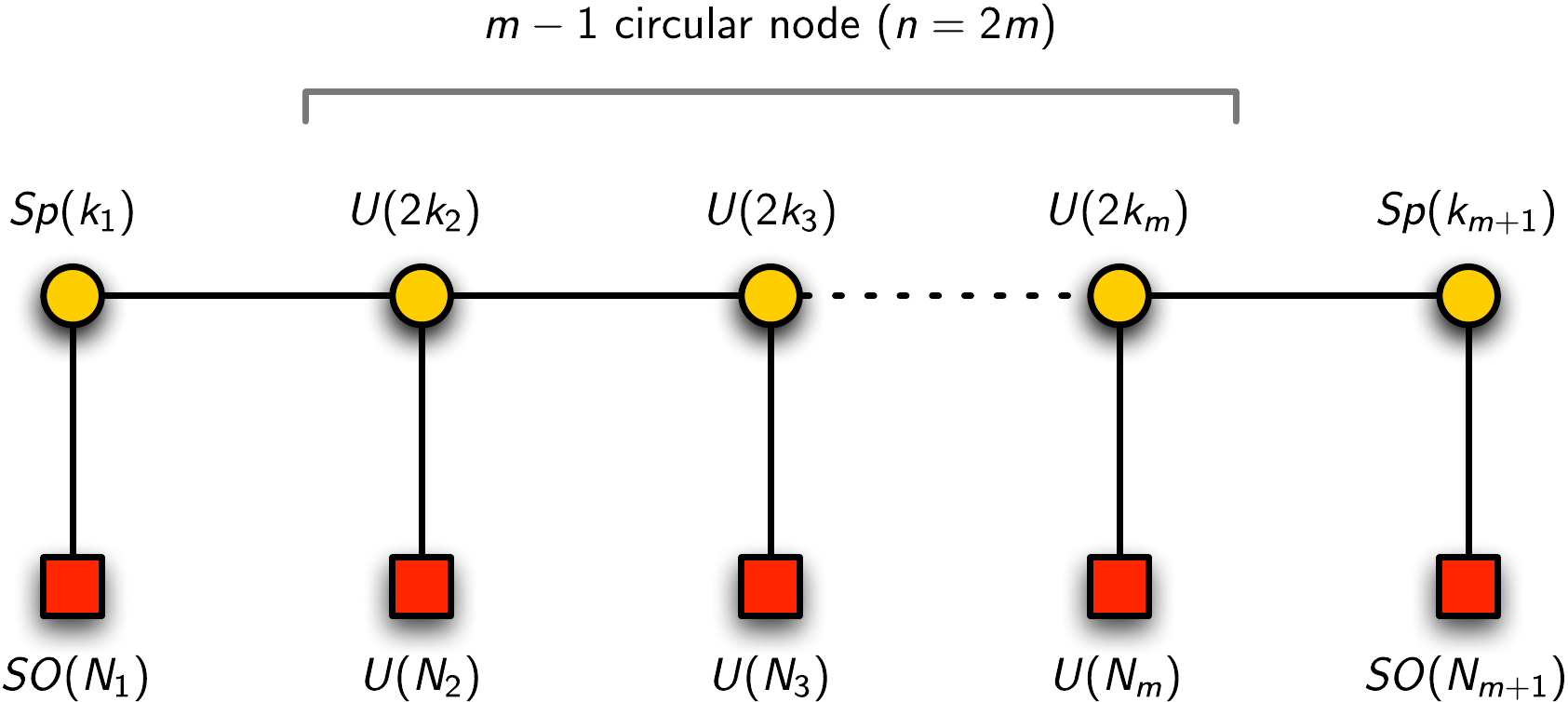}
\caption{The $O/O$ quiver for $SO(N)$ instantons on $\mathbb{C}^2/\mathbb{Z}_{2m}$.  Here, each of $k_2, \ldots, k_{m}$ can take either integral or half-integral value and $N=N_1+ 2N_2 +\ldots+2N_m+N_{m+1}$.  Each line between $Sp(r_1)$ and $U(r_2)$ denotes $2r_1 r_2$ hypermultiplets, each line between $U(r_1)$ and $U(r_2)$ denotes $r_1 r_2$ hypermultiplets, and each line between $Sp(r_1)$ and $SO(r_2)$ denotes $2r_1 r_2$ half-hypermultiplets.}
\label{fig:SO_N_withVSevenn}
\end{center}
\end{figure}

If $k_1 =\cdots =k_{m+1} =k$, the quaternionic dimension of the Higgs branch of \fref{fig:SO_N_withVSevenn} is
\bea
kN+4k^2+ (m-2) 4k^2+4k^2 - (m-1)(2 k)^2 - 2\left[ \frac{1}{2}(2k)(2k+1)\right]= k(N-2)~,
\eea
where $N=N_1+2N_2+\ldots+2N_m+N_{m+1} $.  This is to be expected for $k$ $SO(N)$ instantons on $\mathbb{C}^2/\mathbb{Z}_{2m}$.

\subsubsection{$SO(5)$ instanton on $\BC^2/\BZ_4$: $\vec k=(1,1,1)$ and $\vec N=(1,1,2)$}
\begin{figure}[H]
\begin{center}
\includegraphics[scale=0.7]{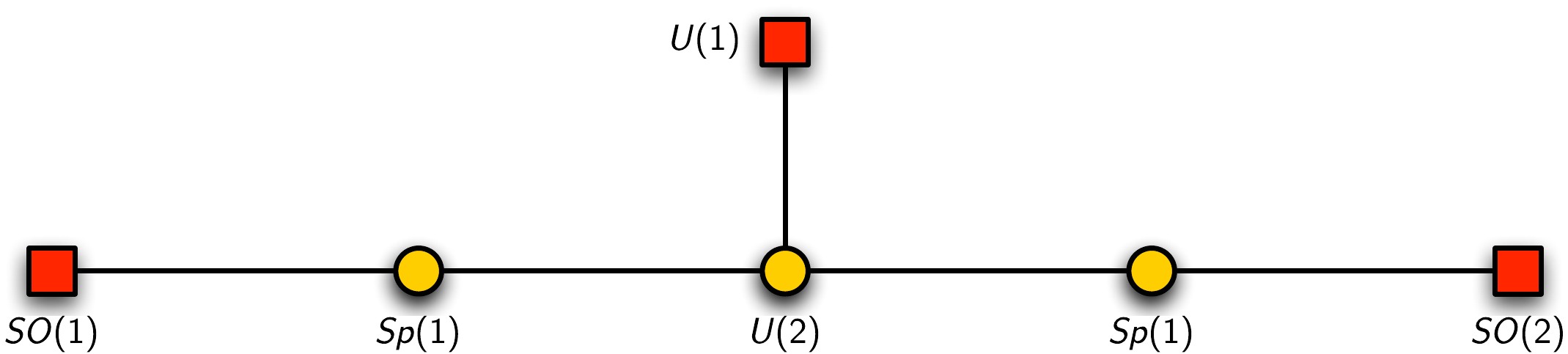}
\caption{The $O/O$ quiver for $SO(5)$ instanton on $\BC^2/\BZ_4$: $\vec k=(1,1,1)$ and $\vec N=(1,1,2)$.}
\label{fig:C2Z4SO_5_N112}
\end{center}
\end{figure}

The quiver of our question is depicted in \fref{fig:C2Z4SO_5_N112}.
Let $(q,z_2)$ be the fugacities for the gauge group $U(2) = U(1) \times SU(2)$, $z_1, z_3$ be those of the gauge groups $Sp(1)$ corresponding to the left and right $Sp(1)$ circular nodes, $b$ be that of the global symmetry $U(1)$, and $c$ be that of the global symmetry $SO(2)$.  The Hilbert series is given by
\bea
g^{(1,1,2)}_{(1,1,1)} (t, b,c)  &= \frac{1}{(2 \pi i)^4} \oint_{|q|=1} \frac{\ud q}{q} \oint_{|z_1|=1} \frac{1-z_1^2}{z_1} \ud z_1 \oint_{|z_2|=1} \frac{1-z_2^2}{z_2} \ud z_2\oint_{|z_3|=1} \frac{1-z_3^2}{z_3} \ud z_3 \times \nn \\
& \quad \frac{\chi_{Q_1}(z_1) \chi_{X_1} (q,z_1,z_2)  \chi_{X_2} (q,z_2,z_3)   \chi_{Q_2} (z_2, b) \chi_{Q_3} (z_3,c)}{\chi_{F}(z_1, z_2,z_3)}~.
\eea
where the subscripts $Q_1$, $X_1$, $X_2$, $Q_2$ and $Q_3$ denote the contribution from the hypermultiplets in the bi-fundamental representation of $SO(1) \times Sp(1)$, those of $Sp(1) \times U(2)$, those of $U(2)\times Sp(1)$, those of $U(1)\times U(2)$, and those of $Sp(1) \times SO(2)$.  The function $\chi_{F}$ takes into account of the $F$-terms.  Explicitly,
\bea
\begin{array}{lll}
\chi_{Q_1}(z_1) &= \PE \left[ t [1]_{z_1} \right]    \\
\chi_{X_1} (q,z_1,z_2)  &=  \PE \left[ t (q+q^{-1}) [1]_{z_1} [1]_{z_2} \right]  \\
\chi_{X_2} (q,z_2,z_3) &=  \PE \left[ t (q+q^{-1}) [1]_{z_2} [1]_{z_3} \right]  \\
 \chi_{Q_2} (q,z_2, b) &= \PE \left[ t  [1]_{z_2} (q b^{-1}+q^{-1} b) \right] \\
  \chi_{Q_3} (z_3,c) &= \PE \left[ t  [1]_{z_3} (c+c^{-1}) \right] \\
\chi_{F} (z_1,z_2,z_3) &= \PE \left[ t^2 ([2]_{z_1}+ [2]_{z_2}+1 +[2]_{z_3})\right] & ~.
\end{array}
\eea
The corresponding unrefined Hilbert series is
\bea \label{C2Z4SO_5_N112}
g^{(1,1,2)}_{(1,1,1)} (t, 1,1) &= \frac{1}{(1- t)^6 (1 + t)^2 (1 + t^2) (1 + t + t^2)^3 (1 + t + t^2 + t^3 + t^4)} \Big(1+2 t^2+3 t^3+ \nn \\
& \qquad 8 t^4+11 t^5+13 t^6+12 t^7+13 t^8+11 t^9+8 t^{10}+3 t^{11}+2 t^{12}+t^{14} \Big) \nn \\
&= 1 + 3 t^2 + 6 t^3 + 12 t^4 + 24 t^5 + 42 t^6 + 68 t^7 + 115 t^8+\ldots~.
\eea
This Hilbert series agrees with that of $Sp(2)$ instanton on $\BC^2/\BZ_4$ ($SS$ quiver): $\vec k=(1,1)$ and $\vec N=(1,1)$; see \eref{C2Z4Sp_2_N11}.

\subsubsection{$SO(6)$ instanton on $\BC^2/\BZ_4$: $\vec k=(1,1,1)$ and $\vec N=(2,0,4)$}
\begin{figure}[H]
\begin{center}
\includegraphics[scale=0.7]{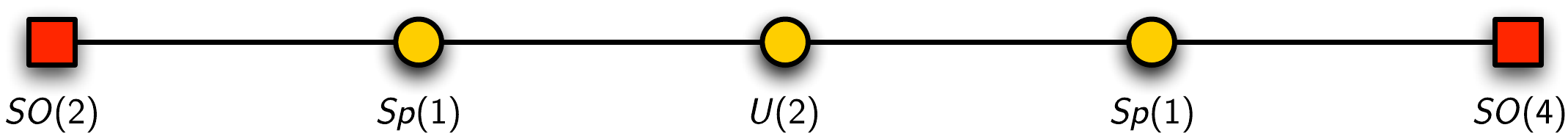}
\caption{The $O/O$ quiver for $SO(6)$ instanton on $\BC^2/\BZ_4$: $\vec k=(1,1,1)$ and $\vec N=(2,0,4)$.}
\label{fig:C2Z4SO_6_N204}
\end{center}
\end{figure}

The quiver of our question is depicted in \fref{fig:C2Z4SO_6_N204}.
Let $(q,z_2)$ be the fugacities for the gauge group $U(2) = U(1) \times SU(2)$, $z_1, z_3$ be those of the gauge groups $Sp(1)$ corresponding to the left and right $Sp(1)$ circular nodes, $a$ be that of the global symmetry $SO(2)$, and $b_1,b_2$ be that of the global symmetry $SO(4)$.  The Hilbert series is given by
\bea
g^{(2,0,4)}_{(1,1,1)} (t, a, b_1,b_2)  &=\frac{1}{(2 \pi i)^4} \oint_{|q|=1} \frac{\ud q}{q} \oint_{|z_1|=1} \frac{1-z_1^2}{z_1} \ud z_1 \oint_{|z_2|=1} \frac{1-z_2^2}{z_2} \ud z_2\oint_{|z_3|=1} \frac{1-z_3^2}{z_3} \ud z_3 \times \nn \\
& \quad \frac{\chi_{Q_1}(a,z_1) \chi_{X_1} (q,z_1,z_2)  \chi_{X_2} (q,z_2,z_3)   \chi_{Q_2} (z_3, b_1,b_2)}{\chi_{F}(z_1, z_2,z_3)}~.
\eea
where the subscripts $Q_1$, $X_1$, $X_2$ and $Q_2$ denote the contribution from the hypermultiplets in the bi-fundamental representation of $SO(2) \times Sp(1)$, those of $Sp(1) \times U(2)$, those of $U(2)\times Sp(1)$ and those of $Sp(1) \times SO(4)$.  The function $\chi_{F}$ takes into account of the $F$-terms.  Explicitly,
\bea
\begin{array}{lll}
\chi_{Q_1}(a,z_1) &= \PE \left[ t (a+a^{-1}) [1]_{z_1} \right]    \\
\chi_{X_1} (q,z_1,z_2)  &=  \PE \left[ t (q+q^{-1}) [1]_{z_1} [1]_{z_2} \right]  \\
\chi_{X_2} (q,z_2,z_3) &=  \PE \left[ t (q+q^{-1}) [1]_{z_2} [1]_{z_3} \right]  \\
 \chi_{Q_2} (z_3, b_1,b_2) &= \PE \left[ t  [1]_{z_3} [1,1]_{b_1,b_2} \right] \\
\chi_{F} (z_1,z_2,z_3) &= \PE \left[ t^2 ([2]_{z_1}+ [2]_{z_2}+1 +[2]_{z_3})\right] & ~.
\end{array}
\eea
After integration, the Hilbert series is given by
\bea
g^{(2,0,4)}_{(1,1,1)} (t, a, b)  &= 1+([2,0]_{\vec b}+[0,2]_{\vec b}+2)t^2+ ([4,0]_{\vec b}+[0,4]_{\vec b}+[2,2]_{\vec b}+2[2,0]_{\vec b}+2[0,2]_{\vec b} \nn \\
& \quad +2[1,1]_{\vec b}(a+a^{-1}) +5) t^4 + \ldots~.
\eea
The plethystic logarithm of this Hilbert series is
\bea
\PL \left[ g^{(2,0,4)}_{(1,1,1)}(t, a, b)  \right] &= ([2,0]_{\vec b} + [0,2]_{\vec b} +2)t^2 +2(a+a^{-1}) [1,1]_{\vec b} t^4 - 4(a+a^{-1}) [1,1]_{\vec b} + \ldots~.
\eea
The corresponding unrefined Hilbert series is
\bea \label{HS:C2Z4SO_6_N204}
g^{(2,0,4)}_{(1,1,1)} (t, 1,1)  &= \frac{1+4 t^2+22 t^4+36 t^6+54 t^8+36 t^{10}+22 t^{12}+4 t^{14}+t^{16}}{(1-t)^8 (1+t)^8 \left(1+t^2\right)^4} \nn \\
&= 1 + 8 t^2 + 52 t^4 + 216 t^6 + 735 t^8 + 2064 t^{10}+ \ldots~.
\eea
Notice that this Hilbert series agrees with that of $SU(4)$ instantons on $\BC^2/\BZ_4$ with $\vec k=(1,1,1,1)$, and $\vec N=(2,0,2,0)$.  

\subsubsection{$SO(6)$ instanton on $\BC^2/\BZ_4$: $\vec k=(1,1,1)$ and $\vec N=(2,1,2)$}
\begin{figure}[H]
\begin{center}
\includegraphics[scale=0.7]{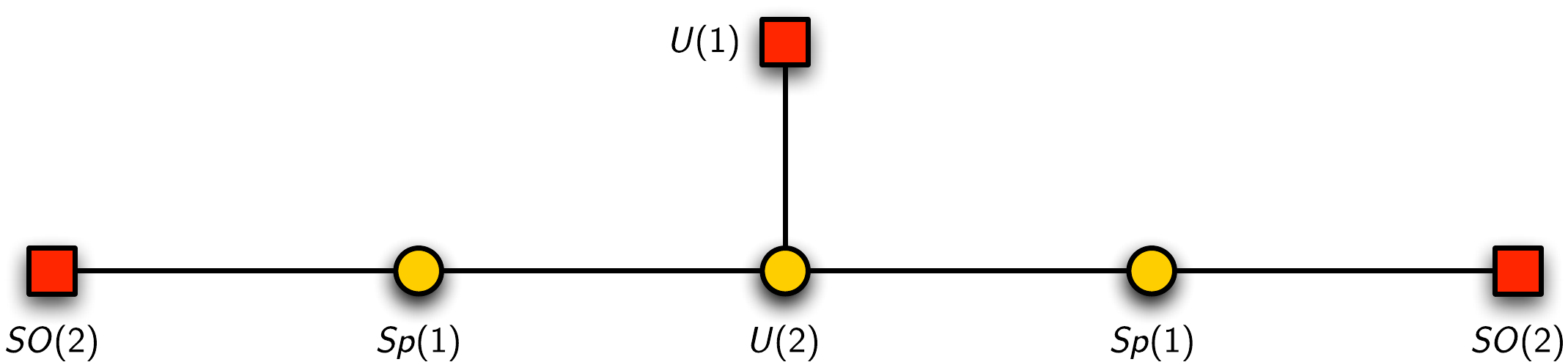}
\caption{The $O/O$ quiver for $SO(6)$ instanton on $\BC^2/\BZ_4$: $\vec k=(1,1,1)$ and $\vec N=(2,1,2)$.}
\label{fig:C2Z4SO_6_N212}
\end{center}
\end{figure}

The quiver of our question is depicted in \fref{fig:C2Z4SO_6_N212}.
Let $(q,z_2)$ be the fugacities for the gauge group $U(2) = U(1) \times SU(2)$, $z_1, z_3$ be those of the gauge groups $Sp(1)$ corresponding to the left and right $Sp(1)$ circular nodes, $a, c$ be that of the global symmetries $SO(2)$ corresponding to left and right square nodes respectively, and $b$ be that of the global symmetry $U(1)$.  The Hilbert series is given by
\bea
g^{(2,1,2)}_{(1,1,1)} (t, a, b,c)  &=\frac{1}{(2 \pi i)^4} \oint_{|q|=1} \frac{\ud q}{q} \oint_{|z_1|=1} \frac{1-z_1^2}{z_1} \ud z_1 \oint_{|z_2|=1} \frac{1-z_2^2}{z_2} \ud z_2\oint_{|z_3|=1} \frac{1-z_3^2}{z_3} \ud z_3 \times \nn \\
& \quad \frac{\chi_{Q_1}(a,z_1) \chi_{X_1} (q,z_1,z_2)  \chi_{X_2} (q,z_2,z_3)  \chi_{Q_2} (q,z_2, b)   \chi_{Q_3} (z_3, c)}{\chi_{F}(z_1, z_2,z_3)}~.
\eea
where the subscripts $Q_1$, $X_1$, $X_2$, $Q_2$, $Q_3$ denote the contribution from the hypermultiplets in the bi-fundamental representation of $SO(2) \times Sp(1)$, those of $Sp(1) \times U(2)$, those of $U(2)\times Sp(1)$, those of $U(2) \times U(1)$, and those of $Sp(1) \times SO(2)$.  The function $\chi_{F}$ takes into account of the $F$-terms.  Explicitly,
\bea
\begin{array}{lll}
\chi_{Q_1}(a,z_1) &= \PE \left[ t (a+a^{-1}) [1]_{z_1} \right]    \\
\chi_{X_1} (q,z_1,z_2)  &=  \PE \left[ t (q+q^{-1}) [1]_{z_1} [1]_{z_2} \right]  \\
\chi_{X_2} (q,z_2,z_3) &=  \PE \left[ t (q+q^{-1}) [1]_{z_2} [1]_{z_3} \right]  \\
 \chi_{Q_2} (q,z_2, b) &= \PE \left[ t  [1]_{z_2} (q b^{-1}+q^{-1}b) \right] \\
 \chi_{Q_3} (z_3, c) &= \PE \left[ t  [1]_{z_3} [1]_{c} \right] \\
\chi_{F} (z_1,z_2,z_3) &= \PE \left[ t^2 ([2]_{z_1}+ [2]_{z_2}+1 +[2]_{z_3})\right] & ~.
\end{array}
\eea
The corresponding unrefined Hilbert series is
\bea \label{HS:C2Z4SO_6_N212}
g^{(2,1,2)}_{(1,1,1)} (t, 1,1,1)  &= \frac{1}{(1 - t)^8 (1 + t)^4 (1 + t^2)^2 (1 + t + t^2)^4 (1 + t + t^2 + t^3 + 
   t^4)} \Big( 1+t \nn \\
   & \quad +3 t^2+7 t^3+18 t^4+33 t^5+51 t^6+69 t^7+93 t^8+110 t^9+120 t^{10} \nn \\
   & \quad +110 t^{11}+ \text{palindrome}+ t^{20} \Big) \nn \\
&= 1+4 t^2+8 t^3+20 t^4+40 t^5+84 t^6+152 t^7+285 t^8+ \ldots~.
\eea
Notice that this Hilbert series agrees with that of $SU(4)$ instantons on $\BC^2/\BZ_4$ with $\vec k=(1,1,1,1)$, and $\vec N=(1,1,1,1)$.  In both cases, the symmetry of the Higgs branch is $S(U(1)^4)$.

\subsection{The $S/S$ quiver for $Sp(N)$ instantons on $\BC^2/\BZ_{2m}$}
\begin{figure}[H]
\begin{center}
\includegraphics[scale=0.7]{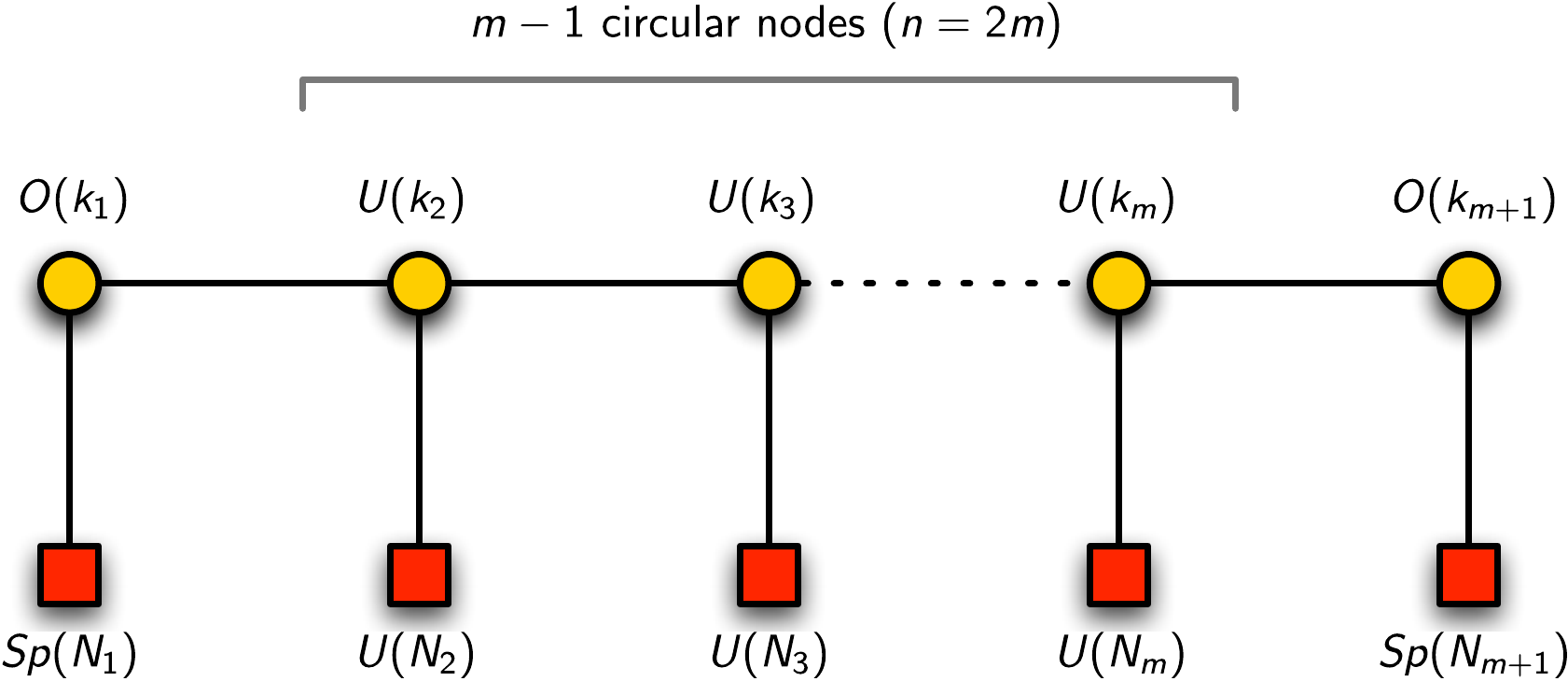}
\caption{The $S/S$ quiver for $Sp(N)$ instantons on $\mathbb{C}^2/\mathbb{Z}_{2m}$. Here, $N=N_1+ \ldots+N_{m}$.}
\label{fig:Sp_N_withVSevenn}
\end{center}
\end{figure}

If $k_1 =\cdots =k_{m+1} =k$, the quaternionic dimension of the Higgs branch of \fref{fig:Sp_N_withVSevenn} is
\bea
k \left(\sum_{i=1}^{m+1} N_i \right)+k^2+ (m-2) k^2+k^2 - (m-1) k^2 - 2\left[ \frac{1}{2}(k)(k-1)\right]
 = k(N+1)~,
\eea
where $N=\sum_{i=1}^{m+1} N_i $.  This is to be expected for $k$ $Sp(N)$ instantons on $\mathbb{C}^2/\mathbb{Z}_{2m}$.

\subsubsection{$Sp(1)$ instanton on $\BC^2/\BZ_4$: $\vec k=(1,1,1)$ and $\vec N=(1,0,0)$}
\begin{figure}[H]
\begin{center}
\includegraphics[scale=0.7]{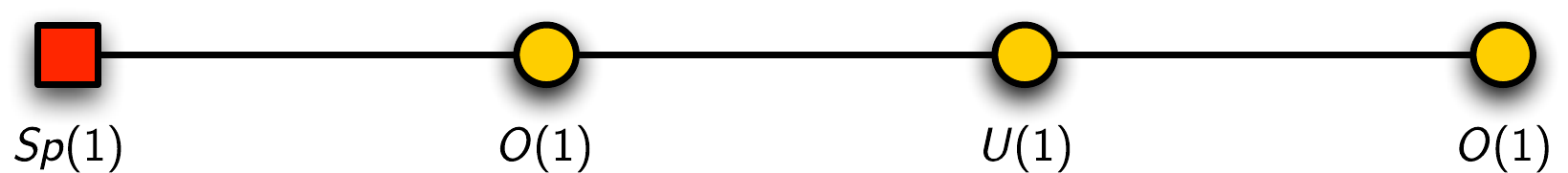}
\caption{The $S/S$ quiver for $Sp(1)$ instanton on $\BC^2/\BZ_4$: $\vec k=(1,1,1)$ and $\vec N=(1,0,0)$.}
\label{fig:C2Z4Sp_1_N100}
\end{center}
\end{figure}

The quiver of our question is depicted in \fref{fig:C2Z4Sp_1_N100}.
Let $z$ be the fugacity for the gauge group $U(1)$.  We take each gauge group $O(1)$ to be $\BZ_2$ and denote their discrete actions by $\omega_1$ and $\omega_2$, with $\omega_1^2= \omega_2^2=1$.  Let $a$ be the fugacity of the global $Sp(1)$.
The Hilbert series is given by
\bea
g^{(1,0,0)}_{(1,1,1)} (t, a)  &=\frac{1}{4} \sum_{\omega_1 = \pm1} \sum_{\omega_2 = \pm1}\oint_{|z|=1} \frac{\ud z}{(2 \pi i)z}  \frac{ \chi_{Q} (a, \omega_1) \chi_{X_1}(\omega_1, z)   \chi_{X_2} (z,\omega_2)}{\chi_{F}}~.
\eea
where the subscripts $Q$, $X_1$ and $X_2$ denote the contribution from the hypermultiplets in the bi-fundamental representation from left to right in \fref{fig:C2Z4Sp_1_N100}.  The function $\chi_{F}$ takes into account of the $F$-terms.  Explicitly,
\bea
\begin{array}{lll}
\chi_Q(a, \omega_1) &=  \PE \left[ t (\omega_1 a^{-1} + \omega_1^{-1} a)\right]   \\
\chi_{X_1} (\omega_1,z) &= \PE \left[ t \omega_1 z \right] \\
\chi_{X_2} (z, \omega_2) &= \PE \left[ t \omega_2 z \right] \\
\chi_{F} &= \PE \left[ t^2 \right] = (1-t^2)^{-1} ~.
\end{array}
\eea
After the integration and summation, the Hilbert series is given by
\bea
g^{(1,0,0)}_{(1,1,1)} (t, a)= (1-t^8) \PE[([2]_a+1) t^2 + t^4] = g_{\BC^2/\BZ_4}(t) \sum_{n=0}^\infty [2m]_a t^{2m}~.
\eea
The plethystic logarithm is given by
\bea
\PL [ g^{(1,0,0)}_{(1,1,1)} (t, a) ] = ([2]_a+1) t^2 + t^4 - t^8~.
\eea
The corresponding unrefined Hilbert series is 
\bea \label{HS:C2Z4Sp_1_N100}
g^{(1,0,0)}_{(1,1,1)} (t, 1) = \frac{1 + t^4}{(1 - t^2)^4}~.
\eea
Note that this Hilbert series agrees with that of $SU(2)$ instanton on $\BC^2/\BZ_4$ with $\vec k = (1, 1, 1, 1)$ and $\vec N = (0,0,0,2)$; see \eref{C2Z4SU2N0002} and \eref{C2Z4SU2N0002ur}.

\subsubsection{$Sp(1)$ instanton on $\BC^2/\BZ_4$: $\vec k=(1,1,1)$ and $\vec N=(0,1,0)$}
\begin{figure}[H]
\begin{center}
\includegraphics[scale=0.7]{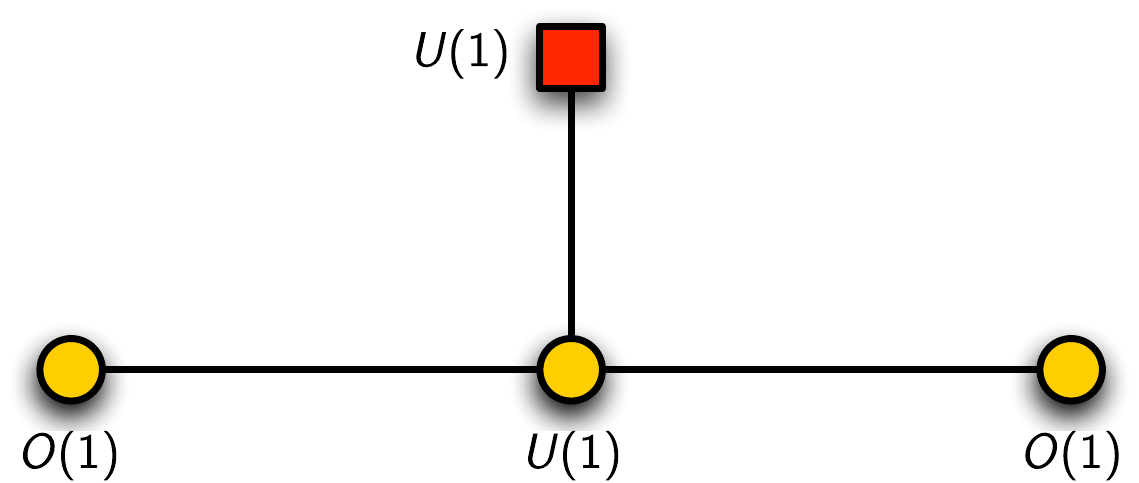}
\caption{The $S/S$ quiver for $Sp(1)$ instanton on $\BC^2/\BZ_4$: $\vec k=(1,1,1)$ and $\vec N=(0,1,0)$.}
\label{fig:C2Z4Sp_1_N010}
\end{center}
\end{figure}

The quiver of our question is depicted in \fref{fig:C2Z4Sp_1_N010}.
Let $z$ be the fugacity for the gauge group $U(1)$.  We take each gauge group $O(1)$ to be $\BZ_2$ and denote their discrete actions by $\omega_1$ and $\omega_2$, with $\omega_1^2= \omega_2^2=1$.  Let $a$ be the fugacity of the global $U(1)$.
The Hilbert series is given by
\bea
g^{(0,1,0)}_{(1,1,1)} (t, a)  &=\frac{1}{4} \sum_{\omega_1 = \pm1} \sum_{\omega_2 = \pm1}\oint_{|z|=1} \frac{\ud z}{(2 \pi i)z}  \frac{ \chi_Q(a, z)  \chi_{X_1}(\omega_1, z)   \chi_{X_2} (z,\omega_2)}{\chi_{F}}~.
\eea
where the subscripts $Q$, $X_1$ and $X_2$ denote the contribution from the hypermultiplets in the bi-fundamental representation from left to right in \fref{fig:C2Z4Sp_1_N100}.  The function $\chi_{F}$ takes into account of the $F$-terms.  Explicitly,
\bea
\begin{array}{lll}
\chi_Q(a, z) &=  \PE \left[ t (z a^{-1} + z^{-1} a)\right]   \\
\chi_{X_1} (\omega_1,z) &= \PE \left[ t \omega_1 z \right] \\
\chi_{X_2} (z, \omega_2) &= \PE \left[ t \omega_2 z \right] \\
\chi_{F} &= \PE \left[ t^2 \right] = (1-t^2)^{-1} ~.
\end{array}
\eea
After the integration and summation, the Hilbert series is given by
\bea
g^{(0,1,0)}_{(1,1,1)} (t, a)= 1+2t^2 + (2[2]_a+3)t^4+(4[2]_a+4)t^6 + (3[4]_a + 5[2]_a+5)t^8+\ldots~.
\eea
The plethystic logarithm is given by
\bea
\PL [ g^{(0,1,0)}_{(1,1,1)} (t, a) ] =2t^2 + (2[2]_a+3)t^4 - (2[2]_a +3) t^8 + \ldots~.
\eea
The corresponding unrefined Hilbert series is 
\bea \label{HS:C2Z4Sp_1_N010}
g^{(0,1,0)}_{(1,1,1)} (t, 1) &= \frac{1 + 4 t^4 + t^8}{(1 - t^2)^4 (1 + t^2)^2} \nn \\
&=1+2 t^2+9 t^4+16 t^6+35 t^8+54 t^{10}+\ldots~.
\eea
Note that this Hilbert series agrees with that of $SU(2)$ instanton on $\BC^2/\BZ_4$ with $\vec k=(1,1,1,1)$, $\vec N=(0,1,0,1)$.

\subsection{The $AA$ quiver for $SO(2N)$ instantons on $\BC^2/\BZ_{2m}$}
\begin{figure}[H]
\begin{center}
\includegraphics[scale=0.7]{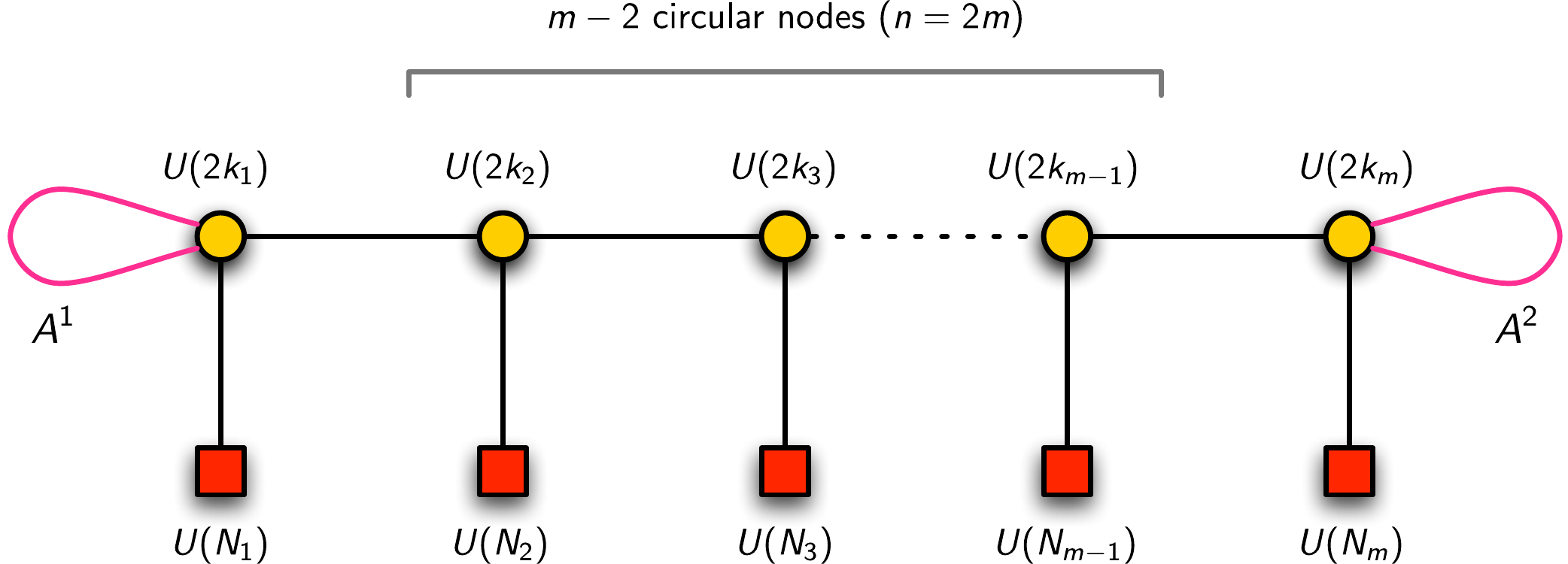}
\caption{The $AA$ quiver for $SO(2N)$ instantons on $\mathbb{C}^2/\mathbb{Z}_{2m}$.  Here, each of $k_1, \ldots, k_{m}$ can take either integral or half-integral value and $2N=2N_1+ \ldots+2N_{m}$.}
\label{fig:SO_N_withNOVSevenn}
\end{center}
\end{figure}

If $k_1 =\cdots =k_{m+1} =k$, the quaternionic dimension of the Higgs branch of \fref{fig:SO_N_withNOVSevenn} is
\bea
k \left(2\sum_{i=1}^{m+1} N_i \right)+(m-1) (4k^2)  + 2\left[ \frac{1}{2}(2k)(2k-1)\right]  - m(4 k^2) 
 = k(2N-2)~,
\eea
where $2N=2\sum_{i=1}^{m+1} N_i $.  This is to be expected for $k$ $SO(2N)$ instantons on $\mathbb{C}^2/\mathbb{Z}_{2m}$, for $N\geq 2$.

\subsubsection{$SO(6)$ instanton on $\BC^2/\BZ_4$: $\vec k=(1,1)$ and $\vec N=(2,1)$}
\begin{figure}[H]
\begin{center}
\includegraphics[scale=0.7]{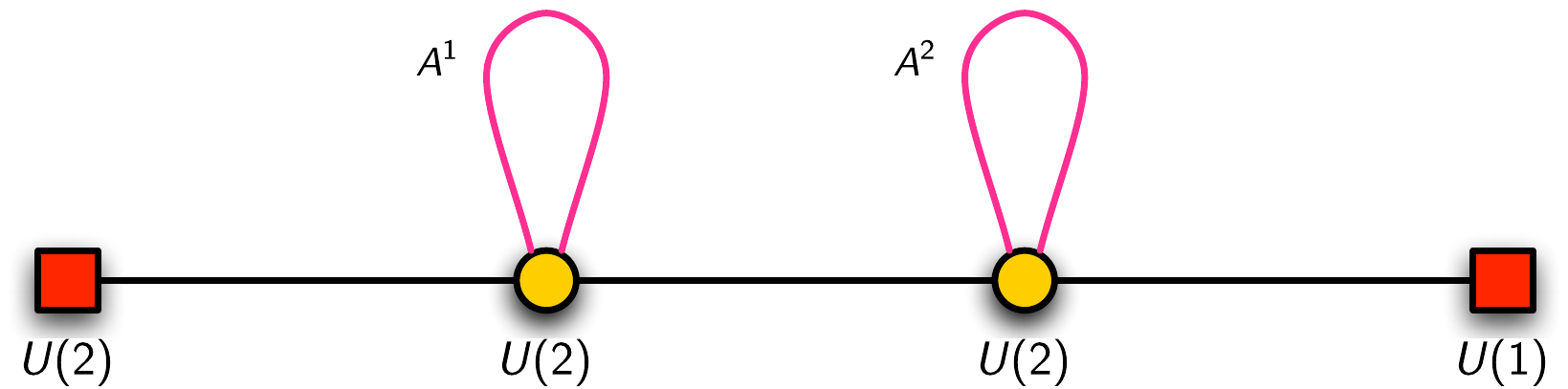}
\caption{The $AA$ quiver for $SO(6)$ instanton on $\BC^2/\BZ_4$: $\vec k=(1,1)$ and $\vec N=(2,1)$.}
\label{fig:C2Z4SO_6_N21}
\end{center}
\end{figure}

The quiver of our question is depicted in \fref{fig:C2Z4SO_6_N21}.
Let $(q_1,z_1)$ and $(q_2,z_2)$ be the fugacities for the gauge groups $U(2)=U(1)\times SU(2)$.  Let $(u,a)$ and $b$ be the fugacities of the global symmetries $U(2)$ and $U(1)$.

The Hilbert series is given by
\bea
&g^{(2,1)}_{(1,1)} (t, u, a,b)  =\oint_{|q_1|=1} \frac{\ud q_1}{(2 \pi i)q_1} \oint_{|q_2|=1} \frac{\ud q_2}{(2 \pi i)q_2}  \oint_{|z_1|=1} \frac{(1-z_1^2) \ud z_1}{(2 \pi i)z_1} \oint_{|z_2|=1} \frac{(1-z_2^2) \ud z_2}{(2 \pi i)z_2} \times \nn \\
&\qquad \frac{ \chi_{Q_1} (u,b,q_1,z_1) \chi_{X} (q_1,z_1,q_2,z_2)\chi_{Q_2} (q_2,z_2,b) \chi_{A_1}(q_1)  \chi_{A_2}(q_2) }{\chi_{F}(z_1,z_2)}~,
\eea
where the subscripts $Q_1$, $X$, $Q_2$ denote the contribution from the hypermultiplets in the bi-fundamental representation from left to right in \fref{fig:C2Z4SO_6_N21}, and the subscripts $A^1$ and $A^2$ denote the hypermultiplets in the antisymmetric representation of each $U(2)$ gauge group.  The function $\chi_{F}$ takes into account of the $F$-terms.  Explicitly,
\bea
\begin{array}{lll}
 \chi_{Q_1} (u,b,q_1,z_1) &=  \PE \left[ t [1]_a (u q_1^{-1} +u^{-1} q_1) [1]_{z_1}\right]   \\
\chi_{X} (q_1,z_1,q_2,z_2)&= \PE \left[ t [1]_{z_1} (q_1 q_2^{-1}+ q_1^{-1} q_2) [1]_{z_2}  \right] \\
\chi_{Q_2} (q_2,z_2,b) &=  \PE \left[ t [1]_{z_2} (b q_2^{-1} +b^{-1} q_2)\right] & \\
\chi_{A^1}(q_1) &= \PE \left[ t(q_1^2 +q_1^{-2}) \right] \\
\chi_{A^2}(q_2) &=  \PE \left[ t(q_2^2 + q_2^{-2}) \right] \\
\chi_{F}(z_1,z_2) &= \PE \left[ t^2 ([2]_{z_1}+1)+ t^2 ([2]_{z_2}+1)\right]
\end{array}
\eea
After computing the integrals, we obtain
\bea
g^{(2,1)}_{(1,1)} (t, u, a,b)  = 1+([2]_a +3) t^2 + \left \{ (u b^{-1}+u^{-1}b) [1]_a + (u^2+u^{-2}) \right\} t^3 +\ldots~.
\eea
The plethystic logarithm of this Hilbert series is
\bea
\PL \left[g^{(2,1)}_{(1,1)} (t, u, a,b)  \right] &= ([2]_a +3) t^2 + \left \{ (u b^{-1}+u^{-1}b) [1]_a + (u^2+u^{-2}) \right\} t^3   \nn \\
&\quad +\left \{ 2(b u + b^{-1} u^{-1}) [1]_a+1 \right \} t^4 -\left \{ 3(b u + b^{-1} u^{-1}) [1]_a+[2]_a +2 \right \} t^4 \nn \\
&\quad - \ldots~.
\eea
The corresponding unrefined Hilbert series is
\bea \label{HS:C2Z4SO_6_N21}
g^{(2,1)}_{(1,1)} (t, 1,1,1)  & = \frac{1}{(1 - t)^8 (1 + t)^6 (1 + t^2)^3 (1 + t + t^2)^3 (1 + t + t^2 + t^3 + 
   t^4)^2} \Big( 1 + 3 t  \nn \\
   & \quad + 9 t^2 + 22 t^3 + 54 t^4 + 114 t^5 + 219 t^6 + 371 t^7 + 
 582 t^8 + 827 t^9  \nn \\
 & \quad + 1092 t^{10}+1323 t^{11}+1493 t^{12}+1548 t^{13}+1493 t^{14}+ \text{palindrome} +t^{26} \Big) \nn \\
&= 1 + 6 t^2 + 6 t^3 + 30 t^4 + 38 t^5 + 114 t^6 + 158 t^7 + 369 t^8+\ldots~. 
 \eea 
This Hilbert series agrees with that of $SU(4)$ instanton on $\BC^2/\BZ_4$ with $\vec k=(1,1,1,1)$, $\vec N=(2,1,1,0)$.  In both cases, the symmetry of the instanton moduli space is $S(U(2) \times U(1) \times U(1))$.

\subsection{The $SS$ quiver for $Sp(N)$ instantons on $\BC^2/\BZ_{2m}$}
\begin{figure}[H]
\begin{center}
\includegraphics[scale=0.7]{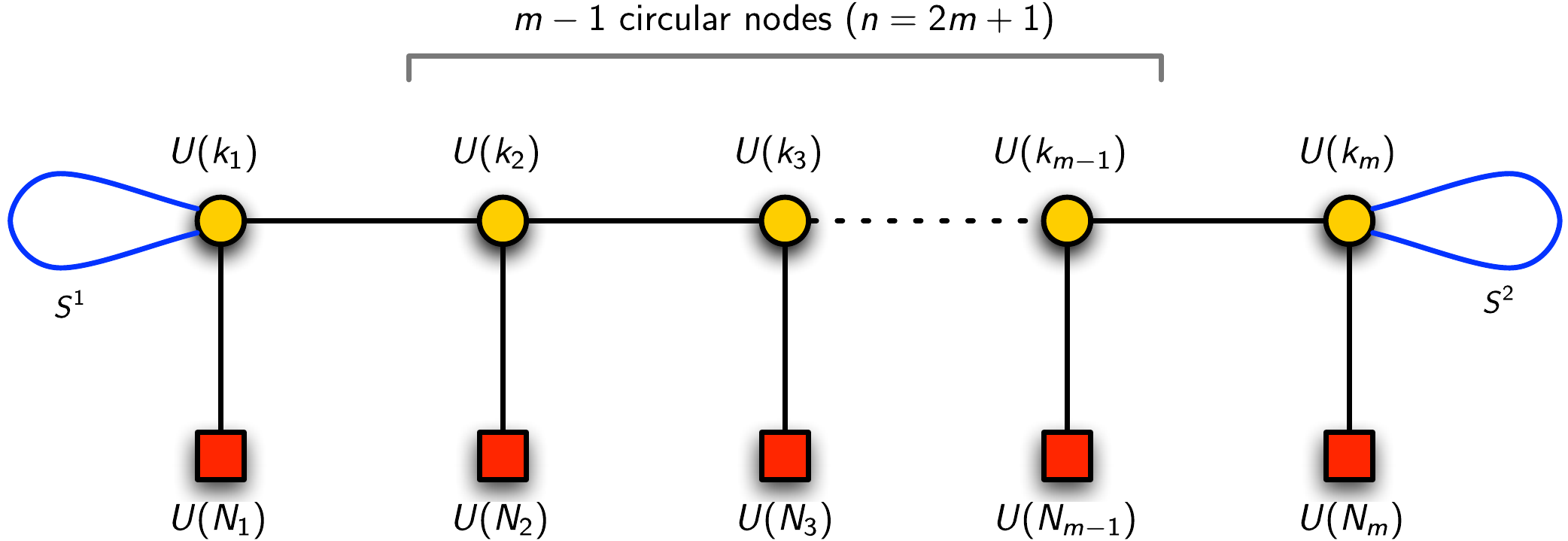}
\caption{The $SS$ quiver for $Sp(N)$ instantons on $\mathbb{C}^2/\mathbb{Z}_{2m}$. Here, $N=2N_1+ \ldots+N_{m}$.}
\label{fig:Sp_N_withNOVSevenn}
\end{center}
\end{figure}
If $k_1 =\cdots =k_{m+1} =k$, the quaternionic dimension of the Higgs branch of \fref{fig:SO_N_withNOVSevenn} is
\bea
k \left(\sum_{i=1}^{m+1} N_i \right)+(m-1) k^2 + 2\left[ \frac{1}{2}(k)(k+1)\right] - m k^2 
 = k(N+1)~,
\eea
where $N=\sum_{i=1}^{m+1} N_i $.  This is to be expected for $k$ $Sp(N)$ instantons on $\mathbb{C}^2/\mathbb{Z}_{2m}$.

\subsubsection{$Sp(1)$ instanton on $\BC^2/\BZ_4$: $\vec k=(1,1)$ and $\vec N=(1,0)$}
\begin{figure}[H]
\begin{center}
\includegraphics[scale=0.7]{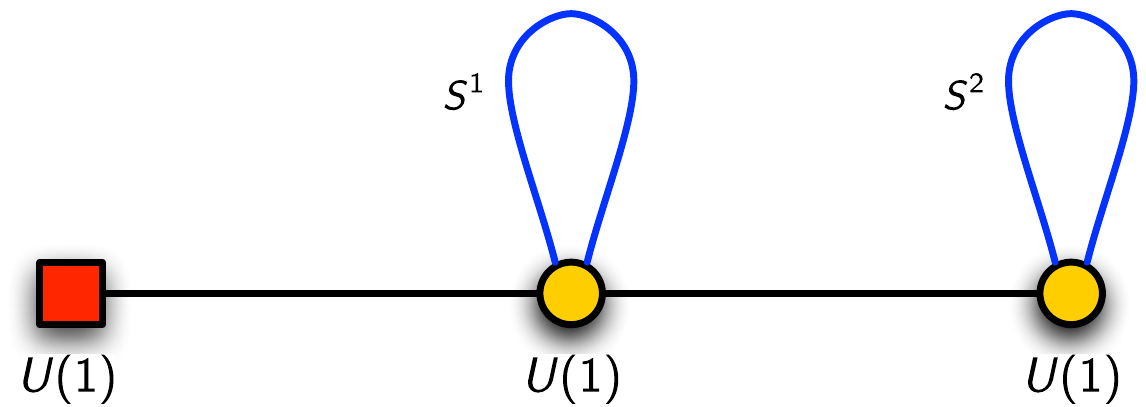}
\caption{The $SS$ quiver for $Sp(1)$ instanton on $\BC^2/\BZ_4$: $\vec k=(1,1)$ and $\vec N=(1,0)$.}
\label{fig:C2Z4Sp_1_N10}
\end{center}
\end{figure}

The quiver of our question is depicted in \fref{fig:C2Z4Sp_1_N10}.
Let $z_1$ and $z_2$ be the fugacities for the gauge groups $U(1)$.  Let $a$ be the fugacity of the global symmetry $U(1)$.  The Hilbert series is given by
\bea
&g^{(1,0)}_{(1,1)} (t, a)  =\oint_{|q_1|=1} \frac{\ud z_1}{(2 \pi i)z_1} \oint_{|z_2|=1} \frac{\ud z_2}{(2 \pi i)z_2}  
 \frac{ \chi_{Q} (a,z_1) \chi_{X} (z_1,z_2)\chi_{S^1}(z_1)  \chi_{S^2}(z_2) }{\chi_{F}}~,
\eea
where the subscripts $Q$, $X$ denote the contribution from the hypermultiplets in the bi-fundamental representation from left to right in \fref{fig:C2Z4Sp_1_N10}, and the subscripts $S^1$ and $S^2$ denote the hypermultiplets in the symmetric representation of each $U(2)$ gauge group.  The function $\chi_{F}$ takes into account of the $F$-terms.  Explicitly,
\bea
\begin{array}{lll}
 \chi_{Q} (a,z_1) &=  \PE \left[ t  (a z_1^{-1} +a^{-1} z_1) \right]   \\
\chi_{X} (z_1,z_2)&= \PE \left[ t (z_1 z_2^{-1}+ z_1^{-1} z_2)  \right] \\
\chi_{S^1}(z_1) &= \PE \left[ t(z_1^2 +z_1^{-2}) \right] \\
\chi_{S^2}(z_2) &=  \PE \left[ t(z_2^2 + z_2^{-2}) \right] \\
\chi_{F} &= \PE \left[ 2t^2 \right]
\end{array}
\eea
The corresponding unrefined Hilbert series is
\bea \label{HS:C2Z4Sp_1_N10}
g^{(1,0)}_{(1,1)} (t, 1) &=\frac{1-t+2 t^3-t^5+t^6}{(1-t)^4 (1+t)^2 \left(1+t+t^2+t^3+t^4\right)} \nn\\
&= 1 + 2 t^2 + 2 t^3 + 5 t^4 + 6 t^5 + 10 t^6 + 12 t^7 + 19 t^8+\ldots~.
\eea
This Hilbert series agrees with that of $SU(2)$ instanton on $\BC^2/\BZ_4$ with $\vec k=(1,1,1,1)$, $\vec N=(1,1,0,0)$.  In both cases, the symmetry of the instanton moduli space is $S(U(1) \times U(1))$.

\subsubsection{$Sp(2)$ instanton on $\BC^2/\BZ_4$: $\vec k=(1,1)$ and $\vec N=(1,1)$}
\begin{figure}[H]
\begin{center}
\includegraphics[scale=0.7]{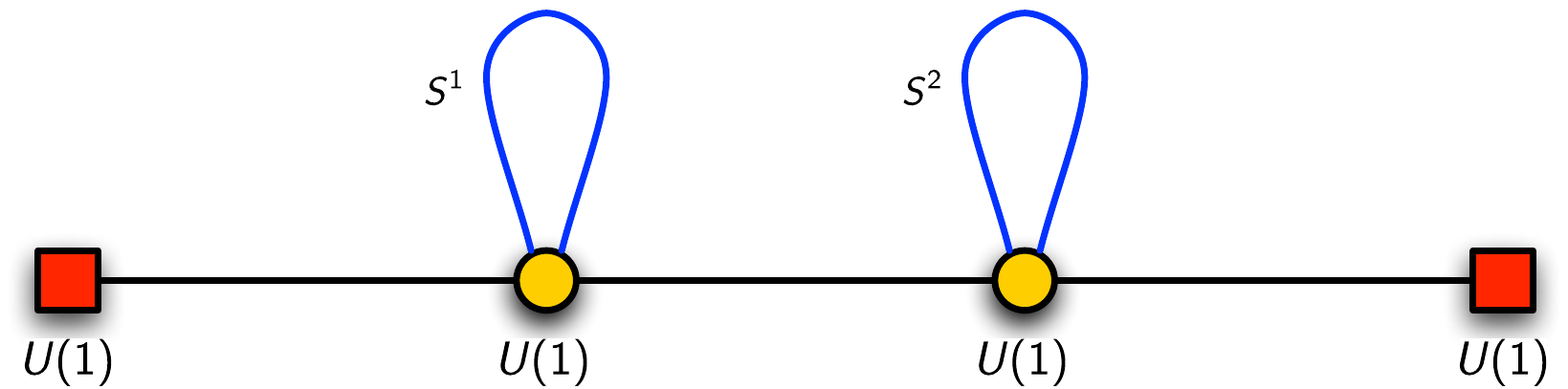}
\caption{The $SS$ quiver for $Sp(2)$ instanton on $\BC^2/\BZ_4$: $\vec k=(1,1)$ and $\vec N=(1,1)$.}
\label{fig:C2Z4Sp_2_N11}
\end{center}
\end{figure}

The quiver of our question is depicted in \fref{fig:C2Z4Sp_2_N11}.
Let $z_1$ and $z_2$ be the fugacities for the gauge groups $U(1)$.  Let $a$ and $b$ be the fugacity of the global symmetries $U(1)$, left and right square nodes in the quiver. The Hilbert series is given by
\bea
&g^{(1,1)}_{(1,1)} (t, a,b)  =\oint_{|q_1|=1} \frac{\ud z_1}{(2 \pi i)z_1} \oint_{|z_2|=1} \frac{\ud z_2}{(2 \pi i)z_2}  
 \frac{ \chi_{Q_1} (a,z_1) \chi_{X} (z_1,z_2)\chi_{S^1}(z_1)  \chi_{S^2}(z_2)  \chi_{Q_2} (a,z_1)}{\chi_{F}}~,
\eea
where the subscripts $Q_1$, $X$ and $Q_2$ denote the contribution from the hypermultiplets in the bi-fundamental representation from left to right in \fref{fig:C2Z4Sp_2_N11}, and the subscripts $S^1$ and $S^2$ denote the hypermultiplets in the symmetric representation of each $U(2)$ gauge group.  The function $\chi_{F}$ takes into account of the $F$-terms.  Explicitly,
\bea
\begin{array}{lll}
 \chi_{Q} (a,z_1) &=  \PE \left[ t  (a z_1^{-1} +a^{-1} z_1) \right]   \\
\chi_{X} (z_1,z_2)&= \PE \left[ t (z_1 z_2^{-1}+ z_1^{-1} z_2)  \right] \\
\chi_{S^1}(z_1) &= \PE \left[ t(z_1^2 +z_1^{-2}) \right] \\
\chi_{S^2}(z_2) &=  \PE \left[ t(z_2^2 + z_2^{-2}) \right] \\
 \chi_{Q} (b,z_2) &=  \PE \left[ t  (b z_2^{-1} +b^{-1} z_2) \right]   \\
\chi_{F} &= \PE \left[ 2t^2 \right]
\end{array}
\eea
The corresponding Hilbert series is
\bea \label{C2Z4Sp_2_N11}
g^{(1,1)}_{(1,1)} (t, a,b)  &= \frac{1}{(1- t)^6 (1 + t)^2 (1 + t^2) (1 + t + t^2)^3 (1 + t + t^2 + t^3 + t^4)} \Big(1+2 t^2+3 t^3+ \nn \\
& \qquad 8 t^4+11 t^5+13 t^6+12 t^7+13 t^8+11 t^9+8 t^{10}+3 t^{11}+2 t^{12}+t^{14} \Big) \nn \\
&= 1 + 3 t^2 + 6 t^3 + 12 t^4 + 24 t^5 + 42 t^6 + 68 t^7 + 115 t^8+\ldots~.
\eea
This Hilbert series agrees with that of $SO(5)$ instanton on $\BC^2/\BZ_4$ with the $O/O$ quiver: $\vec k=(1,1,1)$ and $\vec N=(1,1,2)$; see \eref{C2Z4SO_5_N112}.
\subsection{Summary: Matching of Hilbert series for instantons on $\BC^2/\BZ_n$ ($n \geq 3$)}
We summarise in \tref{tab:match1} matching of Hilbert series for equivalent instantons of isomorphic groups on $\BC^2/\BZ_n$ but very different quiver description, as presented in the preceding subsections.
\ei
\begin{center}
\begin{longtable}{|c|c|c|}
\hline
Quiver & Matches with & Hilbert series\\
\hline
\fref{fig:C2Z3SO_2_wVS}: $SO(2)$ instanton on $\BC^2/\BZ_3$ & $\BC^2/\BZ_3$ & \eref{C2Z3SO_2_N02HS}, \eref{C2Z3HS}
\\ $\vec k =(1,1)$ and $\vec N=(0,2)$& & \\
\hline 
\fref{fig:C2Z3SO_5_wVS}: $SO(5)$ instanton on $\BC^2/\BZ_3$ & \fref{fig:C2Z3Sp_2_N20.pdf}: $Sp(2)$ instanton on $\BC^2/\BZ_3$ & \eref{SO51311}, \eref{Sp22011}
\\ $\vec k =(1,1)$ and $\vec N=(1,3)$ & $\vec k=(1,1)$, $\vec N=(2,0)$ & \\
\hline
\fref{fig:C2Z3SO_5_N21}: $SO(5)$ instanton on $\BC^2/\BZ_3$ & \fref{fig:C2Z3Sp_2_N11.pdf}: $Sp(2)$ instanton on $\BC^2/\BZ_3$ & \eref{C2Z3SO5N21HS}, \eref{HS:C2Z3Sp2N11} \\
$\vec k =(1,1)$ and $\vec N=(2,1)$ &$\vec k=(1,1)$, $\vec N=(1,1)$ &  \\
\hline
\fref{fig:C2Z3Sp_1_N01.pdf}: $Sp(1)$ instanton on $\BC^2/\BZ_3$ &$SU(2)$ instanton on $\BC^2/\BZ_3$ & \eref{HS:C2Z3Sp_1_N01.pdf}, \eref{HS200111unref} \\
$\vec k =(1,1)$ and $\vec N=(0,1)$ & $\vec k=(1,1,1)$ and $\vec N=(2,0,0)$ & \\
\hline
\fref{fig:C2Z3Sp_1_N10.pdf}: $Sp(1)$ instanton on $\BC^2/\BZ_3$ &$SU(2)$ instanton on $\BC^2/\BZ_3$ & \eref{HS:C2Z3Sp_1_N10.pdf} \\
$ \vec k =(1,1)$ and $\vec N=(1,0)$ & $\vec k=(1,1,1)$ and $\vec N=(1,1,0)$ & \\
\hline \hline
\fref{fig:C2Z4SO_5_N112}: $SO(5)$ instanton on $\BC^2/\BZ_4$ & \fref{fig:C2Z4Sp_2_N11}: $Sp(2)$ instanton on $\BC^2/\BZ_4$  & \eref{C2Z4SO_5_N112}, \eref{C2Z4Sp_2_N11} \\
$O/O$:  $\vec k=(1,1,1)$ and $\vec N=(1,1,2)$ & $SS$: $\vec k=(1,1)$, $\vec N=(1,1)$ & \\
\hline
\fref{fig:C2Z4SO_6_N204}: $SO(6)$ instanton on $\BC^2/\BZ_4$ & $SU(4)$ instantons on $\BC^2/\BZ_4$ & \eref{HS:C2Z4SO_6_N204}\\
$O/O$:  $\vec k=(1,1,1)$ and $\vec N=(2,0,4)$ & $\vec k=(1,1,1,1)$ and $\vec N=(2,0,2,0)$ & \\
\hline
\fref{fig:C2Z4SO_6_N212}: $SO(6)$ instanton on $\BC^2/\BZ_4$ & $SU(4)$ instantons on $\BC^2/\BZ_4$ & \eref{HS:C2Z4SO_6_N212}, \eref{HS:C2Z4SU(4)N1111}\\
$O/O$:  $\vec k=(1,1,1)$ and $\vec N=(2,1,2)$ & $\vec k=(1,1,1,1)$ and $\vec N=(1,1,1,1)$ & \\
\hline
\fref{fig:C2Z4Sp_1_N100}: $Sp(1)$ instanton on $\BC^2/\BZ_4$ & $SU(2)$ instanton on $\BC^2/\BZ_4$ & \eref{HS:C2Z4Sp_1_N100}, \eref{C2Z4SU2N0002ur}\\
$S/S$:  $\vec k=(1,1,1)$ and $\vec N=(1,0,0)$ & $\vec k=(1,1,1,1)$ and $\vec N=(0,0,0,2)$ & \\
\hline
\fref{fig:C2Z4Sp_1_N010}: $Sp(1)$ instanton on $\BC^2/\BZ_4$ & $SU(2)$ instanton on $\BC^2/\BZ_4$ & \eref{HS:C2Z4Sp_1_N010} \\
$S/S$:  $\vec k=(1,1,1)$ and $\vec N=(0,1,0)$ & $\vec k=(1,1,1,1)$ and $\vec N=(0,1,0,1)$ & \\
\hline \hline
\fref{fig:C2Z4SO_6_N21}: $SO(6)$ instanton on $\BC^2/\BZ_4$ & $SU(4)$ instantons on $\BC^2/\BZ_4$ & \eref{HS:C2Z4SO_6_N21}\\
$AA$:  $\vec k=(1,1)$ and $\vec N=(2,1)$ & $\vec k=(1,1,1,1)$ and $\vec N=(2,1,1,0)$ & \\
\hline
\fref{fig:C2Z4Sp_1_N10}: $Sp(1)$ instanton on $\BC^2/\BZ_4$ & $SU(2)$ instantons on $\BC^2/\BZ_4$ & \eref{HS:C2Z4Sp_1_N10} \\
$SS$:  $\vec k=(1,1)$ and $\vec N=(1,0)$ & $\vec k=(1,1,1,1)$ and $\vec N=(1,1,0,0)$ & \\
\hline
\caption{Matching between the Hilbert series of the hypermultiplet moduli spaces of different quivers.}
\label{tab:match1}
\end{longtable}%
\end{center}

\section{$SO(N)$ and $Sp(N)$ instantons on $\mathbb{C}^2/\mathbb{Z}_2$} \label{sec:SONC2Z2}
In this section, we focus on the special and degenerate case $n=2$ (\ie~$SO(N)$ and $Sp(N)$ instantons on $\BC^2/\BZ_2$) of the quivers presented in the previous section. 

\subsection{The $O/O$ quiver for $SO(N)$ instantons on $\mathbb{C}^2/\mathbb{Z}_2$}
\begin{figure}[H]
\begin{center}
\includegraphics[scale=0.7]{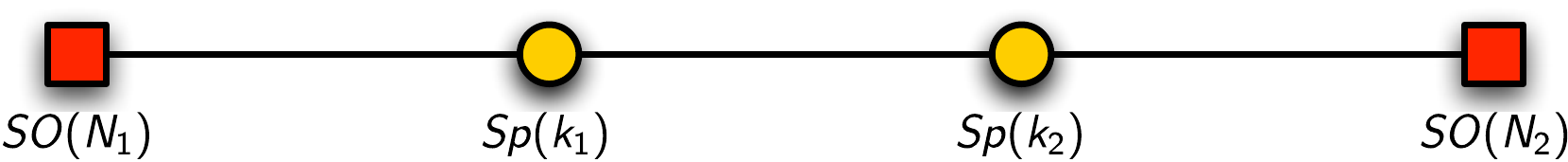}
\caption{The $O/O$ quiver for $SO(N)$ instantons on $\mathbb{C}^2/\mathbb{Z}_2$.  Here, $N_1+N_2=N$. The line between $Sp(k_1)$ and $Sp(k_2)$ gauge groups denote $4k_1k_2$ hypermultiplets (whose scalar components have $8k_1k_2$ complex degrees of freedom), and each line connecting the square node and the circular node denotes $2kN_1$ and $2kN_2$ half-hypermultiplets respectively.}
\label{fig:oneO2NVS}
\end{center}
\end{figure}

This class of theories contains $Sp(k_1)\times Sp(k_2)$ gauge groups, with hypermultiplets in the bi-fundamental representation of $Sp(k_1) \times Sp(k_2)$ and $N$ flavours fundamental half-hypermultiplets distributed among the two gauge group factors. We can represent the distribution of fundamental half-hypermultiplets by $\vec N= (N_1,N_2)$ such that $N_1+N_2=N$, where $N_1$ and $N_2$ are integers.   The 4d $\CN=2$ quiver diagram is depicted in \fref{fig:oneO2NVS}.

\noindent If $k_1 =k_2 =k$, the quaternionic dimension of the Higgs branch of this quiver is
\bea
k N_1 +4k^2 +kN_2 - \frac{1}{2}(2k)(2k+1)- \frac{1}{2}(2k)(2k+1) = k(N_1+N_2-2) = k(N-2)~.
\eea

\subsubsection*{Chiral fields}

Let us denote the $Sp(k_1)$ and $Sp(k_2)$ fundamental indices by $a_1,a_2, \ldots=1,2,..,2k_1$ and $b_1,b_2, \ldots=1,2, \ldots, 2k_2$, respectively. The indices $i_1,j_1$ and $j_1,j_2$ correspond to fundamental representations of $SO(N_1)$ and $SO(N_2)$ respectively, \ie~ $i_1,i_2, \ldots=1,2, \ldots,N_1$ and $j_1, j_2, \ldots=1,2, \ldots,N_2$.   

It is convenient to use $\CN=1$ language in the subsequent computation.  For each of the gauge groups $Sp(k_1)$ and $Sp(k_2)$, there are respectively adjoint scalars in the vector multiplet, denoted by $(S_1)_{a_1 a_2}$ and $(S_2)_{b_1b_2}$.   The bi-fundamental hypermultiplets of $Sp(k_1) \times Sp(k_2)$ contains two chiral multiplets $\{ (X_{12})^{a_1}_{~b_1}, (X_{21})^{b_1}_{~a_1} \}$ transforming as a doublet under a global $SU(2)$ symmetry; let us refer to this as $SU(2)_x$ and denote its fundamental indices by $\alpha, \beta, \alpha_1, \alpha_2, \ldots = 1,2$.  We can therefore denote these chiral multiplets by 
\bea (X^\alpha)^{a_1}_{~b_1} :=\{ (X_{12})^{a_1}_{~b_1}, (X_{21})^{b_1}_{~a_1} \}~. \label{defXSp}
\eea  
The bi-fundamental chiral multiplets under $Sp(k)_1 \times O(2N_1)$ and $Sp(k)_2 \times O(2N_2)$ are respectively denoted by $Q^{i_1}_{a_1}$ and $q^ {j_1}_{b_1}$.  

Since the theory has $\CN=2$ symmetry in four dimensions, it possesses the $R$ symmetry $SU(2)$, under which each chiral multiplet and its complex conjugate transform as a doublet.   In $\CN=1$ language, only one generator of the $SU(2)_R$ symmetry is manifest; let us denote it by $U(1)_t$.  Each chiral multiplet in $X_{a_1 b_1}^\alpha,~ Q^{i_1}_{a_1}, ~ q^{j_1}_{b_1}$ carries charge $+1$ under this $U(1)_t$ symmetry.   On the other hand, since $(S_1)_{a_1 a_2}$ and $(S_2)_{b_1b_2}$ are singlets under $SU(2)_R$ symmetry, they are neutral under $U(1)_t$.

\subsubsection*{Superpotential}
The superpotential of a quiver in this class is
\bea \label{sup:O2NwVS}
W &=   (X_{21})^{b_1}_{~a_1} (S_1)^{a_1}_{~a_2} (X_{12})^{a_2}_{~b_1}- (X_{12})^{a_1}_{~b_1} (S_2)^{b_1}_{~b_2} (X_{21})^{b_2}_{~a_1} \nn \\
& \quad +M^{SO(N_1)}_{i_1 i_2} J^{a_1a_2} J^{a_3 a_4} Q^{i_1}_{a_1} (S_1)_{a_2a_3} Q^{i_2}_{a_4}+M^{SO(N_2)}_{j_1j_2} J^{b_1b_2} J^{b_3 b_4} q^{j_1}_{b_1} (S_2)_{b_2 b_3} q^{j_2}_{b_4} ~,
\eea
where $J^{a_1 a_2}$ and $J^{b_1 b_2}$ is the $2k \times 2k$ symplectic matrix for $Sp(k_1)$ and $Sp(k_2)$:
\bea
J^{ab} = \begin{pmatrix} 0 & {\bf 1}_{m \times m} \\ - {\bf 1}_{m \times m} & 0 \end{pmatrix} \quad \text{for $Sp(m)$}~,
\eea
the indices $a_1,a_2$ and $b_1,b_2$ are raised and lowered by appropriate symplectic matrices, and the symmetric matrices $M^{SO(N_1)}_{i_1 i_2}$ and $M^{SO(N_2)}_{j_1j_2}$ define a quadratic form on the Lie algebra of $SO(N_1)$ and $SO(N_2)$.  

Note that $M^{SO(m)}$ can be taken to be an identity matrix, in which case the Lie algebra consists of anti-symmetric matrices and there is no non-zero diagonal matrix in the Lie algebra.  However, subsequently we shall make use of the diagonal charges from the Cartan subalgebra of $SO(N_1)$ and $SO(N_2)$, $M^{SO(m)}$ should be taken as
\bea \label{SOmquad}
M^{SO(m)}_{ij}= \begin{cases} \begin{pmatrix} 0 & I_d \\ I_d & 0 \end{pmatrix} &\quad \text{for $m=2d$}~,\\  
\begin{pmatrix} 0 & I_d & 0 \\ I_d & 0 &0 \\ 0 & 0 & 1 \end{pmatrix} &\quad \text{for $m=2d+1$}~,
\end{cases}
\eea
where $I_d$ is an $d \times d$ identity matrix.

On the Higgs branch, the vacuum expectation values of $(S_1)_{a_1 a_2}$ and $(S_2)_{b_1b_2}$ are zero.  Therefore, the relevant $F$ terms are
\bea
0=(\CF_1)^{a_2}_{~a_1} &:= \partial_{(S_1)^{a_1}_{~a_2}} W = (X_{12})^{a_2}_{~b_1} (X_{21})^{b_1}_{~a_1} + M^{SO(N_1)}_{i_1 i_2} Q^{i_1 a_2} Q^{i_2}_{a_1}~,  \label{Fterm1} \\
0=(\CF_2)^{b_2}_{~b_1} &:= \partial_{(S_2)^{b_1}_{~b_2}} W =- (X_{21})^{b_2}_{~a_1} (X_{12})^{a_1}_{~b_1} + M^{SO(N_2)}_{j_1 j_2} q^{j_1 b_2} q^{j_2}_{b_1}~. \label{Fterm2}
\eea

\subsubsection*{Symmetry of the Higgs branch}

In summary, the symmetry of the Higgs branch of the theory is $U(1)_t \times SU(2)_x \times SO(N_1) \times SO(N_2)$.  The transformation properties of the various chiral multiplets in the theory, including the $F$-term constraints are summarised in the table below:
\begin{equation}
\begin{array}{c ||  c c| c c  c c  }
& Sp(k_1) & Sp(k_2) & U(1)_t & SU(2)_x  & SO(N_1) & SO(N_2)  \\ \hline
 X_{ab_1}^\alpha&   \Box                      &   \Box  &   1  &   \Box             &    \textbf{1} &  \textbf{1}  \\
 (S_1)_{a_1 a_2}      &    \mathbf{Adj}      &   \textbf{1}     &   0   &   \textbf{1}     &  \textbf{1}   & \textbf{1}    \\
(S_2)_{b_1b_2}     & \textbf{1}               & \mathbf{Adj}  &   0  &   \textbf{1}     &    \textbf{1} &  \textbf{1}  \\
 Q^{i}_a           &   \Box                  &   \textbf{1}  &   1    &   \textbf{1}          &  {\Box}  &  \textbf{1}  \\
q^ {j_1}_{b_1}       &    \textbf{1}            &  \Box            &   1   &    \textbf{1}          &  \textbf{1}    &  {\Box} \\ 
\hline
\mathcal{F}_1      &   \mathbf{Adj}   &   \textbf{1}             &   2   &   \textbf{1}        &  \textbf{1}   &   \textbf{1} \\
  \mathcal{F}_2     &  \textbf{1}             & \mathbf{Adj}        &   2   &    \textbf{1}   &   \textbf{1}  &   \textbf{1}  \\
 \end{array}
 \label{Symm_VS}
 \end{equation}

For a Higgs branch Hilbert Series, the chiral multiplets which contribute are $X_{ab}^i$, $Q^{(1)\alpha}_a $ and $Q^ {(2)\beta}_a $, in addition to the $F$-term constraints. In the following examples, we consider the special case of instanton number $k=1$.

\subsubsection{$SO(2)$ instanton on $\BC^2/\BZ_2$: $\vec k =(1,1)$ and $\vec N= (0,2)$}
It should be pointed out that the space of solutions of the $F$ terms \eref{Fterm1} and \eref{Fterm2}, also known as the $F$-flat space, contain more than one branch and hence can be decomposed further.  The primary decomposition (using {\tt Macaulay2} \cite{mac2} or {\tt STRINGVACUA} \cite{Gray:2008zs}) reveals that the relevant branch of the $F$-flat space are given by the relations:
\bea
Q^1_2 =0 , \quad Q^2_2 = 0, \quad \partial_{S_1} W =0 , \quad \partial_{S_2} W =0~. \label{rel:C2Z2SO2N02}
\eea

Let $x$ be the fugacity of the global symmetry $SU(2)_x$ and $y$ be the fugacity of the global symmetry $SO(2)$.
Using {\tt Macaulay2} \cite{mac2}, we obtain the $F$-flat Hilbert series as
\bea
\CF^\flat(t; z_1,z_2; x,y) &= \Big(1-t^2([2]_{z_1} +[2]_{z_2})+t^3[1]_x[1]_{z_1}[1]_{z_2}-t^4[2]_x \Big) \times \nn \\
& \quad \PE \left[ t [1]_x [1]_{z_1} [1]_{z_2} + t [1]_{z_1} y^{-1}\right]
\eea
Note that the $F$-flat space is $7$ complex dimensional and the corresponding Hilbert series is
\bea
\CF^\flat(t; 1,1; 1,1) &= \frac{1+3 t}{(1-t)^7}~.
\eea
Integrating over the gauge group of the corresponding refined $F$-flat Hilbert series gives
\bea
g^{(0,2)}_{(1,1)}(t,x,y) &= \oint_{|z_1|=1} \frac{1-z_1^2}{(2 \pi i)z_1} \oint_{|z_2|=1} \frac{1-z_2^2}{(2 \pi i)z_2}  \CF^\flat(t; z_1,z_2; x,y)  \nn \\
&= \sum^{\infty}_{n=0} [2n]_x t^{2n}~. \label{C2Z2SO2N02VS}
\eea
This is indeed the Hilbert series of $\BC^2/\BZ_2$, as should be expected for the moduli space of one $SO(2)$ instanton on $\BC^2/\BZ_2$.

The generators in $[2]$ of $SU(2)_x$ can be written as
\bea
G^{\alpha \beta} := \epsilon^{a_1a_2} \epsilon^{b_1b_2} X^\alpha_{a_1 b_1} X^\beta_{a_2b_2} = \tr (X^\alpha \cdot X^\beta) ~, \label{def:genG}
\eea
where $X^\alpha_{ab}$ are defined as in \eref{defXSp}. 
Notice that $G^{\alpha \beta}$ is a rank two symmetric tensor.  The relation at order $t^4$, after taking into account of the relations \eref{rel:C2Z2SO2N02}, is
\bea 
(G^{12})^2- G^{11} G^{22} =0~,  \label{rel:C2Z2n00}
\eea

\subsubsection{$SO(3)$ instanton on $\BC^2/\BZ_2$: $\vec k =(1,1)$ and $\vec N= (0,3)$}
The space of solutions of the $F$ terms \eref{Fterm1} and \eref{Fterm2} contain more than one branch and hence can be decomposed further.  The primary decomposition (using {\tt Macaulay2}) reveals that the relevant branch of the $F$-flat space is defined by
\bea
M^{SO(3)}_{ij} Q^i_1 Q^j_2 + (X_{12})^2_{~1} (X_{21})^1_{~2}- (X_{12})^1_{~2} (X_{21})^2_{~1} =0, \quad \partial_{S_1} W =0 , \quad \partial_{S_2} W =0~.
\eea
and has the unrefined Hilbert series:
\bea
\CF^\flat(t,z_1=1,z_2=1,x=1,y=1) = \frac{(1+t)^3 (1+3 t)}{(1-t)^8}~,
\eea
where $x$ be the fugacity of the global symmetry $SU(2)_x$ and $y$ be the fugacity of the global symmetry $SO(3)$.  Integrating over the gauge group of the corresponding refined $F$-flat Hilbert series gives
\bea
g^{(0,3)}_{(1,1)}(t,x,y) &= \oint_{|z_1|=1} \frac{1-z_1^2}{(2 \pi i)z_1} \oint_{|z_2|=1} \frac{1-z_2^2}{(2 \pi i)z_2}  \CF^\flat(t; z_1,z_2; x,y)  \nn \\
&= \sum^{\infty}_{n=0} [2n]_x t^{2n}~. \label{C2Z2SO3N03VS}
\eea
This is the Hilbert series of $\BC^2/\BZ_2$.

\subsubsection{$SO(5)$ instanton on $\BC^2/\BZ_2$: $\vec k =(1,1)$ and $\vec N =(1,4)$ }
Let $\vec b =(b_1,b_2)$ be the fugacities of the global symmetry $SO(4)$.
The Hilbert series is given by
\bea
g^{(1,4)}_{(1,1)} (t,x,\vec b) = \left(\prod_{m=1}^2 \oint_{|z_m|=1} \frac{\ud z_m}{2\pi i} \frac{1-z_m^2}{z_m} \right) \frac{\chi_Q(t,z_1) \chi_{q}(t,z_2,\vec b) \chi_{X}(t,x,z_1,z_2) }{\chi_{F}(t,z_1,z_2)}~,
\eea
where the contributions from $Q$, $q$, $X$ and the $F$ terms are given respectively by
\bea
\begin{array}{ll}
\chi_Q(t,z_1) = \PE [t  [1]_{z_1}]~, & \quad
\chi_{q}(t,z_2,\vec b) = \PE [t [1]_{z_2} [1,1]_{\vec b}]~, \\
\chi_{X}(t,x,z_1,z_2) = \PE [t[1]_x [1]_{z_1} [1]_{z_2}]~, & \quad
\chi_{F}(t,z_1,z_2) = \PE\left[ t^2 ([2]_{z_1} +[2]_{z_2})\right]~.
\end{array}
\eea
The unrefined Hilbert series is
\bea
g^{(1,4)}_{(1,1)} (t,1,1)&=\frac{1-2 t+6 t^2-2 t^3+t^4}{(1-t)^6 (1+t)^4}~.  \label{C2Z2k1N14VS}
\eea
Notice that $g^{(1,4)}_{(1,1)} (t,1,1,1,1)$ has the correct order for the pole $t=1$ (matches with the expected $8$ complex dimensional Higgs branch) and has a palindromic numerator.  Moreover, it can be seen that the Hilbert series \eref{HS:C2Z2SO6N24} agrees with that of the $S/S$ quiver for $Sp(2)$ instanton on $\BC^2/\BZ_2$: $\vec k=(1,1), \vec N=(1,1)$.

It is instructive to look at the first few terms in the power series of $g^{(2,4)}_{(1,1)} (t,x,a,b_1,b_2) $:
\bea
g^{(1,4)}_{(1,1)} (t,x,b_1,b_2)  &= 1+\left( [2;0,0]+ [0;2,0]+[0;0,2]\right)t^2 + [1;1,1] t^3 +([4;0,0]+ \nn \\
& \quad [0;4,0]+[0;0,4] +[0;2,2]+[2;2,0]+[2;0,2])t^4+\ldots~,
\eea
where the notation $[n_x; n_{b_1},n_{b_2}]$ denotes a representation of $SU(2)_x\times SO(4)_{\vec b}$; here and henceforth the subscripts indicate the corresponding fugacities. The PL of the above result is
\bea
\PL\left[ g^{(1,4)}_{(1,1)} (t,x,b_1,b_2) \right] &=\left( [2;0,0]+ [0;2,0]+[0;0,2]\right)t^2 + [1;1,1] t^3  -3t^4-\ldots~.
\eea
At order $t^2$, the 9 generators consist of 
\bea 
\begin{array}{ll}
\text{$\tr(X^{\alpha_1} \cdot X^{\alpha_2})$} \qquad \qquad& \text{in $[2;0,0]$}~, \\
\text{$m^{j_1j_2}= q^{j_1}_{b_1} q^{j_2}_{b_2} \epsilon^{b_1b_2}$ (with $j_1,j_2 =1,2,3,4$)} \qquad \qquad& \text{in $[0;2,0]+[0;0,2]$}~.
\end{array}
\eea
The generators at order $t^3$ consist of 
\bea 
(B^{\alpha})^{j_1}=X^\alpha_{a_1 b_1} Q_{a_2} q^{j_1}_{b_2} \epsilon^{a_1 a_2} \epsilon^{b_1 b_2} \qquad \qquad& \text{in $[1;1,1]$}~.
\eea

\subsubsection{$SO(5)$ instanton on $\BC^2/\BZ_2$: $\vec k =(1,1)$ and $\vec N =(2,3)$ }
Let $a$ and $b$ be the fugacity of the global symmetries $SO(2)$ and $SO(3)$ respectively.
The Hilbert series is given by
\bea
g^{(2,3)}_{(1,1)} (t,x,a,b) = \left(\prod_{m=1}^2 \oint_{|z_m|=1} \frac{\ud z_m}{2\pi i} \frac{1-z_m^2}{z_m} \right) \frac{\chi_Q(t,a,z_1) \chi_{q}(t,z_2,b) \chi_{X}(t,x,z_1,z_2) }{\chi_{F}(t,z_1,z_2)}~,
\eea
where the contributions from $Q$, $q$, $X$ and the $F$ terms are given respectively by
\bea
\begin{array}{ll}
\chi_Q(t,a,z_1) = \PE [t (a+a^{-1}) [1]_{z_1}]~, & \quad
\chi_{q}(t,z_2, b) = \PE [t [1]_{z_2} [2]_b]~, \\
\chi_{X}(t,x,z_1,z_2) = \PE [t[1]_x [1]_{z_1} [1]_{z_2}]~, & \quad
\chi_{F}(t,z_1,z_2) = \PE\left[ t^2 ([2]_{z_1} +[2]_{z_2})\right]~.
\end{array}
\eea
The unrefined Hilbert series for each component is
\bea \label{C2Z2k1N23VS}
g^{(2,3)}_{(1,1)} (t,1,1,1)&=\frac{1-t+5 t^2+4 t^3+4 t^4+4 t^5+5 t^6-t^7+t^8}{(1-t)^6 (1+t)^2 \left(1+t+t^2\right)^3}~.
\eea
Notice that $g^{(2,3)}_{(1,1)} (t,1,1,1,1)$ has the correct order for the pole $t=1$ (he expected $6$ complex dimensional Higgs branch) and has a palindromic numerator.
Observe also that this Hilbert series agrees with that of $Sp(2)$ instanton ($SS$ quiver): $k=1$ and $N=2$.

It is instructive to look at the first few terms in the power series of $g^{(2,3)}_{(1,1)} (t,x,a,b) $:
\bea
g^{(2,3)}_{(1,1)} (t,x,a,b) &= 1+\left( [2;0;0]+ [0;0;0]+[0;0;2]\right)t^2 + [1;1;2] t^3 + \ldots~.
\eea
where the notation $[n_x; n_a;n_b]$ denotes a representation of $SU(2)_x \times SO(2)_a \times SO(3)_{b}$. The PL of the above result is
\bea
\PL\left[ g^{(2,3)}_{(1,1)} (t,x,a,b)\right] &=\left( [2;0;0]+ [0;0;0]+[0;0;2]\right)t^2 + [1;1;2] t^3 -2t^4\ldots~.
\eea
At order 2, the generators consists of 
\bea  \label{gensSO5H++order2}
\begin{array}{ll}
\text{$\tr(X^{\alpha_1} \cdot X^{\alpha_2})$} \qquad  \qquad & \text{in $[2;0;0]$}~,  \\
\text{$M^{i_1 i_2}= Q^{i_1}_{a_1} Q^{i_2}_{a_2} \epsilon^{a_1 a_2}$ (with $i_1, i_2=1,2$)} \qquad  \qquad&\text{in $[0;0;0]$}~, \\
\text{$m^{j_1j_2}= q^{j_1}_{b_1} q^{j_2}_{b_2} \epsilon^{b_1b_2}$ (with $j_1,j_2 =1,2,3$)} \qquad  \qquad& \text{in $[0;0;2]$}~.
\end{array}
\eea
The generators at order $t^3$ consist of 
\bea  \label{gensSO5H++order3}
(B^{\alpha})^{i_1 j_1}=X^\alpha_{a_1 b_1} Q^{i_1}_{a_1} q^{j_1}_{b_2} \epsilon^{a_1 a_2} \epsilon^{b_1 b_2} \qquad  \qquad& \text{in $[1;1;2]$}~.
\eea

\subsubsection{$SO(5)$ instantons on $\BC^2/\BZ_2$: $\vec k =(2,2)$ and $\vec N =(2,3)$ }
Let $a$ and $b$ be the fugacity of the global symmetries $SO(2)$ and $SO(3)$ respectively.
The Hilbert series is given by
\bea
g^{(2,3)}_{(2,2)} (t,x,a,b) &=  \oint_{|z_1|=1} \frac{\ud z_1}{(2\pi i)z_1} \oint_{|z_2|=1} \frac{\ud z_2}{(2\pi i)z_2} (1-z_1^2)(1-z_2)(1-z_1^2 z_2^{-1}) (1-z_1^2 z_2^{-2})   \nn \\
& \qquad \frac{\chi_Q(t,a,z_1) \chi_{q}(t,z_2,b) \chi_{X}(t,x,z_1,z_2) }{\chi_{F}(t,z_1,z_2)}~,
\eea
where the contributions from $Q$, $q$, $X$ and the $F$ terms are given respectively by
\bea
\begin{array}{ll}
\chi_Q(t,a,z_1) = \PE [t (a +a^{-1} ) [1,0]_{z_1}]~, & \quad
\chi_{q}(t, z_2, b) = \PE [t [1,0]_{z_2} [2]_b]~, \\
\chi_{X}(t,x,z_1,z_2) = \PE [t[1]_x [1,0]_{z_1} [1,0]_{z_2}]~, & \quad
\chi_{F}(t,z_1,z_2) = \PE\left[ t^2 ([2,0]_{z_1} +[2,0]_{z_2})\right]~.
\end{array}
\eea
The corresponding unrefined Hilbert series is
\bea \label{C2Z3k2N23VS}
g^{(2,3)}_{(2,2)} (t,1,1,1) &= \frac{1}{(1- t)^{12} (1 + t)^4 (1 + t^2)^2 (1 + t + t^2)^6 (1 + t + t^2 + t^3 +
    t^4)^3} \times \nn \\
& \quad \Big( 1+t+6 t^2+12 t^3+42 t^4+93 t^5+214 t^6+415 t^7+790 t^8+1348 t^9 \nn \\
& \quad +2156 t^{10}+3133 t^{11}+4275 t^{12}+5392 t^{13}+6416 t^{14}+7078 t^{15}+7352 t^{16} \nn \\
& \quad +7078 t^{17}+6416 t^{18} + \text{palindrome} + t^{32} \Big) \nn \\
&= 1+7 t^2+12 t^3+45 t^4+108 t^5+271 t^6+ \ldots~.
\eea
Notice that this Hilbert series agrees with that of two $Sp(2)$ instantons ($SS$ quiver): $k=2$ and $N=2$.
The first few terms in the plethystic logarithm is 
\bea
\PL\left[ g^{(2,3)}_{(2,2)} (t,x,a,b)\right] &=\left( [2;0;0]+ [0;0;0]+[0;0;2]\right)t^2 + [1;1;2] t^3 + ([4;0;0]+[2;0;0] \nn \\
& \quad +[2;0;2]) t^4 + [3;1;2] t^5+\ldots~.
\eea

\subsubsection{$SO(6)$ instanton on $\BC^2/\BZ_2$: $\vec k =(1,1)$ and $\vec N =(2,4)$ }
Let $a$ be the fugacity of the global symmetry $SO(2)$ and $\vec b =(b_1,b_2)$ be the fugacities of the global symmetries $SO(4)$.
The Hilbert series is given by
\bea
g^{(2,4)}_{(1,1)} (t,x,a,\vec b) = \left(\prod_{m=1}^2 \oint_{|z_m|=1} \frac{\ud z_m}{2\pi i} \frac{1-z_m^2}{z_m} \right) \frac{\chi_Q(t,a,z_1) \chi_{q}(t,z_2,\vec b) \chi_{X}(t,x,z_1,z_2) }{\chi_{F}(t,z_1,z_2)}~,
\eea
where the contributions from $Q$, $q$, $X$ and the $F$ terms are given respectively by
\bea
\begin{array}{ll}
\chi_Q(a,z_1) = \PE [t (a+a^{-1}) [1]_{z_1}]~, & \quad
\chi_{q}(z_2,\vec b) = \PE [t [1]_{z_2} [1,1]_{\vec b}]~, \\
\chi_{X}(x,z_1,z_2) = \PE [t[1]_x [1]_{z_1} [1]_{z_2}]~, & \quad
\chi_{F}(z_1,z_2) = \PE\left[ t^2 ([2]_{z_1} +[2]_{z_2})\right]~.
\end{array}
\eea
The unrefined Hilbert series is
\bea
& g^{(2,4)}_{(1,1)} (t,1,1,1,1) \nn \\
&=\frac{1+6 t^2+12 t^3+18 t^4+24 t^5+34 t^6+24 t^7+18 t^8+12 t^9+6 t^{10}+t^{12}}{(1-t)^8 (1+t)^4 \left(1+t+t^2\right)^4}~. \label{HS:C2Z2SO6N24}
\eea
Notice that $g^{(2,4)}_{(1,1)} (t,1,1,1,1)$ has the correct order for the pole $t=1$ (matches with the expected $8$ complex dimensional Higgs branch) and has a palindromic numerator.  Moreover, it can be seen that the Hilbert series \eref{HS:C2Z2SO6N24} agrees with that of $SU(4)$ instanton on $\BC^2/\BZ_2$ with $\vec k=(1,1), \vec N=(2,2)$ given by \eref{HS:C2Z2SU4N22}.

It is instructive to look at the first few terms in the power series of $g^{(2,4)}_{(1,1)} (t,x,a,b_1,b_2) $:
\bea
g^{(2,4)}_{(1,1)} (t,x,a,b_1,b_2)  &= 1+\left( [2;0,0]+ [0;2,0]+[0;0,2]+1\right)t^2 + [1;1,1](a+a^{-1}) t^3 + \nn \\
& \quad ([4;0,0]+[2;0,0]+[0;0,0]+[0;4,0]+[0;0,4] \nn \\
&\quad +[0;2,2]+[2;2,0]+[2;0,2]+[0;2,0]+[0;0,2])t^4+\ldots~,
\eea
where the notation $[m_1;n_1,n_2]$ denotes a representation of $SU(2)_x \times SO(4)_{\vec b}$; here and henceforth the subscripts indicate the corresponding fugacities. The PL of the above result is
\bea
\PL\left[ g^{(2,4)}_{(1,1)} (t,x,a,b_1,b_2) \right] &=\left( [2;0,0]+ [0;2,0]+[0;0,2]+[0;0,0]\right)t^2+ [1;1,1](a+a^{-1})t^3 \nn \\
& \quad -3t^4-3 [1;1,1](a+a^{-1})t^5-\ldots~.
\eea
At order 2, the 10 generators consists of 
\bea  \label{gensSO6H++order2}
\begin{array}{ll}
\text{$\tr(X^{\alpha_1} \cdot X^{\alpha_2})$} \qquad & \text{in $[2;0,0]$}~,  \\
\text{$M^{i_1 i_2}= Q^{i_1}_{a_1} Q^{i_2}_{a_2} \epsilon^{a_1 a_2}$ (with $i_1, i_2=1,2$)} \qquad &\text{in $[0;0,0]$}~, \\
\text{$m^{j_1j_2}= q^{j_1}_{b_1} q^{j_2}_{b_2} \epsilon^{b_1b_2}$ (with $j_1,j_2 =1,2,3,4$)} \qquad & \text{in $[0;2,0]+[0;0,2]$}~.
\end{array}
\eea
The generators at order $t^3$ consist of 
\bea  \label{gensSO6H++order3}
(B^{\alpha})^{i_1 j_1}=X^\alpha_{a_1 b_1} Q^{i_1}_{a_1} q^{j_1}_{b_2} \epsilon^{a_1 a_2} \epsilon^{b_1 b_2}~.
\eea

\subsubsection{$SO(6)$ instanton on $\BC^2/\BZ_2$: $\vec k =(1,1)$ and $\vec N =(3,3)$ }
Let $a$ and $b$ be the fugacity of the global symmetries $SO(3)$ corresponding to the left and right node respectively.
The Hilbert series is given by
\bea
g^{(3,3)}_{(1,1)} (t,x,a,b) = \left(\prod_{m=1}^2 \oint_{|z_m|=1} \frac{\ud z_m}{2\pi i} \frac{1-z_m^2}{z_m} \right) \frac{\chi_Q(t,a,z_1) \chi_{q}(t,z_2,b) \chi_{X}(t,x,z_1,z_2) }{\chi_{F}(t,z_1,z_2)}~,
\eea
where the contributions from $Q$, $q$, $X$ and the $F$ terms are given respectively by
\bea
\begin{array}{ll}
\chi_Q(t,a,z_1) = \PE [t [2]_a [1]_{z_1}]~, & \quad
\chi_{q}(t,z_2, b) = \PE [t [1]_{z_2} [2]_b]~, \\
\chi_{X}(t,x,z_1,z_2) = \PE [t[1]_x [1]_{z_1} [1]_{z_2}]~, & \quad
\chi_{F}(t,z_1,z_2) = \PE\left[ t^2 ([2]_{z_1} +[2]_{z_2})\right]~.
\end{array}
\eea
The unrefined Hilbert series for each component is
\bea
g^{(3,3)}_{(1,1)} (t,1,1,1,1)&=\frac{1-2 t+8 t^2+5 t^4+12 t^5+5 t^6+8 t^8-2 t^9+t^{10}}{(1-t)^8 (1+t)^2 \left(1+t+t^2\right)^4}~.
\eea
Notice that $g^{(3,3)}_{(1,1)} (t,1,1,1,1)$ has the correct order for the pole $t=1$ (he expected $8$ complex dimensional Higgs branch) and has a palindromic numerator. 

It is instructive to look at the first few terms in the power series of $g^{(3,3)}_{(1,1)} (t,x,a,b) $:
\bea
g^{(3,3)}_{(1,1)} (t,x,a,b) &= 1+\left( [2;0;0]+ [0;2;0]+[0;0;2]\right)t^2 + [1;2;2] t^3 + ([4;0;0]+[0;4;0]+[0;0;4]+\nn \\
& \quad [0;0;0]+[2;2;0]+[2;0;2] +[0;2;2]) t^4+ \ldots~.
\eea
where the notation $[n_x; n_a;n_b]$ denotes a representation of $SU(2)_x \times SO(3)_a \times SO(3)_{b}$. The PL of the above result is
\bea
\PL\left[ g^{(3,3)}_{(1,1)} (t,x,a,b)\right] &=\left( [2;0;0]+ [0;2;0]+[0;0;2]\right)t^2 + [1;2;2] t^3 -2t^4\ldots~.
\eea
At order 2, the 10 generators consists of 
\bea  \label{gensSO61133order2}
\begin{array}{ll}
\text{$\tr(X^{\alpha_1} \cdot X^{\alpha_2})$} \qquad & \text{in $[2;0;0]$}~,  \\
\text{$M^{i_1 i_2}= Q^{i_1}_{a_1} Q^{i_2}_{a_2} \epsilon^{a_1 a_2}$ (with $i_1, i_2=1,2,3$)} \qquad &\text{in $[0;2;0]$}~, \\
\text{$m^{j_1j_2}= q^{j_1}_{b_1} q^{j_2}_{b_2} \epsilon^{b_1b_2}$ (with $j_1,j_2 =1,2,3$)} \qquad & \text{in $[0;0;2]$}~.
\end{array}
\eea
The generators at order $t^3$ consist of 
\bea  \label{gensSO61133order3}
(B^{\alpha})^{i_1 j_1}=X^\alpha_{a_1 b_1} Q^{i_1}_{a_1} q^{j_1}_{b_2} \epsilon^{a_1 a_2} \epsilon^{b_1 b_2}~.
\eea

\subsubsection{$SO(8)$ instanton on $\BC^2/\BZ_2$: $\vec k =(1,1)$ and $\vec N =(2,6)$}
Let $a$ be the fugacity of the global symmetry $SO(2)$ and $\vec b= (b_1,b_2,b_3)$ be the fugacities of the global symmetry $SO(6)$.  The Hilbert series is given by
\bea
g^{(2,6)}_{(1,1)} (t,x,a,\vec b) =\left(\prod_{m=1}^2 \oint_{|z_m|=1} \frac{\ud z_m}{2\pi i} \frac{1-z_m^2}{z_m} \right) \frac{\chi_Q(t,a,z_1) \chi_{q}(t,z_2,b) \chi_{X}(t,x,z_1,z_2) }{\chi_{F}(t,z_1,z_2)}~,
\eea
where the contributions from $Q$, $q$, $X$ and the $F$ terms are given respectively by
\bea
\begin{array}{ll}
\chi_Q(t,a,z_1) = \PE [t (a+a^{-1}) [1]_{z_1}]~, & \quad
\chi_{q}(t,z_2, \vec b) = \PE [t [1]_{z_2} [1,0,0]_{\vec b}]~, \\
\chi_{X}(t,x,z_1,z_2) = \PE [t[1]_x [1]_{z_1} [1]_{z_2}]~, & \quad
\chi_{F}(t,z_1,z_2) = \PE\left[ t^2 ([2]_{z_1} +[2]_{z_2})\right]~.
\end{array}
\eea
Let us report the unrefined Hilbert series as follows:
\bea \label{C2Z3k1N26VS}
g^{(2,6)}_{(1,1)} (t,1,1,{\bf 1})& =\frac{1}{(1 - t)^{12} (1 + t)^8 (1 + t + t^2)^6} \Big(1+2 t+14 t^2+44 t^3+123 t^4+272 t^5\nn \\ 
&\quad +546 t^6+886 t^7+1259 t^8+1544 t^9 +1678 t^{10}+1544 t^{11} \nn \\
& \quad +\text{palindrome}+t^{20} \Big)~.
\eea
it can be seen that the Hilbert series \eref{HS:C2Z2SO6N24} agrees with that of one $SO(8)$ instanton on $\BC^2/\BZ_2$ ($AA$ quiver): $k=1$ and $N=4$.

The PL of $g^{(2,6)}_{(1,1)}(t,x,a,\vec b)$ is 
\bea
\PL\left[ g^{(2,6)}_{(1,1)} (t,x,a,\vec b)\right]&=([2;0;0,0,0]+[0;0;0,1,1]+1)t^2 +[1;1;1,0,0]t^3 \nn \\
& \qquad -([0;0;0,1,1]+2)t^4- \ldots~.
\eea
At order 2, there are $19$ generators: 
\bea  \label{mesonH++SO8}
\begin{array}{ll}
\text{$\tr(X^{\alpha_1} \cdot X^{\alpha_2})$} \qquad & \qquad  \text{in $[2;0,0]$}~,  \\
\text{$M^{i_1 i_2}= Q^{i_1}_{a_1} Q^{i_2}_{a_2} \epsilon^{a_1 a_2}$ (with $i_1, i_2=1,2$)} \qquad & \qquad  \text{in $[0;0;0,0,0]$}~, \\
\text{$m^{j_1j_2}= q^{j_1}_{b_1} q^{j_2}_{b_2} \epsilon^{ab}$ (with $j_1,j_2 =1,\ldots,6$)} \qquad & \qquad  \text{in $[0;0;0,1,1]$}~.
\end{array}
\eea
while, at order 3, there are $24$ generators in the representation $[1;1;1,0,0]$:
\bea  \label{gensSO8H++order3}
(B^{\alpha})^{i_1 j_1}=X^\alpha_{a_1 b_1} Q^{i_1}_{a_1} q^{j_1}_{b_2} \epsilon^{a_1 a_2} \epsilon^{b_1 b_2}~,
\eea
as expected.

\subsubsection{Summary: The $O/O$ quiver with $\vec k=(1,1)$ and $\vec N = (N_1, N_2)$}
Let us summarise the generators of the moduli space of one $SO(N_1+N_2)$ instanton on $\BC^2/\BZ_2$ with $O/O$ quivers.  

Without gauging any parity symmetry, the symmetry of the Higgs branch is $SU(2) \times SO(N_1) \times SO(N_2)$.  
The generators at order 2 are
\bea  
\begin{array}{lll}
\text{$\tr(X^{\alpha_1} \cdot X^{\alpha_2})$} \qquad & \qquad \text{(with $\alpha,\beta=1,2$)}  \qquad & \qquad  \text{in $[2; ~{\bf singlet}; ~{\bf singlet}]$}~,  \\
\text{$M^{i_1 i_2}= Q^{i_1}_{a_1} Q^{i_2}_{a_2} \epsilon^{a_1 a_2}$} \qquad & \qquad \text{(with $i_1, i_2=1,\ldots,N_1$)} \qquad & \qquad  \text{in $[0;~{\bf Adj};~ {\bf singlet}]$}~, \\
\text{$m^{j_1j_2}= q^{j_1}_{b_1} q^{j_2}_{b_2} \epsilon^{b_1 b_2}$} \qquad & \qquad \text{(with $j_1,j_2 =1,\ldots,N_2$)} \qquad & \qquad  \text{in $[0;~{\bf singlet};~ {\bf Adj}]$}~.
\end{array}
\eea
At order 3, the generators are
\bea
(B^{\alpha})^{i_1 j_1}=X^\alpha_{a_1 b_1} Q^{i_1}_{a_1} q^{j_1}_{b_2} \epsilon^{a_1 a_2} \epsilon^{b_1 b_2} \qquad  \text{in $[1;~1,0,\ldots,0; ~1,0,\ldots,0]$}~.
\eea
Thus, the total number of generators are
\bea
3+\frac{1}{2}N_1(N_1-1) +\frac{1}{2}N_2(N_2-1)+2N_1N_2~.
\eea

\subsection{The $AA$ quiver for $SO(2N)$ instantons on $\BC^2/\BZ_2$}
\begin{figure}[H]
\begin{center}
\includegraphics[scale=0.7]{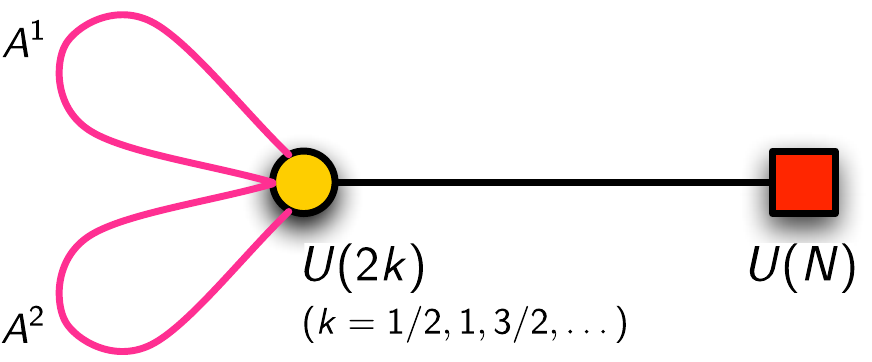}
\caption{The $AA$ quiver for $k$ $SO(2N)$ instanton on $\mathbb{C}^2/\mathbb{Z}_2$.  Here, $k$ can take a half-integral or an integral value, so that the rank of $U(2k)$ can be odd or even. The loops labelled by $A^1$ and $A^2$ denotes two hypermultiplets, each transforms as the rank 2-antisymmetric tensor under $U(2k)$ gauge group. There is a global symmetry $SU(2)$ in which the hypermultiplets $A^1, A^2$ transform as a doublet.  The line between the circular and the square node denote $N$ flavour fundamental hypermultiplets.}
\label{fig:oneO2NnoVS}
\end{center}
\end{figure}

Let us examine the $AA$ quiver for the construction of $k$ $SO(2N)$ instantons on $\BC^2/\BZ_2$.  This quiver contains the gauge group $U(2k)$ with $2$ flavours of antisymmetric hypermultiplets and $N$ flavours fundamental hypermultiplets.  The quiver diagram is depicted in \fref{fig:oneO2NnoVS}.  Note that the $SO(2N)$ symmetry is broken to $U(N)$ in this model.

Let us denote the $U(2k)$ fundamental indices by $a,b, \ldots=1,2,..,2k$. The indices $i, j, \ldots$ correspond to fundamental representations of $U(N)$, with $i, j, \ldots=1,2, \ldots,N$.   

It is convenient to use $\CN=1$ language in the subsequent computation.  We denote by $Q^i_a$ and $\tQ^a_i$ the bi-fundamental chiral multiplets in $U(2k) \times U(N)$ and the ones in $U(N) \times U(2k)$, respectively.  We denote also by $A^\alpha$ and $\widetilde{A}^\alpha$ (with $\alpha, \beta=1,2$) the chiral multiplets in the antisymmetric and its conjugate representations of the gauge group $U(2k)$.  Note that there is an $SU(2)$ global symmetry under which each of $A^\alpha$ and $\widetilde{A}^\alpha$ transform as a doublet; we refer to this $SU(2)$ global symmetry as $SU(2)_x$.

The transformation rules of the various chiral multiplets, including the $F$-term constraints are summarised in the table below:
\begin{equation}
\begin{array}{c | c | c  c c c c  }
& U(2k) & U(N)  & U(1)_t & SU(2)_x  \\ \hline
 (A^{\alpha})^{ab}                &   [0,1,0,\ldots,0]_{+2}          &    [0,\ldots,0]_0  &   1  &   [1]        \\
 (\widetilde{A}^{\alpha})_{ab}                    &  [0,\ldots,0,1,0]_{-2}    & [0,\ldots,0]_0 &   1  &   [1]       \\
  Q^{i}_a                   &    [1,0,\ldots,0]_{+1}       &  [0,\ldots,0,1]_{-1}      &   1    &   [0]   \\
\tQ_{i}^a                  & [0,\ldots,0,1]_{-1}           &   [1,0,\ldots,0]_{+1}       &   1   &    [0] \\ 
 \varphi^a_{~b}        &   [1,0,\ldots,0,1]_{0}      &  [0,\ldots,0]_0 &   0   &  [0]    \\
 \hline \hline
\mathcal{F}^a_{b}           &    [1,0,\ldots,0,1]_{0}          &  [0,\ldots,0]_0     &   2   &   [0] \\
 \end{array}
 \label{Symm_withoVS}
 \end{equation}
 The superpotential is given by
 \begin{equation} \label{WnoVS}
 W=(\widetilde{A}_{\alpha})^{ a b} \varphi^c_{~b} (A^\alpha)_{ca}+ {\tQ}_{i}^a \varphi^{b}_{~a} Q^{i}_b
  \end{equation}
The $F$-terms corresponding to the Higgs branch is given by
\bea
0 = \partial_{\varphi^c_{~b}} W = (\widetilde{A}_{\alpha})^{ a b}  (A^\alpha)_{ca}+ {\tQ}_{i}^b Q^{i}_c = \epsilon_{\alpha \beta} (\tA^\alpha)^{ab} (A^\beta)_{ca}+ {\tQ}_{i}^b Q^{i}_c~.
\eea

\subsubsection{$SO(2)$ instanton on $\BC^2/\BZ_2$: $k=1$ and $N=1$}
It should be pointed out that the space of solutions of the $F$ terms, also known as the $F$-flat space, contain more than one branch and hence can be decomposed further.  The primary decomposition (using {\tt Macaulay2} \cite{mac2} or {\tt STRINGVACUA} \cite{Gray:2008zs}) reveals that the relevant branch of the $F$-flat space are given by the relations:
\bea
\tQ = 0, \qquad \epsilon_{\alpha \beta} (\tA^\alpha)^{ab} (A^\beta)_{ca}= 0~.
\eea
Let $(q,z)$ be the gauge symmetry $U(2) = U(1) \times SU(2)$, $x$ be the fugacity of $SU(2)_x$ and $y$ be the fugacity of the $U(1)$ flavour symmetry.  Using {\tt Macaulay2} \cite{mac2}, we obtain the unrefined $F$-flat Hilbert series as
\bea
\CF^\flat(t;q,z;x,y) &= (1-t^2) \chi_Q(t,q,z,y) \chi_{A}(t,q,x) \chi_{\tA}(t,q,x) \nn \\
&=\frac{1-t^2}{\left(1-\frac{t}{q^2 x}\right) \left(1-\frac{q^2 t}{x}\right) \left(1-\frac{t x}{q^2}\right) \left(1-q^2 t x\right) \left(1-\frac{q t}{y z}\right) \left(1-\frac{q t z}{y}\right)}~.
\eea
where the contributions from the chiral fields $Q$, $A$, $\tA$ and the $F$ terms are, respectively,
\bea
\begin{array}{ll}
\chi_Q(t,q,z,y) &= \PE [t q [1]_z y^{-1})]~, \\
\chi_{A}(t,q,x) &= \PE [ t (x+x^{-1}) q^2 ]~,\\
\chi_{\tA}(t,q,x) &= \PE [ t (x+x^{-1}) q^{-2} ]~.
\end{array}
\eea
Integrating over the gauge group of the corresponding refined $F$-flat Hilbert series gives
\bea \label{C2Z2k1N1noVS}
g^{N=2}_{k=1}(t,x,y) &= \left(\oint_{|q|=1} \frac{\ud q}{(2 \pi i) q} \oint_{|z|=1} \ud z\frac{1-z^2}{(2 \pi i)z}\right) \CF^\flat(t;q,z;x,y) \nn \\
&= \sum^{\infty}_{n=0} [2n]_x t^{2n} = (1-t^4) \PE [[2]_x t^2]~.
\eea
This is indeed the Hilbert series of $\BC^2/\BZ_2$, as should be expected for the moduli space of one $SO(2)$ instanton on $\BC^2/\BZ_2$.  The generators in $[2]$ of $SU(2)_x$ can be written as
\bea
G^{\alpha \beta} =  \tr({\tA^\alpha A^\beta})~.
\eea
Using the $F$-flat conditions, we see that $G^{12} = G^{21}$ and the relation at order $4$ is
\bea
G^{11} G^{22} - (G^{12})^2 =0 ~.
\eea

\subsubsection{$SO(6)$ instanton on $\BC^2/\BZ_2$: $k=1$ and $N=3$}
Let $x$ and $(u, \vec y)$ be the fugacities of $SU(2)_x$ and the flavour symmetry $U(3)=U(1)\times SU(3)$ such that the characters of fundamental and anti-fundamental representations of $U(3)$ can be written as
\bea
u [1,0]_{\vec y} = u (y_1 + y_2 y_1^{-1} + y_2^{-1} )~, \qquad u^{-1} [0,1]_{\vec y} = u^{-1} (y_1^{-1} + y_2^{-1} y_1 + y_2 )~.
\eea
The Hilbert series are given by
\bea \label{SO6noVSpm}
g^{N=3}_{k=1} (t,x,u,\vec y)&= \left(\oint_{|q|=1} \frac{\ud q}{(2 \pi i) q} \oint_{|z|=1} \ud z\frac{1-z^2}{(2 \pi i)z}\right) \frac{\chi_Q(t,q,z,u,\vec y)\chi_{\tQ}(t,q,z,u, \vec y) \chi_A(t,q,x) \chi_{\tA} (t,q,x)}{\chi_{F}(t,z_1,z_2)}~,
\eea
where the contributions from the chiral fields $Q$, $\tQ$, $A$, $\tA$ and the $F$ terms are given respectively by
\bea
\begin{array}{ll}
\chi_Q(t,q,z,u,\vec y) &= \PE [t q [1]_z u^{-1} [0,1]_{\vec y}]~, \\
\chi_{\tQ}(t,q,z,u, \vec y) &= \PE [t q^{-1} [1]_{z} u [1,0]_{\vec y}]~, \\
\chi_{A}(t,q,x) &= \PE [ t (x+x^{-1}) q^2 ]~,\\
\chi_{\tA}(t,q,x) &= \PE [ t (x+x^{-1}) q^{-2} ]~,\\
\chi_{F}(t,z_1,z_2) &= \PE\left[ t^2 ([2]_z +1)\right]~.
\end{array}
\eea
Evaluating the integrals, we obtain
\bea
& g^{N=3}_{k=1} (t,x,u,\vec y)= 1+([2;0,0]+[0;1,1]+[0;0,0])t^2+ (u^2 [1;0,1]+u^{-2}[1;1,0])t^3 \nn \\
& \quad + ([4;0,0]+[2;0,0]+[2;1,1] +[0;2,2]+[0;1,1]+[0;0,0] )t^4 + \ldots~.
\eea
The unrefined Hilbert series is
\begin{equation} \label{C2Z2k1N3noVS}
g^{N=3}_{k=1}(t,1,1,{\bf 1})=\frac{1+2 t+9 t^2+24 t^3+50 t^4+76 t^5+108 t^6+120 t^7+108 t^8+\text{palindrome}+t^{14}}{(1-t)^8 (1+t)^6 \left(1+t+t^2\right)^4}~.
\end{equation}
The plethystic logarithm of $g^{N=3}_{k=1} (t,x,u,\vec y)$ is
\bea
\PL [g^{N=3}_{k=1} (t,x,u,\vec y)] &= ([2;0,0]+[0;1,1]+[0;0,0])t^2+ (u^2 [1;0,1]+u^{-2}[1;1,0])t^3 \nn \\
& \quad - ([0;1,1]+2)t^4 - \ldots~.
\eea
The generators at order 2 are
\bea
\begin{array}{ll}
\text{$G^{\alpha \beta} =\tr ( A^\alpha \widetilde{A}^\beta)$} \qquad & \qquad  \text{in \quad $[2;0,0]+[0;0,0]$}~,  \\
\text{$M^{i}_{j}= Q^{i}_{a} \tQ^{a}_{j}$} \qquad & \qquad  \text{in \quad $[0;1,1]$}~,
\end{array}
\eea
subject to a relation coming from the $F$ terms:
\bea
\epsilon_{\alpha \beta} G^{\alpha \beta}+  M^i_i = 0~.
\eea
The generators at order 3 are
\bea
\begin{array}{ll}
\text{$(B^\alpha)^{ij}  =(\widetilde{A}^\alpha)^{ab} Q^i_a Q^j_b$} \qquad & \qquad  \text{in \quad $u^{2} [1;0,1]$}~, \\
\text{$(\tB^\alpha)_{ij} =(A^\alpha)_{ab} \tQ^a_i \tQ^b_j$} \qquad & \qquad  \text{in \quad$u^{-2} [1;1,0]$}~.
\end{array}
\eea

\subsubsection{$SO(6)$ instantons on $\BC^2/\BZ_2$: $k=3/2$ and $N=3$}
Let $(q, z_1, z_2)$ be the fugacities of the gauge group $U(3)=U(1) \times SU(3)$, $x$ and $(u, \vec y)$ be the fugacities of $SU(2)_x$ and the flavour symmetry $U(3)=U(1)\times SU(3)$ respectively.  Then, the Hilbert series are given by
\bea 
g^{N=3}_{k=3/2} (t,x,u,\vec y)&= \left(\oint_{|q|=1} \frac{\ud q}{(2 \pi i) q} \oint_{|z_1|=1} \frac{\ud z_1}{(2 \pi i)z_1}  \oint_{|z_2|=1} \frac{\ud z_2}{(2\pi i)z_2} (1-z_1 z_2)(1-z_1^2 z_2^{-1})(1-z_2^2 z_1^{-1}) \right) \times \nn \\
& \quad  \frac{\chi_Q(t,q,z,u,\vec y)\chi_{\tQ}(t,q,z,u, \vec y) \chi_A(t,q,x) \chi_{\tA} (t,q,x)}{\chi_{F}(t,z_1,z_2)}~,
\eea
where the contributions from the chiral fields $Q$, $\tQ$, $A$, $\tA$ and the $F$ terms are given respectively by
\bea
\begin{array}{ll}
\chi_Q(t,q,z,u,\vec y) &= \PE [t q [1,0]_{\vec z} u^{-1} [0,1]_{\vec y}]~, \\
\chi_{\tQ}(t,q,z,u, \vec y) &= \PE [t q^{-1} [0,1]_{\vec z} u [1,0]_{\vec y}]~, \\
\chi_{A}(t,q,x) &= \PE [ t (x+x^{-1}) q^2 [0,1]_{\vec z}]~,\\
\chi_{\tA}(t,q,x) &= \PE [ t (x+x^{-1}) q^{-2} [1,0]_{\vec z} ]~,\\
\chi_{F}(t,z_1,z_2) &= \PE\left[ t^2 ([1,1]_z +1)\right]~.
\end{array}
\eea
The corresponding unrefined Hilbert series is
\bea \label{C2Z2k32N3noVS}
g^{N=3}_{k=3/2} (t,1,1,\vec 1) & = \frac{1}{(1-t)^{12} (1+t)^{10} \left(1+t^2\right)^5 \left(1+t+t^2\right)^6} \Big( 1+4 t+17 t^2+54 t^3+175 t^4 \nn \\
& \quad +470 t^5+1164 t^6+2506 t^7+5008 t^8+9010 t^9+15054 t^{10}+22874 t^{11} \nn \\
& \quad +32297 t^{12}+41858 t^{13}+50631 t^{14}+56488 t^{15}+58818 t^{16}+56488 t^{17}  \nn \\
& \quad + \text{palindrome} +t^{32} \Big) \nn \\
&= 1 + 12 t^2 + 12 t^3 + 101 t^4 + 132 t^5 + 622 t^6 + 900 t^7 + 3050 t^8 + \ldots~.
\eea
Notice that this is in agreement with the Hilbert series of $3/2$ $SU(4)$-instantons on $\BC^2/\BZ_2$: $\vec k =(2,1)$ and $\vec N=(3,1)$.

The plethystic logarithm of the Hilbert series $g^{N=3}_{k=3/2} (t,x,u,\vec y)$ is
\bea
\PL \left[g^{N=3}_{k=3/2} (t,x,u,\vec y)  \right] = ([2;0,0]+[0;1,1]+1)t^2 +(u^2 [1;0,1]+u^{-2} [1;1,0])t^3 +( [2;1,1]-1)t^4- \ldots~.
\eea

\subsubsection{$SO(8)$ instanton on $\BC^2/\BZ_2$: $k=1$ and $N=4$}
Let $x$ and $(u,\vec y)$ be the fugacities of $SU(2)_x$ and the flavour symmetry $U(4)=U(1)\times SU(4)$ such that the characters of fundamental and anti-fundamental representations of $U(4)$ can be written as
\bea
u [1,0,0]_{\vec y} &= u (y_1 + y_2 y_1^{-1} + y_3 y_2^{-1}+ y_3^{-1} )~, \nn \\ 
u^{-1} [0,0,1]_{\vec y} &= u^{-1} (y_1^{-1} + y_2^{-1} y_1 + y_3^{-1} y_2 + y_3 )~.
\eea
The even and odd components of the Hilbert series are given by
\bea \label{SO8noVSpm}
g^{N=4}_{k=1} (t,u,\vec y)&= \left(\oint_{|q|=1} \frac{\ud q}{(2 \pi i) q} \oint_{|z|=1} \ud z\frac{1-z^2}{(2 \pi i)z}\right)\frac{\chi_Q(t,q,z,u,\vec y) \chi_{\tQ}(t,q,z,u, \vec y) \chi_{A}(t,q,x) \chi_{\tA}(t,q,x)}{\chi_{F}(t,z_1,z_2)}~,
\eea
where the contributions from the chiral fields $Q$, $\tQ$, $A$, $\tA$ and the $F$ terms are given respectively by
\bea
\begin{array}{ll}
\chi_Q(t,q,z,u,\vec y) &= \PE [t q [1]_z u^{-1} [0,0,1]_{\vec y}]~, \\
\chi_{\tQ}(t,q,z,u, \vec y) &= \PE [t q^{-1} [1]_{z} u [1,0,0]_{\vec y}]~, \\
\chi_{A}(t,q,x) &= \PE [ t (x+x^{-1}) q^2 ]~,\\
\chi_{\tA}(t,q,x) &= \PE [ t (x+x^{-1}) q^{-2} ]~,\\
\chi_{F}(t,z_1,z_2) &= \PE\left[ t^2 ([2]_z +1)\right]~.
\end{array}
\eea
Evaluating the integrals, we obtain
\bea
g^{N=4}_{k=1} (t,x,u,\vec y) &= 1+([2;0,0,0]+[0;1,0,1]+[0;0,0])t^2+ (u^2 [1;0,1,0]+u^{-2}[1;0,1,0])t^3 \nn \\
& \quad + ([4;0,0,0]+[2;0,0,0]+[2;1,0,1] +[0;2,0,2]+[0;0,2,0]+[0;1,0,1] \nn \\
& \quad +[0;0,0,0] )t^4 + \ldots~.
\eea
The unrefined Hilbert series is
\bea \label{C2Z3SO8k1N4noVS}
g^{N=4}_{k=1} (t,1,1,{\bf 1})&=\frac{1}{(1 - t)^{12} (1 + t)^8 (1 + t + t^2)^6} \Big(1+2 t+14 t^2+44 t^3+123 t^4+272 t^5+546 t^6+ \nn \\
& \quad 886 t^7+1259 t^8+1544 t^9+1678 t^{10}+1544 t^{11}+\text{palindrome}+t^{20} \Big)~.
\eea
The plethystic logarithm of $g^{N=4}_{k=1} (t,x,u,\vec y)$ is
\bea
\PL [g^{N=4}_{k=1} (t,x,u,\vec y)] &= ([2;0,0,0]+[0;1,0,1]+[0;0,0,0])t^2+ (u^2 [1;0,1,0]+u^{-2}[1;0,1,0])t^3 \nn \\
& \quad - ([0;1,0,1]+2)t^4 - \ldots~.
\eea
The generators at order 2 are
\bea
\begin{array}{ll}
\text{$G^{\alpha \beta} =\tr ( A^\alpha \widetilde{A}^\beta)$} \qquad & \qquad  \text{in \quad $[2;0,0,0]$}~,  \\
\text{$M^{i}_{j}= Q^{i}_{a} \tQ^{a}_{j}$} \qquad & \qquad  \text{in \quad $[0;1,0,1]+[0;0,0,0]$}~,
\end{array}
\eea
subject to a relation coming from the $F$ terms:
\bea
\epsilon_{\alpha \beta} G^{\alpha \beta}+  M^i_i = 0~.
\eea
The generators at order 3 are
\bea
\begin{array}{ll}
\text{$(B^\alpha)^{ij}  =(\widetilde{A}^\alpha)^{ab} Q^i_a Q^j_b$} \qquad & \qquad  \text{in \quad $u^{2} [1;0,1,0]$}~, \\
\text{$(\tB^\alpha)_{ij} =(A^\alpha)_{ab} \tQ^a_i \tQ^b_j$} \qquad & \qquad  \text{in \quad$u^{-2} [1;0,1,0]$}~.
\end{array}
\eea

\subsubsection{Summary: One $SO(2N)$ instanton on $\BC^2/\BZ_2$, the $AA$ quiver}
Without gauging any parity symmetry, the symmetry of the Higgs branch is $SU(2) \times U(N)$.  The generators at order 2 are
\bea
\begin{array}{ll}
\text{$G^{\alpha \beta} =\tr ( A^\alpha \widetilde{A}^\beta)$} \qquad & \qquad  \text{in \quad $[2;0,\ldots,0]$}~,  \\
\text{$M^{i}_{j}= Q^{i}_{a} \tQ^{a}_{j}$} \qquad & \qquad  \text{in \quad $[0;1,0,\ldots,0,1]+[0;0,\ldots,0]$}~,
\end{array}
\eea
subject to a relation coming from the $F$ terms:
\bea
\epsilon_{\alpha \beta} G^{\alpha \beta}+  M^i_i = 0~.
\eea
The generators at order 3 are
\bea
\begin{array}{ll}
\text{$(B^\alpha)^{ij}  =(\widetilde{A}^\alpha)^{ab} Q^i_a Q^j_b$} \qquad & \qquad  \text{in \quad $u^{2} [1;0,1,0, \ldots,0]$}~, \\
\text{$(\tB^\alpha)_{ij} =(A^\alpha)_{ab} \tQ^a_i \tQ^b_j$} \qquad & \qquad  \text{in \quad$u^{-2} [1;0,\ldots,0,1,0]$}~.
\end{array}
\eea
%

\subsection{The $S/S$ quiver for $Sp(N)$ instantons on $\mathbb{C}^2/\mathbb{Z}_2$}
\begin{figure}[H]
\begin{center}
\includegraphics[scale=0.7]{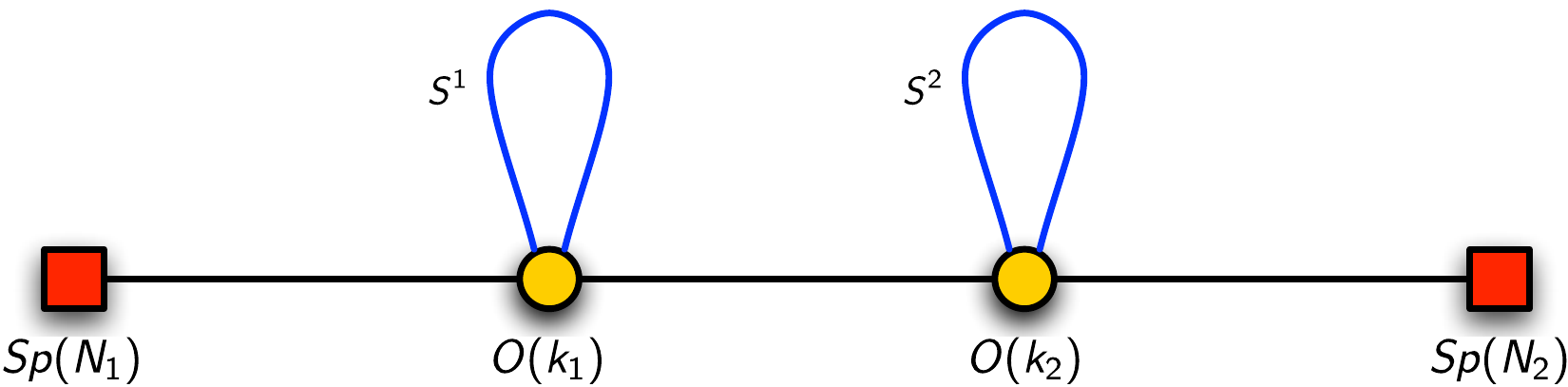}
\caption{The $S/S$ quiver for $Sp(N)$ instantons on $\mathbb{C}^2/\mathbb{Z}_2$.  Here, $N_1+N_2=N$. The line between $O(k_1)$ and $O(k_2)$ gauge groups denote $k_1k_2$ hypermultiplets (whose scalar components have $2k_1k_2$ complex degrees of freedom), and each line connecting the square node and the circular node denotes $2kN_1$ and $2kN_2$ half-hypermultiplets respectively.}
\label{fig:SpNVS}
\end{center}
\end{figure}

The $S/S$ quiver for $Sp(N)$ instantons on $\BC^2/\BZ_2$ is depicted in \fref{fig:SpNVS}. From the quiver diagram, we see that the Higgs branch is
\bea
& k_1 N_1 + k_2 N_2 +k_1 k_2- \frac{1}{2}k_1(k_1-1)- \frac{1}{2}k_2(k_2-1) \nn \\
&= k_1 \left(N_1 +\frac{1}{2} \right)+ k_2\left(N_2 +\frac{1}{2} \right) - \frac{1}{2} (k_1-k_2)^2~.
\eea
quaternionic dimensional.  When $k_1=k_2 =k$, this expression gives $k(N+1)$, as should be expected for the quaternionic dimension of $k$ $Sp(N)$ instantons on $\BC^2/\BZ_2$.

In what follows, we denote the half-hypermultiplets in the bi-fundamental representations of $Sp(N_1)\times O(k_1)$ and $O(k_2) \times Sp(N_2)$ by $Q$ and $q$ and the hypermultiplet in the the bi-fundamental representation of $O(k_1) \times O(k_2)$ by $X$.

\paragraph{The special case of $\vec k=(1,1)$.}
The gauge symmetries are $O(1) \times O(1)$.  In this case, we regard them as $\BZ_2 \times \BZ_2$.
Let $\omega_1, \omega_2 = \pm 1$ be the action of the gauge groups $O(1)$ corresponding to the left and the right circular nodes.  Let $\vec a$ be the fugacity of the flavour symmetry $Sp(N_1)$ and $\vec b$ be the fugacity of the flavour symmetry $Sp(N_2)$. The contribution from $Q$ and $X$ to the Hilbert series is respectively
\bea
g^{(N_1,N_2)} _{(1,1)}(t,x,\vec a, \vec b)=& \frac{1}{4} \sum_{\omega_1,\omega_2=\pm 1} \chi_Q(t, \omega_1,  \vec a)\chi_X(t, \omega_1, \omega_2, x) \chi_q(t, \omega_2,  \vec b)~,
\eea
where $\chi_X$, $\chi_Q$ and $\chi_q$ denote contributions from $X$, $Q$ and $q$:
\bea
\chi_X(t, \omega_1, \omega_2, x) &= \PE\left[ z w t \left(x+1/x\right)\right]~, \nn \\
\chi_Q(t, \omega_1,  \vec a) &=  \PE\left[ z\;t \left[1,0,0,...,0\right]_{\vec a}\right]~, \nn \\ 
\chi_q(t, \omega_2,  \vec b) &=  \PE\left[ w\;t \left[1,0,0,...,0\right]_{\vec b}\right]~.
\eea

\subsubsection{$Sp(N)$ instanton on $\BC^2/\BZ_2$: $\vec k =(1,1)$ and $\vec N=(N,0)$}
The Hilbert series is given by
\bea
\label{C2Z2Spk11N0}
g^{(N,0)}_{(1,1)}(t,x,\vec a) &= \frac{1}{4} \sum_{\omega_1, \omega_2=\pm 1} \chi_Q(t, \omega_1, \vec a)\chi_X(t, \omega_1, \omega_2, x)  \nn \\
&= (1-t^4) \PE[ [2]_x t^2] \sum_{n=0}^\infty [2n,0,\ldots,0]_{\vec a} t^{2n}  \nn \\
&=g_{\mathbb{C}^2/\mathbb{Z}_2} (t,x) \times \widetilde{g}_{1, Sp(N), \BC^2} (t,\vec a)
\eea
where $g_{\mathbb{C}^2/\mathbb{Z}_2} (t,x)$ is the Hilbert series of $\mathbb{C}^2/\mathbb{Z}_2$ and $ \widetilde{g}_{1, Sp(N), \BC^2} (t,\vec a)$ is the Hilbert series of the reduced moduli space of one $Sp(N)$ instanton on $\mathbb{C}^2$.  Notice that, for $N=1$, this agrees with the Hilbert series for one $SU(2)$ instanton on $\BC^2/\BZ_2$ discussed in Section \ref{sec:pureinstfeatures}.

\subsubsection{$Sp(N+1)$ instanton on $\BC^2/\BZ_2$: $\vec k=(1,1)$ and $\vec N = (1, N)$}
The Hilbert series is given by
\bea
g^{(1,N)}_{(1,1)}(t,x,\vec a, \vec b)&= \frac{1}{4} \sum_{\omega_1, \omega_2=\pm 1} \PE\left[ \omega_1 t [1]_{\vec a} \right] \PE\left[ \omega_1 \omega_2 t [1]_x\right]\PE\left[ \omega_2 t \left[1,0,0,...,0\right]_{\vec b}\right] \nn \\
&= 1+ ([2;0;0]+[0;2;0]+[0;0;2,0,\ldots]) t^2 + [1;1;1,0,\ldots,0]t^3 \nn \\
& \quad + \ldots~.
\eea
The plethystic logarithm is given by
\bea
g^{(1,N)}_{(1,1)}(t,x,\vec a, \vec b)&= ([2;0;0]+[0;2;0]+[0;0;2,0,\ldots]) t^2 + [1;1;1,0,\ldots,0]t^3 \nn \\
& \quad -( [0,1,0,\ldots,0]+[0,2,0,\ldots,0]+3)t^4-\ldots~.
\eea
The unrefined Hilbert series for $N=1$ and $N=2$ are
\bea
g^{(1,1)}_{(1,1)}(t,1,\vec 1, \vec 1)&=\frac{1-2 t+6 t^2-2 t^3+t^4}{(1-t)^6(1+t)^4}~, \label{C2Z2Spk11N11noVS} \\
g^{(1,2)}_{(1,1)}(t,1,\vec 1, \vec 1)&=\frac{(1+t^2)(1-2 t+10t^2-2 t^3+t^4)}{(1-t)^8(1+t)^6}~.
\eea

\subsection{The $SS$ quiver for $Sp(N)$ instantons on $\mathbb{C}^2/\mathbb{Z}_2$}
\begin{figure}[H]
\begin{center}
\includegraphics[scale=0.7]{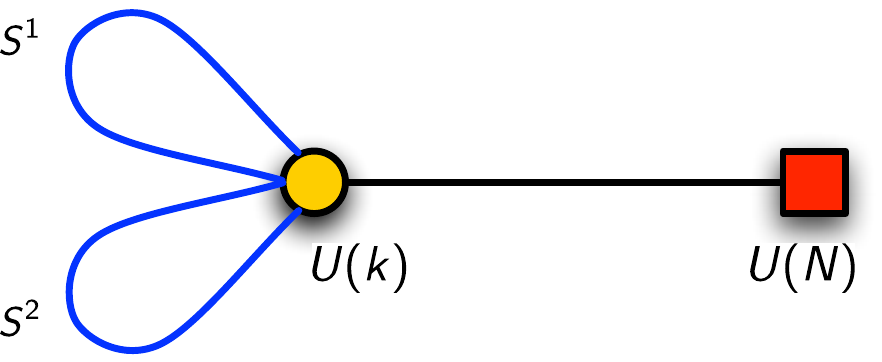}
\caption{The $SS$ quiver for $k$ $Sp(N)$ instanton on $\mathbb{C}^2/\mathbb{Z}_2$.  Here, $k$ takes an integral value.  The loops labelled by $S^1$ and $S^2$ denotes two hypermultiplets, each transforms as the rank 2-symmetric tensor under $U(k)$ gauge group. There is a global symmetry $SU(2)$ under which the hypermultiplets $S^1, S^2$ transform as a doublet.  The line between the circular and the square node denote $N$ flavour fundamental hypermultiplets.}
\label{fig:SpNnoVS}
\end{center}
\end{figure}

Let us examine the $SS$ quiver, depicted in \fref{fig:SpNnoVS}, for $k$ $Sp(N)$ instantons on $\BC^2/\BZ_2$.  This quiver contains the gauge group $U(2k)$ with $2$ flavours of antisymmetric hypermultiplets and $N$ flavours fundamental hypermultiplets.  Note that the $Sp(N)$ symmetry is broken to $U(N)$ in this model.

From the quiver diagram, we see that the Higgs branch is
\bea
kN + \frac{1}{2}k(k+1)+ \frac{1}{2}k(k+1)-k^2 = k(N+1)
\eea
quaternionic dimensional, as should be expected for $k$ $Sp(N)$ instantons on $\BC^2/\BZ_2$.

Let us denote the $U(k)$ fundamental indices by $a,b=1,2,..,k$. The indices $i, j$ correspond to fundamental representations of $U(N)$, with $i, j=1,2, \ldots,N$.   

It is convenient to use $\CN=1$ language in the subsequent computation.  We denote by $Q^i_a$ and $\tQ^a_i$ the bi-fundamental chiral multiplets in $U(k) \times U(N)$ and the ones in $U(N) \times U(k)$, respectively.  We denote also by $S^\alpha$ and $\widetilde{S}^\alpha$ (with $\alpha=1,2$) the chiral multiplets in the symmetric and its conjugate representations of the gauge group $U(k)$.  Note that there is an $SU(2)$ global symmetry under which each of $S^\alpha$ and $\widetilde{S}^\alpha$ transform as a doublet; we refer to this $SU(2)$ global symmetry as $SU(2)_x$.

The transformation rules of the various chiral multiplets, including the $F$-term constraints are summarised in the table below:
\begin{equation}
\begin{array}{c | c | c  c c c c  }
& U(k) & U(N)  & U(1)_t & SU(2)_x  \\ \hline
 (S^{\alpha})^{ab}                &   [2,0,\ldots,0]_{+2}          &    [0,\ldots,0]_0  &   1  &   [1]        \\
 (\widetilde{S}^{\alpha})_{ab}                    &  [0,\ldots,0,2]_{-2}    & [0,\ldots,0]_0 &   1  &   [1]       \\
  Q^{i}_a                   &    [1,0,\ldots,0]_{+1}       &  [0,\ldots,0,1]_{-1}      &   1    &   [0]   \\
\tQ_{i}^a                  & [0,\ldots,0,1]_{-1}           &   [1,0,\ldots,0]_{+1}       &   1   &    [0] \\ 
 \varphi^a_{~b}        &   [1,0,\ldots,0,1]_{0}      &  [0,\ldots,0]_0 &   0   &  [0]    \\
 \hline \hline
\mathcal{F}^a_{b}           &    [1,0,\ldots,0,1]_{0}          &  [0,\ldots,0]_0     &   2   &   [0] \\
 \end{array}
 \end{equation}
 The superpotential is given by
 \begin{equation}
 W=(\widetilde{S}_{\alpha})^{ a b} \varphi^c_{~b} (S^\alpha)_{ca}+ {\tQ}_{i}^a \varphi^{b}_{~a} Q^{i}_b
  \end{equation}
The $F$-terms corresponding to the Higgs branch is given by
\bea
0 = \partial_{\varphi^c_{~b}} W = (\widetilde{S}_{\alpha})^{ a b}  (S^\alpha)_{ca}+ {\tQ}_{i}^b Q^{i}_c = \epsilon_{\alpha \beta} (\widetilde{S}^\alpha)^{ab} (S^\beta)_{ca}+ {\tQ}_{i}^b Q^{i}_c~.
\eea  

\subsubsection{$Sp(1)$ instanton on $\mathbb{C}^2/\mathbb{Z}_2$: $k=1$ and $N=1$}
Let $q$ be fugacity of the gauge fugacity of $U(1)$, $x$ be the fugacity of the global symmetry $SU(2)_x$ and $y$ be the fugacity of the global symmetry $U(1)$.  The Hilbert series is
\bea
g^{N=1}_{k=1}(t,x,y) &= \oint_{|q|=1} \frac{\ud q}{(2 \pi i) q}  \frac{\chi_{Q} (t,q,z,y) \chi_{\tQ} (t,z,y) \chi_{S} (t,q,z,x) \chi_{\tS} (t,q,z,x)}{\chi_F (t,z)}~,
\eea
where
\bea
\begin{array}{llll}
\chi_{Q} (t,q,z,y) &= \PE \left[t q y^{-1} \right]~,  &\qquad \chi_{\tQ} (t,z,y) &= \PE \left[t q^{-1} y \right]~, \\
\chi_{S} (t,q,z,x) &=  \PE[t q^2 [1]_x  ]~,  &\qquad \chi_{\tS} (t,q,z,x) &=  \PE[t q^{-2} [1]_x ]~, \\
\chi_{F} (t,z) &= \PE[t^2 ]~. & &
\end{array}
\eea
The corresponding unrefined Hilbert series is 
\bea \label{C2Z2Spk1N1noVS}
g^{N=1}_{k=1}(t,1, 1) &=  \frac{1 + 2 t^2 + 2 t^3 + 
 2 t^4 + t^6}{(1 - t)^4 (1 + t)^2 (1 + t + t^2)^2}~.
\eea
The plethystic logarithm of $g^{N=1}_{k=1}(t,x,u, y)$ is
\bea
\PL \left[ g^{N=1}_{k=1}(t,x, y) \right] 
&= ([2]_x+1)t^2 + (y^2+y^{-2})[1]_xt^3 -t^4  \nn \\
& \quad - (y^2+y^{-2})[1]_x t^5 -\ldots~.
\eea

\subsubsection{$Sp(1)$ instantons on $\mathbb{C}^2/\mathbb{Z}_2$: $k=2$ and $N=1$}
Let $(q,z)$ be fugacity of the gauge fugacity of $U(2)=U(1) \times SU(2)$, $x$ be the fugacity of the global symmetry $SU(2)_x$ and $y$ be the fugacity of the global symmetry $U(1)$.  The Hilbert series is given by
\bea
g^{N=1}_{k=2}(t,x,y) &= \left(\oint_{|q|=1} \frac{\ud q}{(2 \pi i) q} \oint_{|z|=1} \ud z\frac{1-z^2}{(2 \pi i)z}\right) \times \nn \\
& \qquad  \frac{\chi_{Q} (t,q,z,y) \chi_{\tQ} (t,z,y) \chi_{S} (t,q,z,x) \chi_{\tS} (t,q,z,x)}{\chi_F (t,z)}~,
\eea
where
\bea
\begin{array}{llll}
\chi_{Q} (t,q,z,y) &= \PE \left[t q [1]_z  y^{-1} \right]~,  &\qquad \chi_{\tQ} (t,z,y) &= \PE \left[t q^{-1} [1]_z y \right]~, \\
\chi_{S} (t,q,z,x) &=  \PE[t q^2 [1]_x [2]_z ]~,  &\qquad \chi_{\tS} (t,q,z,x) &=  \PE[t q^{-2} [1]_x [2]_z ]~, \\
\chi_{F} (t,z) &= \PE[t^2 ([2]_z +1) ]~. & &
\end{array}
\eea
The corresponding unrefined Hilbert series is 
\bea \label{C2Z2k2N1noVS}
g^{N=1}_{k=2}(t,1, 1) &=  \frac{1}{(1-t)^8 (1 + t)^4 (1 + t^2)^2 (1 + t + t^2)^2 (1 + t + t^2 + t^3 + t^4)^2} \Big( 1 + 2 t^2 + 2 t^3  \nn \\
& \quad + 9 t^4 + 10 t^5 + 15 t^6 + 18 t^7 + 28 t^8 + 
 26 t^9 + 34 t^{10} + 26 t^{11} + \text{palindrome} +t^{20} \Big) \nn \\
&= 1 + 4 t^2 + 4 t^3 + 18 t^4 + 24 t^5 + 58 t^6 + 92 t^7 + 181 t^8 + \ldots~.
\eea
The plethystic logarithm of $g^{N=1}_{k=2}(t,x, y)$ is
\bea
\PL \left[ g^{N=1}_{k=2}(t,x, y) \right] 
&= ([2]+1)t^2 + (y^2+y^{-2})[1]t^3 +([4]+[2]) t^4  \nn \\
& \quad + (y^2+y^{-2})[3]t^5 - ([2]+1)t^6-\ldots~.
\eea

\subsubsection{$Sp(2)$ instanton on $\mathbb{C}^2/\mathbb{Z}_2$: $k=1$ and $N=2$}
Let $z$ be fugacity of the gauge fugacity of $U(1)$, $x$ be the fugacity of $SU(2)_x$ and $(u, y)$ be the fugacity of $U(2) = U(1) \times SU(2)$.  The Hilbert series is given by
\bea
g^{N=2}_{k=1}(t,x,u, y) = \oint_{|z|=1} \frac{\ud q}{(2 \pi i) z} \frac{\chi_{Q} (t,z,u,y) \chi_{\tQ} (t,z,u,y) \chi_{S} (t,z,x) \chi_{\tS} (t,z,x)}{\chi_F (t)}~,
\eea
where
\bea
\begin{array}{llll}
\chi_{Q} (t,z,u,y) &= \PE \left[ z [1]_y u^{-1} \right]~,  &\qquad \chi_{\tQ} (t,z,u,y) &= \PE \left[ z^{-1} [1]_y u^{1} \right]~, \\
\chi_{S} (t,z,x) &=  \PE[t [1]_x z^2 ]~,  &\qquad \chi_{\tS} (t,z,x) &=  \PE[t [1]_x z^{-2} ]~, \\
\chi_{F} (t) &= \PE[t^2] = (1-t^2)^{-1}~. & &
\end{array}
\eea
The corresponding unrefined Hilbert series is 
\bea \label{C2Z2Spk1N2noVS}
g^{N=2}_{k=1}(t,1,1, 1) =  \frac{1-t+5 t^2+4 t^3+4 t^4+4 t^5+5 t^6-t^7+t^8}{(1-t)^6 (1+t)^2 \left(1+t+t^2\right)^3}~.
\eea
The plethystic logarithm of $g^{N=2}_{k=1}(t,x,u, y)$ is
\bea
\PL \left[ g^{N=2}_{k=1}(t,x,u, y) \right] = ([2;0]+[0;2]+1)t^2 + (u^2+u^{-2})[1;2]t^3 -2 t^4 -\ldots~.
\eea

\subsubsection{$Sp(2)$ instantons on $\mathbb{C}^2/\mathbb{Z}_2$: $k=2$ and $N=2$}
Let $(q,z)$ be fugacity of the gauge fugacity of $U(2)=U(1) \times SU(2)$, $x$ be the fugacity of the global symmetry $SU(2)_x$ and $(u, y)$ be the fugacity of the global symmetry $U(2) = U(1) \times SU(2)$.  The Hilbert series is given by
\bea
g^{N=2}_{k=2}(t,x,u, y) &= \left(\oint_{|q|=1} \frac{\ud q}{(2 \pi i) q} \oint_{|z|=1} \ud z\frac{1-z^2}{(2 \pi i)z}\right) \times \nn \\
& \qquad  \frac{\chi_{Q} (t,q,z,u,y) \chi_{\tQ} (t,q,z,u,y) \chi_{S} (t,q,z,x) \chi_{\tS} (t,q,z,x)}{\chi_F (t,z)}~,
\eea
where
\bea
\begin{array}{llll}
\chi_{Q} (t,q,z,u,y) &= \PE \left[t q [1]_z  [1]_y u^{-1} \right]~,  &\qquad \chi_{\tQ} (t,z,u,y) &= \PE \left[t q^{-1} [1]_z [1]_y u \right]~, \\
\chi_{S} (t,q,z,x) &=  \PE[t q^2 [1]_x [2]_z ]~,  &\qquad \chi_{\tS} (t,q,z,x) &=  \PE[t q^{-2} [1]_x [2]_z ]~, \\
\chi_{F} (t,z) &= \PE[t^2 ([2]_z +1) ]~. & &
\end{array}
\eea
The corresponding unrefined Hilbert series is 
\bea  \label{C2Z2Spk2N2noVS}
g^{N=2}_{k=2}(t,1,1, 1) &=  \frac{1}{(1-t)^{12} (1+t)^4 \left(1+t^2\right)^2 \left(1+t+t^2\right)^6 \left(1+t+t^2+t^3+t^4\right)^3} \Big( 1+t+6 t^2 \nn \\
& \quad +12 t^3+42 t^4+93 t^5+214 t^6+415 t^7+790 t^8+1348 t^9+2156 t^{10}+3133 t^{11} \nn \\
& \quad +4275 t^{12}+5392 t^{13}+6416 t^{14}+7078 t^{15}+7352 t^{16}+7078 t^{17}  \nn \\
& \quad + \text{palindrome} +t^{32} \Big) \nn \\
&= 1 + 7 t^2 + 12 t^3 + 45 t^4 + 108 t^5 + 271 t^6 + 624 t^7 + 
 1382 t^8 + \ldots~.
\eea
The plethystic logarithm of $g^{N=2}_{k=2}(t,x,u, y)$ is
\bea
\PL \left[ g^{N=2}_{k=2}(t,x,u, y) \right] &= ([2;0]+[0;2]+1)t^2 + (u^2+u^{-2})[1;2]t^3 +([4;0]+[2;2]+[2;0]) t^4  \nn \\
& \quad + (u^2+u^{-2})[3;2]t^5 - (2[2;0]+[0;2]+1)t^6 -\ldots~.
\eea

\subsection{Summary: Matching of Hilbert series for instantons on $\BC^2/\BZ_2$}
We summarise in \tref{tab:match2} matching of Hilbert series for equivalent instantons of isomorphic groups on $\BC^2/\BZ_2$ but very different quiver description, as presented in the preceding subsections.

\begin{center}
\begin{longtable}{|c|c|c|}
\hline
Quiver & Matches with & Hilbert series\\
\hline
$SO(2)$ instanton on $\BC^2/\BZ_2$ & $\BC^2/\BZ_2$  & \eref{C2Z2SO2N02VS}
\\ $O/O$: $\vec k =(1,1)$ and $\vec N=(0,2)$&  & \\
\hline
$SO(5)$ instanton on $\BC^2/\BZ_2$ & $Sp(2)$ instanton on $\BC^2/\BZ_2$  & \eref{C2Z2k1N14VS}, \eref{C2Z2Spk11N11noVS}
\\ $O/O$: $\vec k =(1,1)$ and $\vec N=(1,4)$& $S/S$: $\vec k=(1,1)$ and $\vec N=(1,1)$ & \\
\hline
$SO(5)$ instanton on $\BC^2/\BZ_2$ & $Sp(2)$ instanton on $\BC^2/\BZ_2$  & \eref{C2Z2k1N23VS}, \eref{C2Z2Spk1N2noVS}
\\ $O/O$: $\vec k =(1,1)$ and $\vec N=(2,3)$& $SS$: $k=1$ and $N=2$ & \\
\hline
$SO(5)$ instantons on $\BC^2/\BZ_2$ & $Sp(2)$ instantons on $\BC^2/\BZ_2$  &\eref{C2Z3k2N23VS}, \eref{C2Z2Spk2N2noVS}
\\ $O/O$: $\vec k =(2,2)$ and $\vec N=(2,3)$& $SS$: $k=2$ and $N=2$ & \\
\hline
$SO(6)$ instanton on $\BC^2/\BZ_2$ & $SU(4)$ instanton on $\BC^2/\BZ_2$ with  & \eref{HS:C2Z2SO6N24}
\\ $O/O$: $\vec k =(1,1)$ and $\vec N=(2,4)$& $\vec k=(1,1)$ and $\vec N=(2,4)$ & \\
\hline
$SO(8)$ instanton on $\BC^2/\BZ_2$ & $SO(8)$ instanton on $\BC^2/\BZ_2$ & \eref{C2Z3k1N26VS}, \eref{C2Z3SO8k1N4noVS}
\\ $O/O$: $\vec k =(1,1)$ and $\vec N=(2,6)$& $AA$: $k=1$ and $N=4$ & \\
\hline
$Sp(1)$ instanton on $\BC^2/\BZ_2$ & $SU(2)$ instanton on $\BC^2/\BZ_2$ with & \eref{C2Z2Spk11N0}, \eref{C2Z2SUk110N}
\\ $S/S$: $k =(1,1)$ and $N=(1,0)$&  $\vec k =(1,1)$ and $\vec N=(0,2)$ & \\
\hline \hline
$SO(2)$ instanton on $\BC^2/\BZ_2$ & $\BC^2/\BZ_2$ & \eref{C2Z2k1N1noVS}
\\ $AA$: $k =1$ and $N=1$&  & \\
\hline
$SO(6)$ instanton on $\BC^2/\BZ_2$ & $SU(4)$ instanton on $\BC^2/\BZ_2$ with  & \eref{C2Z2k1N3noVS}
\\ $AA$: $k =1$ and $N=3$&  $\vec k=(1,1)$ and $\vec N=(3,1)$& \\
\hline
$SO(6)$ instantons on $\BC^2/\BZ_2$ & $SU(4)$ instantons on $\BC^2/\BZ_2$ with & \eref{C2Z2k32N3noVS}
\\ $AA$: $k =3/2$ and $N=3$&  $\vec k=(2,1)$ and $\vec N=(3,1)$~\footnote{Note that this is not pure instantons; according to \eref{betafn}, $\vec \beta =(1,3)$.} & \\
\hline
$Sp(1)$ instanton on $\BC^2/\BZ_2$ & $SU(2)$ instanton on $\BC^2/\BZ_2$ with & \eref{C2Z2Spk1N1noVS}, \eref{C2Z2SUk11N11}
\\ $SS$: $k =1$ and $N=1$&  $\vec k =(1,1)$ and $\vec N=(1,1)$ & \\
\hline
$Sp(1)$ instantons on $\BC^2/\BZ_2$ & $SU(2)$ instanton on $\BC^2/\BZ_2$ with & \eref{C2Z2k2N1noVS}
\\ $SS$: $k =2$ and $N=1$&  $\vec k =(2,2)$ and $\vec N=(1,1)$ &  \\
\hline
\caption{Matching of Hilbert series of different quivers for instantons on $\BC^2/\BZ_2$.}\label{tab:match2}
\end{longtable}
\end{center}

\section{The hybrid configurations} \label{sec:hybrids}

\begin{figure}[H]
\centering
\includegraphics[scale=.45]{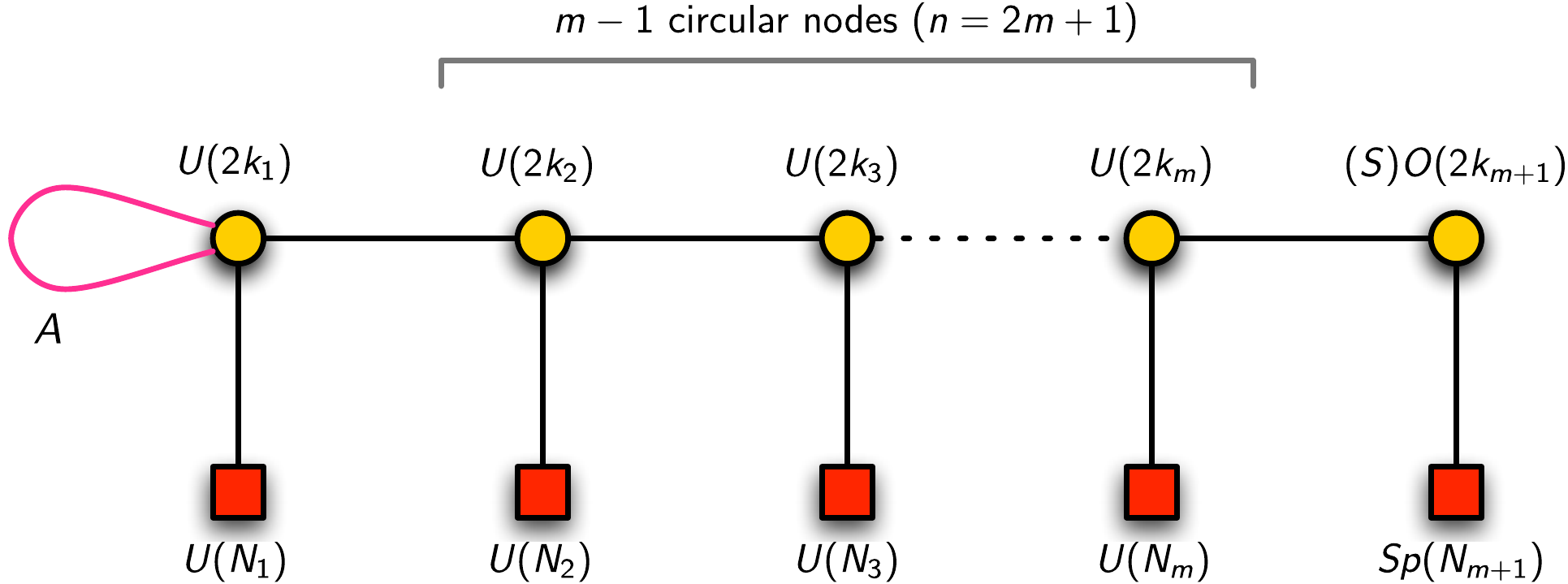} \\~\\
\includegraphics[scale=.45]{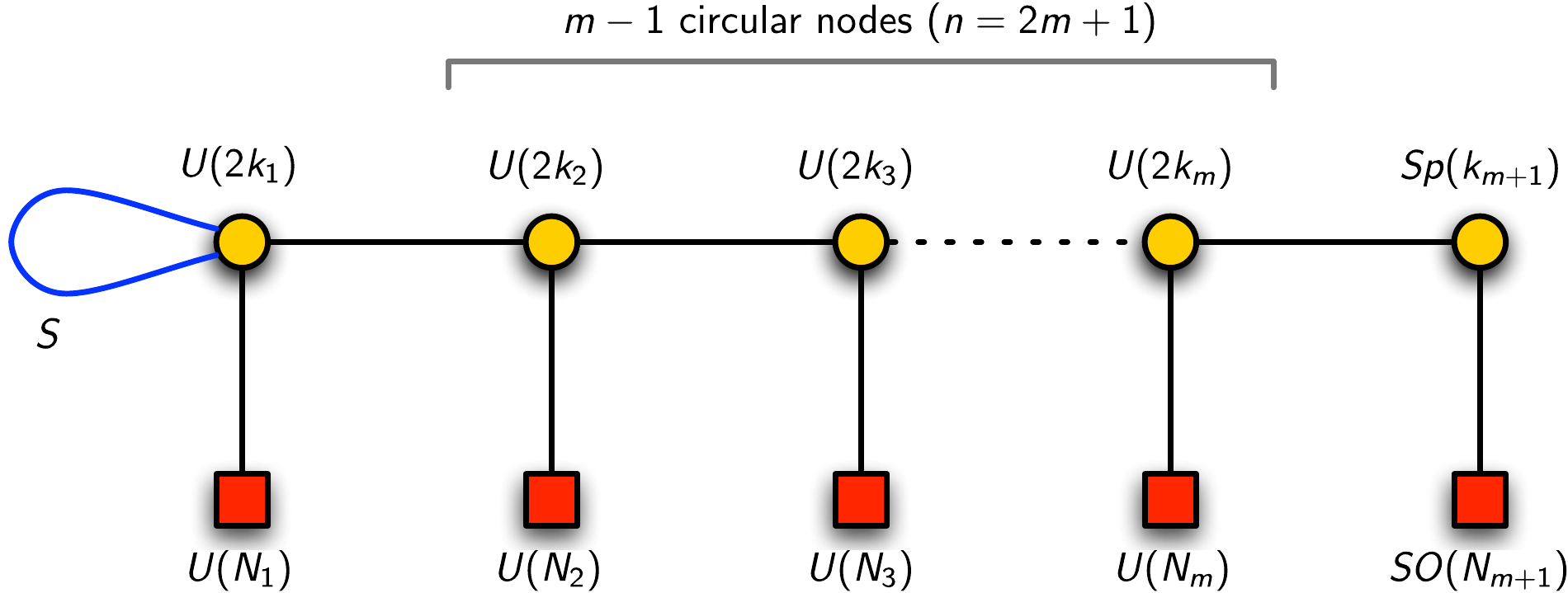}
\caption{The hybrids between the quivers for $SO(N)$ and $Sp(N)$ instantons on $\BC^2/\BZ_{2m+1}$. For the top quiver, $k_1, k_2, \ldots, k_m, k_{m+1}$ can be either integral or half-integral and for the bottom quiver, $k_1, k_2, \ldots, k_m$ can be either integral or half-integral.  The notation $(S)O$ indicates that the gauge symmetry can be taken to be (special) orthogonal group; these different choices yield different moduli spaces as discussed below \eref{relHplusg1011}. Below, we take this group to be special orthogonal.}
\label{fig:hybridA}
\end{figure}

\begin{figure}[H]
\centering
\includegraphics[scale=.45]{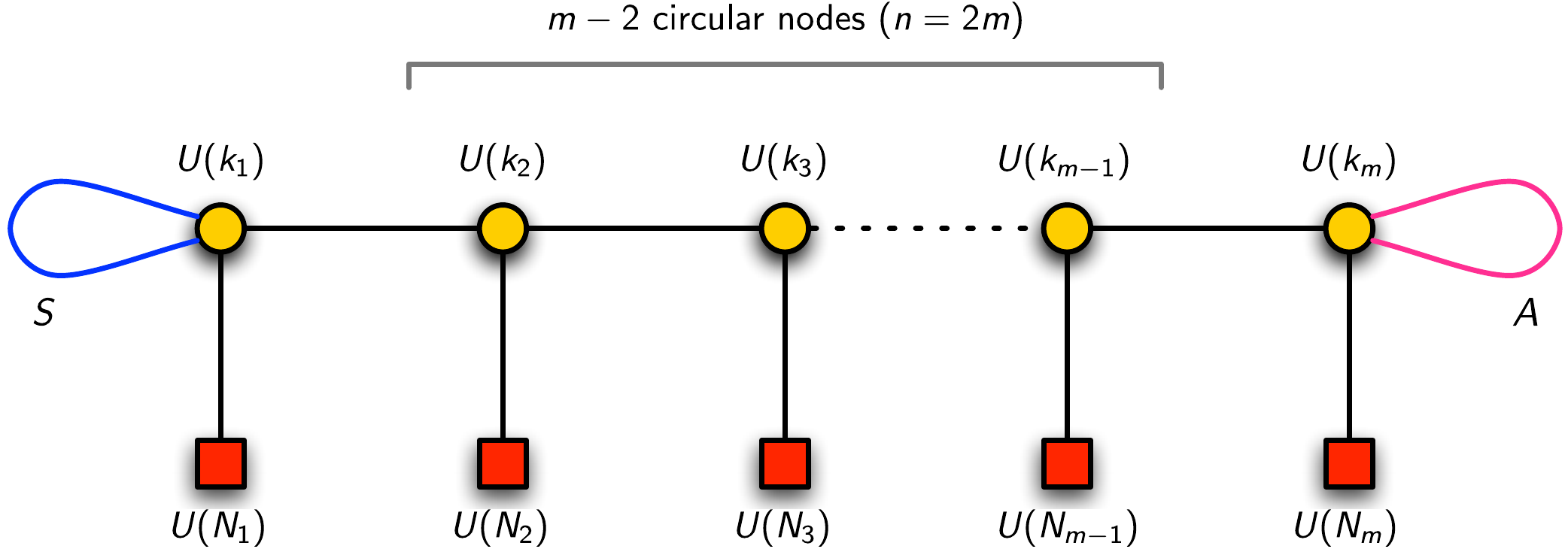} \\~\\
\includegraphics[scale=.45]{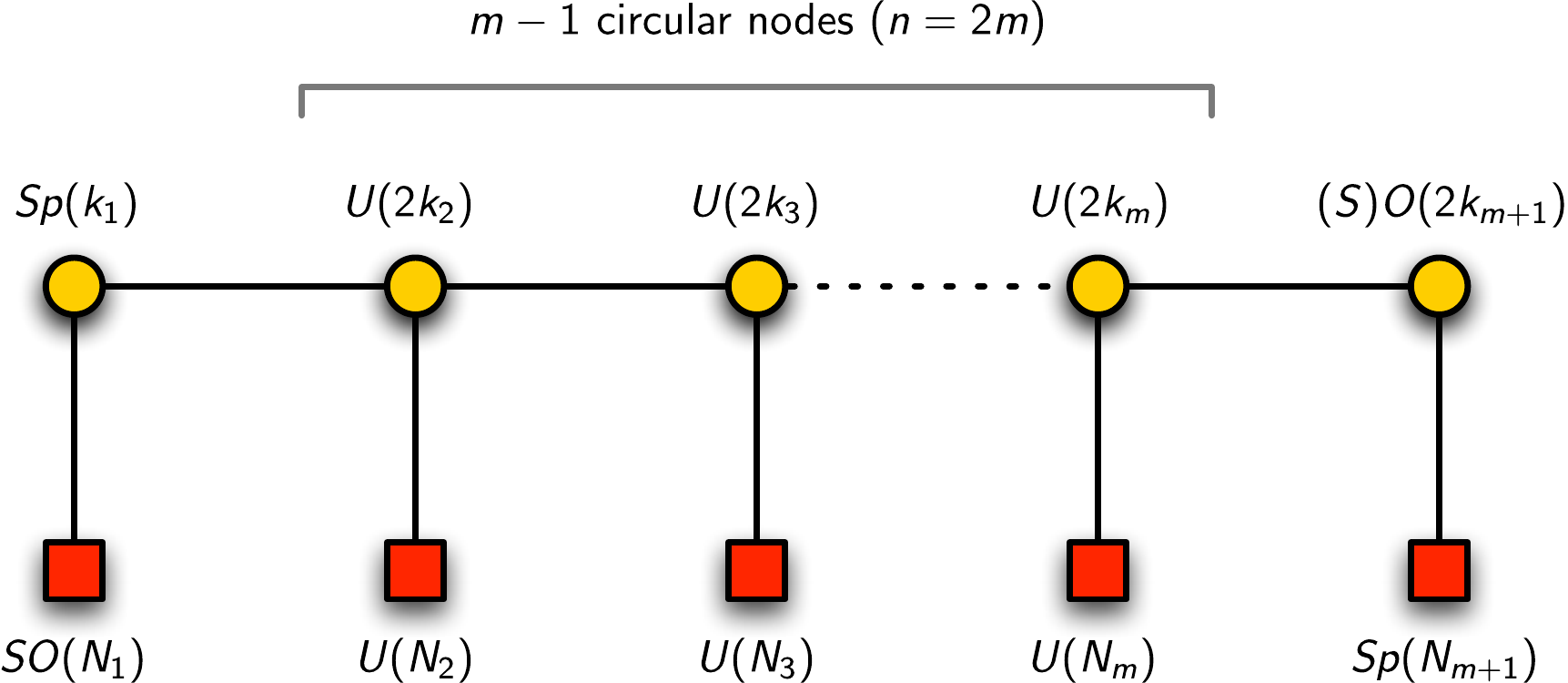}
\caption{The $SA$ hybrid (top) and the $O/S$ hybrid (bottom). For the bottom quiver,  $k_2, \ldots, k_m, k_{m+1}$ can be either integral or half-integral.}
\label{fig:hybridB}
\end{figure}

In addition to the quivers presented in Figures \ref{fig:SO_N_oddn}, \ref{fig:Sp_N_oddn}, \ref{fig:SO_N_withVSevenn}, \ref{fig:Sp_N_withVSevenn}, \ref{fig:SO_N_withNOVSevenn} and \ref{fig:Sp_N_withNOVSevenn}, we can construct four more quivers that are `hybrids' of such quivers in a similar fashion to \cite{Hanany:2001iy}.  
\bi
\item For $\BC^2/\BZ_{2m+1}$, we present the hybrid between the quivers for $SO(N)$ and $Sp(N)$ instantons in \fref{fig:hybridA}.
\item For $\BC^2/\BZ_{2m}$, there are two hybrids that can be formed:
\bi
\item The top diagram of \fref{fig:hybridB} depicts the hybrid between the $SS$ and $AA$ quivers.  We refer to this as the $SA$ hybrid.
\item The bottom diagram of \fref{fig:hybridB} depicts the $O/S$ hybrid.
\ei
\ei

If $k_1 =k_2 =\cdots = k_{m+1} =k$, the dimension of the Higgs branch of each quiver in \fref{fig:hybridA} is $k N$, with $N=2\sum_{i=1}^{m+1} N_i$.

If $k_1 =k_2 =\cdots = k_m = k_{m+1} =k$, the dimension of the Higgs branch of each quiver in \fref{fig:hybridB} is $k N$, with $N=\sum_{i=1}^{m} N_i$ for the top quiver and $N= N_1 + 2N_2 + \cdots + 2N_{m-1}$ for the bottom quiver.

\begin{figure}[H]
\begin{center}
\includegraphics[scale=0.7]{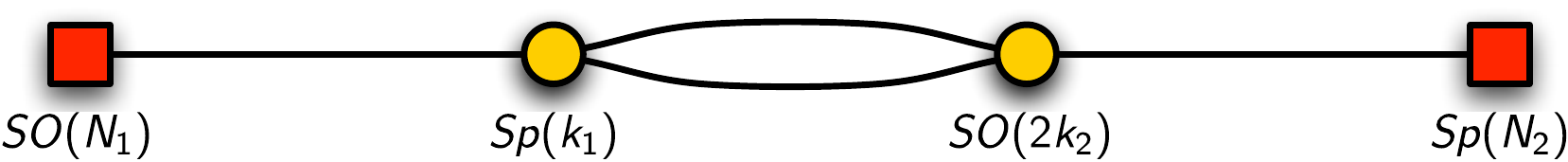}
\caption{The $O/S$ hybrid with $m=1$.  Here, $k_1$ takes an integral value, whereas $k_2$ can either take {\bf integral} or {\bf half-integral} values.  Each line between $Sp(r_1)$ and $SO(r_2)$ denotes $2r_1r_2$ half-hypermultiplets. Hence the total complex degrees of freedom of the bi-fundamental fields between $Sp(k_1)$ and $SO(2k_2)$ is $8 k_1 k_2$.}
\label{fig:SOSphybrid}
\end{center}
\end{figure}

\begin{figure}[H]
\begin{center}
\includegraphics[scale=0.7]{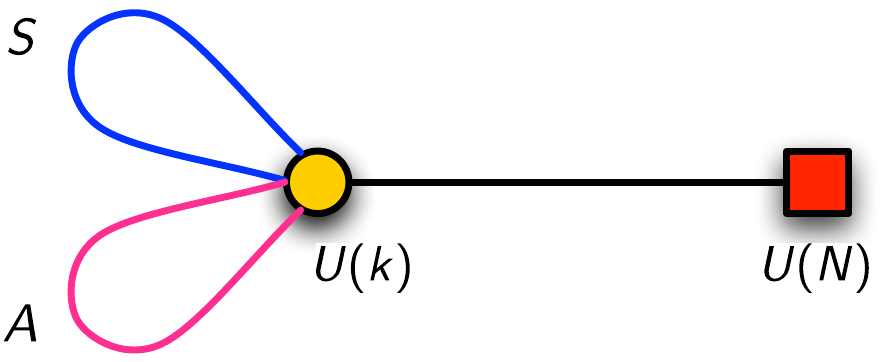}
\caption{The $SA$ hybrid with $m=1$.  The loops labelled by $A$ and $S$ denotes hypermultiplets transforming as the rank 2-antisymmetric and rank 2-symmetric tensors of the gauge group $U(k)$, respectively.}
\label{fig:SUNinstAS}
\end{center}
\end{figure}

Subsequently, we focus on the special cases of \fref{fig:hybridB} with $m=1$.  The corresponding quivers are depicted in Figures \ref{fig:SOSphybrid} and \ref{fig:SUNinstAS}. It should be pointed out that the symmetry present on the Higgs branch of each of these quivers is similary to that present in the string backgrounds studied in \cite{Chaudhuri:1995fk, Chaudhuri:1995bf,Dabholkar:1996pc}. As briefly described in appendix \ref{classification}, these are related by dualities to elliptic brane/orientifold containing two opposite charged O-planes, instead of two equally charged ones.

\subsection{The $O/S$ hybrid with $m=1$}
If $k_1=k_2=k$, we see that the quaternionic dimension of the Higgs branch is
\bea
 kN_1 + 2k^2 + 2k^2 + 2kN_2 - \frac{1}{2}(2k)(2k+1)- \frac{1}{2}(2k)(2k-1) = k (N_1+2N_2) = kN~,
\eea
where $N = N_1+2N_2$.

Let us denote the half-hypermultplets in the bi-fundamental representation of $SO(N_1) \times Sp(k_1)$ by $Q^{i_1}_{a_1}$ and those of $SO(2k_2) \times Sp(N_2)$ by $q^{j_1}_{b_1}$, where $i, i_1, i_2,\ldots=1, \ldots, N_1$ are the indices for $SO(N_1)$, $a_1,a_2,\ldots=1,\ldots, 2k_1$ are the indices for $Sp(k_1)$, $b_1,b_2,\ldots=1,\ldots, 2k_2$ are the indices for $O(2k_2)$ and $j, j_1,j_2,\ldots=1,\ldots, 2N_2$ are the indices for $Sp(N_2)$.   We also denote by $\varphi_1$ and $\varphi_2$ the scalars in the vector multiplets of the gauge group $Sp(k_1)$ and $SO(2k_2)$ respectively.
The half-hypermultiplets in the bi-fundamental representation of $Sp(k_1) \times SO(2k_2)$ are denoted by $(X^\alpha)^{a_1 b_1}$, where $\alpha, \beta, \alpha_1, \alpha_2, \ldots=1,2$ corresponding to a global symmetry which we shall refer to as $SU(2)_x$.  


\paragraph{The superpotential.}
In 4d $\CN=1$ fomalism, the superpotential is
\bea
W &= M^{SO(N_1)} _{i_1 i_2} Q^{i_1}_{a_1} Q^{i_2}_{a_2} (\varphi_1)^{a_1a_2} + J^{Sp(N_2)}_{j_1j_2} q^{j_1}_{b_1} q^{j_2}_{b_2} (\varphi_2)^{b_1b_2} + \epsilon_{\alpha_1\alpha_2}M^{SO(2k_2)}_{b_1b_2} (\varphi_1)_{a_1 a_2} (X^{\alpha_1})^{a_1 b_1} (X^{\alpha_2})^{a_2 b_2}\nn \\
&\quad +  \epsilon_{\alpha_1\alpha_2} J^{Sp(k_1)}_{a_1 a_2}(\varphi_2)_{b_1b_2} (X^{\alpha_1})^{a_1 b_1} (X^{\alpha_2})^{a_2 b_2}~, \label{supSOSphybrid}
\eea
where $M^{SO(m)}$ can be chosen to be \eref{SOmquad} in order that the Cartan subalgebra is the subalgebra of diagonal matrices and $J^{Sp(m)}$ is the symplectic matrix for $Sp(m)$.
With such choices of quadratic forms, the field $(\varphi_1)^{a_1a_2}$ takes the following form: 
\bea
\varphi_1^{a_1 a_2} = \begin{pmatrix} A & B \\ C & D \end{pmatrix}~,
\eea
where the off-diagonal blocks $B$ and $C$ are symmetric and the diagonal blocks $A$ and $D$ are negative transposes of each other; on the other hand, $(\varphi_2)^{b_1b_2}$ takes the following form:
\bea
\varphi_2^{b_1b_2} = \begin{pmatrix} A' & B' \\ C' & D' \end{pmatrix}~,
\eea
where the off-diagonal blocks $B'$ and $C'$ are anti-symmetric and the diagonal blocks $A'$ and $D'$ are negative transposes of each other.


\subsubsection{The case of $\vec k =(1,1)$ and $\vec N =(1,0)$}
%

Since both $\epsilon_{b_1b_2}$ and $M^{SO(2)}_{b_1b_2}$ are invariant tensors of $SO(2)$, in this special case the superpotential is
\bea
W&=  \varphi_1^{a_1 a_2} Q_{a_1} Q_{a_2} + \epsilon_{\alpha_1 \alpha_2} (\varphi_1)_{a_1a_2} \epsilon_{b_1 b_2} (X^{\alpha_1})^{a_1b_1} (X^{\alpha_2})^{a_2b_2} \nn \\
& \quad + \epsilon_{\alpha_1 \alpha_2} \epsilon_{a_1a_2} \varphi_2 \epsilon_{b_1 b_2} (X^{\alpha_1})^{a_1b_1} (X^{\alpha_2})^{a_2b_2}~.
\eea
where we take $(\varphi_2)_{b_1b_2} =\varphi_2 \epsilon_{b_1 b_2}$ with $\varphi_2$ a complex scalar field.  

Let $z_1$ and $z_2$ be the fugacities for the gauge symmetries $Sp(1)$ and $SO(2)$ respectively.  Then, the Hilbert series is
\bea
g^{(1,0)}_{(1,1)} (t,x,b) = \left( \oint_{|z_1|=1} \frac{(1-z_1^2) \ud z_1}{(2 \pi i) z_1}   \oint_{|z_2|=1} \frac{\ud z_2}{(2 \pi i) z_2} \right) \frac{\chi_X(t,z_1, z_2, x) \chi_Q(t,z_1)}{\chi_F(t^2,z_1)} ~,
\eea
where the contributions from the fields $q$, $X$ and the $F$ terms are given by
\bea
\chi_Q(t,z_1) &= \PE[t [1]_{z_1} ]~, \nn \\
\chi_X(t,z_1, z_2, x) &= \PE[ t [1]_{z_1} (z_2+z_2^{-1})  [1]_x ]~, \nn \\
\chi_F(t,z_1) &= \PE[ t^2 + t^2 [2]_{z_1} ]~.
\eea
Computing the above integral, we obtain 
\bea
g^{(1,0)}_{(1,1)} (t,x,b) = (1-t^4) \PE [[2]_x t^2]  = \sum_{m=0}^\infty [2m]_x t^{2m}~.
\eea
Notice that this is the Hilbert series of $\BC^2/\BZ_2$.  The generators of the moduli space are
\bea \label{genHplusg1011}
G^{\alpha_1 \alpha_2} = (X^{\alpha_1})^{a_1b_1} (X^{\alpha_2})^{a_2b_2} \epsilon_{a_1 a_2} \epsilon_{b_1b_2}
\eea
Note that $G^{\alpha_1 \alpha_2}$ is symmetric under the exchange of $\alpha_1$ and $\alpha_2$, so it transform as a triplet under $SU(2)_x$.\footnote{Note that the $F$ terms $\partial_{\varphi_2} W=0$ imply that the gauge invariant combination $\widehat{G}^{\alpha \beta} := (X^\alpha)^{a_1 b_1} (X^\beta)^{a_2 b_2} \epsilon_{a_1 a_2} M^{SO(2)}_{b_1b_2}$ vanishes.} Using the $F$-terms, one find that the relation at order 4 is
\bea
\det G= G^{11}G^{22}-(G^{12})^2 = 0~. \label{relHplusg1011}
\eea

We emphasise again that in the above analysis the gauge group in the right circular node of the quiver is taken to be $SO(2)$, not $O(2)$.  The former group possesses $\epsilon_{b_1b_2}$ as an invariant tensor, whereas the latter group does not.  Hence it is clear that the global symmetry $SU(2)_x$ of $\BC^2/\BZ_2$ is broken.

\subsubsection{The case of $\vec k =(1,1)$ and $\vec N =(0,1)$}
By examining the $F$-terms arising from the superpotential in \eref{supSOSphybrid} using {\it primary decomposition} of {\tt Macaulay2} \cite{mac2}, we find that the $F$-flat space has two branches.  
\bi
\item Branch 1 is defined by the following homogeneous relations:
\bea
\begin{array}{ll}  
\epsilon_{j_1j_2} Q^{j_1}_1 Q^{j_2}_2 + 2 \epsilon_{a_1 a_2} (X^1)^{a_1 1}(X^2)^{a_2 2} =0~, &\qquad
\epsilon_{j_1j_2} Q^{j_1}_1 Q^{j_2}_2 + 2 \epsilon_{\alpha_1 \alpha_2} (X^{\alpha_1})^{11}(X^{\alpha_2})^{22} =0~, \\
\epsilon_{a_1 a_2} (X^1)^{a_1 b_1}(X^1)^{a_2 b_2}=0~, & \qquad \epsilon_{a_1 a_2} (X^2)^{a_1 b_1}(X^2)^{a_2 b_2}=0~, \\
\epsilon_{\alpha_1 \alpha_2} (X^{\alpha_1})^{1b_1}(X^{\alpha_2})^{1b_2}=0~, & \qquad \epsilon_{\alpha_1 \alpha_2} (X^{\alpha_1})^{2b_1}(X^{\alpha_2})^{2b_2}=0~, \\
\epsilon_{b_1b_2} (X^1)^{1b_1}(X^2)^{2b_2}=0~. & 
\end{array}
\eea
\item Branch 2 is defined by the following homogeneous relations:
\bea \label{defbr20111}
\begin{array}{ll}  
\epsilon^{b_1b_2} \epsilon_{j_1j_2}q^{j_1}_{b_1} q^{j_2}_{b_2} =0~, & \qquad  \\
\epsilon_{\alpha_1 \alpha_2} (X^{\alpha_1})^{1b_1}(X^{\alpha_2})^{1b_2}=0~, & \qquad \epsilon_{\alpha_1 \alpha_2} (X^{\alpha_1})^{2b_1}(X^{\alpha_2})^{2b_2}=0~, \\
\epsilon_{\alpha_1 \alpha_2} (X^{\alpha_1})^{a_11}(X^{\alpha_2})^{a_21}=0~, & \qquad \epsilon_{\alpha_1 \alpha_2} (X^{\alpha_1})^{a_1 2}(X^{\alpha_2})^{a_2 2}=0~, \\
\epsilon_{\alpha_1 \alpha_2} (X^{\alpha_1})^{12}(X^{\alpha_2})^{21}=0~, & \qquad \epsilon_{\alpha_1 \alpha_2} (X^{\alpha_1})^{11}(X^{\alpha_2})^{22}=0~.
\end{array}
\eea
\ei

\subsubsection*{Branch 1: Hilbert series}
The $F$-flat Hilbert series of Branch 1 is
\bea
\fflat_1 (t;z_1,z_2;x,b) &= \Big[ 1-t^2 (1+[2]_x+[2]_{z_1}) +t^3 [1]_x [1]_{z_1} [1]_{z_2} + t^4( [1]_x +[2]_{z_1} -[2]_{z_2} ) \nn \\
& \qquad - t^5  [1]_x [1]_{z_1} [1]_{z_2} + t^6 [2]_{z_2} \Big] \times \PE \left[ t [1]_{z_2} ([1]_x [1]_{z_1} + [1]_b) \right]~,
\eea
where we wrote $SO(2)$ characters in terms of $z_2$ in the $SU(2)$ character notation. The corresponding unrefined Hilbert series is
\bea
\fflat_1 (t;z_1=1,z_2=1;x=1,b=1)  = \frac{1 + 4t + 3t^2}{(1 - t)^8}~.
\eea
Integrating over the gauge symmetries, we obtain
\bea
g^{(0,1)}_{(1,1), {\rm Br. 1}}(t,b) &= \frac{1}{(2 \pi i)^2}\oint_{|z_1|=1} \frac{\ud z_1}{z_1} (1-z_1^2)\oint_{|z_2|=1} \frac{\ud z_2}{z_2} \fflat_1 (t;z_1,z_2;x,b) \nn\\
&= 1+ (1+[2]_b)t^2 + (1+3[2]_b +[4]_b)t^4 + (1+ 3[2]_b +3[4]_b +[6]_b)t^6 \nn \\
& \qquad +(1 + 3[2]_b + 5 [4]_b + 3 [6]_b + [8]_b) t^8 + \ldots~.
\eea
The corresponding unrefined Hilbert series is
\bea
g^{(0,1)}_{(1,1), {\rm Br. 1}}(t,b=1) =  \frac{1+2 t^2+6 t^4+2 t^6+t^8}{\left(1-t^2\right)^4 \left(1+t^2\right)^2}~.
\eea
The plethystic logarithm is
\bea
\PL [ g^{(0,1)}_{(1,1), {\rm Br. 1}}(t,b)  ] = (1+[2]_b)t^2 +(2[2]_b -1) t^4 - \ldots~.
\eea

The generators at order 2 in the triplet and the singlet of $Sp(1)$ are respectively
\bea
G^{j_1j_2}_{[2]} = M_{SO(2)}^{b_1b_2} q^{j_1}_{b_1} q^{j_2}_{b_2}~, \qquad G^{j_1j_2}_{[0]} = \epsilon^{b_1b_2} q^{j_1}_{b_1} q^{j_2}_{b_2}~.
\eea
The generators at order 4 are
\bea
H^{j_1 j_2}_{[2]} &=  \epsilon_{a_1 a_2} ( X^\alpha)^{a_1 b_1} (X^\beta)^{a_2 b_2} q^{j_1}_{b_1} q^{j_2}_{b_2}~, \nn \\
\widehat{H}^{j_1 j_2}_{[2]} &=  \epsilon_{a_1 a_2} ( X^\alpha)^{a_1 b_1} (X^\beta)^{a_2 b_2} q^{j_1}_{b_1}  q^{j_1b_3}M^{SO(2)}_{b_2b_3}~,
\eea
where we define $q^{j_1b_3}$ as
\bea
q^{j_1b_3}  = \epsilon^{b_3b_4} q^{j_1}_{b_4}~.
\eea

\subsubsection*{Branch 2: Hilbert series of $(\BC^2/\BZ_2)^2$}
The $F$-flat Hilbert series of Branch 2 is
\bea
\fflat_2 (t;z_1,z_2;x,b) &= \Big[ 1-t^2 (2+[2]_x+[2]_{z_1}) +t^3 [1]_x [1]_{z_1} [1]_{z_2} + t^4( [2]_{z_1} +[2]_{z_1} -[2]_{x}) \nn \\
& \qquad - t^5  [1]_x [1]_{z_1} [1]_{z_2} + t^6 [2]_{x} \Big] \times \PE \left[ t [1]_{z_2} ([1]_x [1]_{z_1} + [1]_b) \right]~,
\eea
The corresponding unrefined Hilbert series is
\bea
\fflat_2 (t;z_1=1,z_2=1;x=1,b=1)  = \frac{1 + 4t + 3t^2}{(1 - t)^8}~.
\eea
Integrating over the gauge symmetries, we obtain
\bea
g^{(0,1)}_{(1,1), {\rm Br. 2}}(t,x,b) &= \frac{1}{(2 \pi i)^2}\oint_{|z_1|=1} \frac{\ud z_1}{z_1} (1-z_1^2)\oint_{|z_2|=1} \frac{\ud z_2}{z_2} \fflat_2 (t;z_1,z_2;x,b) \nn\\
&= (1-t^4)^2 \PE[ ([2]_b+[2]_x) t^2 ]~.
\eea
This is the Hilbert series of $(\BC^2/\BZ_2)^2$.  

Let us consider the generators if the moduli space.  The relevant gauge invariant combinations are
\bea
\begin{array}{ll}
G_2^{j_1j_2} = M_{SO(2)}^{b_1b_2} q^{j_1}_{b_1} q^{j_2}_{b_2}~, & \qquad \widehat{G}_2^{j_1j_2} = \epsilon^{b_1b_2} q^{j_1}_{b_1} q^{j_2}_{b_2}~, \\
H_2^{\alpha_1 \alpha_2} = \epsilon_{a_1 a_2} \epsilon_{b_1b_2} (X^{\alpha_1})^{a_1 b_1}  (X^{\alpha_2})^{a_2 b_2}~, & \qquad 
\widehat{H}_2^{\alpha_1 \alpha_2} = \epsilon_{a_1 a_2} M^{SO(2)}_{b_1b_2} (X^{\alpha_1})^{a_1 b_1}  (X^{\alpha_2})^{a_2 b_2}~
\end{array}
\eea
Observe that $G_2^{j_1j_2}$ and $H_2^{\alpha_1 \alpha_2}$ are rank-two symmetric tensors and hence each has three independent components; on the other hand, $\widehat{G}_2^{j_1j_2}$ and $\widehat{H}_2^{\alpha_1 \alpha_2}$ are rank-two antisymmetric tensors and hence each has one independent component.  The first and the last line of \eref{defbr20111} set
\bea
\widehat{G}^{j_1j_2}_2 =0~, \qquad \widehat{H}^{\alpha_1 \alpha_2}_2 = 0~. \label{HAvanish}
\eea
Thus, the generators of each factor $\BC^2/\BZ_2$ in $(\BC^2/\BZ_2)^2$ are $G_2^{j_1j_2}$ and $S_2^{\alpha_1 \alpha_2}$; the former satisfy
\bea
(G_2^{12})^2= G_2^{11} G_2^{22} + \left( \frac{1}{2} \epsilon_{j_1j_2} \widehat{G}_2^{j_1j_2} \right)^2  = G_2^{11} G_2^{22} ~.
\eea
and, using the conditions \eref{defbr20111} and \eref{HAvanish}, we find that $H_2^{\alpha_1 \alpha_2}$ satisfies
\bea
(H_2^{12})^2= H_2^{11} H_2^{22}~.
\eea

\subsubsection{The case of $\vec k =(1,1)$ and $\vec N =(1,1)$}
The Hilbert series is
\bea
g^{(1,1)}_{(1,1)} (t,x,b) &= \left( \oint_{|z_1|=1} \frac{(1-z_1^2) \ud z_1}{(2 \pi i) z_1}   \oint_{|z_2|=1} \frac{\ud z_2}{(2 \pi i) z_2} \right)  \times \nn \\
& \qquad  \frac{ \chi_Q(t,z_1) \chi_X(t,z_1, z_2, x) \chi_q(t,z_2,b)}{\chi_F(t,z_1)} ~,
\eea
where the contributions from the fields $Q$, $q$, $X$ and the $F$ terms are given by
\bea
\chi_Q(t,z_1) &= \PE[t [1]_{z_1} ]~, \nn \\
\chi_q(t,z_2,b) &= \PE[t (z_2+z_2^{-1}) [1]_b ]~, \nn \\
\chi_X(t,z_1, z_2, x) &= \PE[ t [1]_{z_1} (z_2+z_2^{-1})  [1]_x ]~, \nn \\
\chi_F(t,z_1) &= \PE[ t^2 + t^2 [2]_{z_1} ]~.
\eea
The integration gives
\bea
g^{(1,1)}_{(1,1)} (t,x,b) &= 1+(1+ [0; 2] + [2; 0])t^2 +2[1;1]t^3 + (1+ 3 [0; 2] + [0; 4] + [2; 0] + [2; 2]  \nn \\
& \quad + [4; 0])t^4+(2 [1; 1] + 2 [1; 3] + 2 [3; 1])t^5+ \ldots~,
\eea
where the notation $[n_1;n_2]$ denotes a representation of $SU(2)_x \times Sp(1)$.
The corresponding unrefined Hilbert series is
\bea
g^{(1,1)}_{(1,1)} (t,x=1,b=1) &=\frac{1-2 t+6 t^2-2 t^3+6 t^4-2 t^5+6 t^6-2 t^7+t^8}{(1-t)^6 (1+t)^4 \left(1+t^2\right)^2}~.
\eea
The plethystic logarithm of the Hilbert series is
\bea
\PL [g^{(1,1)}_{(1,1)} (t,x,b) ] = ([0;0]+ [0; 2] + [2; 0])t^2 +2[1;1]t^3+(2[0;2]-2)t^4 - \ldots~.
\eea
The generators at order $2$ are
\bea
G_{[2;0]}^{\alpha_1 \alpha_2} &= (X^{\alpha_1})^{a_1b_1} (X^{\alpha_2})^{a_2b_2} \epsilon_{a_1 a_2} \epsilon_{b_1b_2}~, \quad
G_{[0;2]}^{j_1j_2} = M_{SO(2)}^{b_1b_2} q^{j_1}_{b_1} q^{j_2}_{b_2}~, \nn \\
G_{[0;0]}^{\alpha_1 \alpha_2} &= (X^{\alpha_1})^{a_1b_1} (X^{\alpha_2})^{a_2 b_2}  \epsilon_{a_1a_2} M^{SO(2)}_{b_1b_2}~.
\eea
The generators at order $3$ are
\bea
G_{[1;1]}^{\alpha j_1} &= (X^\alpha)^{a_1 b_1}  Q_{a_1} q^{j_1}_{b_1}~, \qquad \widehat{G}_{[1;1]}^{\alpha j_1} = (X^\alpha)^{a_1 b_1}  Q_{a_1} q^{j_1b_2} M^{SO(2)}_{b_1b_2}~,
\eea
where we define $q^{j_1b_3}$ as
\bea
q^{j_1b_2}  = \epsilon^{b_2b_1} q^{j_1}_{b_1}~.
\eea
The generators at order $4$ are
\bea
H^{j_1 j_2}_{[0;2]} &=  \epsilon_{a_1 a_2} ( X^\alpha)^{a_1 b_1} (X^\beta)^{a_2 b_2} q^{j_1}_{b_1} q^{j_2}_{b_2}~, \nn \\
\widehat{H}^{j_1 j_2}_{[0;2]} &=  \epsilon_{a_1 a_2} ( X^\alpha)^{a_1 b_1} (X^\beta)^{a_2 b_2} q^{j_1}_{b_1}  q^{j_2b_3}M^{SO(2)}_{b_2b_3}~,
\eea

\subsubsection{The case of $\vec k =(1,1)$ and $\vec N =(2,1)$}
The Hilbert series is
\bea
g^{(2,1)}_{(1,1)} (t,x,a, b) &= \left( \oint_{|z_1|=1} \frac{(1-z_1^2) \ud z_1}{(2 \pi i) z_1}   \oint_{|z_2|=1} \frac{\ud z_2}{(2 \pi i) z_2} \right)  \times \nn \\
& \qquad  \frac{ \chi_Q(t,z_1,a) \chi_X(t,z_1, z_2, x) \chi_q(t,z_2,b)}{\chi_F(t,z_1)} ~,
\eea
where the contributions from the fields $Q$, $q$, $X$ and the $F$ terms are given by
\bea
\chi_Q(t,z_1,a) &= \PE[t [1]_{z_1}[1]_a ]~, \nn \\
\chi_q(t,z_2,b) &= \PE[t (z_2+z_2^{-1}) [1]_b ]~, \nn \\
\chi_X(t,z_1, z_2, x) &= \PE[ t [1]_{z_1} (z_2+z_2^{-1})  [1]_x ]~, \nn \\
\chi_F(t,z_1) &= \PE[ t^2 + t^2 [2]_{z_1} ]~.
\eea
The Hilbert series admits the following character expansion
\bea
g^{(2,1)}_{(1,1)} (t,x,a,b) &= 1+(2+ [0;2] + [2; 0])t^2 +2[1;1](a+a^{-1})t^3 + (3 + 4 [0; 2] + [0; 4]  \nn \\
& \quad + [2; 0] + [2; 2] + [2; 0](a^2+1+a^{-2}) + [4; 0])t^4+ \ldots~,
\eea
where the notation $[n_1;n_2]$ denotes a representation of $SU(2)_x \times Sp(1)_b$.
The corresponding unrefined Hilbert series is
\bea
g^{(2,1)}_{(1,1)} (t,x=1,a=1,b=1) &=\frac{1}{(1-t)^8 (1 + t)^4 (1 + t^2)^2 (1 + t + t^2)^3}\Big(1 - t + 6 t^2 + 7 t^3 + 16 t^4 + 17 t^5  \nn \\
& \quad + 33 t^6 + 26 t^7 + 33 t^8 + \text{palindrome} + t^{14} \Big)~.
\eea
The plethystic logarithm of the Hilbert series is
\bea
\PL [g^{(2,1)}_{(1,1)} (t,x,a,b) ] &=(2+ [0;2] + [2; 0])t^2 +2[1;1](a+a^{-1})t^3+( 2 [0; 2] + [2; 0](a^2+1+a^{-2}) \nn \\
& \qquad -2 -[2;0])t^4 - \ldots~.
\eea

\paragraph{The generators.} Let $\alpha, \beta$ be the $SU(2)_x$ indices, $i, i_1, i_2, \ldots$ be the $SO(2)_a$ indices and $j,j_1, j_2, \ldots$ be the $Sp(1)_b$ indices.
The generators at order $2$ are
\bea
\begin{array}{ll}
G_{[2;0]}^{\alpha \beta} = (X^\alpha)^{a_1 b_1} (X^\beta)^{a_2 b_2} \epsilon_{a_1 a_2} \epsilon_{b_1b_2}~, &\quad
H_{[0;2]}^{j_1j_2} = M_{SO(2)}^{b_1b_2} q^{j_1}_{b_1} q^{j_2}_{b_2}~, \\
\widehat{G}_{[0;0]}^{\alpha \beta} = (X^\alpha)^{a_1 b_1} (X^\beta)^{a_2 b_2} M^{SO(2)}_{b_1b_2} \epsilon_{a_1 a_2}~, &\quad 
\widehat{H}_{[0;0]}^{i_1 i_2} = \epsilon^{a_1 a_2} Q^{i_1}_{a_1} Q^{i_2}_{a_2}~.
\end{array}
\eea
The generators at order $3$ are
\bea
G_{[1;1](a+a^{-1})}^{\alpha i_1 j_1} &= (X^\alpha)^{a_1 b_1}  Q^{i_1}_a q^{j_1}_{b_1}~, \qquad \widehat{G}_{[1;1](a+a^{-1})}^{\alpha i_1 j_1} = (X^\alpha)^{a_1 b_1}  Q^{i_1}_a q^{j_1b_2} M^{SO(2)}_{b_1b_2}~,
\eea
where we define $q^{j_1b_3}$ as
\bea
q^{j_1b_2}  = \epsilon^{b_2b_1} q^{j_1}_{b_1}~.
\eea
The generators at order $4$ are
\bea
\begin{array}{ll}
H^{j_1 j_2}_{[0;2]} =  \epsilon_{a_1 a_2} ( X^\alpha)^{a_1 b_1} (X^\beta)^{a_2 b_2} q^{j_1}_{b_1} q^{j_2}_{b_2}~,
& \widehat{H}^{j_1 j_2}_{[0;2]} =  \epsilon_{a_1 a_2} ( X^\alpha)^{a_1 b_1} (X^\beta)^{a_2 b_2} q^{j_1}_{b_1}  q^{j_2b_3}M^{SO(2)}_{b_2b_3}~,\\
 G^{\alpha \beta i j}_{[2;0](a^2+1+a^{-2})} = ( X^\alpha)^{a_1 b_1} (X^\beta)^{a_2 b_2} Q^{i}_{a} Q^{j}_{b} M^{SO(2)}_{b_1b_2}~. &
\end{array}
\eea

\subsection{The $SA$ hybrid with $m=1$}

From the quiver diagram, we see that the Higgs branch is
\bea
kN + \frac{1}{2}k(k+1)+ \frac{1}{2}k(k-1)-k^2 = kN
\eea
quaternionic dimensional.

\subsubsection{The case of $k=1$: General set-up}
If $k=1$, the antisymmetric hypermultiplet $A$ is absent from the quiver. 
Let us write the hypermultiplets in terms of the chiral multiplets as follows: The symmetric hypermultiplet $S$ as $S, \widetilde{S}$ and the fundamental hypermultiplet $Q$ as $Q^i, \tQ_i$, with $i=1, \ldots, N$.  The $F$ term is given by
\bea
\widetilde{S} S+ \tQ_i Q^i =0~. \label{Ftermk1ASZ2}
\eea

We then compute the Hilbert series of the Higgs branch.  Let $z$ be the fugacity the gauge group $U(1)$.
The global symmetry is $U(1)_S \times U(N)$, where $U(N)$ corresponds to the flavour node in the quiver and $U(1)_S$ is associated with the symmetric hypermultiplet in such a way that $S$ and $\widetilde{S}$ carry charges $+1$ and $-1$ under $U(1)_S$ respectively.

Let $(u, \vec y)$ be the fugacities of the flavour group $U(N) = U(1) \times SU(N)$, and $s$ be the fugacity of the global symmetry $U(1)_S$.  The chiral fields $S$, $\widetilde{S}$, $Q^i$ and $\tQ_i$ correspond to the fugacities $t s z^2$, $t s^{-1} z^{-2}$, $t z u^{-1} [0,0,\ldots,1]_{\vec y}$ and $t z^{-1} u [1,0,\ldots,0]_{\vec y}$.
The Hilbert series is given by
\bea
g^{N}_{k=1; \text{AS}}(t, s, u, \vec y) &= \frac{1}{2 \pi i}\oint_{|z|=1} \frac{\ud z}{z} (1-t^2) \PE \left[ t (s z^2+s^{-1} z^{-2}) \right]  \times \nn \\
& \qquad \PE \left[ t z u^{-1} [0,0,\ldots,1]_{\vec y}+t z^{-1} u [1,0,\ldots,0]_{\vec y} \right]~.
\eea

\paragraph{Unrefined Hilbert series.} Setting $q=1$ and $y_i=1$ for all $i=1, \ldots, N-1$, we obtain the unrefined Hilbert series.  Let us report them for a few values of $N$ below:
\bea
g^{N=1}_{k=1; \text{AS}}(t, 1,1, \vec 1) &= \frac{1-t^6}{\left(1-t^2\right)\left(1-t^3\right)^2} = \frac{1-t+t^2}{(1-t)^2 \left(1+t+t^2\right)}~, \nn \\
g^{N=2}_{k=1; \text{AS}}(t, 1,1, \vec 1) &= \frac{1+2 t^2+4 t^3+2 t^4+t^6}{(1-t)^4 (1+t)^2 \left(1+t+t^2\right)^2}~, \nn \\
g^{N=3}_{k=1; \text{AS}}(t, 1,1, \vec 1) &= \frac{1+t+6 t^2+15 t^3+21 t^4+18 t^5+21 t^6+\text{palindrome}+t^{10}}{(1-t)^6 (1+t)^4 \left(1+t+t^2\right)^3}~, \nn \\
g^{N=4}_{k=1; \text{AS}}(t, 1,1, \vec 1) &= \frac{1+2 t+13 t^2+40 t^3+86 t^4+132 t^5+194 t^6+220 t^7+194 t^8+\text{palindrome}+t^{14}}{(1-t)^8 (1+t)^6 \left(1+t+t^2\right)^4}~.
\eea

\subsubsection{The case of $k=1$ and $N=1$} 
The Hilbert series is
\bea
g^{N=1}_{k=1; \text{AS}}(t, q) = (1-t^6) \PE \left[ t^2 + t^3 (q^2 + q^{-2}) \right] = g_{\BC^2/\BZ_3} (t, q^{2/3})~,
\eea
where the fugacity $q$ is defined as
\bea
q^2= s u^2~, \label{qdef}
\eea
and $g_{\BC^2/\BZ_3}$ is defined as in \eref{C2Z3HS}. Indeed, the hypermultiplet moduli space is $\BC^2/\BZ_3$.  The generator at order 2 can be written as
\bea 
G_2= \widetilde{S} S = -\tQ Q 
\eea 
and the two generators at order 3 can be written as
\bea
(G_3)_1 = S \tQ^2~, \qquad  (G_3)_2 = \widetilde{S} Q^2~.
\eea
The relation at order $6$ is therefore
\bea
(G_3)_1 (G_3)_2 - G_2^3 =0~.
\eea  

\subsubsection{The case of $k=1$ and $N=2$}
 The Hilbert series admits the following character expansion:
\bea
&g^{N=2}_{k=1; \text{AS}}(t, q, y)  \nn \\
&= \frac{1}{1-t^2} \Bigg \{ \sum_{n_2=0}^\infty [2n_2]_y t^{2n_2}  + \sum_{n_2=0}^\infty \sum_{n_3=0}^\infty (q^{2n_3+2} + q^{-2n_3-2}) [2n_2+2n_3+2]_y t^{2n_2+3n_3 +3} \Bigg \} ~,
\eea
where the fugacity $q$ is defined as in \eref{qdef}.
The plethystic logarithm is given by
\bea
\PL \left[ g^{N=2}_{k=1; \text{AS}}(t, q, y) \right] = ([2]_y+1) t^2 + \left\{ (q^2+q^{-2}) [2]_y \right \}t^3  -t^4 - \ldots~.
\eea
This indicates that the generators at order $t^2$ are
\bea
G_2 = \widetilde{S} S~, \qquad  M^i_j = Q^i \tQ_j~, \label{genG2AS}
\eea
where the $F$ term \eref{Ftermk1ASZ2} impose the condition $M^i_i +G_2 =0$.
The generators at order $t^3$ are
\bea
\tB_{ij} = S \tQ_i \tQ_j~, \qquad  B^{ij} = \widetilde{S} Q^i Q^j~. \label{genBAS}
\eea
Note that both $B^{ij}$ and $\tB_{ij}$ are symmetric under the exchange of $i$ and $j$.

\subsubsection{The case of $k=1$ and $N=3$} 
The Hilbert series admits the following character expansion:
\bea
g^{N=3}_{k=1; \text{AS}}(t, q,  \vec y) &= \frac{1}{1-t^2} \Bigg \{ \sum_{n_2=0}^\infty [n_2,n_2]_{\vec y} t^{2n_2}  + \sum_{n_2=0}^\infty \sum_{n_3=0}^\infty (q^{2n_3+2}[n_2+2n_3+2,n_2]_{\vec y}  \nn \\
& \qquad \qquad + q^{-2n_3-2} [n_2,n_2+2n_3+2]_{\vec y}) t^{2n_2+3n_3 +3} \Bigg \} ~,
\eea
The plethystic logarithm is given by
\bea
\PL \left[ g^{N=3}_{k=1; \text{AS}}(t, q, \vec  y) \right] = ([1,1]_y+1) t^2 + \left\{ q^2 [2,0]_y +q^{-2} [0,2]_y\right \}t^3  -( [1,1]_y +1) t^4 - \ldots~.
\eea
Note that the generators of the moduli space take the same form as in \eref{genG2AS} and \eref{genBAS} with the indices $i, j=1,2,3$.

\subsubsection{The case of $k=1$ and $N=4$} 
The Hilbert series admits the following character expansion:
\bea
g^{N=4}_{k=1; \text{AS}}(t, q,  \vec y) &= \frac{1}{1-t^2} \Bigg \{ \sum_{n_2=0}^\infty [n_2,0, n_2]_{\vec y} t^{2n_2}  + \sum_{n_2=0}^\infty \sum_{n_3=0}^\infty (q^{2n_3+2}[n_2+2n_3+2,0, n_2]_{\vec y}  \nn \\
& \qquad \qquad + q^{-2n_3-2} [n_2,0, n_2+2n_3+2]_{\vec y}) t^{2n_2+3n_3 +3} \Bigg \} ~,
\eea
The plethystic logarithm is given by
\bea
\PL \left[ g^{N=3}_{k=1; \text{AS}}(t, q, \vec y) \right] &= ([1,0,1]_{\vec y}+1) t^2 + \left\{ q^2 [2,0,0]_{\vec y} +q^{-2} [0,0,2]_{\vec y}\right \}t^3   \nn \\
& \qquad -( [1,0,1]_{\vec y}+[0,2,0]_{\vec y} +1) t^4 - \ldots~.
\eea
Note that the generators of the moduli space take the same form as in \eref{genG2AS} and \eref{genBAS} with the indices $i, j=1,\ldots,4$.

\subsubsection{Character expansion for $k=1$ and general $N$} 
For $k=1$ and general $N$, the Hilbert series can be written as
\bea
g^{N}_{k=1; \text{AS}}(t, q,\vec y) &= \frac{1}{1-t^2} \Bigg \{ \sum_{n_2=0}^\infty [n_2,0, \ldots,0, n_2]_{\vec y} t^{2n_2}  + \sum_{n_2=0}^\infty \sum_{n_3=0}^\infty (q^{2n_3+2}[n_2+2n_3+2,0, \ldots,0, n_2]_{\vec y}  \nn \\
& \qquad \qquad + q^{-2n_3-2} [n_2,0, \ldots,0, n_2+2n_3+2]_{\vec y}) t^{2n_2+3n_3 +3} \Bigg \} ~,
\eea
where the fugacity $q$ is defined as in \eref{qdef}.

\subsubsection{The case of $k=2$: General set-up}
The global symmetry is $U(1)_S \times U(1)_A \times U(N)$, where $U(N)$ corresponds to the flavour node in the quiver, $U(1)_S$ is associated with the symmetric hypermultiplet in such a way that $S$ and $\widetilde{S}$ carry charges $+1$ and $-1$ under $U(1)_S$ respectively, and $U(1)_A$ is associated with the antisymmetric hypermultiplet in such a way that $A$ and $\widetilde{A}$ carry charges $+1$ and $-1$ under $U(1)_A$ respectively.

Let $(z_1, z_2)$ be the fugacities of the gauge group $U(2)=U(1) \times SU(2)$, $(u, \vec y)$ be the fugacities of the flavour group $U(N) = U(1) \times SU(N)$, $s$ be the fugacity of the global symmetry $U(1)_S$, and $a$ be the fugacity of the global symmetry $U(1)_A$.  
The Hilbert series is given by
\bea
g^{N}_{k=2; \text{AS}}(t, s, a,u, \vec y) &= \oint_{|z_1|=1} \frac{\ud z_1}{(2 \pi i)z_1} \oint_{|z_2|=1} \frac{\ud z_2}{(2 \pi i)z_2} (1-z_2^2)  \times \nn\\ 
&\quad \frac{ \chi_{S,\tS}(t,z_1,z_2,s) \chi_{A,\tA}(t,z_1, z_2, a) \chi_{Q,\tQ}(t,z_1,z_2,u, \vec y)}{\chi_F(t,z_2)} ~,
\eea
where the contributions from the chiral fields $S$, $\widetilde{S}$, $A$, $\tA$ , $Q$, $\tQ$ and $F$-terms are
\bea
\chi_{S,\tS}(t,z_1,z_2,s) &= \PE[t (s z_1^2 [2]_{z_2} + s^{-1} z_1^{-2} [2]_{z_2})]~, \nn \\
\chi_{A,\tA}(t,z_1,z_2,a) &= \PE[t (a z_1^2+a^{-1} z_1^{-2}) ]~, \nn \\
\chi_{Q,\tQ}(t,z_1, z_2, u, \vec y) &= \PE[ t (z_1^{-1} [1]_{z_2} u [1,0,\ldots,0,0]_{\vec y}+ z_1 [1]_{z_2} u^{-1} [0,0,\ldots,0,1]_{\vec y})  ]~, \nn \\
\chi_F(t,z_2) &= \PE[ t^2 (1+[2]_{z_2}) ]~.
\eea


\paragraph{Unrefined Hilbert series.} Setting $s=a=u=1$ and $y_i=1$ for all $i=1, \ldots, N-1$, we obtain the unrefined Hilbert series.  Let us report them for a few values of $N$ below:
\bea
g^{N=1}_{k=2; \text{AS}}(t, 1,1,1, \vec 1) &= \frac{1-t+2 t^3-t^5+t^6}{(1-t)^4 (1+t)^2 \left(1+t+t^2+t^3+t^4\right)}~, \nn \\
g^{N=2}_{k=2; \text{AS}}(t, 1,1,1, \vec 1) &=\frac{1}{(1 - t)^8 (1 + t)^4 (1 + t + t^2)^4 (1 + t + t^2 + t^3 + t^4)^2} \times \nn \\
& \quad \Big( 1+2 t+4 t^2+10 t^3+25 t^4+48 t^5+77 t^6+108 t^7+142 t^8+172 t^9 \nn \\
& \quad +186 t^{10}+172 t^{11}+ \text{palindrome}+t^{20} \Big)~, \nn\\
g^{N=3}_{k=2; \text{AS}}(t, 1,1,1, \vec 1) &=\frac{1}{(1-t)^{12} (1+t)^6 \left(1+t+t^2\right)^6 \left(1+t+t^2+t^3+t^4\right)^3} \times \nn \\
& \quad \Big( 1+3 t+10 t^2+34 t^3+103 t^4+268 t^5+620 t^6+1273 t^7+2359 t^8 \nn \\
& \quad +3982 t^9+6139 t^{10}+8653 t^{11}+11245 t^{12}+13555 t^{13}+15179 t^{14} \nn \\
& \quad +15764 t^{15}+15179 t^{16}+ \text{palindrome}+t^{30} \Big)~.
\eea

The plethystic logarithm of the refined Hilbert series is given by
\bea
&\PL [ g^{N}_{k=2; \text{AS}}(t, s, a,u, \vec y) ]= (2+[1,0,\ldots,0,1]_{\vec y})t^2 +  (s u^2 [2,0,\ldots,0]_{\vec y}+ a u^2 [0, \ldots, 0,1]_{\vec y}+{\rm c.c.}) t^3 + \nn \\
& \qquad (sa^{-1} [1,1]_{\vec y} + s^2 a^{-2} + {\rm c.c.})t^4+ \Big \{ (s^{-1}a^{2}u^2 [2,0,\ldots,0]_{\vec y} + s^{-2}au^{-2} [1,0,\ldots,0]_{\vec y} + {\rm c.c.})- \nn \\
& \qquad (s u^2 [0,1]_{\vec y} + a u^2 [0,1]_{\vec y} +{\rm c.c}) \Big \} t^5 + \ldots~,
\eea
where `c.c.' denotes the character of the complex conjugate representation.

The generators at order $2$ are
\bea
Q^i_a \tQ^a_j~, \qquad A^{ab} \tA_{ba}~, \qquad S^{ab} \tS_{ba}~.
\eea
The generators at order $3$ are
\bea
S^{ab} Q^i_a Q^j_b~, \qquad A^{ab} Q^i_a Q^j_b~, \qquad \tS_{ab} \tQ^a_i \tQ^b_j~, \qquad \tA_{ab} \tQ^a_i \tQ^b_j~.
\eea
The generators at order $4$ are
\bea
S^{ab}\tA_{bc}Q^i_a\tQ^c_j~, \quad \tS_{ab} A^{bc} \tQ^a_i Q^j_c~, \quad S^{a_1 a_2} S^{a_3 a_4}  \tA_{a_1 a_2} \tA_{a_3 a_4}~, \quad \tS_{a_1 a_2} \tS_{a_3 a_4}  A^{a_1 a_2} A^{a_3 a_4}~.
\eea
The generators at order $5$ are
\bea
\tS_{ab} A^{ac} A^{bd} Q^i_c Q^j_d~, \qquad S^{ab} \tA_{ac} \tA_{bd} \tQ^{c}_i \tQ^d_j~.
\eea

\section*{Acknowledgements}
A.~D.'s  work is supported by the National Science Foundation under Grant Number PHY-0969020. A.~D., A.~H., N.~M., R-K.~S. would like to thank the Simons Center for Geometry and Physics for the hospitality during the completion of this paper. N.~M.~ thanks Raffaele Savelli and I\~naki Garc\'ia-Etxebarria for a number of useful discussions. The work of N.~M. is supported by a research grant of the Max Planck Society and is based upon work supported in part by the National Science Foundation under Grant No. PHY-1066293 and the hospitality of the Aspen Center for Physics.
D.R-G warmly thanks the Technion for warm hospitality while this work was in progress. He would like to specially thank O.Bergman for very useful conversations. D. R G is partially supported by the research grants MICINN-09-FPA2009-07122 and MEC-DGI-CSD2007-00042, as well as by the Ramon y Cajal fellowship RyC-20011- 07593.\\

\begin{appendix}

\section{Constructing bundles on ALF/ALE spaces}\label{classification}

We review the construction of topologically non-trivial rank $N$ bundles on ALE spaces mostly following the presentation of \cite{Witten:2009xu}. For further details we refer to \cite{Witten:2009xu} and references therein. 

\subsection{The  ALE/ALF space as a hyperK\"ahler quotient}

In the singular limit, the ALE space can be thought as $\mathbb{C}^2$ modded out by $(z_1,\,z_2)\,\sim\,(\omega\,z_1,\,\omega\,z_2)$ where  $\omega^n=1$. However, following \cite{Witten:2009xu}, it turns out to be more useful to think of the geometry as a limit of the ALF space. To that matter, we construct the ALF space as a hyperKaheler quotient. We start by considering $n$ copies of $\mathbb{R}^4$ times $\mathbb{R}^3\times S^1$. For each $\mathbb{R}^4$ we write the metric as

\begin{equation}
d\vec{X}_i^2=d\rho^2_i+\rho^2_i\,d\Omega_{3,\,i}^2=d\rho^2_i+\frac{\rho^2_i}{4}\,d\Omega_{2,\,i}^2+\frac{\rho^2_i}{4}\,(d\psi_i+\cos\theta_i\,d\phi_i)^2=\frac{d\vec{r}_i^2}{r_i}+r_i\,(d\psi_i+\vec{\omega}_i\cdot d\vec{r}_i)^2
\end{equation}
where we have done $r_i=2\,\sqrt{\rho_i}$. Then, the metric of the $\prod_{i=1}^n\, \mathbb{R}_i^4\times\mathbb{R}^3\times S^1$ is

\begin{equation}
ds^2=\sum_{i=1}^n\,\frac{d\vec{r}_i^2}{r_i}+\,r_i\,(d\psi_i+\vec{\omega}_i\cdot d\vec{r}_i)^2+d\vec{X}+\lambda^2\,d\theta^2
\end{equation}
being $\lambda$ the radius of the $S^1$ parametrized by $\theta$ and $d\vec{X}^2$ the flat euclidean metric on $\mathbb{R}^3$.

The space $\mathbb{R}^3\times S^1$, being itself hyperKahler, admits a triple of symplectic structures. These can be constructed explicitly introducing the following matrices

\begin{equation}
\mathcal{J}_1=\left(\begin{array}{cccc}0 & 1 & 0 & 0 \\ -1 & 0 & 0 & 0 \\ 0 & 0 & 0 & 1 \\ 0 & 0 & -1 & 0\end{array}\right)\quad \mathcal{J}_2=\left(\begin{array}{cccc}0 & 0 & -1 & 0 \\ 0 & 0 & 0 & 1 \\ 1 & 0 & 0 & 0 \\ 0 & -1 & 0 & 0\end{array}\right)\quad \mathcal{J}_3=\left(\begin{array}{cccc}0 & 0 & 0 & 1 \\ 0 & 0 & 1 & 0 \\ 0 & -1 & 0 & 0 \\ -1 & 0 & 0 & 0\end{array}\right)
\end{equation}
satisfying $\mathcal{J}_i^2=-1$ and $\mathcal{J}_1\,\mathcal{J}_2=\mathcal{J}_3$. Then, the  associated three symplectic 2-forms read

\begin{equation}
J_1=dX_1\wedge dX_2+dX_3\wedge d\theta\qquad J_2=dX_3\wedge dX_1+dX_2\wedge d\theta\qquad J_3=dX_2\wedge dX_3+dX_1\wedge d\theta
\end{equation}
Since the three symplectic structures remain invariant under shifts in $\theta$, the vector field $\partial_{\theta}$ generates a symplectomorphism. The associated moment maps can be taken to be  simply $\vec{\mu}^{\,\theta}=\lambda\, \vec{X}$.

Likewise each $\mathbb{R}^4$ is a hyperKahler space. Introducing complex coordinates $(z_1,\,z_2)$, the three symplectic forms can be constructed out of the three Pauli matrices satisfying $\mathcal{J}_i^2=-1$ and $\mathcal{J}_1\,\mathcal{J}_2=\mathcal{J}_3$. In this case, the associated symplectic structures 
are

\begin{equation}
J_1=d\bar{z}_2\wedge dz_1+d\bar{z}_1\wedge dz_2 \qquad 
J_2=i\,d\bar{z}_2\wedge dz_1-i\,d\bar{z}_1\wedge dz_2 \qquad 
J_3=d\bar{z}_1\wedge dz_1-d\bar{z}_2\wedge dz_2
\end{equation}

Introducing real coordinates on each $\mathbb{R}^4$ as

\begin{equation}
z_1^i=\sqrt{2\,r_i}\,e^{i\frac{\psi_i-\phi_i}{2}}\,\cos\frac{\theta_i}{2}\qquad z_2^i=\sqrt{2\,r_i}\,e^{i\frac{\psi_i+\phi_i}{2}}\,\sin\frac{\theta_i}{2}
\end{equation}
the metric of the $i$-th $\mathbb{R}^4$ becomes 

\begin{equation}
ds^2_i=\frac{d\vec{r}_i^2}{|\vec{r}_i|}+|\vec{r}_i|\,(d\psi_i+\vec{\omega}_i\cdot d\vec{r}_i)
\end{equation}
It is easy to see that shifts in $\psi_i$ leave the three symplectic structures invariant. It is easy to show that the associated moment maps are simply $\vec{\mu}^{\,\psi_i}_i=\vec{r}_i$.

The ALF space arises as the symplectic quotient of $\prod_{i=1}^n\mathbb{R}^4_i\times \mathbb{R}^3\times S^1$ by the group generated by simultaneous shifts of each $\psi_i$ and $\theta$. In order to construct it, note that the moment maps under the simultaneous shift of $\theta$ and $\psi_i$ are, upon adding integration constants $\vec{x}_i$, just $\vec{\mu}_i=\vec{r}_i+\lambda\,\vec{X}+\vec{x}_i$. Hence, the zero-level set $\vec{\mu}^{-1}(0)$ is $-\lambda\,\vec{X}=\vec{r}_i+\vec{x}_i$.  Thus, writing $\vec{r}=-\lambda\,\vec{X}$, the metric on the zero level set becomes

\begin{equation}
ds^2=\Big( \sum_i\,\frac{1}{|\vec{r}-\vec{x}_i|} +\frac{1}{\lambda^2}\Big)\,d\vec{r}^2+\sum_i|\vec{r}-\vec{x}_i|\,(d\psi_i+\vec{\omega}_i\cdot d\vec{r})^2+\lambda^2\,d\theta^2
\end{equation}
We still need to divide by the $U(1)$ actions. Following \cite{Witten:2009xu}, this amounts to set 

\begin{equation}
|\vec{r}-\vec{x}_i|\,(d\psi_i+\vec{\omega}_i\cdot d\vec{r})+\lambda^2\,d\theta=0
\end{equation}
Using these conditions one can eliminate $d\theta$ and $d\psi_i$, trading it for the invariant $\chi=\sum_i\psi_i-\theta$. This wat we find the ALF space, whose metric is

\begin{equation}
ds^2=\Big( \sum_i\,\frac{1}{|\vec{r}-\vec{x}_i|} +\frac{1}{\lambda^2}\Big)\,d\vec{r}^2+\Big( \sum_i\,\frac{1}{|\vec{r}-\vec{x}_i|} +\frac{1}{\lambda^2}\Big)^{-1}\,\Big(d\chi+\sum_i\,\vec{\omega}_i\cdot d\vec{r}\Big)^2
\end{equation}

The ALE space emerges in the limit $\lambda\rightarrow \infty$, when the metric becomes

\begin{equation}
ds^2=\Big( \sum_i\,\frac{1}{|\vec{r}-\vec{x}_i|}\Big)\,d\vec{r}^2+\Big( \sum_i\,\frac{1}{|\vec{r}-\vec{x}_i|} \Big)^{-1}\,\Big(d\chi+\sum_i\,\vec{\omega}_i\cdot d\vec{r}\Big)^2
\end{equation}
For generic $\vec{x}_i$, we can recognize here the blow-up of the $\mathbb{C}^2/\mathbb{Z}_n$. The singular limit corresponds, without loss of generality, to setting all $\vec{x}_i=0$.

\subsection{Topology of the quotient space}

\begin{figure}[H]
\centering
\includegraphics[scale=.5]{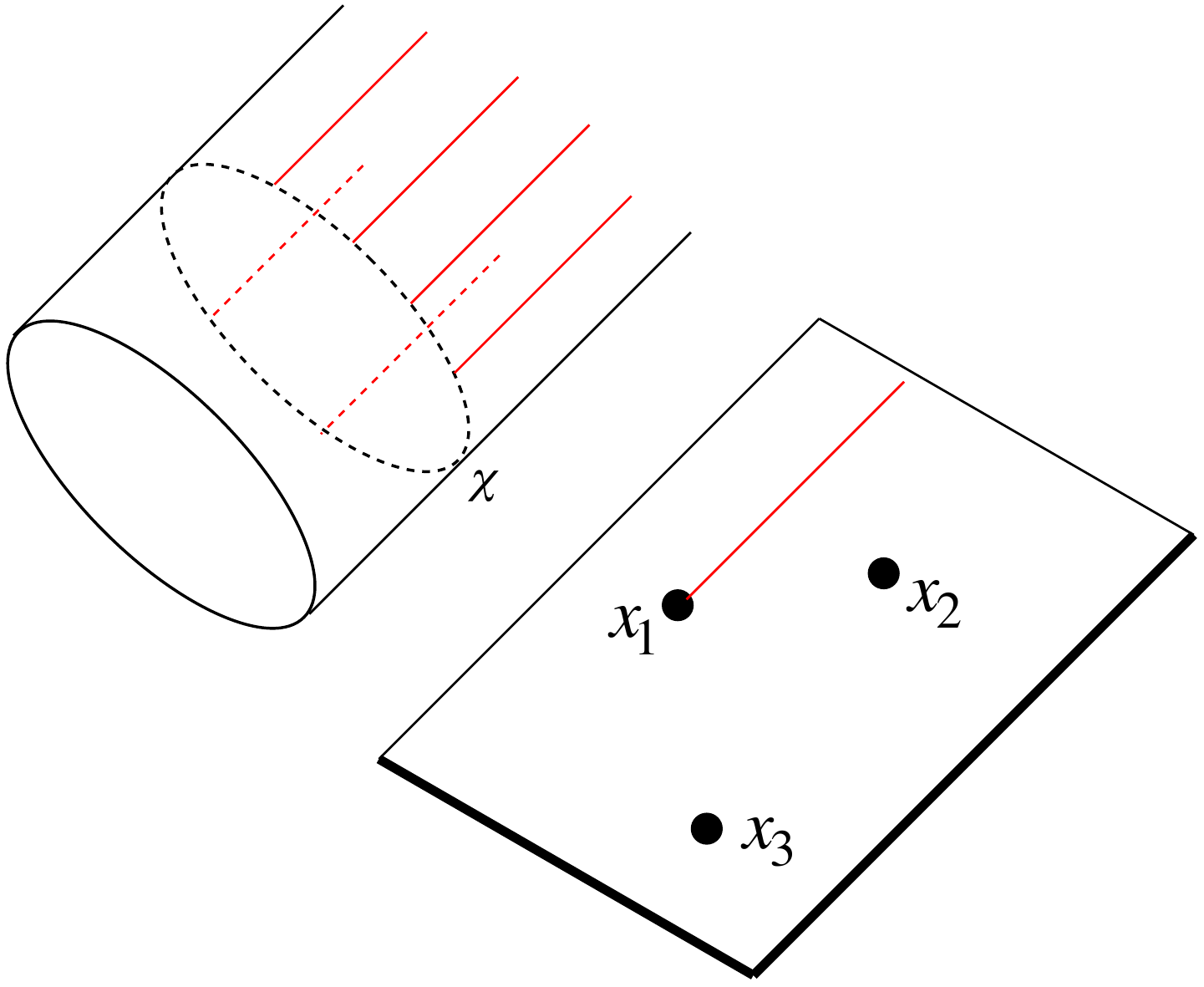}
\caption{Cartoon of the ALF $C_i$'s. Taking a segment from the $\vec{x}_i$ center off to infinity times the $S^1$ we find the $C_i$ --shown in red--}
\label{ALFcycles}
\end{figure}

The ALF manifold is basically $\mathbb{R}^3\times S^1$ with $n$ marked centers $\vec{x}_i\in \mathbb{R}^3$. We can consider a segment starting at each $\vec{x}_i$ and going all the way off to infinity --obviously without intersecting other segments-- and the circle, parametrized by $\chi$, on top of it. This defines sort of a cigar --topologically a disk-- which we will denote by $C_i$. We illustrate this in \fref{ALFcycles}.

\begin{figure}[H]
\centering
       \begin{minipage}[b]{0.4\textwidth}
               \centering
               \includegraphics[scale=0.5]{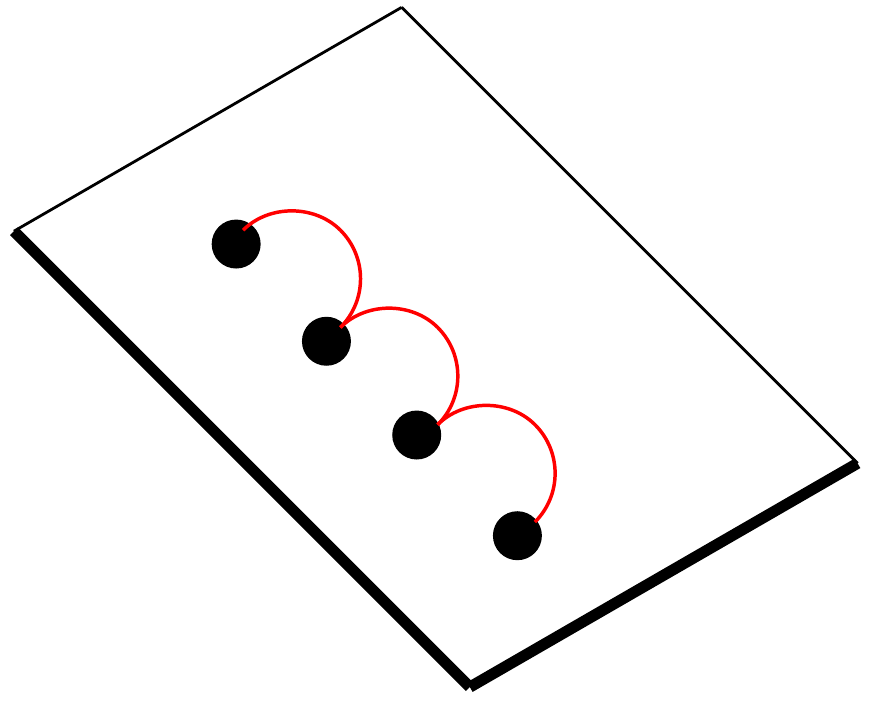}
               \caption{}
               \label{cycles2}
       \end{minipage}
       \hspace{1cm}
       \begin{minipage}[b]{0.4\textwidth}
               \centering
               \includegraphics[scale=0.5]{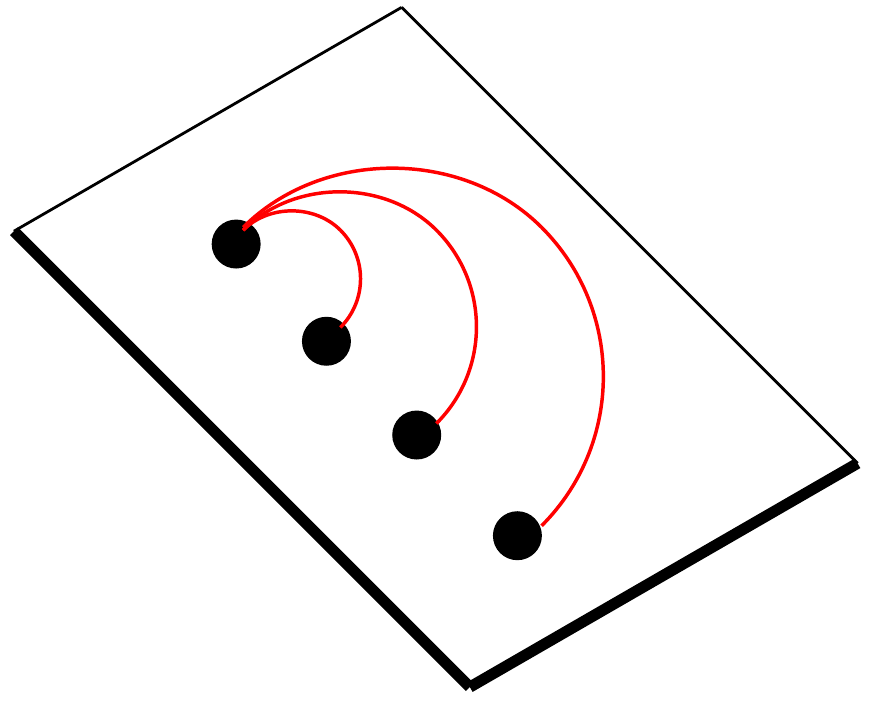}
               \caption{}
               \label{cycles1}
       \end{minipage}
             \caption{Different choices for the segments in $\mathbb{R}^3$ defining the compact two-cycles.}
             \label{choicescycles}
\end{figure}

We can take $C_i$ and $C_j$ and glue them to form $C_{ij}$, which has the topology of a two-sphere. This two-cycle is homotopically equivalent to taking the segment $\vec{x}_i-\vec{x}_j$ together with the $S^1 $ on top. As obviously there are $n-1$ such two-cycles we have that $H_2(ALF,\,\mathbb{Z})=\mathbb{Z}^{n-1}$. Note that we could have chosen different definitions for the two-cycles. For example, instead of taking the segment homotopically equivalent to the difference $\vec{x}_i-\vec{x}_j$ --depicted in \fref{cycles2}-- we could have taken the segments $\vec{x}_i-\vec{x}_1$, as shown in \fref{cycles1}.

On the other hand, $H_2(ALF,\,\mathbb{Z})$ is dual to $H^2(ALF,\,\mathbb{Z})$ in the sense that there is a natural pairing between an element $b$ in $H^2(ALF,\,\mathbb{Z})$ and an element in $H_2(ALF,\,\mathbb{Z})$ which assigns to each $C_{ij}$ a number $b_{ij}$. Since $C_{i_1i_2}+C_{i_2i_3}+C_{i_3i_1}$ is trivial in homology, we should have that $b_{i_1i_2}+b_{i_2i_3}+b_{i_3i_1}=0$, which is solved by itroducing $n$ integers $b_i$ so that $b_{ij}=b_i-b_j$. Obviously this is defined up to overall shifts $b_i\rightarrow b_i+b$. Thus $H^2(ALF,\,\mathbb{Z})$ is spanned by integer valued sequences $\{b_i\}$ modulo an overall shift. Note that, due to the overall shift, there are really $n-1$ independent $b_i$'s as expected.

\subsection{Line bundles on ALF spaces}

We we reviewed the construction of the ALF space as a hyperKahler quotient. As described, this proceeds in two steps whereby one first restricts to the zero level set of the moment maps and then one divides by the $U(1)$'s. Hence, this construction shows that $\vec{\mu}^{-1}(0)$ can be thought as a $U(1)^n$ bundle over the ALF space. In particular, the corresponding Riemannian connections associated to each $U(1)$ automatically provide $n$ gauge fields on the ALF space, or, alternatively, $n$ complex line bundles. These gauge fields can easily be explicitly constructed. To that matter one picks a certain $U(1)$ --say the $j$-th-- and divides the zero level $\vec{\mu}^{-1}(0)$ by all the $U(1)$ actions but the $j$-th. One then finds a metric which is just the one of the ALF space plus

\begin{equation}
A\,\Big(D\psi_j-\frac{D\chi}{|\vec{r}-\vec{x}_j|\,(\frac{1}{\lambda^2}+\sum_i\,\frac{1}{|\vec{r}-\vec{x}_i|})}\Big)^2
\end{equation}
where

\begin{equation}
A=|\vec{r}-\vec{x}_j|\,\frac{\frac{1}{\lambda^2}+\sum_i\,\frac{1}{|\vec{r}-\vec{x}_i|}}{\frac{1}{\lambda^2}+\sum_{i\ne j}\,\frac{1}{|\vec{r}-\vec{x}_i|}}\qquad D\psi_i=d\psi_i+\vec{\omega}_j\cdot d\vec{r}\qquad D\chi=d\chi+\sum_i\vec{\omega}_j\cdot d\vec{r}
\end{equation}
Re-defining $\psi_i$ as $\psi_i=2\tilde{\psi}_i$ so that $\tilde{\psi}_i$ has $2\pi$ length, this extra piece in the metric can be re-written as

\begin{equation}
4\,A\,\Big(d\tilde{\psi}_j-\frac{1}{2}\frac{D\chi}{|\vec{r}-\vec{x}_j|\,(\frac{1}{\lambda^2}+\sum_i\,\frac{1}{|\vec{r}-\vec{x}_i|})}+\frac{1}{2}\vec{\omega}_j\cdot d\vec{r}\Big)^2
\end{equation}
As anticipated, this Riemannian connection provides a natural connection for a $U(1)_j$ line bundle on the ALF space. In other words, we can now borrow the Riemannian connection from here, divide over the $j$-th $U(1)$ and write down the following gauge field on the ALF space

\begin{equation}
\Lambda_j=-\frac{1}{2}\frac{D\chi}{|\vec{r}-\vec{x}_j|\,(\frac{1}{\lambda^2}+\sum_i\,\frac{1}{|\vec{r}-\vec{x}_i|})}+\frac{1}{2}\vec{\omega}_j\cdot d\vec{r}
\end{equation}
One can check that the curvature $B_j=d\Lambda_j$ associated to this connection is in fact anti-selfdual, and it integrates to an integer on the $C_i$ ``cycles"

\begin{equation}
\int_{C_i}\,\frac{B_j}{2\,\pi}=\delta_{ij}
\end{equation}
Thus, the first Chern class of the $j$-th bundle is given by the sequence $(0,\cdots,\,1,\cdots,\,0)$ where the 1 is in the $j$-th position.

The tensor product $\mathcal{L}_{\star}$ is topologically trivial. Upon adding the exact form $d\chi$, the corresponding connection $\Lambda$ can be taken to be

\begin{equation}
\Lambda=\frac{1}{2}\,\frac{D\chi}{1+\lambda^2\,\sum_i\frac{1}{|\vec{r}-\vec{x}_i|}}
\end{equation}
Note that at infinity $\Lambda\sim \frac{d\chi}{2}$, and so it has a trivial $2\pi$ holonomy on the $S^1$. This is in contrast to the individual $\Lambda_i$'s, which vanish asymptotically. Hence, each individual $\Lambda_i$ has zero monodromy at infinity --although non-trivial topology as the first Chern class is non-vanishing--. Coming back to the overall bundle $\Lambda$, in addition to being topologically trivial it has trivial monodromy $2\,\pi$ However, since the one-form $\Lambda$ is well-defined everywhere, we can imagine multiplying it by some $t$, so that the integral over the circle at infinity is $2\pi\,t$. We will denote the corresponding bundle by $\mathcal{L}_{\star}^t$. Thus, putting it all together, one has that the general form for an anti-selfdual bundle over the ALF space is

\begin{equation}
\label{ALFbundle}
\mathcal{L}=\mathcal{L}_{\star}^t\otimes\,\Big(\otimes_{i=1}^n\,\mathcal{L}_i^{n_i}\,\Big)
\end{equation}
where the powers $n_i$ of each individual bundle $\mathcal{L}_i$ encode the first Chern class and $t$ encodes the holonomy around the $S^1$ at infinity.

\subsection{Branes and self-dual bundles on ALF spaces}

\begin{figure}[H]
\centering
\includegraphics[scale=.6]{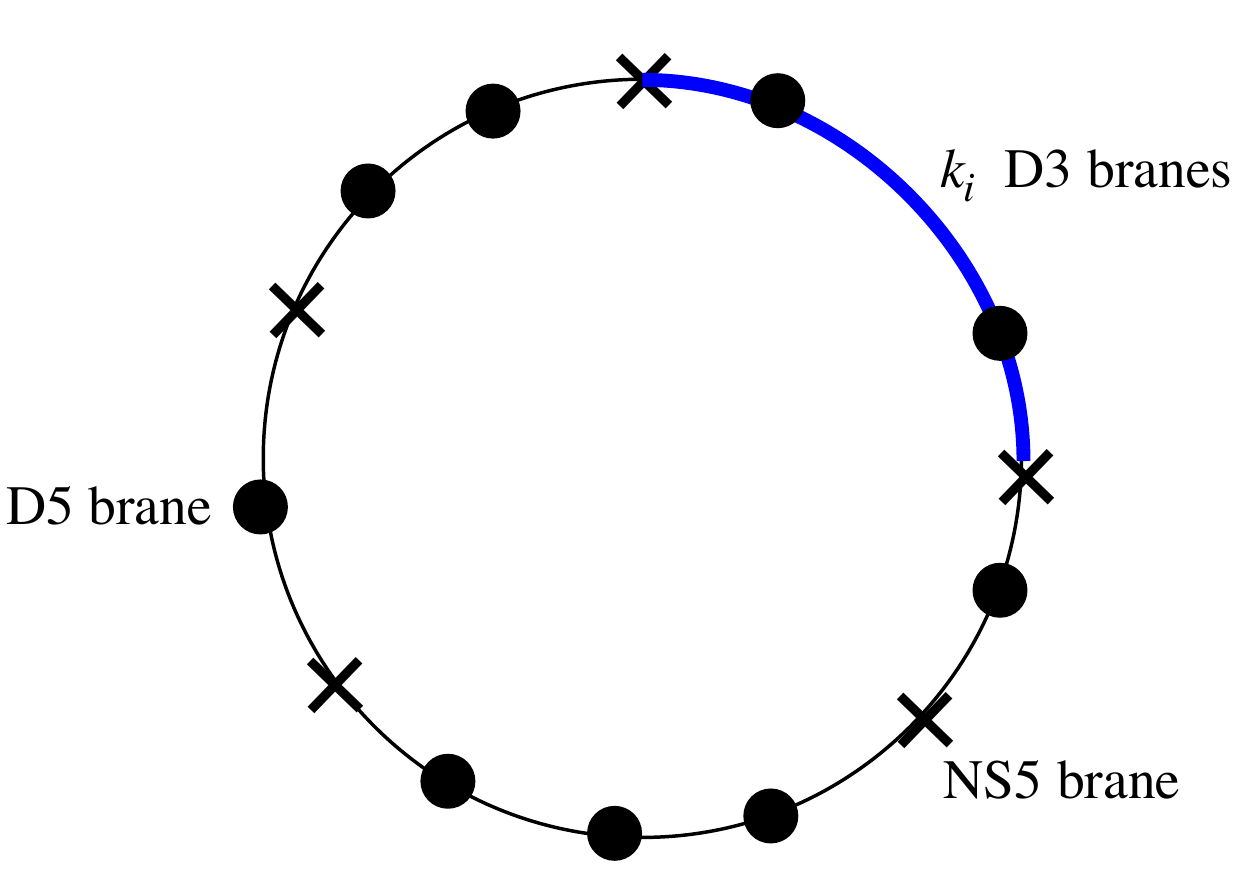}
\caption{Arrangement of the configuration on the $S^1$ parametrized by $x^3$. In this figure, each cross denotes an NS5 brane and each dot denotes a D5-branes.  A segment of curve between two adjacent crosses denotes an interval of D3 branes stretched between the two adjacent NS5 branes.  In particular, we highlight the $i$-th segment in blue which contains $k_i$ D3 branes.}
\label{Cartoon}
\end{figure}

Let us momentarily switch gears and consider the brane configuration including D5, NS5 and D3 in Type IIB on $\mathbb{R}^{1,\,2}\,\times S^1\times \mathbb{R}^3\times \mathbb{R}^3$ depicted in (\ref{Tdual}). As it will become clear shortly, this system will allow us to realize the desired topologically non-trivial bundles on ALF/ALE spaces.

\begin{equation}
\begin{array}{l c |  c c c c c c c c c}
& 0 & 1 & 2 & \raisebox{.5pt}{\textcircled{\raisebox{-.9pt} {3}}}  & 4 &5 & 6 & 7 & 8 & 9 \\ \hline
\mathrm{D5} & \times & \times & \times &  & \times & \times  & \times &  & &\\
\mathrm{NS5} & \times & \times & \times &  &  &   &  & \times &\times &\times \\
\mathrm{D3} & \times & \times & \times & \times & & & & & &\\
\end{array}
\label{Tdual}
\end{equation}
The circled $x^3$ stands for the fact that $x^3$ is an $S^1$. The locations of the $n$ NS5 branes on $x^3$ will be denoted by $y_i$, while those of the $N$ $D5$ branes will be denoted by $s_{\sigma}$. Furthermore, we have $k_i$ D3 branes in each segment between NS5. We stress that we assume a standard presentation where the configuration arranged so that no D3 brane ends on a D5 brane. Besides, we will denote by $N_i$ the number of D5 in each segment between NS5's. A cartoon of the configuration can be seen in \fref{Cartoon}

Note that the theory describing this brane configuration is precisely the one used in section \ref{U(N)inst}.

It will be useful to introduce, following \cite{Witten:2009xu}, the linking numbers for the fivebranes as in \cite{Hanany:1996ie}. There is one subtlety, however, relative to the fact that the configuration lives on a circle. One way to circumvent this problem is by selecting a basepoint in the circle and cut it open into a segment, so that we can use the standard linking number definition

\begin{itemize}
\item NS5 linking number: for the $i$-th NS5  ($i=1\cdots n$) we define its linking number $\ell_i$ as the number of D3 to the right $k_{i_>}$ minus the number of D3 to the left $k_{i_<}$ minus the number of D5 to the right $N_{i_>}$ 

\begin{equation}
\ell_i = k_{i_>}-k_{i_<}-N_{i_>}
\end{equation}

\item D5 linking number: for the $\sigma$-th NS5  ($\sigma=1\cdots N$) we define its linking number $\tilde{\ell}_{\sigma}$ as the number of D3 to the right $k_{\sigma_>}$ minus the number of D3 to the left $k_{\sigma_<}$ plus the number of NS to the left $n_{\sigma_<}$ 

\begin{equation}
\tilde{\ell}_{\sigma} = k_{\sigma_>}-k_{\sigma_<}-n_{\sigma_>}
\end{equation}

\end{itemize}


It is clear that these linking numbers do depend on the choice of basepoint. We can remove such dependence by considering the relative linking numbers. For the NS5, the relative linking number $\Delta\ell_i$ is just $\ell_{i+1}-\ell_i$, which simplifies into
\begin{equation}
\Delta\ell_i=k_{i+1}+k_{i-1}-2\,k_i+N_i~.
\end{equation}
This actually equals to the negative one-loop beta function coefficient of the $U(k_i)$ gauge group.  

Note now that upon $T$-duality along  $S^1$, the configuration (\ref{Tdual}) maps into the system of D2 branes wrapped on $\BR^{1,2}$ and supported at a point in the ALF/ALE space with D6 branes that fill $\BR^{1,2} \times \text{ALF/ALE space}$. To be specific, the D3 brane going all the way around the circle become the regular D2 brane, while those stretching between NS's become fractional D2 branes. In turn, the D5 brane map to the D6 branes. The crucial observation \cite{Douglas:1995bn} is that, from the point of view of the D6 branes, the D2 branes appear as topologically non-trivial, instantonic, worldvolume gauge field configurations. On the other hand, from the point of view of the D2 branes such D2/D6 bound state is described by the Higgs branch of the worldvolume D2 theory, which then allows to realize the desired instanton bundle as the Higgs branch of a gauge theory.  This is the construction used in section \ref{U(N)inst} to compute the instanton moduli space Hilbert series.

Furthermore, upon T-duality, we are interested on the line bundles associated to D5 branes in (\ref{Tdual}). Denoting the bundle associated to the D5 at the position $s_{\sigma}$ on the circle by $\mathcal{R}_{s_{\sigma}}$, a topologically non-trivial bundle on the ALF space $V$ of rank $N$ can be written as

\begin{equation}
V=\bigoplus_{\sigma=1}^{N}\,\mathcal{R}_{s_{\sigma}}
\end{equation}
%

However, as the bundle $V$ describes D2 branes dissolved in D6 branes, it has to be anti-selfdual. This in turn implies that each $\mathcal{R}_{s_{\sigma}}$ is of the form of eq. (\ref{ALFbundle}). Moreover, the power $t$ of the $\mathcal{L}_{\star}$ is naturally related to the possition of the D5 brane on the circle. This can be understood performing a T-duality, so that the monodromy is dual to the position of the D5 on the circle --whose radius will be denoted by $R$--. Since the monodromy is encoded in the $\mathcal{L}_{\star}$ bundle --recall that the $\mathcal{L}_{i}$ are trivial at infinity-- it follows that 

\begin{equation}
\label{Rsbundletrial}
\mathcal{R}_{s_{\sigma}}=\mathcal{L}_{\star}^{\frac{s_{\sigma}}{2\pi R}}\,\otimes\,\Big(\otimes_{i=1}^n\,\mathcal{L}_i^{n_i}\Big)
\end{equation}

In order to further proceed, recall that the topological data is encoded in the integers $n_i$, which we need to determine. To that matter, we now recall that the D5 is linked to the NS5 such that when one D5 crosses one NS5, one D3 is created \cite{Hanany:1996ie}. In order to determine the consequences of such requirement, note that due to eq. (\ref{Rsbundletrial}) whenever the D5 goes all the way around the circle $s_{\sigma}$ shifts by $2\pi\,R$, and so we should have that $\mathcal{R}_{s}\rightarrow \mathcal{R}_s\otimes\mathcal{L}_{\star}$. This can be achieved if  each time we cross an NS5 we have $\mathcal{R}_s\rightarrow \mathcal{R}_s\otimes \mathcal{L}_i^{-1}$.  Then going around the circle we accumulate the desired factor of $\mathcal{L}_{\star}\sim \mathcal{L}_{\star}^{-1}=\otimes \mathcal{L}_i^{-1}$Thus, crossing iteratively NS5 it follows that 

\begin{equation}
\label{Rsbundle}
\mathcal{R}_{s_{\sigma}}=\mathcal{L}_{\star}^{\frac{s_{\sigma}}{2\pi R}}\,\otimes\,\Big( \otimes_{i,\,s_{\sigma}\geq y_i}\,\mathcal{L}_i^{-1}\Big)
\end{equation}
In words, this means that each time we cross the $i$-th NS5 we collect a factor of $\mathcal{L}_i^{-1}$, and so the $\mathcal{R}_{s_{\sigma}}$ is the product of all the $\mathcal{L}_i^{-1}$ crossed --from a certain basepoint-- to reach the particular D5.

Having determined the form of the $\mathcal{R}_{s_{\sigma}}$ bundles, we can now determine the first Chern class of $V$. On general grounds $c_1(V)=\sum_{\sigma}\,c_1(\mathcal{R}_{s_{\sigma}})$, and so we just need to compute $c_1(\mathcal{R}_{s_{\sigma}})$. Since $\mathcal{L}_{\star}$ is topologically trivial, for these matters we can think of $\mathcal{R}_{s_{\sigma}}$ as

\begin{equation}
\mathcal{R}_{s_{\sigma}}\sim\otimes_{i,\,s_{\sigma}\geq y_i}\,\mathcal{L}_i^{-1}
\end{equation}
As described above, $c_1(\mathcal{L}_i)$ corresponds to a vector of the form $(0,\cdots,0,1,0\cdots,0)$ with the 1 in the $i$-th position. Hence, $\mathcal{R}_{s_{\sigma}}$ will contribute with a $-1$ in the $i$-th position for all $i$ such that $y_i\leq s_{\sigma}$. Thus, in the $i$-th position we will have as many -1 as D5 to the right if the $i$-th NS5, that is, precisely the D5 contribution to the NS5 linking number $\ell_i$. Thus, the first Chern class equals the linking number $\ell_i$. Note that strictly speaking in this discussion we have omitted fractional branes. A more through analysis \cite{Witten:2009xu} shows that indeed the first Chern class can be indeed associated to the linking numbers $\ell_i$.

This presentation corresponds to the choice of two-cycles as in \fref{cycles1}. The choice of basepoint corresponds to the choice of say the $\vec{x}_1$ center to form the other $n-1$ segments. In the language of the line bundles, this corresponds to the choice of $B$ field gauge trivializing  $\mathcal{R}_0$. However, we might as well consider the generic sequence $b_i$ and, instead of trivializing one of the entries by a $B$-field gauge transformation, consider defining the gauge-independent sequence $\Delta b_i=b_{i+1}-b_i$. It is clear that these $\Delta b_i$ correspond to the relative linking numbers above. This would correspond to use the cycles described in \fref{cycles2}, where, instead if using directly the $\mathcal{L}_i$ we would use 
\bea 
\mathcal{T}_i=\mathcal{L}_{i+1}\otimes\mathcal{L}_i^{-1}~. \label{deflinebundT}
\eea 
In this case, the $c_1$ correspond to the relative linking number as in \cite{Fucito:2004ry}. Note that there are really $n-1$ independent linking numbers corresponding to the $n-1$ independent 2-cycles. However, we can introduce the $n$-th $\mathcal{T}_n$ such that $c_1(\mathcal{T}_n)=0$. Hence, the topological data can be taken to be

\begin{equation}
c_1=\sum_i\,\Delta\ell_i\,c_1(\mathcal{T}_i)\qquad c_2=\sum_i\,\Delta\ell_i\,c_2(\mathcal{T}_i)+\frac{1}{n}\,\sum\,k_i
\end{equation}
Clearly there are $n-1$ relative linking numbers

\subsection{The ALE/orbifold Limit}

As discussed above, the ALE limit corresponds to $\lambda\rightarrow \infty$. Obviously the topology remains the same. However there is a crucial difference as the asymptotic behavior of the individual connections is different. Asymptotically, the one-forms $\Lambda_i$ become

\begin{equation}
\Lambda_j=-\frac{1}{2}\frac{D\chi}{|\vec{r}-\vec{x}_j|\,\sum_i\,\frac{1}{|\vec{r}-\vec{x}_i|}}+\frac{1}{2}\vec{\omega}_j\cdot d\vec{r}
\end{equation}
And so at infinity $\Lambda_i\sim \frac{d\chi}{2\,n}$. On the other hand $\Lambda$ goes to zero. Thus, in the ALE limit the monodromy of the individual $\mathcal{L}_i$ is $e^{i\,\frac{2\pi}{n}}$, while the monodromy of the overall bundle vanishes. This means that when describing generic bundles in the ALE space  $c_1$, $c_2$ and the rank of the bundle will not be enough to fully specify the bundle since we will need to supply information about the monodromies.

Making use of the results in the previous subsection, and since the $\mathcal{L}_{\star}$ bundle vanishes uniformly in the ALE limit, we can read immediately the monodromy at infinity from the expression of the individual $\mathcal{R}_{s_{\sigma}}$ in (\ref{Rsbundle}). Since the monodromy of each $\mathcal{L}_i$ is just $e^{i\frac{2\pi}{n}}$, the monodromy of  $\mathcal{R}_{s_{\sigma}}$ is $e^{i\frac{2\pi\,j}{n}}$, being $j$ the number of NS5 jumped until arriving to the corresponding D5. Obviously there are $N_j$ of them, and so the $N_j$ thus label the holonomy at infinity.

Since the topological data is the same as in the ALF case, and we have just learned how to determine the monodromy in the ALE limit, we can now fully characterize instantons on an ALE space. The Chern classes are given by

\begin{equation}
c_1=\sum_i\,\Delta\ell_i\,c_1(\mathcal{T}_i)\qquad c_2=\sum_i\,\Delta\ell_i\,c_2(\mathcal{T}_i)+\frac{1}{n}\,\sum\,k_i
\end{equation}
Note that the instanton number defined as

\begin{equation}
k=\frac{1}{n}\,\sum\,k_i
\end{equation}
does not generically coincide with the second Chern class.

Furthermore, the rank of the bundle is \cite{Bianchi:1996zj}

\begin{equation}
{\rm rank}\,V=N=\sum N_i
\end{equation}
while the array of integers $\vec{N}$ specifies the holonomy at infinity as described above.

\subsection{Orthogonal and symplectic groups} \label{app:SOSpbrane}

As argued above, the construction in (\ref{Tdual}) is mapped under T-duality along $S^1$ to a system containing D6 branes wrapping the ALE space with dissolved D2 branes on their worldvolume. Indeed, those arise as non-trivial configurations for the wolrdvolume gauge field on the D6 brane along the ALE directions. As, on general grounds, the worldvolume gauge field is a section of a $U(N)$ bundle, this provides a construction of non-trivial $U(N)$ bundles as reviewed above.

Furthermore, this also suggests a way to construct and classify the bundles with orthogonal and symplectic groups. Indeed, starting with the D2-D6 system we can add orientifold 6-planes parallel to the D6 branes with no further breaking of supersymmetry. In particular, we concentrate on the addition of either $O6^-$, $\widetilde{O6}^-$ or $O6^+$ plane. When the D6 branes coincide with the orientifold, the worldvolume symmetry on the D6-branes is, respectively, $O(2N)$, $O(2N+1)$ or $Sp(N)$. 

Upon T-duality along $S^1$, we end up with a brane configruation as in (\ref{TdualO}):
\begin{equation}
\begin{array}{l c |  c c c c c c c c c}
& 0 & 1 & 2 & \raisebox{.5pt}{\textcircled{\raisebox{-.9pt} {3}}}  & 4 &5 & 6 & 7 & 8 & 9 \\ \hline
\text{D5} & \times & \times & \times &  & \times & \times  & \times &  & &\\
\text{O5} & \times & \times & \times &  & \times & \times  & \times &  & &\\
\text{NS5} & \times & \times & \times &  &  &   &  & \times &\times &\times \\
\text{D3} & \times & \times & \times & \times & & & & & &\\
\end{array}
\label{TdualO}
\end{equation}
where O5 stands for the corresponding class of O5 plane. Note that T-duality on an orientifold worldvolume coordinate leaves behind an inversion symmetry. Hence in particular in the $x^3$ circle we will have two orientifolds located at opposite positions in the circle, so that the orientifold projection demands the left half to mirror the right half of the $x^3$ circle. In particular this implies that the distribution of NS5 branes must be symmetric so that depending on wether we have an even or odd number $n$ of NS5 branes we will have different possibilities.  Note that a similar orientifold actions were also considered in the recent papers \cite{GarciaEtxebarria:2012qx, Garcia-Etxebarria:2013tba} in the context of $4d$ $\CN=1$ quiver gauge theories.

\subsubsection{$\BC^2/\BZ_n$ with odd $n=2m+1$}
\begin{figure}[H]
\centering
\includegraphics[scale=.4]{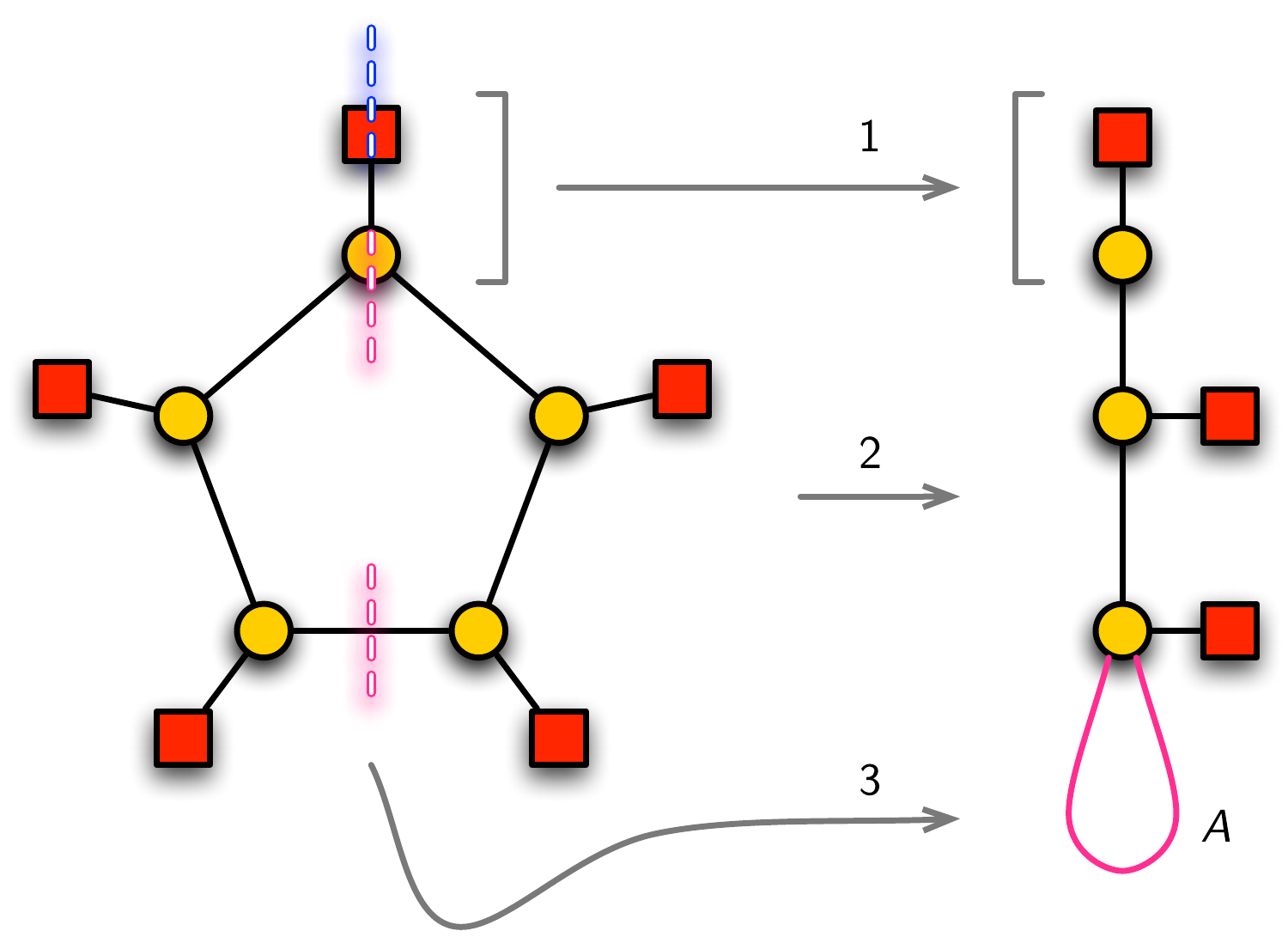} \hspace{2cm}
\includegraphics[scale=.4]{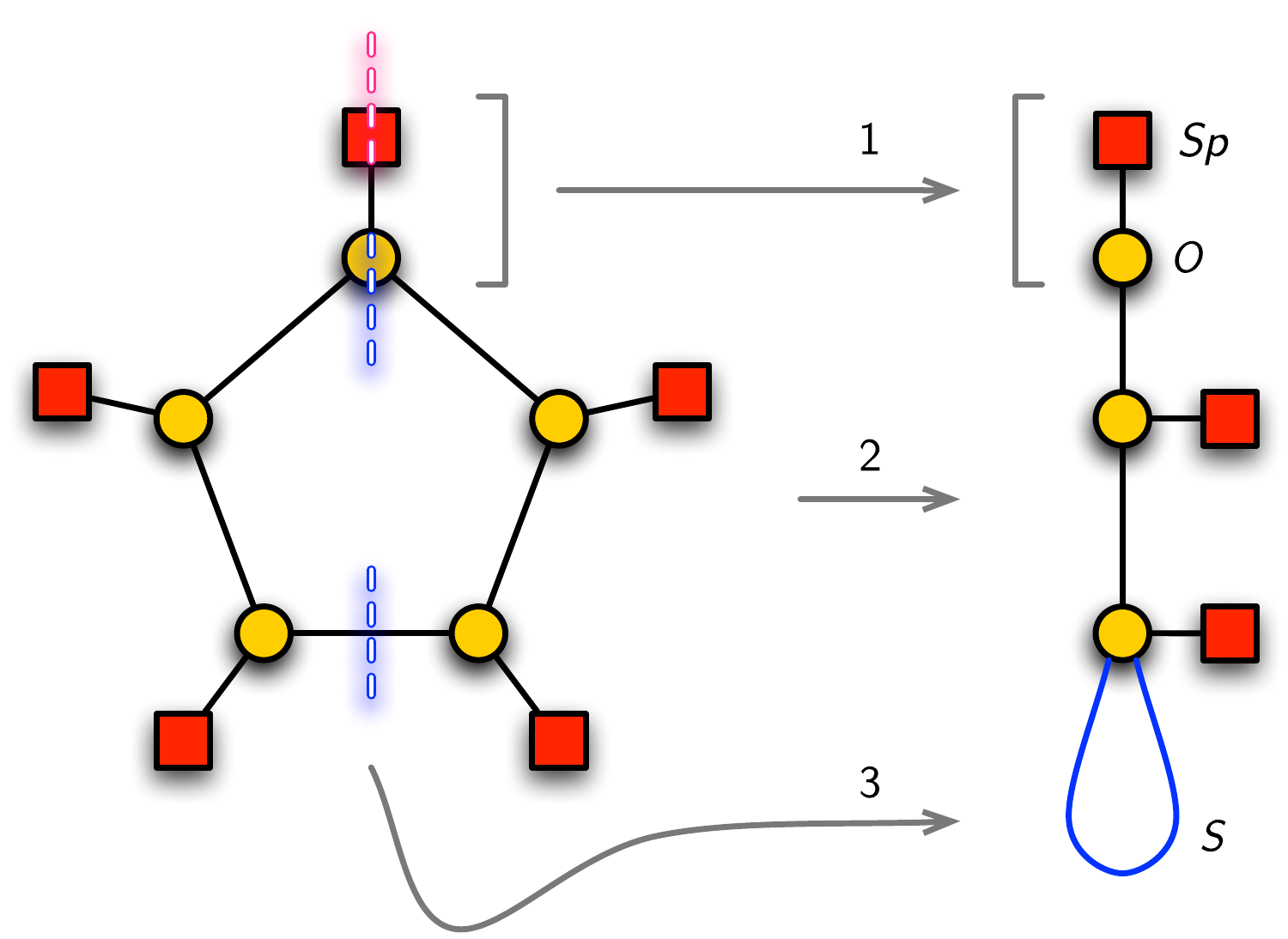}
\caption{{\bf Orientifold action} on the KN quiver to obtain the quivers for $SO(N)$ (left) and $Sp(N)$ (right) instantons on $\BC^2/\BZ_{2m+1}$; here $m=2$.  The nodes without label denote unitary groups $U(k_i)$ or $U(N_i)$, and the shaded notes denote the unitary groups $U(2k_i)$ whose the parameters $k_i$ can be integral or half-integral.  The red and blue dashed lines denote the action of $O5^+/\widetilde{O5}^+$ and $O5^-/\widetilde{O5}^-$ respectively. The label $A$ and $S$ denote the hypermultiplets in the antisymmetric and symmetric representations respectively. Actions in the parts labelled by 1, 2 and 3 are explained below.}
\label{fig:SOSpNoddorb}
\end{figure}

Two possible orientifold actions on the brane configuration corresponding to the KN quiver is depicted in \fref{fig:SOSpNoddorb}. 
For the pedagogical reason, \fref{fig:SOSpNoddorb} demonstrate explicitly the case of $\BC^2/\BZ_5$. However, a generalisation to any $\BC^2/\BZ_{2m+1}$ is straightforward; the results of the latter are given by Figures \ref{fig:SO_N_oddn} and \ref{fig:Sp_N_oddn}.
For each diagram in \fref{fig:SOSpNoddorb}, there are three different types of the action:
\ben
\item The action of $O5^+/\widetilde{O5}^+$ ({\it resp.} $O5^-/\widetilde{O5}^-$) on the unitary gauge node yields a symplectic ({\it resp.} even/odd orthogonal) gauge node. Since the orientifold plane alternates its charge as it crosses an NS5 brane, if the gauge node is symplectic ({\it resp.} orthogonal), then the adjacent flavour node is orthogonal ({\it resp.} symplectic).
\item The orientifold projection maps the left and right halves of the circle to a linear quiver.  Note that in the left diagram, we denote by shaded nodes the gauge groups $U(2k_1), U(2k_2), \ldots$ such that $k_1, k_2, \ldots$ can either be integral or half-integral.
\item In this part of the left diagram, one NS5 brane stuck on an orientifold plane.  The massless fundamental hypermultiplet is therefore projected to the hypermultiplets in the antisymmetric ({\it resp.} symmetric) representation when the orientifold plane is $O5^+/\widetilde{O5}^+$ ({\it resp.} $O5^-/\widetilde{O5}^-$).
\een
It should be noted that the choice between $O5^+$ and $\widetilde{O5}^+$ or between $O5^-$ and $\widetilde{O5}^-$ should be as follows:
\begin{quote}
If the flavour node is of $D$ type ({\it resp.} $B$ type), the orientifold projection on that flavour node comes from $O5^-$ ({\it resp.} $\widetilde{O5}^+$).  Consequently, the orientifold action on the adjacent gauge node comes from $O5^+$ ({\it resp.} $\widetilde{O5}^-$).
\end{quote}
%
%
%

\subsubsection{$\BC^2/\BZ_n$ with even $n=2m$}
\begin{figure}[H]
\centering
\includegraphics[scale=.4]{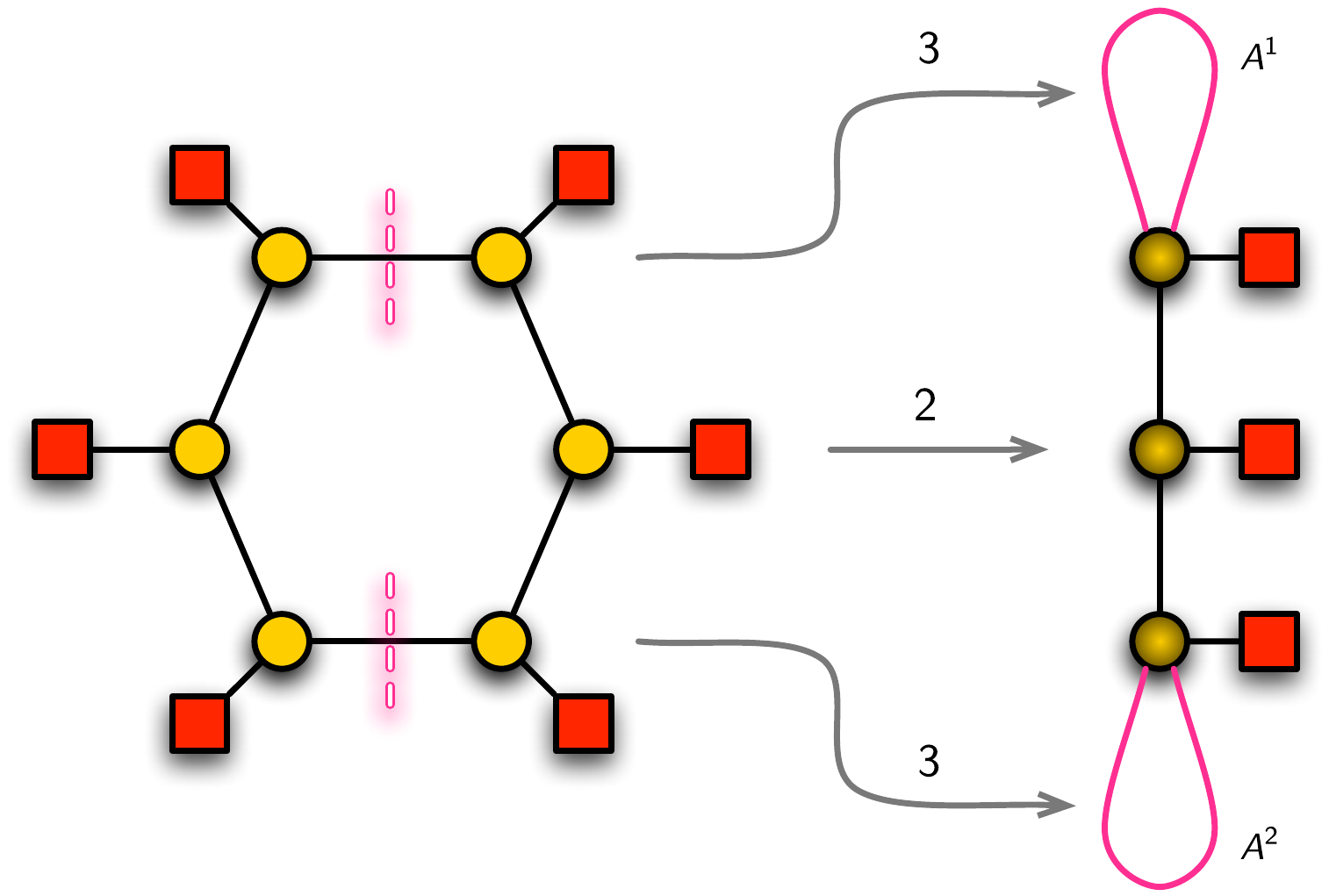} \hspace{1.2cm}
\includegraphics[scale=.4]{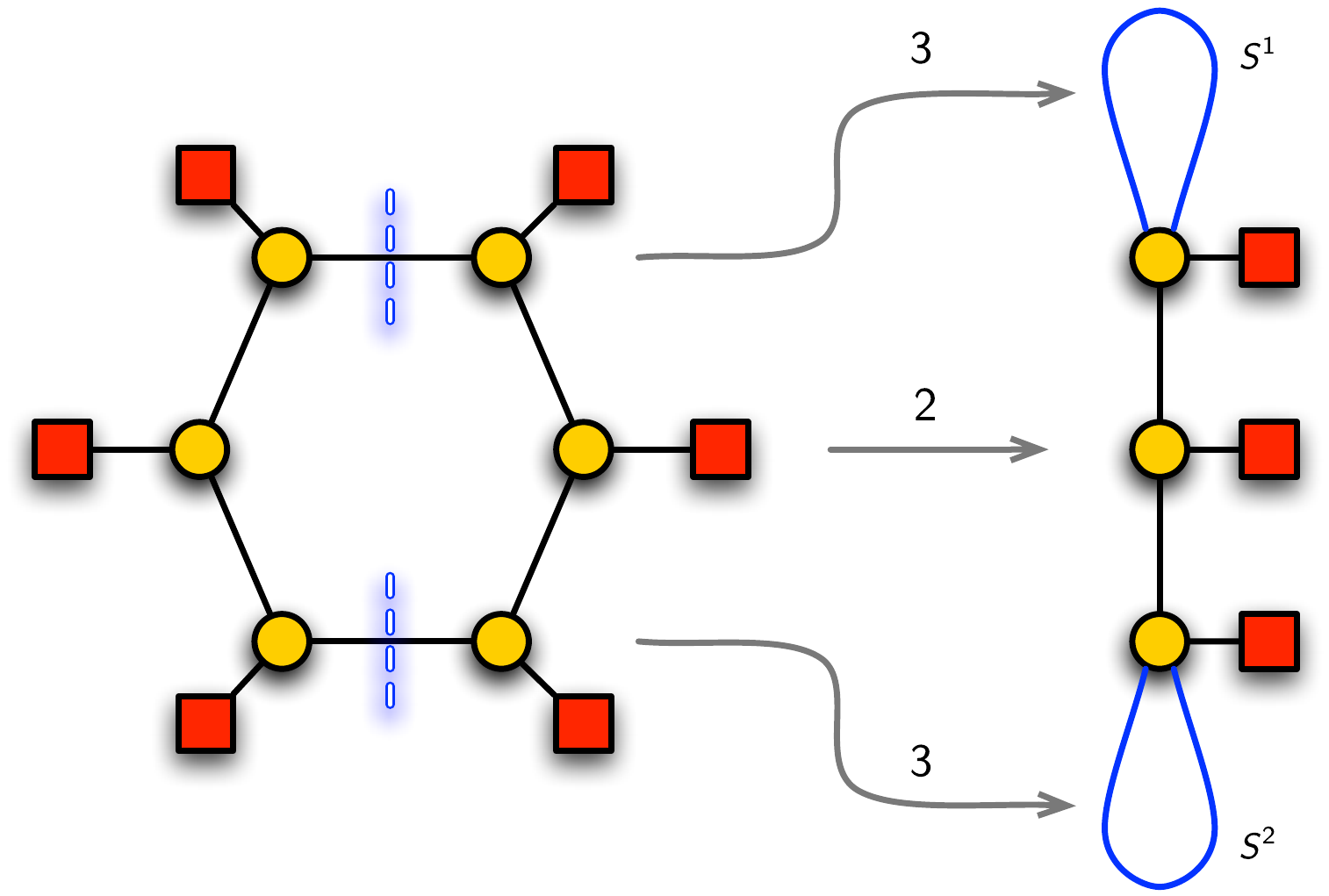}
\caption{Orientifold action on the KN quiver to obtain the $AA$ quiver for $SO(2N)$ (left) and the $SS$ quiver for $Sp(N)$ (right) instantons on $\BC^2/\BZ_{2m}$; here $m=3$.  The nodes without label denote unitary groups $U(k_i)$ or $U(N_i)$, and the shaded notes denote the unitary groups $U(2k_i)$ whose the parameters $k_i$ can be integral or half-integral.  The red and blue dashed lines denote the action of $O5^+/\widetilde{O5}^+$ and $O5^-/\widetilde{O5}^-$ respectively. The label $A$ and $S$ denote the hypermultiplets in the antisymmetric and symmetric representations respectively. Actions in the parts labelled by 1 and 2 are explained above.}
\label{fig:SOSpNevenorbVS}
\end{figure}

\begin{figure}[H]
\centering
\includegraphics[scale=.4]{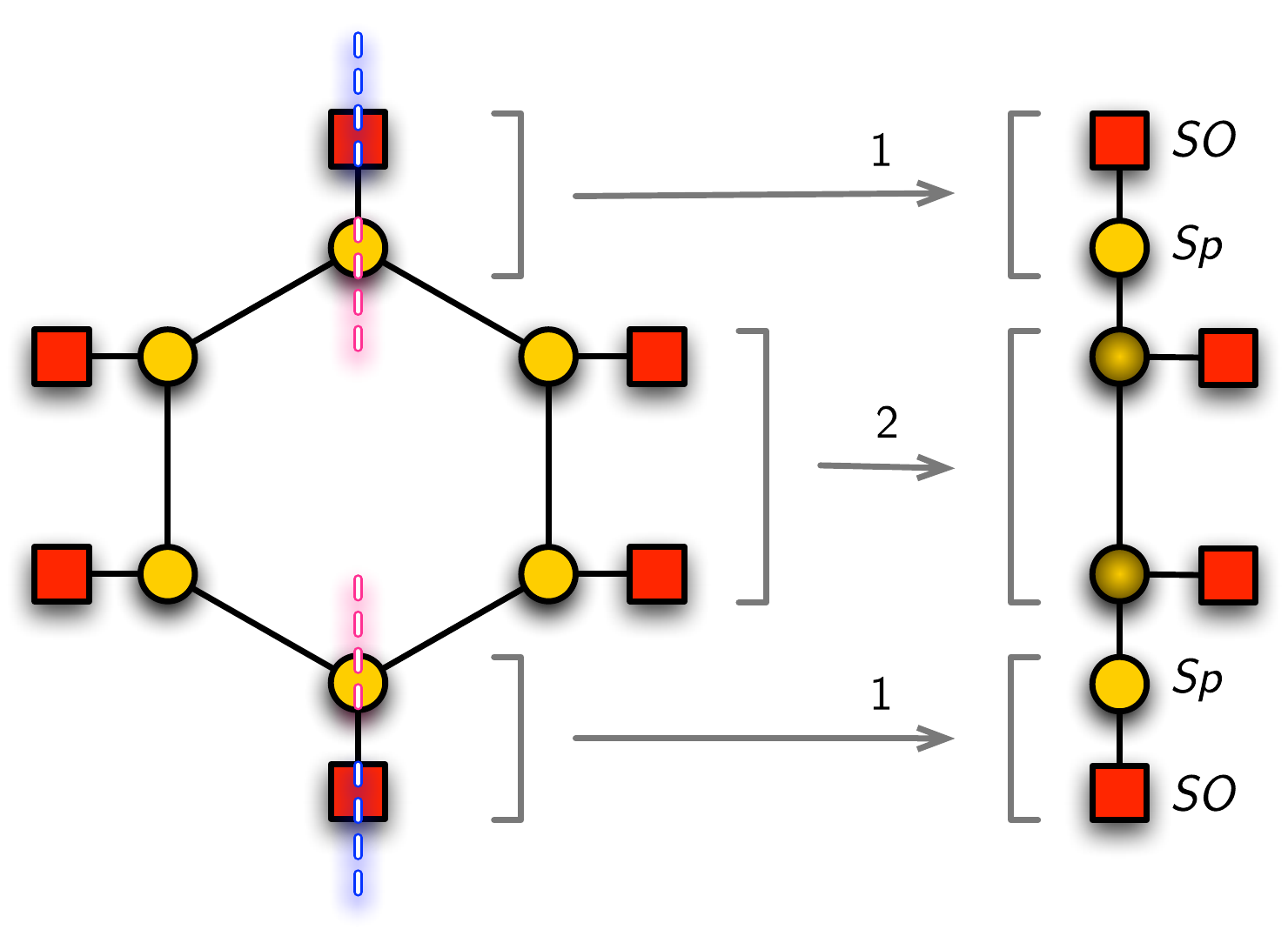} \hspace{1.2cm}
\includegraphics[scale=.4]{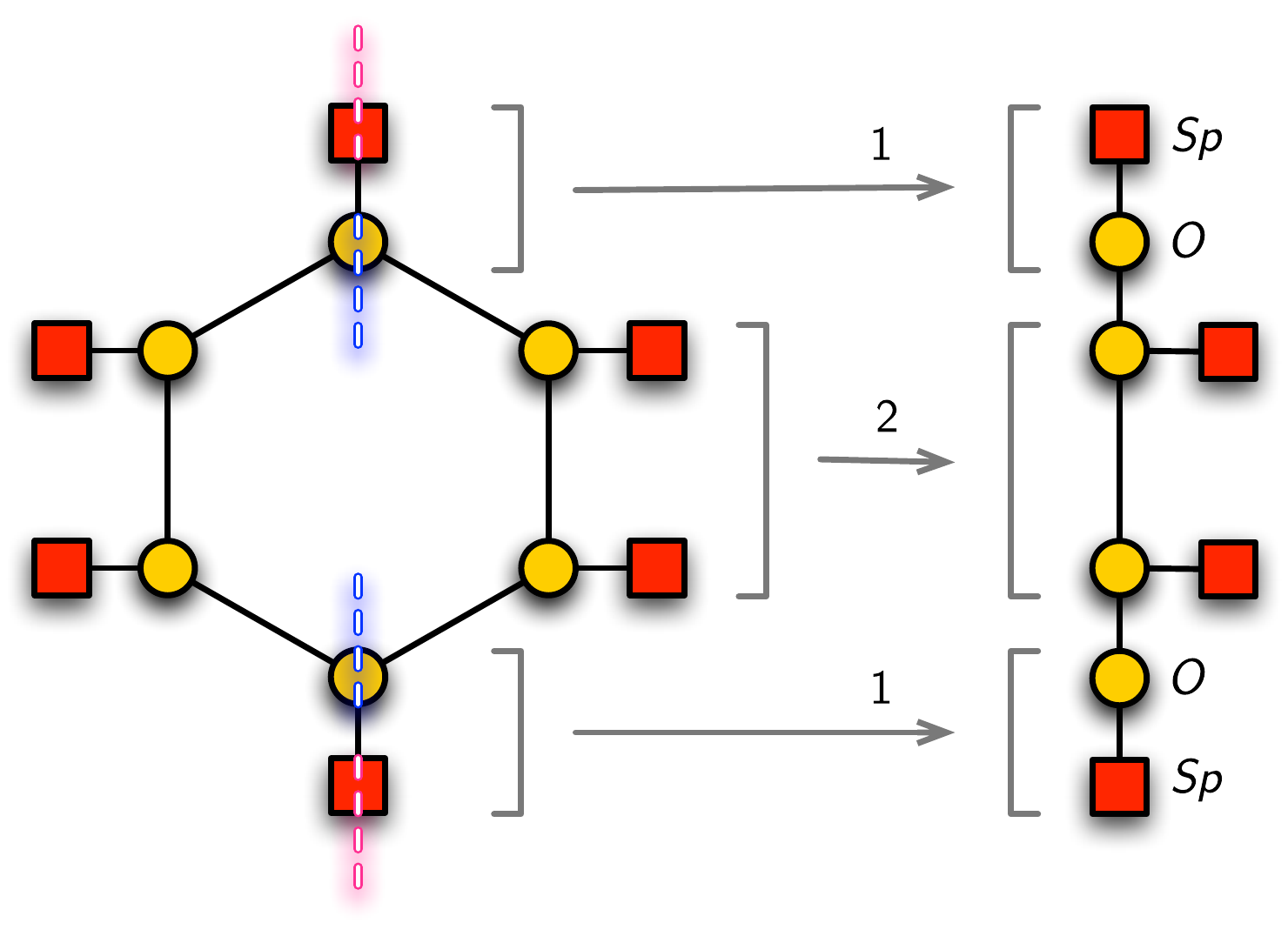}
\caption{Orientifold action on the KN quiver to obtain the $O/O$ quiver for $SO(N)$ (left) and the $S/S$ quiver for $Sp(N)$ (right) instantons on $\BC^2/\BZ_{2m}$; here $m=3$.  The nodes without label denote unitary groups $U(k_i)$ or $U(N_i)$, and the shaded notes denote the unitary groups $U(2k_i)$ whose the parameters $k_i$ can be integral or half-integral.  The red and blue dashed lines denote the action of $O5^+/\widetilde{O5}^+$ and $O5^-/\widetilde{O5}^-$ respectively. The label $A$ and $S$ denote the hypermultiplets in the antisymmetric and symmetric representations respectively. Actions in the parts labelled by 1 and 2 are explained above.}
\label{fig:SOSpNevenorbnoVS}
\end{figure}

For even $n$, there are two possibilities of symmetrically distributing the NS5 branes: we can either have one NS5 stuck at each orientifold or none; these are depicted in Figures \ref{fig:SOSpNevenorbnoVS} and \ref{fig:SOSpNevenorbVS} respectively. We refer to such configurations as $O/O$ or $S/S$ (sometimes called in the literature \emph{vector structure case}) and $AA$ or $SS$ (sometimes referred to as \emph{no vector structure case}) respectively.  For the pedagogical reason, \fref{fig:SOSpNoddorb} demonstrate explicitly the case of $\BC^2/\BZ_6$. However, a generalisation to any $\BC^2/\BZ_{2m}$ is straightforward; the results of the latter are given by Figures \ref{fig:SO_N_withVSevenn}, \ref{fig:Sp_N_withVSevenn}, \ref{fig:SO_N_withNOVSevenn} and \ref{fig:Sp_N_withNOVSevenn}.
The three prescriptions described above can still be applied in order to determine the results of orientifold projections.

%
%
%
%
%
%
%
%

\subsubsection{The hybrid configurations} \label{sec:hybridbranes}
In the six configurations presented earlier in Figures \ref{fig:SOSpNoddorb}, \ref{fig:SOSpNevenorbVS} and \ref{fig:SOSpNevenorbnoVS}, we take the orietifold planes that acts on the `antipodal' gauge nodes to be of the same charges.  However, there are also possibilities to take them to be of the opposite charges; we list all four of them below. These configurations are related to the Dabholkar-Park orientifold with orientifold group $\{1,\,S\,\Omega\}$, being $S$ a half-circle shift \cite{Dabholkar:1996pc}. Such configurations are referred to as the {\it hybrid configurations}.  

As depicted below, as a result of the orientifold action, there is an ambiguity between the orthogonal gauge group and the special orthogonal gauge group; we denote this ambiguity by $(S)O$ in the diagram below.  Since it is not immediate that these hybrid quivers can be associated with a known instanton moduli space, we shall not resolve such an ambiguity in this paper. 

\begin{figure}[H]
\centering
\includegraphics[scale=.35]{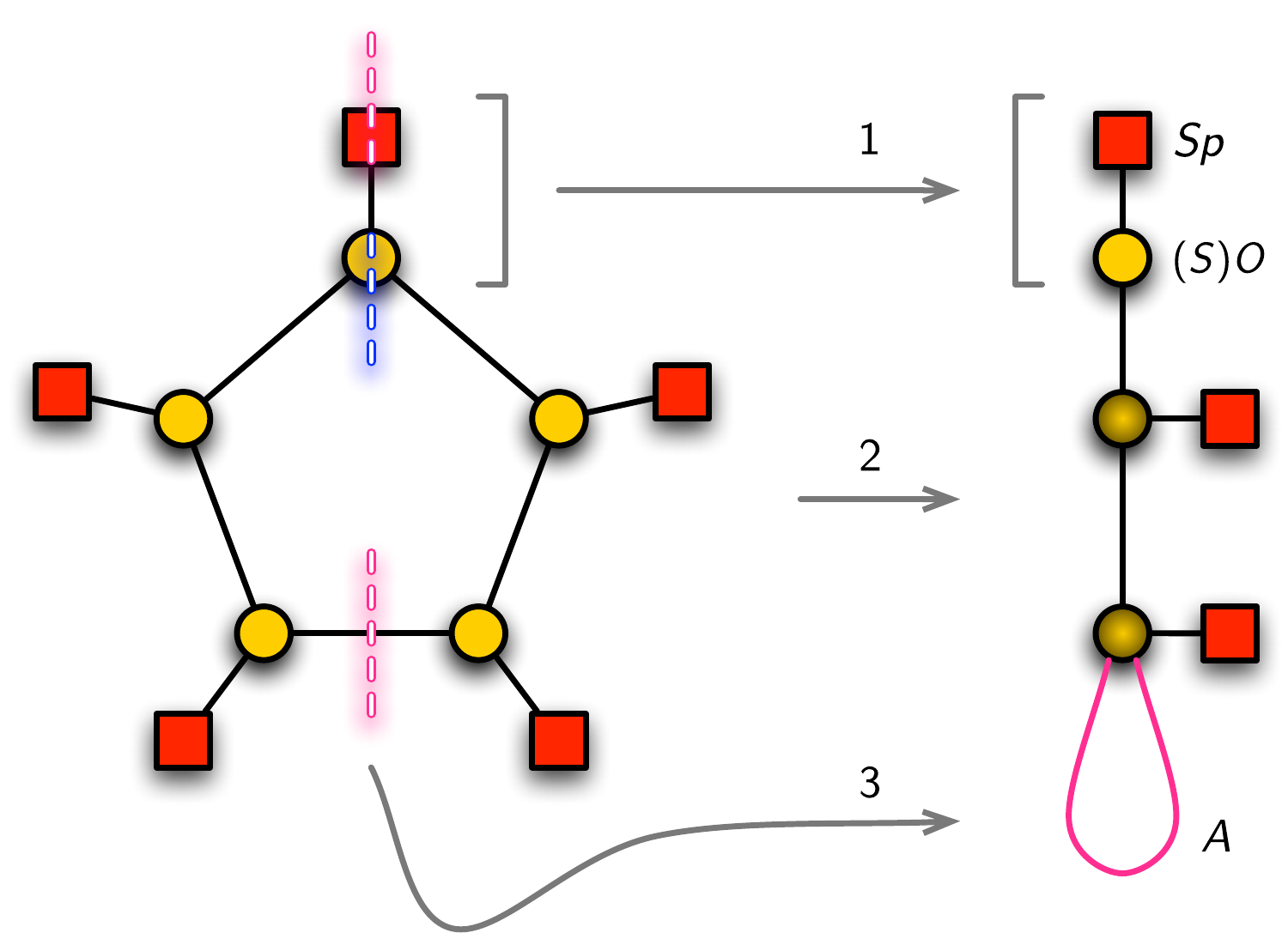} \hspace{0.5cm}
\includegraphics[scale=.35]{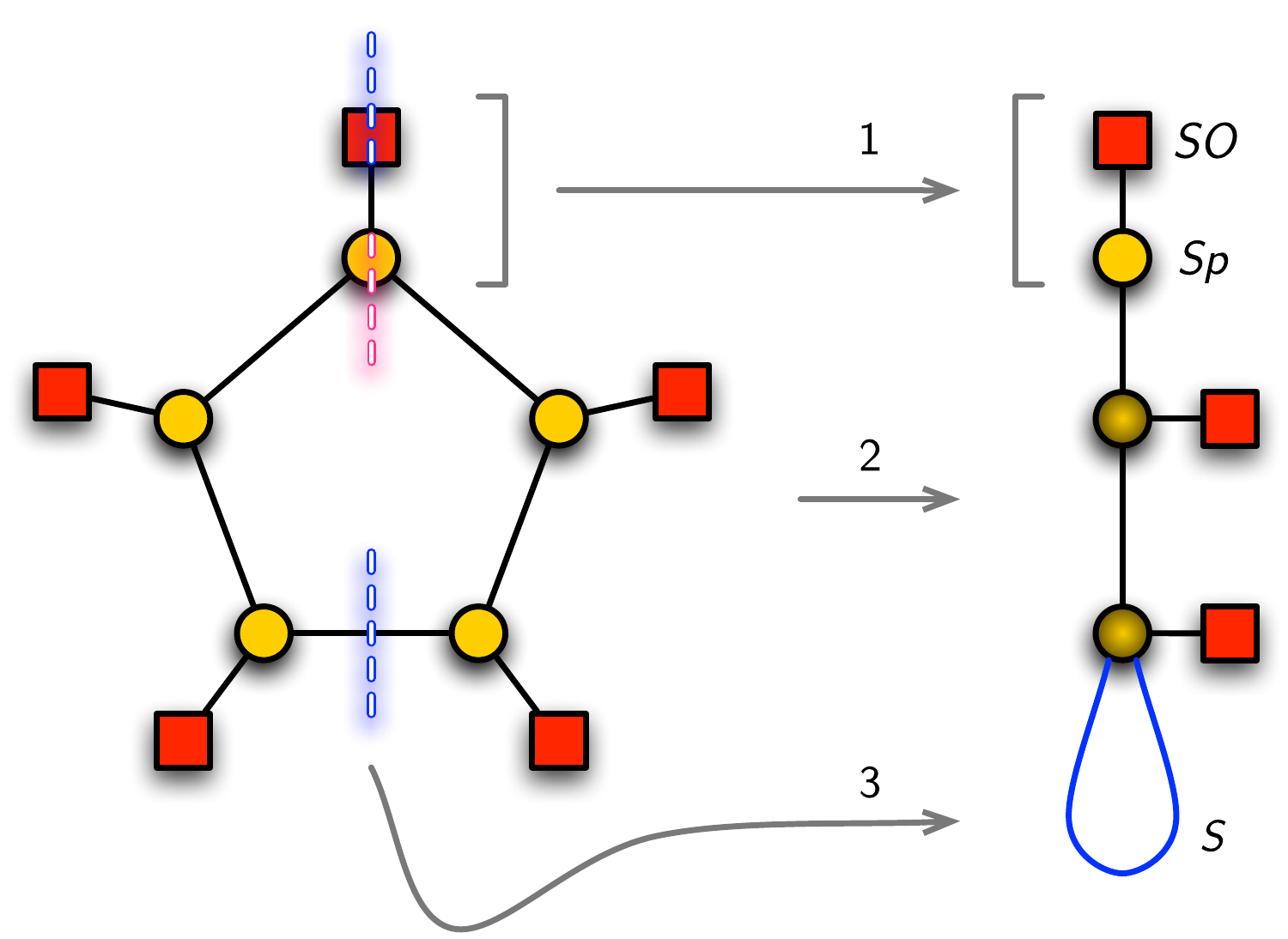}
\caption{The hybrids between the quivers for $SO(N)$ and $Sp(N)$ instantons on $\BC^2/\BZ_{2m+1}$.  The nodes without label denote unitary groups $U(k_i)$ or $U(N_i)$, and the shaded notes denote the groups $U(2k_i)$ or $(S)O(2k_i)$ whose the parameters $k_i$ can be integral or half-integral.  The red and blue dashed lines denote the action of $O5^+/\widetilde{O5}^+$ and $O5^-/\widetilde{O5}^-$ respectively. Actions in the parts labelled by 1, 2 and 3 are explained above.}
\label{fig:hybridAbrane}
\end{figure}

\begin{figure}[H]
\centering
\includegraphics[scale=.4]{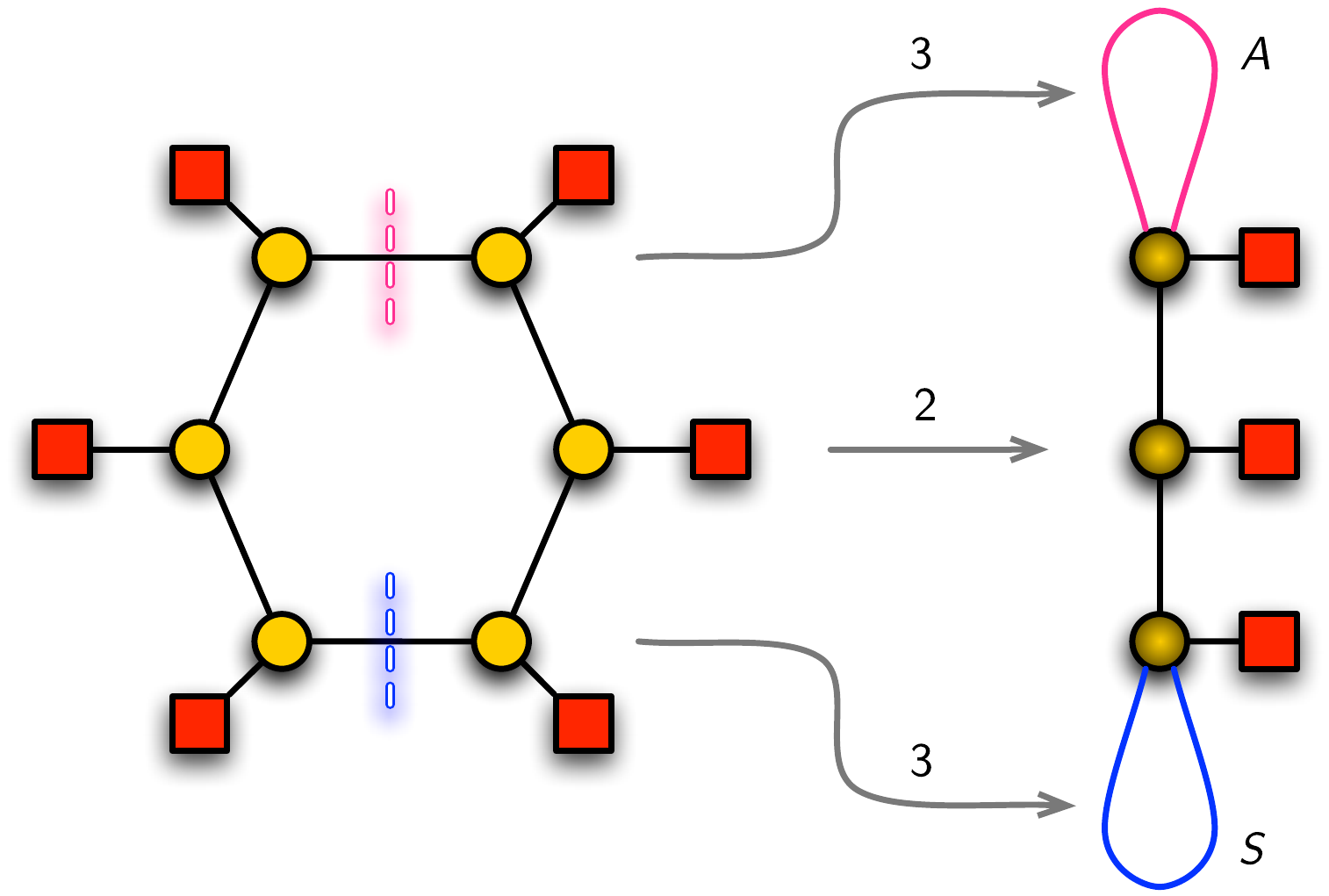} \hspace{0.5cm}
\includegraphics[scale=.35]{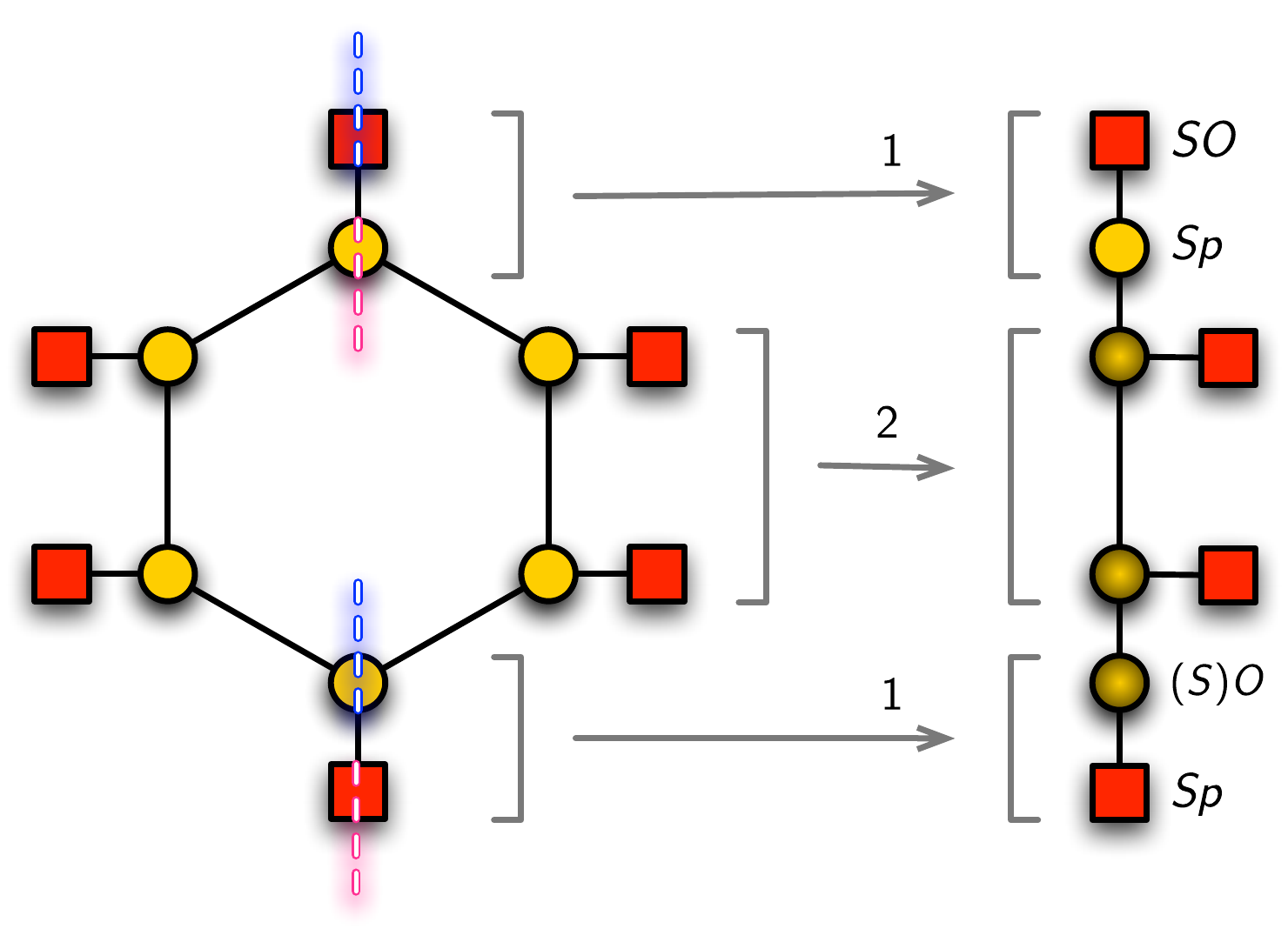}
\caption{The $SA$ hybrid (left) and the $O/S$ hybrid (right). The nodes without label denote unitary groups $U(k_i)$ or $U(N_i)$, and the shaded notes denote the groups $U(2k_i)$ or $(S)O(2k_i)$ whose the parameters $k_i$ can be integral or half-integral.  The red and blue dashed lines denote the action of $O5^+/\widetilde{O5}^+$ and $O5^-/\widetilde{O5}^-$ respectively. Actions in the parts labelled by 1, 2 and 3 are explained above.}
\label{fig:hybridBbrane}
\end{figure}

For the pedagogical reason, Figures \ref{fig:hybridAbrane} and \ref{fig:hybridBbrane} demonstrate explicitly the case of $\BC^2/\BZ_5$ and $\BC^2/\BZ_6$. However, a generalisation to any $\BC^2/\BZ_{2m+1}$ is straightforward; the results of the latter are given by Figures \ref{fig:hybridA} and \ref{fig:hybridB}.

\section{Rank zero pure instantons on $\mathbb{C}^2/\mathbb{Z}_2$ and holography } \label{holo}
In this appendix, we focus on the case of pure `instantons' with $\vec{N}=(0,\,0)$ in \fref{UNC2modZ2}.  

\subsection{The case of $\vec k =(1,1)$ and $\vec N=(0,0)$}
The superpotential \eref{WSUNC2Z2} simplifies to 
\bea
W=(\varphi_{2}-\varphi_{1})\,(X_{12}^1\,X_{21}^2+X_{21}^1\,X_{12}^2)~.
\eea
The relevant $F$-flat space of our interest is parametrised by $X^\alpha_{12}, X^\alpha_{21}$, with $\alpha=1,2$, subject to the defining relation:
\bea
0=\partial_{\varphi_{1}} W=\partial_{\varphi_{2}} W  = X_{12}^1\,X_{21}^2+X_{21}^1\,X_{12}^2~.
\eea
The Hilbert series of the $F$-flat space is
\bea
\fflat (t,x,z_1,z_2) = (1-t^2){\PE}\Big[ t [1]_x\,(z_1^{-1}z_2+z_1 z_2^{-1})\,\Big]
\eea
The Hilbert series for the hypermultiplet moduli space is given by
\bea
g^{(0,0)}_{(1,1)} (t,x) &= \oint_{|z_1|=1} \frac{\ud z_1}{(2 \pi i)z_1}  \oint_{|z_2|=1} \frac{\ud z_2}{(2 \pi i)z_2}\fflat (t,x,z_1,z_2) \nn \\
&= (1-t^4) \PE[ [2]_x t^2] =\sum_{n=0}^\infty [2n]_x t^{2n}~.
\eea
This is indeed the Hilbert series of $\BC^2/\BZ_2$.  

Observe that the gauge symmetry $U(1) \times U(1)$ is not completely broken at a generic point on the hypermultiplet moduli space. In fact, the diagonal gauge group $S(U(1)\times U(1))$, corresponding to the massless combination $(\varphi_{1}+\varphi_{2})$, remains unbroken on this space.  This can be seen as follows: The mass matrix $m_{ij}$ (with $i_1, i_2=1,2$) for the adjoint fields $\varphi_1, \varphi_2$ satisfies
\bea
m_{ij} \varphi_i \varphi^*_j &= |\partial_{X^1_{12}} W|^2+|\partial_{X^2_{12}}  W|^2+|\partial_{X^1_{21}} W|^2+|\partial_{X^2_{21}}  W|^2 \nn\\
&= 4|\varphi_1 - \varphi_2|^2 = 4(|\varphi_1|^2 + |\varphi_2|^2 - \varphi_1 \varphi_2^* - \varphi_2 \varphi_1^*)~,
\eea
and so the mass matrix is
\bea
m_{ij} = \begin{pmatrix} 4 & -4 \\ -4 & 4 \end{pmatrix}~,
\eea
whose eigenvalues are $8$ and $0$ with respect to eigenvectors $(-1,1)$ and $(1,1)$.

\subsection{Rank zero instantons and the gauge/gravity duality }
Rank zero $k$ pure instantons are described by the Higgs branch of a $U(k)\times U(k)$ theory, which corresponds to $k$ regular D3 branes probing $\mathbb{C}\times \BC^2/\mathbb{Z}_2$. For large $k$ the backreaction of the D3 branes generates, in the near-brane region, an $AdS_5\times S^5/\mathbb{Z}_2$ geometry. Then, on general grounds, IIB SUGRA on such space should be dual to the gauge theory describing $k$ rank zero instantons. More explicitly, the geometric background, in global coordinates for $AdS_5$, is 

\begin{equation}
ds^2=-(1+\frac{r^2}{L^2})\,dt^2+\frac{dr^2}{(1+\frac{r^2}{L^2})}+r^2\,d\Omega_3^2+L^2\,\Big(\sin^2\alpha\,d\chi^2+\frac{\cos^2\alpha}{4}\,g_5^2+\frac{\cos^2\alpha}{4}\,d\Omega_2^2\Big)
\end{equation}
There is also a RR 4-form field strength

\begin{equation}
C^{(4)}=\frac{r^4}{L}\,dt\wedge d\Omega_3
\end{equation}
The quantization condition demands in this case (we set $\ell_s=1$)

\begin{equation}
L^4=8\,\pi\,N\
\end{equation}

We would like to study the Higgs branch of the gauge theory holographically (see also \cite{Bergman:2012qh}). To that matter, we consider a dual giant graviton D3 brane wrapping the $S^3$ in $AdS$ and moving along $\psi$, $\chi$ and $\phi$ \cite{Mandal:2006tk,Martelli:2006vh,Biswas:2006tj}. The action is 

\begin{equation}
S_{\text{DBI}}=-T_3\,\Omega_3\,\int\,r^3\,\sqrt{(1-\frac{r^2}{L^2})}\,\sqrt{1-\frac{L^2}{(1+\frac{r^2}{L^2})}\,(\sin^2\alpha\,\dot{\chi}^2+\frac{\cos^2\alpha}{4}\,(\dot{\psi}+\cos\theta\,\dot{\phi})^2+\frac{\cos^2\alpha}{4}\,\sin^2\theta\,\dot{\phi}^2) } 
\end{equation}
where we denoted by $\Omega_3$ the angular volume of the 3-sphere, and 

\begin{equation}
S_{CS}=T_3\,\Omega_3\,\int\,\frac{r^4}{L}
\end{equation}
Let us concentrate on branes not moving on the $\chi$ direction. After a bit of algebra, one finds that the hamiltonian $\mathcal{H}$ is minimized at $\alpha=0$, \textit{i.e.} at the origin of the $\mathbb{C}$, factor and

\begin{eqnarray}
a)\qquad P_{\phi}=P_{\psi}\,\cos\theta\qquad r=\frac{\sqrt{2}}{L\,\sqrt{T_3\,\Omega_3}}\,\sqrt{P_{\psi}} \quad \rightarrow\quad \mathcal{H}=2\,\frac{P_{\psi}}{L} \\ \nonumber \\ 
b)\qquad P_{\psi}=P_{\phi}\,\cos\theta \qquad r=\frac{\sqrt{2}}{L\,\sqrt{T_3\,\Omega_3}}\,\sqrt{P_{\phi}}\quad \rightarrow\quad  \mathcal{H}=2\,\frac{P_{\phi}}{L} 
\end{eqnarray}
Following \cite{Mandal:2006tk, Martelli:2006vh}, we should proceed to geometrically quantize the phase space for these branes. To that matter, we can follow a shortcut and note that the symplectic one-forms are given by $\nu=P_{\phi}\,d\phi+P_{\psi}\,d\psi$, which read

\begin{equation}
a)\quad \nu=\frac{L\,\mathcal{H}}{2}\,(d\psi+\cos\theta\,d\phi)\qquad b)\quad \nu=\frac{L\,\mathcal{H}}{2}\,(d\phi+\cos\theta\,d\phi)
\end{equation}
Upon recalling that $L\,\mathcal{H}/2=L^2\,T_3\,\Omega_3\,r^2=\frac{N}{L^2}\,r^2$ we recognize the local integration of the Kahler form for the $\mathbb{C}^2/\mathbb{Z}_2$ space in the two orthogonal complex structures. Thus, both a) and b) describe the same operators but in orthogonal complex structures, corresponding to the freedom to write our operators using the R-symmetry doublets or the $SU(2)_M$ doublets. While the later are holomorphic in $\mathcal{N}=1$ formalism, the former are not.

Since $\cos\theta\in[-1,\,1]$, we have that, for one of the a), $|P_{\phi}|\leq |P_{\psi}|$. Recall that $P_{\psi}$ is the $U(1)\in SU(2)_M$. Hence this corresponds to describing our operators using the non-holomorphic doublets adapted to see the $SU(2)_R$. In turn, in the other complex structure corresponding to case b) we have $|P_{\psi}|\leq |P_{\phi}|$, which corresponds to using the doublets adapted to see the $SU(2)_M$. These are the $X_{12}^i$ and $X_{21}^i$. In order to make the correspondence more explicit, let us denote as $Q_M$ the $U(1)\in SU(2)_M$ and as $Q_R$ the $U(1)\in SU(2)_R$. \footnote{Note that the period of the coordinate $\psi$ in the unorbifolded case is $4\pi$, as opposed to the $\mathbb{Z}_2$ orbifold at hand. Since in this case $\psi$ is canonically normalized, the Killing vector corresponding to $Q_M$ is $\partial_{\psi}$. Hence the charges of the mesons are $\{1,\,0,\,-1\}$.} Then, the smallest operators are

\begin{equation}
\begin{array}{c | c | c | c }
 & Q_M & Q_R & \Delta \\ \hline
 X_{12}^1\,X_{21}^1 & 1& 1 & 2\\
 X_{12}^1\,X_{21}^2 & 0&  1&2\\
 X_{12}^2\,X_{21}^2 & -1&  1 & 2\\
\end{array}
\end{equation}
Hence, upon identifying $Q_M=\frac{P_{\phi}}{L}$, $Q_R=\frac{P_{\psi}}{L}$ and taking into account the standard identification in global $AdS$ $\Delta=L\,\mathcal{H}$ we recover the gravity findings 

\begin{equation}
|Q_M|\leq |Q_R|\qquad \Delta=2\,Q_R
\end{equation}
Needless to say that considering larger operators follows the same pattern, thus recovering the counting anticipated by the Hilbert series.

Since we have found that the phase space for these branes is $\mathbb{C}^2/\mathbb{Z}_2$, the geometric quantization of this phase space \cite{Mandal:2006tk, Martelli:2006vh} will correspond to the geometric quantization of the $\mathbb{C}^2/\mathbb{Z}_2$ space. Indeed, the operators dual to these branes are the ones counted by the Hilbert series described above, as briefly indicated above.

Note that strictly speaking the gravity computation is dual to a gauge theory with $SU(k)$ nodes instead of $U(k)$. Indeed, as it is well-known, the $U(1)$ part of the $U(k)$ gauge groups decouple in the IR and become global baryonic symmetries. However, the dual giants we have considered do not carry baryonic quantum numbers and hence do not explore the baryonic directions corresponding to the $U(1)$'s. Thus, for them it is as if the dual theory was $U(k)$.

It is worth to stress that the $AdS_5\times S^5/\mathbb{Z}_2$ is a singular space with an $S^1$ worth of singularities. Explicitly, this singular locus corresponds to $\alpha=0$. Remarkably, although $\alpha=0$ corresponds to a fixed circle of singularities, the behavior of the branes at hand is perfectly regular. \footnote{A more dramatic example appears in \cite{Bergman:2012qh}, where the Higgs branch sits at a locus where both curvature and dilaton diverge and yet the dual giants probing the Higgs branch are perfectly well behaved.}

\subsection{Generic $k$ rank zero pure instantons}

The fact that the large $k$ and the $k=1$ give the same result for the Hilbert series suggests that the rank zero and arbitrary $k$ pure instanton Hilbert series should be the $\nu$-insterted PE of the $k=1$ \cite{Benvenuti:2006qr}. This is because the coefficient of $\nu^k$ in the expansion of
\bea
\PE \Big[ g_{\BC^2/\BZ_2}(t,x) \nu \Big]= \PE \Big[ \frac{1-t^4}{(1-t^2)(1-t^2 x^2)(1-t^2 x^{-2})} \nu \Big]
\eea
should describe the phase space of the $k$ instanton, and as shown in \cite{Benvenuti:2006qr}, the $k\rightarrow \infty$ coefficient indeed coincides with the $k=1$ coefficient. Note that the direct evaluation of the Hilbert series for arbitrary $k$ --and hence the explict check of the above conjecture-- is not obvious, since the gauge group is not fully Higgsed while, being non-abelian, the simple argument we used does not directly hold. Nevertheless one can compute the Hilbert series  by numerical means \cite{mac2} for the first few $k$'s and convince oneself that they behaved as expected.

\end{appendix}

\bibliographystyle{ytphys}
\bibliography{ref}

\end{document}